\RequirePackage{ifpdf}
\documentclass[hyper,12pt,A4paper]{article}
\pdfoutput=1
\usepackage{jheppub}
\usepackage{graphicx}
\usepackage{latexsym,amsmath,amsfonts,amssymb}
\usepackage{mathrsfs}
\usepackage[makeroom]{cancel}
\usepackage[utf8]{inputenc} % added by Vivek on Nov 4
\usepackage{bbm}
\usepackage{bm}
\usepackage{subfigure}
\usepackage{paralist}
\usepackage{url}

\newcommand{\vol}{\mathrm{vol}}

\numberwithin{equation}{section}

\newcommand{\nn}{\nonumber}

\newcommand{\be}{\begin{equation}} 
\newcommand{\ee}{\end{equation}}
\newcommand{\bea}{\begin{equation} \begin{aligned}} \newcommand{\eea}{\end{aligned} \end{equation}}

\newcommand{\bit}{\begin{itemize}} 
\newcommand{\eit}{\end{itemize}}

\newcommand{\cF}{\mathcal{F}}
\newcommand{\cG}{\mathcal{G}}
\newcommand{\cH}{\mathcal{H}}
\newcommand{\cI}{\mathcal{I}}

\newcommand{\cK}{\mathcal{K}}

\newcommand{\cT}{\mathcal{T}}

\newcommand{\cV}{\mathcal{V}}

\newcommand{\fg}{\mathfrak{g}}

\newcommand{\Z}{\mathbb{Z}}
\newcommand{\C}{\mathbb{C}}
\newcommand{\R}{\mathbb{R}}

\newcommand{\wh}{\widehat }

\newcommand{\CC}{\mathcal{C}}

\newcommand{\CK}{\mathcal{K}}

\newcommand{\CP}{\mathcal{P}}

\newcommand{\GG}{\mathbf{G}}

\newcommand{\GH}{\mathbf{H}}

\newcommand{\p}{\partial}

\newcommand{\bE}{{\bf E}}
\renewcommand{\P}{{\mathbb{P}}}

\newcommand{\MG}{{\mathbf X}} %M-theory singualrity
\newcommand{\TY}{{\mathbf Y}} %IIA resolved toric variety (Vivek)
\newcommand{\FT}{{\mathcal{T}_\MG}} %5d THEORY
\newcommand{\FF}{\mathcal{F}} %prepotential
\newcommand{\NL}{\text{NL}} % non-Lagrangian superscript (Nov 16, 2019)
\newcommand{\fthreehalf}{\frac{3}{2}}
\newcommand{\fhalf}{\frac{1}{2}}
\title{Rank-two 5d SCFTs from M-theory at isolated toric singularities: a systematic study 
}

\author[\natural]{Vivek Saxena}

\affiliation[\natural]{C.N. Yang Institute for Theoretical Physics, Stony Brook University\\Stony Brook, NY 11794-3840, USA}

\preprint{YITP-SB-19-41}
\emailAdd{vivek.saxena@stonybrook.edu}

\abstract{We carry out a detailed exploration of the deformations of rank-two five-dimensional superconformal field theories (SCFTs) $\mathcal{T}_{\mathbf{X}}$,  which are geometrically engineered by M-theory on the space transverse to isolated toric Calabi-Yau (CY) threefold singularities $\mathbf{X}$. Deformations of 5d $\mathcal{N}=1$ SCFTs can lead to ``gauge-theory phases,'' but also to ``non-gauge-theoretic phases,'' which have no known Lagrangian interpretation. In previous work, a technique relying on fiberwise M-theory/type IIA duality was developed to associate a type IIA background to any resolution of $\mathbf{X}$ which admits a suitable projection of its toric diagram. The type IIA background consists of an A-type ALE space fibered over the real line, with stacks of coincident D6-branes wrapping 2-cycles in the ALE resolution. In this work, we combine that technique with some elementary ideas from graph theory, to analyze mass deformations of $\mathcal{T}_{\mathbf{X}}$ when $\mathbf{X}$ is a isolated toric CY$_3$ singularity of rank-two (that is, it has two compact divisors). We explicitly derive type IIA descriptions of all isolated rank-two CY$_3$ toric singularities. We also comment on the renormalization group flows in the extended parameter spaces of these theories, which frequently relate distinct geometries by flowing to theories with lower flavor symmetries, including those that describe non-gauge-theoretic phases.}

\begin{document}

\maketitle

\section{Introduction}
Five-dimensional $\mathcal{N}=1$ supersymmetric gauge theories are curious entities in the landscape of string/M-theory compactifications, interpolating between their more extensively-studied four- and six- dimensional cousins. Being non-renormalizable, they are intrinsically ill-defined as quantum field theories, and yet, they are interesting systems to study as they have well-defined ultraviolet (UV) completions via string/M-theory \cite{Seiberg:1996bd,Aharony:1997ju,Aharony:1997bh}. The basic tool for studying them in the context of this paper is geometric engineering in M-theory \cite{Morrison:1996xf,Douglas:1996xp,Intriligator:1997pq,Witten:1996qb}. (For reviews of geometric engineering, see \cite{Katz:1996fh,Katz:1997eq}.) Specifically, five-dimensional gauge theories with $\mathcal{N}=1$ supersymmetry can be geometrically engineered via the decoupling limit of M-theory on a local Calabi-Yau (CY) threefold $\MG$. In the limit where all the K\"{a}hler moduli of the threefold shrink to zero size, one gets a five-dimensional superconformal field theory (SCFT) in the UV along the spacetime transverse to $\MG$:
\be\label{Mtheory X3 SCFT intro}
\text{M-theory on}\:\; \R^{1,4} \times \MG \qquad \longleftrightarrow \qquad \FT \equiv {\rm SCFT}(\MG)~.
\ee
The geometric engineering analysis in this paper is largely based on \cite{Closset:2018bjz}, where a program to study the mass deformations of these SCFTs was initiated, focusing on the properties \textit{away} from the conformal point. See also \cite{Cherkis:2014vfa,Hayashi:2015fsa,Hayashi:2015zka,Hayashi:2015vhy,Hayashi:2016abm,Hayashi:2017jze,DelZotto:2017pti,Xie:2017pfl,Jefferson:2017ahm,Hayashi:2017btw,Alexandrov:2017mgi,Hayashi:2018bkd,Jefferson:2018irk,Assel:2018rcw,Hayashi:2018lyv,Bhardwaj:2018yhy,Bhardwaj:2018vuu,Apruzzi:2018nre,Hayashi:2019yxj,Hayashi:2019fsa,Apruzzi:2019vpe,Apruzzi:2019opn,Apruzzi:2019enx,Bhardwaj:2019jtr,Bhardwaj:2019fzv} for recent developments. Alternative constructions of 5d SCFTs rely on $(p,q)$-web diagrams in type IIB \cite{Aharony:1997ju,Aharony:1997bh,Bergman:2013aca,Zafrir:2014ywa,DeWolfe:1999hj,Benini:2009gi,Hayashi:2014hfa}, which are dual to the M-theory geometry when $\MG$ is toric. The existence of 5d UV fixed points is also motivated by the AdS/CFT correspondence \cite{Brandhuber:1999np,Bergman:2012kr,Passias:2012vp,Lozano:2012au,Bergman:2013koa,Apruzzi:2014qva,Bergman:2015dpa,Kim:2015hya,DHoker:2016ujz,Gutperle:2018vdd,Bergman:2018hin,Fluder:2019szh,Uhlemann:2019ypp}.

Five-dimensional SCFTs are strongly coupled \cite{Chang:2018xmx}, and do not admit any marginal deformations \cite{Cordova:2016xhm,Cordova:2016emh}, but do admit (relevant) flavor current deformations (with mass dimension one). Therefore a deformation of the UV SCFT can lead to an infrared (IR)-free gauge theory. In the geometric engineering approach, a relevant deformation of the 5d $\mathcal{N}=1$ SCFT $\FT$ of \eqref{Mtheory X3 SCFT intro} is equivalent to a crepant resolution of the singularity:
\begin{align}
\pi_{\ell} &: \wh{\MG}_{\ell} \longrightarrow \MG ~,	
\end{align}
which yields a smooth (or at the very least, less singular) local CY$_{3}$--fold $\wh{\MG}_{\ell}$. Different crepant resolutions are related by flop transitions. Under suitable conditions satisfied by $\wh{\MG}_{\ell}$ that were spelled out in \cite{Closset:2018bjz} for the toric case, the resulting geometry gives rise to a five-dimensional $\mathcal{N}=1$ supersymmetric gauge theory.

%Five-dimensional $\mathcal{N}=1$ gauge theories have a moduli space of vacua characterized by a Coulomb branch, and a Higgs branch (like four-dimensional $\mathcal{N}=2$ theories \cite{Seiberg:1994aj,Seiberg:1994rs}). 

In this paper, we apply the methods developed in \cite{Closset:2018bjz} to analyze deformations of $\FT$ when $\MG$ is a ``rank-two'' isolated toric singularity. The latter refers to the fact that $\MG$ is described by a two-dimensional toric diagram with two interior points (that is, two compact divisors). There is a well-known classification of two-dimensional convex toric diagrams with one interior point \cite{Hori:2003ic,Cox2011} (``rank-one'' in this terminology). A classification of lattice polygons with two-interior points was given by \cite{Wei2012,Xie:2017pfl}, a subset of which describe \textit{isolated} canonical singularities of toric CY$_3$-folds. The advantage of working with isolated singularities is that there is a one-to-one correspondence between crepant resolutions of these singularities and chambers of the corresponding gauge theories that they engineer. This feature is unfortunately lost in the non-isolated case. Nevertheless, we remark that the non-isolated case is extremely important, for instance, for engineering five-dimensional $T_{N}$ theories \cite{Benini:2009gi}.\footnote{A discussion of non-isolated singularities and 5d $T_{N}$ theories has also appeared in \protect{\cite{Closset:2018bjz}}.}
%%%%%%%%%%%%%%%
 \begin{figure}[t]
\centering
\subfigure{\includegraphics[width=6in]{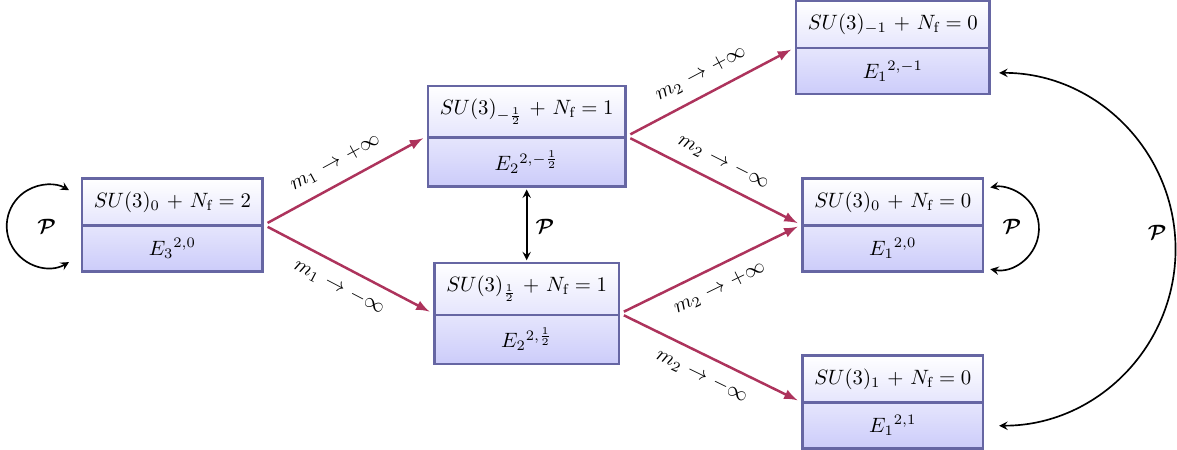}}\\$\,$\\
\subfigure{\includegraphics[width=6in]{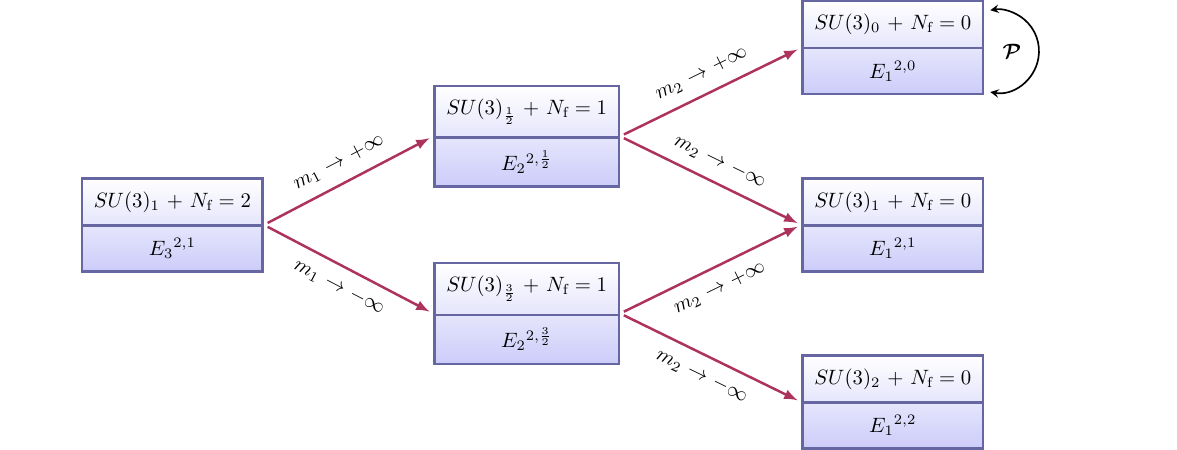}}
\caption{RG flows connecting the different rank-two toric singularities discussed in this paper, the corresponding gauge-theory phases, and their relations under parity (denoted by $\bm{\mathcal{P}})$. Non-gauge-theoretic phases are not shown in the figure, but do arise as discussed in the main text. {\bf Top:} Rank-two theories stemming from the $E_{3}{}^{2,0}$ (``beetle'') singularity. {\bf Bottom:} Rank-two theories stemming from the $E_{3}{}^{2,1}$ singularity. There is a $\bm{\mathcal{P}}$-transformed version of this flow, which is not shown here.\label{fig:RG flow chain}}
 \end{figure} 
 %%%%%%%%%%
 
In Figure \ref{fig:RG flow chain}, we show a map relating the distinct gauge-theory deformations of some of the toric singularities that appear in this paper. A recurrent theme in this paper is that theories with large flavor symmetry can flow to theories with lower flavor symmetry, an operation, which, in geometric engineering, can be understood as a combination of flop transitions in the extended parameter space of the Calabi-Yau geometry, followed by the decoupling of certain divisors in the geometry by blowing up the K\"{a}hler volumes of certain compact curves (thereby rendering them non-compact). As we explain in various examples, this can be understood as a geometric version of renormalization group (RG) flow, because the operation of decoupling divisors corresponds (rather directly in gauge-theory phases) to integrating out some massive degrees of freedom. We remark here that starting from the two geometries on the extreme left in the two RG flows indicated in Figure \ref{fig:RG flow chain}, that is, the toric geometries labeled $E_{3}{}^{2,0}$ and $E_{3}{}^{2,1}$, one can obtain all other isolated toric rank-two singularities, including three singularities which have no Lagrangian description.

On the Coulomb branch, a 5d $\mathcal{N}=1$ gauge theory is characterized by a one-loop exact cubic prepotential \cite{Seiberg:1996bd,Intriligator:1997pq} $\mathcal{F}_{\text{ft}}(\varphi;h_{0,s},m_{\alpha})$, which is a function of the Coulomb branch vevs $\varphi$, masses $m_{\alpha}$ and (inverse) gauge couplings $h_{0,s}$. In the M-theory engineering, the geometric prepotential is given by a triple intersection form on the CY$_3$-fold $\wh{\MG}_{\ell}$ \cite{Cadavid:1995bk,Ferrara:1996hh}, denoted by $\mathcal{F}_{\text{geo}}(\nu_a,\mu_j)$, where $\nu_a$ and $\mu_j$ are K\"{a}hler moduli of $\wh{\MG}_{\ell}$. In \cite{Closset:2018bjz}, the issue of matching geometry to field theory after turning on generic mass parameters and gauge couplings was discussed, such that the prepotentials in both descriptions match: $\mathcal{F}_{\text{ft}}(\varphi;h_{0,s},m_{\alpha}) = 	\mathcal{F}_{\text{geo}}(\nu_a,\mu_j)$, once an appropriate map between geometry and field theory is determined. Under this map, the K\"{a}hler parameter $\mu_{j}$ are interpreted as mass deformations of $\FT$, whereas the K\"{a}hler moduli $\nu_{a}$ are, in general, some combinations of Coulomb branch vevs and mass deformations. In \cite{Closset:2018bjz}, several rank-one examples of $\FT$, and a particular rank-two example (which the authors named the ``beetle geometry'') were discussed, along with their mass deformations which lead to gauge-theory phases.

A crucial new ingredient introduced in \cite{Closset:2018bjz} was a type IIA background for the five-dimensional theory, obtained by a circle reduction of the M-theory setup. Specifically, this involves the choice of an abelian subgroup $U(1)_{M}\subset U(1)^3$ of the toric action on $\wh{\MG}_{\ell}$, and a subsequent reduction to type IIA string theory along this $U(1)_{M}$, treated as the ``M-theory circle'' \cite{Aganagic:2009zk,Benini:2009qs,Jafferis:2009th,Benini:2011cma}, resulting in the duality:
\begin{align}
 \text{M-theory on } \R^{1,4}\times \wh{\MG}_{\ell}	~\longleftrightarrow~ \text{Type IIA string theory on } \R^{1,4}\times \mathcal{M}_{5} ~,
\end{align}
where the transverse five-dimensional space $\mathcal{M}_{5} \cong \wh{\MG}_{\ell}/U(1)_{M}$, is, in fact, a resolved $A_{M-1}$ singularity (a hyperK\"{a}hler ALE space) fibered over the real line (parametrized by $r_0 \in \mathbb{R}$). The ALE resolution contains exceptional $\P^1$s or ``2-cycles,'' which are wrapped at specific values of $r_{0}$ by D6-branes, which engineer gauge groups if they wrap compact 2-cycles (with inverse gauge couplings $\frac{1}{g^2} = \text{vol}(\P^1)$), and flavor groups if they wrap non-compact 2-cycles \cite{Douglas:1996sw,Sen:1997kz}. However, the existence of such a type IIA description in the first place relies on whether the toric diagram of $\wh{\MG}_{\ell}$ admits a ``vertical reduction.'' By viewing a toric diagram as an undirected graph as we briefly explain in this paper, this requirement can be reinterpreted as a condition on the collapsibility of a graph under a sequence of edge reductions. Remarkably, this criterion also distinguishes between toric diagrams that correspond to gauge-theory phases and those that do not. 

The details of the fibration, which can be recovered from the ``Type IIA reduction'' of the gauged linear sigma model (GLSM) associated with the toric $\wh{\MG}$ \cite{Witten:1993yc,Aganagic:2009zk}, are specified by volumes of exceptional $\P^1$s in the ALE resolution, which are piecewise linear functions of $\{r_0\} \simeq \mathbb{R}$, from which one can extract BPS masses of W-bosons, perturbative hypermultiplets, instantons, and tensions of monopole strings, etc. We carry out the type IIA analysis of \cite{Closset:2018bjz} for rank-two isolated toric singularities.

A subtle point that arises in the study of mass deformations is the parity anomaly \cite{Seiberg:2016rsg,Witten:2016cio,AlvarezGaume:1984dr,Closset:2012vp}. In \cite{Closset:2018bjz} this issue was revisited in the context of 5d $\mathcal{N}=1$ gauge  theories, and used to motivate a slightly modified version of the Coulomb-branch prepotential, one that is consistent with the requirement of predicting only integer-quantized (mixed) Chern-Simons levels on the Coulomb branch. This also plays a role in the analysis of this paper, as we use the modified prepotential which is consistent with the so-called ``$U(1)_{-\frac{1}{2}}$ quantization scheme,'' for the effective Chern-Simons levels \cite{Closset:2012vp,Closset:2018ghr,Closset:2019hyt}. This is a fine point and while it may not play a role in classifying SCFTs, it is nevertheless worth emphasizing and will be important for future studies of gauging flavor symmetries.

%Prior to \cite{Closset:2018bjz}, the equality \eqref{eq:intro match prepot} had only explicitly been checked in the literature in the neighborhood of the conformal point where mass parameters and couplings are turned off. 
%A subtlety worth mentioning here is that the geometric prepotential is known only up to some ambiguously-defined contributions from triple-intersection numbers of non-compact divisors. So the equality \eqref{eq:intro match prepot} should really be expected to hold up to such contributions. Note that these terms play no role in studies probing the Coulomb moduli space of the SCFT (where all mass deformations -- which, in the prepotential are coefficients of these ill-defined triple-intersection numbers -- are turned off). Fortunately, such contributions are independent of the Coulomb moduli ($\varphi$), so they do not affect the matching of BPS quantities, but their precise interpretation may be relevant to curved-space compactifications of such five-dimensional theories.

The number of distinct ($SL(2,\Z)$-inequivalent) two-dimensional toric diagrams increases rapidly with the rank. A natural extension will be to address higher-rank toric singularities. We anticipate that a classification of toric diagrams of rank $>2$ will be more involved, although restricting to isolated singularities as a first-order step should make the problem tractable. We expect graph-theoretic techniques to be even more useful in these higher-rank cases. It might also be interesting to interpret the type IIA geometry in terms of calibrated solutions in low-energy supergravity, potentially including orientifolds to describe gauge theories with SO/Sp gauge groups. But this will require leaving the toric realm, and we leave this as another avenue for future work.

This paper is organized as follows. In section \ref{sec:5d review}, we give a brief review of 5d $\mathcal{N}=1$ theories and geometric engineering, commenting on various features such as the prepotential, the parity anomaly, the BPS states on the Coulomb branch, and the M-theory and type-IIA approaches. We also motivate the use of graph theoretic-methods for operations such as enumerating crepant resolutions and characterizing allowed type IIA reductions. In section \ref{sec:rank two section}, we discuss in detail each rank-two isolated toric singularity. For every singularity with a gauge-theory phase, we derive the corresponding type IIA description, and use it to match the M-theory description with the field theory description. Along the way, we also discuss the role of walls in moduli space, and geometric transitions to non-Lagrangian phases. We also give several examples of RG flows in the extended parameter space of these toric geometries, which lead to geometries (theories) with fewer external vertices (lower flavor symmetry). Finally, the appendices contain results relevant for intermediate computations, including geometric and field-theory prepotentials, instanton masses and triple-intersection numbers.

\section{5d $\mathcal{N}=1$ theories and M-theory on a CY$_3$ singularity\label{sec:5d review}}
In this section, we give a lightning review of five-dimensional $\mathcal{N}=1$ supersymmetric field theories and geometric engineering. For more detailed reviews, see \cite{Seiberg:1996bd,Intriligator:1997pq,Closset:2018bjz} and references therein. In this paper, we focus on the Coulomb-branch physics. (See \cite{Cremonesi:2015bld,Ferlito:2017xdq,Cabrera:2018jxt} for some recent work on the Higgs branch.) 
\subsection{Review of 5d $\mathcal{N}=1$ gauge theories}
These theories have eight real supercharges.\footnote{\label{foot:EuclideanFootnote}Recall that the minimal spinor of $\text{Spin}(1,4)$ is a \textit{symplectic} Majorana spinor.}  We will assume that the gauge group $\GG$ is compact and connected, and factorizable into a product of simple factors $\GG_s$, i.e. $\GG = \prod_{s}\GG_s $. The Lie algebra of $\GG$ is $\fg = \text{Lie}(\GG)$.  The two on-shell multiplets of (rigid) 5d $\mathcal{N}=1$ supersymmetry are:\footnote{The tensor multiplet plays no role in our discussion, but on the Coulomb branch one can, of course, dualize a tensor to an abelian vector.} (i) the vector multiplet $\cV$, consisting of a \textit{real} scalar $\varphi$, a gauge field $A_{\mu}$, and gaugini $\lambda$ and $\widetilde{\lambda}$, all valued in the adjoint of $\fg$, and (ii) the hypermultiplet consisting of four real scalars and their fermionic superpartners. The vector multiplet is coupled to matter fields in the hypermultiplets $\cH$ in some representation $\mathfrak{R}$ of the gauge group which is in general reducible.\footnote{In this paper, since we restrict to rank-2 theories with unitary gauge groups, the gauge group will be $\GG = SU(3)$ or $\GG = U(3)$, and $\mathfrak{R}$ will be the fundamental representation of $\GG$.} The R-symmetry group in Lorentzian signature is $SU(2)_{R}$.  
\paragraph{The 5d supersymmetry algebra and central charges.} The most general (i.e. centrally extended) $\mathcal{N}$-extended Poincar\'{e} superalgebra in $d = 1+4$ dimensions has an $Sp(\mathcal{N}) \cong USp(2\mathcal{N})$ R-symmetry, and has the form \cite{Cremmer:1980gs,Ferrara:1999si}:
\begin{align}
\{Q_{\alpha}^A, Q_{\beta}^B\} &= (\gamma^{\mu}C)_{\alpha\beta} P_{\mu}\Omega^{AB} + (\gamma^{\mu}C)_{\alpha\beta} Z_{\mu}^{\circ[AB]} + C_{\alpha\beta} Z^{[AB]} + (\gamma^{\mu\nu}C)_{\alpha\beta} Z_{\mu\nu}^{(AB)}  ~,	\label{eq:5d superpoincare}
\end{align}
where $\alpha, \beta$ here are spinor indices (only for the purposes of this equation, and not to be confused with their use elsewhere in this paper), $\mu, \nu$ are five-dimensional spacetime indices, $C$ is the charge conjugation matrix, $\Omega$ denotes the symplectic form, and (crucially) $Z$'s denote \textit{real} central charges. The indices $A, B$ range over $A, B = 1, \ldots, 2\mathcal{N}$. Here $Z_{\mu}^{\circ[AB]}$ and $Z^{[AB]}$ are antisymmetric, and $Z_{\mu}^{\circ[AB]}$ is symplectic traceless: $\Omega_{AB}Z_{\mu}^{\circ[AB]} = 0$. The central charges $Z_{\mu}^{\circ[AB]}$ and $Z_{\mu\nu}^{(AB)}$ contribute to strings and membranes, respectively, whereas $Z^{[AB]}$ contributes to particle states that enter \eqref{eq:mass BPS particles} (see below). For the purposes of this paper, since we focus on 5d SCFTs, we restrict to $\mathcal{N}=1$ in \eqref{eq:5d superpoincare}.
%
%\paragraph{Moduli space.} These theories have a Coulomb branch along which the $SU(2)_{R}$ symmetry acts freely and the scalars in the vector multiplet get a vacuum expectation value (vev) thus breaking the gauge group $\GG$ to its maximal torus, and a Higgs branch where instead the scalars in the hypermultiplet get vevs and the gauge group is generically broken completely. In this paper, we exclusively focus on the Coulomb-branch physics.

\paragraph{Flavor symmetries.} Five-dimensional gauge theories have a nontrivial global symmetry $\GG_F \times SU(2)_{R}$, where the $SU(2)_{R}$ is the R-symmetry group introduced above, and $\GG_F$ is the ``flavor'' symmetry group, which in turn, can be further decomposed as $\GG_F = \GG_{\mathcal{H}} \times \prod_{s}U(1)_{T_s}$, that is, a group ($\GG_{\mathcal{H}}$) that hypermultiplets transform under, and a product of ``$U(1)$ topological factors,''\footnote{These are due to the ``topological symmetry'' which is associated with a conserved current $j_{T_s} = \frac{1}{8\pi^2} F^{(s)} \wedge F^{(s)}$ (where $F^{(s)} = dA^{(s)} - i A^{(s)} \wedge A^{(s)}$), which is conserved due to the Bianchi identity.} one for each simple factor $\GG_s$ in $\GG$. More precisely, each hypermultiplet transforms under a representation of $\GG_{\mathcal{H}} \times \GG$.

%We denote by $\cV_F$ background vector multiplets for the flavor symmetry $\GG_\mathcal{H}$ in \eqref{eq:flavor group decomp 2}, and background abelian vector multiplets $\cV_{T_s}$ for each topological symmetry factor. A supersymmetric background for $\cV_F$ is obtained by switching on constant vacuum expectation values for the real scalar $\varphi_F \in \cV_F$. These five-dimensional real masses are denoted by $\langle \varphi_F \rangle \equiv \mu = (m_{\alpha}, h_{0,s})$, where $m = (m_{\alpha})$ denote ``flavor masses'' for the hypermultiplet flavor group, $\alpha= 1, \ldots, \text{rank}(\GG_{\mathcal{H}})$ runs over a maximal torus of $\GG_{\mathcal{H}}$. The masses associated with the topological symmetries are $h_0 = (h_{0,s})$. Note that the $U(1)_{T_s}$ mass term is the Yang-Mills Lagrangian for the vector multiplet with gauge group $\GG_s$, with $h_{0} = \frac{8\pi^2}{g^2}$ denoting the 5d inverse gauge coupling \cite{Seiberg:1996bd}.

\paragraph{5d parity anomaly, Chern-Simons terms and the $U(1)_{-\frac{1}{2}}$ quantization scheme.} A detailed discussion of the 5d parity anomaly can be found in \cite{Closset:2018bjz}. (See also \cite{Redlich:1983dv,Redlich:1983kn,Niemi:1983rq,AlvarezGaume:1984nf,Closset:2012vp} for original discussions in the three-dimensional setting.) %In this context, it is useful to consider the Euclidean version of the theory. 
%Here, parity refers to an inversion of \textit{one} of the spacetime coordinates of $\R^5$, say $x^5$, so it is given by $\mathcal{P}: x^5 \rightarrow -x^5$. 
Five-dimensional gauge theories suffer from a ``parity anomaly,'' which is the statement that parity and gauge invariance cannot simultaneously be preserved. Here, the term ``gauge invariance'' is used in a general sense to include all potential background gauge symmetries, including flavor symmetries. If we preserve background gauge invariance (so that a gauge field can be made dynamical), we must accept non-conservation of parity.

There are three sources of parity violation: the first is an explicit Chern-Simons term in the low-energy effective action, which for a $U(1)$ gauge field $A_{\mu}$ in five dimensions, is of the form,
\begin{align}
S_{\text{CS}} &= \frac{i k}{24\pi^2}\int_{\mathcal{M}_{5}}\left(A \wedge F \wedge F	+ \cdots\right) ~. \label{eq:CS term}
\end{align}
on an oriented Riemannian five-manifold $\mathcal{M}_{5}$, where the integrand is understood to include terms needed for a supersymmetric completion. Such a CS term is well-defined only if the CS level $k$ is integer quantized, $k \in \Z$.\footnote{Note that for any simple Lie group with a nonzero cubic index, i.e. for $\mathfrak{g}_{s} = \text{Lie}(\GG_{s}) = \mathfrak{su}(N)$ with $N > 2$, one can have explicit non-abelian supersymmetric CS terms.}
 The second source is a parity-odd contact term in the three-point function of the conserved current $j^\mu$ of a $U(1)$ symmetry acting on a fermion $\psi$ charged under a background $U(1)$ gauge field.\footnote{This term is of the form $\frac{i\kappa} {24\pi^2}\epsilon_{\mu_1 \mu_2 \mu_3 \mu_4 \mu_5} p^{\mu_4} q^{\mu_5} \subset \langle j_{\mu_1}(p)j_{\mu_2}(q)j_{\mu_3}(-p-q)\rangle$.} An explicit CS \eqref{eq:CS term} shifts the \textit{effective} CS level $\kappa$ to $\kappa + k$. But since $k$ is integer-quantized, only the non-integer part of the Chern-Simons contact term, $\kappa \,\, \text{ (mod 1)}$, is physical and it is what probes the presence of a parity-violating term in the effective action. For a collection of Dirac fermions $\psi^i$ with $U(1)$ charges $Q_i \in \Z$, we find that, 
\begin{align}
\kappa &= -\frac{1}{2}\sum_{i}Q_{i}^3 + k ~.	 \label{eq:massless Dirac collection}
\end{align}
where the integer $k$ is due to a scheme ambiguity which, in this case, corresponds simply to adding an explicit $U(1)$ CS term \eqref{eq:CS term} with integer coefficient $k$ to the action \cite{Closset:2019hyt}. For every Dirac fermion in the gauge theory, we should specify a ``quantization scheme'' that is consistent with gauge invariance: this requires specifying all the CS contact terms $\kappa$, for both dynamical and background gauge fields, which corresponds to a scheme choice for $k$ in \eqref{eq:massless Dirac collection}, to remove the integer-valued ambiguity. One such scheme is the so-called ``$U(1)_{-\frac{1}{2}}$ quantization scheme,''\footnote{Any other quantization scheme is related to this one by a shift of $\kappa$ by an integer.} \cite{Closset:2012vp} which declares,
\begin{align}
   \kappa_{\psi} &= -\frac{1}{2}~ , \quad \text{ for a massless free fermion.} \label{eq:U(1) half quantization}
\end{align}
We choose this quantization scheme for every 5d $\mathcal{N}=1$ hypermultiplet. The final source of parity violation is a mass term for a Dirac fermion $\psi$ in the Lagrangian, $\delta \mathcal{L}_{m} = i m \overline{\psi} \psi$ (for $m \in \mathbb{R}$), which explicitly breaks parity. In the limit $|m| \rightarrow \infty$, one can integrate out $\psi$. As shown in \cite{Witten:1996qb,Bonetti:2013ela}, this shifts the parity-odd contact term by $\delta \kappa = -\frac{1}{2}\text{sign}(m)$. The generalization of \eqref{eq:massless Dirac collection} to a collection of \textit{massive} Dirac fermions of masses $m_i \in \mathbb{R}$ and charges $Q_i \in \Z$ is:
\begin{align}
\kappa_{\psi} &= -\frac{1}{2}\sum_{i}Q_{i}^3\,\text{sign}(m_i) \, \,.\label{eq:massive Dirac fermion collection U(1) half quantization}
\end{align}
The notation ``$\GG_{\kappa}$'' where $\GG$ is the gauge group and $\kappa \in \frac{1}{2}\Z$ is frequently used in the literature, and here $\kappa$ denotes the effective Chern-Simons level as in \eqref{eq:massive Dirac fermion collection U(1) half quantization}.

\subsection{The prepotential on the Coulomb branch}\label{sec:new prepotential}
We consider the low-energy effective field theory on the Coulomb branch, where vacuum expectation values of the adjoint scalar, $\langle {\bm{\varphi}}\rangle = \text{diag}(\varphi_a) = (\varphi_1, \ldots, \varphi_{\text{rk}(G)})$ break the gauge group $\GG$ down to a maximal torus $\GH$ times the Weyl group:
\begin{align}
\GG &\longrightarrow \GH \rtimes W_{\GG}~, \qquad \GH \cong \prod_{a=1}^{\text{rk}(\GG)}U(1)_{a} ~. \label{eq:gauge group breaking on CB}
\end{align}
Here, $\bm{\varphi} = (\varphi_{a})$ denotes the set of low-energy Coulomb-branch scalars which reside in abelian vector multiplets $\mathcal{V}_{a}$, and $\bm{\mu} = (m, h_0)$ denotes the set of real flavor masses and inverse gauge couplings.

The low-energy effective field theory on the Coulomb branch is an $\mathcal{N}=1$ supersymmetric gauge theory that is completely determined by a one-loop exact \textit{cubic} prepotential $\mathcal{F}(\bm{\varphi}, \bm{\mu})$ \cite{Cadavid:1995bk,Seiberg:1996bd,Ferrara:1996hh,Intriligator:1997pq}. The one-loop contribution arises from integrating out W-bosons and massive hypermultiplets at a generic point on the Coulomb branch. The widely prevalent expression for the prepotential is due to \cite{Intriligator:1997pq}; we refer to it as the ``IMS prepotential.'' However, this prepotential can lead to non-integer mixed flavor-gauge effective CS levels. To cure this discrepancy, the authors of \cite{Closset:2018bjz} proposed an alternate expression for the prepotential which corrects the IMS expression essentially by adding explicit ``half-integer CS levels'' on the Coulomb branch in order to cancel the parity anomalies by restoring background gauge invariance under the flavor group. 
The prepotential proposed in \cite{Closset:2018bjz} is:
\begin{align}
\mathcal{F}(\bm{\varphi},\bm{\mu}) &= \frac{1}{2}h_{0,s}K_{s}^{ab}\varphi_{a}\varphi_{b} + \frac{k^{abc}}{6}\varphi_{a}\varphi_{b}\varphi_{c} + \frac{1}{6}\sum_{\alpha\in\Delta}\Theta\left(\alpha(\varphi)\right)\left(\alpha(\varphi)\right)^3\nonumber\\
&\qquad - \frac{1}{6}\sum_{\omega}\sum_{\rho\in\mathfrak{R}}\Theta\left(\rho(\varphi) + \omega(m)\right)\left(\rho(\varphi) + \omega(m)\right)^3 \, \, ,	\label{eq:our prepot}
\end{align}
where a sum over repeated indices ($s$, and $a$, $b$, $c$) is understood, and our notation is summarized in Table \ref{tbl:prepot notation}. 
\begin{table}[ht]
\centering
 \begin{tabular}{l c l}
    $s$ & : & index ranging over the simple gauge group factors  \\
    $a, b, c$ &: & indices ranging over $\text{rk}(\GG)$, i.e. $1 \leq a \leq \text{rk}(\GG)$ \\
 	$h_{0,s} = \frac{8\pi^2}{g_s}$ & : & inverse gauge coupling for gauge group factor $\GG_s$ \\
 	$K_{s}^{ab}$ & : & Killing forms of the simple factors $\mathfrak{g}_s$ \\
 	$d_{s}^{abc}$ & : & cubic Casimir of $\mathfrak{g}_{s}$ \\
 	$k^{abc} = k_{s} d_{s}^{abc}$ & : & Chern-Simons coefficient in the prepotential \\
 	$\alpha = (\alpha^a)$ & : & a typical root of $\mathfrak{g}$ \\
 	$\Delta$ & : & set of nonzero roots (adjoint weights) of $\mathfrak{g}$ \\
 	$\alpha(\varphi) = \alpha^a \varphi_a$ & : & natural pairing between Coulomb vevs and adjoint weights \\ % do I need to define alpha or is it obvious?
 	$\omega^{\alpha}$ & : & flavor weights (weights of the repn.($\GG_{\mathcal{H}}$))\\
 	$\omega(m) = \omega^\alpha m_\alpha$ & : & natural pairing between hyper masses and flavor weights \\
 	$\rho^{a}$ &: & gauge weights (weights of repn.($\GG$) that the hyper transforms under)\\
 	$\rho(\varphi) = \rho^a\varphi_a$ & : & natural pairing between Coulomb vevs and gauge weights
 \end{tabular}	
 \caption{Notation for symbols appearing in the prepotential.\label{tbl:prepot notation}}
\end{table}

\noindent By contrast, the IMS prepotential reads:
\begin{align}
\mathcal{F}_{\text{IMS}}(\bm{\varphi},\bm{\mu}) &= \frac{1}{2}h_{0,s}K_{s}^{ab}\varphi_{a}\varphi_{b} + \frac{k^{abc}_{\text{eff}}}{6}\varphi_a \varphi_b \varphi_c + \frac{1}{12}\sum_{\alpha \in\Delta}\left|\alpha(\varphi)\right|^3 \nonumber\\
&\qquad -\frac{1}{12} \sum_{\omega}\sum_{\rho\in\mathfrak{R}}\left|\rho(\varphi) + \omega(m)\right|^3 ~.\label{eq:IMS prepot}
\end{align}
The function $\Theta(x)$ appearing in \eqref{eq:our prepot} is the Heaviside step function defined by:
\begin{align}
\Theta(x) &= \left\{\begin{array}{ll}
          1, & \text{if } x \geq 0~, \\
          0, & \text{if } x < 0 \,.
 \end{array}\right.	
\end{align}
The prepotential \eqref{eq:our prepot} yields the correct result for a hypermultiplet in the ``$U(1)_{-\frac{1}{2}}$ quantization,'' that was introduced above. For a single hypermultiplet coupled to a $U(1)$ vector multiplet containing the real scalar $\varphi$, the contribution to the prepotential is $\mathcal{F}_{\cH} = -\frac{1}{6}\Theta(\varphi)\varphi^3$.  A $U(1)$ Chern-Simons term at level $k$ contributes, on the other hand, $\mathcal{F}_{U(1)_{k}}(\varphi) = \frac{k}{6}\varphi^3$. Therefore the hypermultiplet contribution $\mathcal{F}_{\cH}$ reproduces the correct decoupling limits for both signs of the real mass $\varphi$. 
Comparing \eqref{eq:our prepot} and \eqref{eq:IMS prepot}, one finds that terms of order $\varphi^3$ are the same once one correctly maps the CS levels, via $k^{abc}_{\text{eff}} = k^{abc} - \frac{1}{2}\sum_{\rho,\omega}\rho^{a}\rho^{b}\rho^{c}$, but there is a difference in the lower-order terms (i.e. terms of order $\varphi^2$ and $\varphi$). Specifically, at a generic point on the Coulomb branch, the theory is gapped and therefore the Chern-Simons contact terms $\kappa$ should all be integer-quantized. This must be true not just for the gauge CS levels, but also for mixed (gauge)$^2-$flavor, (gauge)$-$(flavor)$^2$ and (flavor)$^3$ CS levels, etc. More explicitly, \textit{all} the following Chern-Simons levels must be integer quantized at a generic point on the Coulomb branch:
\begin{align}
\small \left. \begin{array}{l l l}\kappa^{abc} = \p_{\varphi_a}\p_{\varphi_b}\p_{\varphi_c}\FF, & \kappa^{ab\alpha} = \p_{\varphi_a}\p_{\varphi_b}\p_{m_{\alpha}}\FF, & \kappa^{a\alpha\beta} = \p_{\varphi_a}\p_{m_\alpha}\p_{m_{\beta}}\FF,	\\
\kappa^{abs} = \p_{\varphi_a}\p_{\varphi_b}\p_{h_{0,s}}\FF, & \kappa^{a s s'} = \p_{\varphi_a}\p_{h_{0,s}}\p_{h_{0,s'}}\FF, & \kappa^{\alpha\beta\gamma} = \p_{m_\alpha}\p_{m_\beta}\p_{m_\gamma}\FF, \\
\kappa^{\alpha\beta s} = \p_{m_\alpha}\p_{m_\beta}\p_{h_{0,s}}\FF, & \kappa^{\alpha s s'} = \p_{m_\alpha}\p_{h_{0,s}}\p_{h_{0,s'}}\FF, & \kappa^{s s' s''} = \p_{h_{0,s}}\p_{h_{0,s'}}\p_{h_{0,s''}}\FF, \\
\kappa^{a \alpha s} = \p_{\varphi_a}\p_{m_\alpha}\p_{h_{0,s}}\FF & &
\end{array} \right\} \in \Z \,.
\end{align}
One finds that \eqref{eq:our prepot} indeed produces produces integer-quantized effective CS levels but \eqref{eq:IMS prepot} does not. For a detailed discussion, including a derivation of \eqref{eq:our prepot}, see \cite{Closset:2018bjz}. Henceforth, we will work exclusively with \eqref{eq:our prepot}.

\subsection{BPS objects on the Coulomb branch}
On the Coulomb branch of 5d $\mathcal{N}=1$ gauge theories, there are half-BPS particles and strings, which saturate suitable BPS bounds relating their masses (or tensions) to central charges in the supersymmetry algebra.
\paragraph{BPS particles.} The masses of BPS particles are given by the absolute value of the (real) central charge of the 5d $\mathcal{N}=1$ Poincar\'{e} superalgebra \eqref{eq:5d superpoincare}:
\begin{align}
M &= |Q^a \varphi_a + Q_{F}^{\alpha}m_{\alpha} + Q_{F}^{s}h_{0,s}| ~, \label{eq:mass BPS particles}
\end{align}
where $Q^a$ are gauge charges, $Q_F^\alpha$ are the $\GG_{\mathcal{H}}$ flavor charges and $Q_F^s$ are $U(1)_{T_s}$ instanton charges. All charges are integer-quantized. (Also see Table \ref{tbl:prepot notation}.) The three categories of BPS particles of interest here are:
\begin{itemize}
	\item \textbf{W-bosons} $W_\alpha$, associated with the roots $\alpha \in \mathfrak{g}$ of the gauge algebra, with masses:
	\begin{align}
	   M(W_\alpha) &= \alpha(\varphi) ~. \label{eq:mass W boson}	
	\end{align}
   \item \textbf{Hypermultiplets} $\mathcal{H}_{\rho,\omega}$, transforming in a representation of $\GG \times \GG_{\mathcal{H}}$ with gauge charges $Q^a = \rho^a$ and flavor charges $Q_F^{\alpha} = \omega^{\alpha}$, with masses:
   \begin{align}
       M(\mathcal{H}_{\rho,\omega}) &= \rho(\varphi) + \omega(m) ~. \label{eq:mass hypermultiplets}	
   \end{align}
   \item \textbf{Instantonic particles}: these are BPS particles charged under topological symmetries such that $Q_F^s \neq 0$ in \eqref{eq:mass BPS particles}. They are really solitonic particles in five dimensions, being uplifts of four-dimensional $\GG$-instantons. The procedure to compute the instanton masses is outlined in Appendix \ref{sec:field theory prepotentials}. The results of these computations appear in Tables \ref{tbl:u3 nf1 instantons} and \ref{tbl:u3 nf2 instantons} for the models discussed in this paper. 
\end{itemize}
\paragraph{BPS monopole strings.} The 5d $\mathcal{N}=1$ gauge theory has real codimension-3 objects which are BPS monopole strings, which are five-dimensional uplifts of 4d $\mathcal{N}=2$ monopoles. %They are labeled by GNO-quantized magnetic fluxes threading any $S^2$ surrounding them inside the ambient $\mathbb{R}^5$. For every $U(1)_{a}$ in the maximal torus of the gauge group $\GG$, there is a string with $U(1)_a$ magnetic flux $\mathfrak{m}_a = 1$. 
The tension of a monopole string is given by the first derivative of the prepotential with respect to the Coulomb modulus \cite{Seiberg:1996bd}:
\begin{align}
T_{a}(\bm{\varphi}, \bm{\mu}) &= \frac{\partial\FF}{\partial\varphi_{a}}	 ~, \label{eq:BPS monopole string tension} \quad \text{for }\quad  a = 1, \ldots, \text{rk}(\GG) ~.
\end{align}  

\subsection{M-theory on a CY$_3$ singularity\label{sec:m-theory-on-cy3}}
In this paper, we consider geometric engineering of 5d $\mathcal{N}=1$ theories that live on the spacetime transverse to M-theory on a local Calabi-Yau three-fold (CY$_3$) $\MG$, an isolated canonical singularity. This is motivated by the conjectured correspondence,
\begin{align}
\text{M-theory on } \mathbb{R}^{1,4} \times \MG \quad  &\longleftrightarrow \quad \mathcal{T}_{X}\,\,\,\text{ SCFT on } \mathbb{R}^{1,4}	 ~ .\label{eq:conjectured correspondence 2}
\end{align}
We give a brief recap of some relevant terminology from singularity theory.\footnote{We refer the mathematically inclined reader to \protect{\cite{Hartshorne1977,Griffiths1994,Closset:2009sv,Ishii2014,Xie:2017pfl}} and references therein.} For an irreducible variety $\MG$, a resolution of singularities of $\MG$ is a proper morphism $\pi: \widehat{\MG} \rightarrow \MG$ such that $\wh{\MG}$ is smooth and irreducible, and $\pi$ induces an isomorphism of varieties $\pi^{-1}(\MG/\wh{\MG}) = \MG/\wh{\MG}$. A projective normal variety $\MG$ such that its canonical class $K_{\MG}$ is $\mathbb{Q}$-Cartier has the property that $K_{\wh{\MG}} = \pi^{*}K_{\MG} + \sum_{i}a_{i}\bE_i$ where the sum (over $i$) is over irreducible exceptional divisors, and the $a_i$'s are rational numbers called the discrepancies. Such a variety is called $\mathbb{Q}$-Gorenstein. The singular variety $\MG$ is said to have canonical singularities if $a_{i} \geq 0$ for all $i$, in which case it is called a Gorenstein canonical singularity.\footnote{The case of strict equality $a_i > 0$ for all $i$ is called a terminal singularity, in which case the variety $\MG$ is called a Gorenstein terminal singularity. Terminal singularities imply that any subsequent resolution changes the canonical class.}

In the case of a generic CY$_3$ singularity $\MG$, a crepant resolution exists:
\begin{align}
\pi &: \wh{\MG} \longrightarrow \MG, \qquad \pi^{*}K_{\MG} = K_{\wh{\MG}} \,\, ,	
\end{align}
yielding a smooth local CY threefold $\wh{\MG}$.\footnote{A singular Calabi-Yau is always Gorenstein. Its singularities are either Gorenstein canonical or $\mathbb{Q}$-factorial Gorenstein terminal. See, for example, \protect{\cite{alg-geom/9606016}}.} %Strictly speaking, the resolution $\wh{\MG}$ is a partial resolution such that it has $\mathbb{Q}$-factorial terminal singularities (i.e. any subsequent resolution changes the canonical class).
A 5d $\mathcal{N}=1$ field theory can be obtained in the decoupling limit of an M-theory compactification on a \textit{compact} CY $_3$ threefold $Y$, by scaling the volume of $Y$ to infinity, while keeping finite the volumes of a collection of holomorphic 2-cycles and holomorphic 4-cycles which intersect within $Y$. This makes the five-dimensional Planck mass infinitely large, thereby decoupling gravity. The requirement of intersecting 2- and 4-cycles ensures that we get an interacting SCFT from the local model $\wh{\MG}$. %(Here $Y \xrightarrow{M_{pl}\rightarrow \infty} \wh{\MG}$.) 

Recall that divisors are complex codimension-1 hypersurfaces (elements of $H_4(\wh{\MG},\Z)$), whereas compact curves are complex dimension-1 hypersurfaces (elements of $H_2(\wh{\MG},\Z)$). The exceptional set $\pi^{-1}(0)$ (with $0 \in \MG$ denoting the isolated singularity) contains a certain number, say $n_{4} \equiv r \geq 0$ of compact divisors, called the ``rank'' of $\MG$. This number is the rank of the SCFT Coulomb branch, that is, $r = \dim \mathcal{M}^{C}_{\mathcal{T}_{\MG}}$. In addition to compact divisors, the resolved space $\wh{\MG}$ contains compact curves which may intersect the exceptional divisors non-trivially. Let $\CC^{\textbf{a}}$ be a basis of compact holomorphic 2-cycles in $H_{2}(\wh{\MG}, \Z)$. Note that the two-cycles $\CC^{\textbf{a}}$ are Poincar\'{e} dual to either compact divisors (in the exceptional set) or to non-compact divisors. Let $n_2 \equiv r + f = \dim H_{2}(\wh{\MG}, \Z)$, with $f \geq 0$ being a non-negative integer. Then, $r$ is the number of compact divisors and $f$ is the number of non-compact divisors. 

Let $D_{k}$ denote a typical divisor (compact or noncompact). We choose some basis of $n_2$ divisors $\{D_k\}_{k=1}^{n_2}$ and collect the intersection numbers of divisors and curves in a (square) matrix denoted by $\textbf{Q}^{\textbf{a}}{}_{k}$:
\begin{align}
   \textbf{Q}^{\textbf{a}}{}_{k} &\equiv \CC^{\textbf{a}}\cdot D_{k}\,\,, \qquad \det \textbf{Q} \neq 0 \, \,. \label{eq:Qintnos}
\end{align}
Let $J$ denote the K\"{a}hler form of $\wh{\MG}$ (a representative of the cohomology class $H^{1,1}(\wh{\MG})$) and let $S$ denote the Poincar\'{e} dual K\"{a}hler class, \footnote{As $\wh{\MG}$ is local, invoking Poincar\'{e} duality entails the use of cohomology with compact support \protect{\cite{hatcher2002algebraic}}.} which can be written as a linear combination of divisors over $\R$:
\begin{align}
S &= \sum_{k=1}^{n_2} \lambda^{k}D_{k} = \sum_{j=1}^{f}\mu^j D_{j} + \sum_{a=1}^{r}\nu^{a}\bE_a \, \, . \label{eq:ess}
\end{align} 
Here we have decomposed $\{D_k\}_{k=1}^{n_2}$ into a set of $r$ compact divisors denoted by $\bE_a$ (where $a = 1, \ldots, r$), and $f$ non-compact divisors, denoted by $D_j$ (where $j = 1, \ldots, f$). The K\"{a}hler volumes of a compact curve $\CC^{\textbf{a}}$ in $\wh{\MG}$ are given by:
\begin{align}
\xi^{\textbf{a}}(\mu, \nu) &= \int_{	C^{\textbf{a}}}J = \CC^{\textbf{a}}\cdot S = \textbf{Q}^{\textbf{a}}{}_{k} \lambda^{k} = \textbf{Q}^{\textbf{a}}{}_{j}\mu^j + \textbf{Q}^{\textbf{a}}{}_{a}\nu^{a} \geq 0 \, \, . \label{eq:nef}
\end{align}
Incidentally, the inequalities of the form \eqref{eq:nef} for all basis curves are also sometimes called the Nef conditions \cite{Xie:2017pfl} in the literature. The curves $\CC^{\textbf{a}}$  generators of the Mori cone. It is clear that the parameters $\mu^k \in \R$ and $\nu^a \in \R$ in \eqref{eq:ess} are, respectively, the K\"{a}hler moduli of two-cycles dual to non-compact four-cycles and compact four-cycles. They play an important role in developing the geometry--field-theory dictionary. In particular, $\mu$'s are mass parameters and couplings (which we collectively refer to as ``K\"{a}hler parameters,'' for they are nondynamical), whereas $\nu$'s involve a combination of dynamical fields (the Coulomb branch scalar vevs, i.e. $\varphi$'s) and in general, also the masses and couplings.\footnote{The important basis-independent feature is that the $\mu$'s \textit{never} depend on Coulomb branch scalars.}

The low-energy 5d $\mathcal{N}=1$ field theory, for generic values of the K\"{a}hler parameters, is an abelian theory with gauge group $U(1)^{r} \cong H^{2}(\wh{\MG},\R)/H^{2}(\wh{\MG},\Z)$. In the geometric engineering picture, the $U(1)$ gauge fields arise from periods of the M-theory 3-form $C_{(3)}$ over the curves $\CC^a$ dual to compact divisors, i.e. $A^{(a)}_{U(1)} = \int_{\CC^a}C_{(3)}$ (where $a = 1, \ldots, r$). The exact prepotential for this abelian gauge theory can be computed from the geometry using the following expression:\footnote{The minus sign is simply a matter of convention, chosen in \protect{\cite{Closset:2018bjz}}.} 
\begin{align}
\FF(\mu, \nu) &= -\frac{1}{6}\int_{\wh{\MG}}J \wedge J \wedge J = -\frac{1}{6}S\cdot S \cdot S \, \, .	\label{eq:geometric prepot}
\end{align}
The prepotential involves triple-intersection numbers of $\wh{\MG}$, specifically those of the form (dropping the dot for brevity) $D_{i}D_{j}\bE_{a}$, $D_{i}E_{a}\bE_{b}$, $\bE_{a}\bE_{b}\bE_{c}$, and $D_{i}D_{j}D_{k}$. But since $\wh{\MG}$ is noncompact, the triple-intersection numbers involving three noncompact divisors ($D_i D_j D_k$) are not well-defined. In subsequent computations of the geometric prepotential \eqref{eq:geometric prepot}, we ignore such contributions to the prepotential, and we refer to the result as the ``compact part'' of the prepotential. We refer the reader to Appendix B of \cite{Closset:2018bjz} for a more detailed discussion of this point. 
 
 \paragraph{BPS states from geometry.} The BPS states from geometric engineering are:
 \begin{itemize}
 	\item Electrically charged BPS particles, from M2-branes wrapping holomorphic (compact) 2-cycles $\CC^a$. These have masses given by the K\"{a}hler volumes \eqref{eq:nef}, and 
 	\item (Dual) magnetically charged BPS monopole strings, from M5-branes wrapping holomorphic surfaces (compact 4-cycles) $\bE_a$. These have tensions by the K\"{a}hler volumes of the compact divisors:
      \begin{align}
 	           T_{a}(\mu,\nu) &\equiv -\partial_{\nu^a}\mathcal{F}(\mu,\nu) = \frac{1}{2}\int_{\bE_a}J\wedge J = \vol(\bE_a) \,\, . \label{eq:tension from geometry}
       \end{align}
 \end{itemize}
% \paragraph{Metric on the Coulomb branch.} The metric on the Coulomb branch can be computed from the geometric prepotential as the volume of a curve at the intersection of two exceptional divisors:
% \begin{align}
% \tau_{ab}(\mu, \nu) &= \partial_{\nu^a}\partial_{\nu^b}\mathcal{F}(\mu, \nu) = \text{vol}(\bE_a\cdot \bE_b)	\,\,.
% \end{align}
\paragraph{The extended parameter space.} Given an isolated canonical CY$_3$ singularity $\MG$, there can be several birationally equivalent resolutions $\pi_{\ell}: \wh{\MG}_{\ell} \rightarrow \MG$, each of which is a local Calabi-Yau 3-fold with the same singular limit. The collection of all such $\wh{\MG}_{\ell}$ constitutes, for a given singularity $\MG$, the set of all crepant resolutions.
For a particular $\wh{\MG}_{\ell}$, the K\"{a}hler cone is given by the set of all positive K\"{a}hler forms:
\begin{align}\hspace{-0.35in}
\cK(\wh{\MG}_{\ell}{\setminus}\MG)	&= \left\{[J]\in H^{1,1}(\wh{\MG}_{\ell})\cap H^{2}(\widehat{\MG}_{\ell},\R) \left|\right. \int_{\CC}J = S\cdot \CC > 0\,\, \forall \text{ hol. curves } \CC \in \wh{\MG}_{\ell}\right\}
\end{align}
The parameter space of all massive deformations of 5d SCFTs obtained from M-theory is given by the extended K\"{a}hler cone, which is the closure of the union of all compatible K\"{a}hler cones: $\CP_{\cT_{\MG}} = \wh{\CK}(\MG) = \left\{\bigcup_{l}\CK\left(\wh{\MG}_{\ell}{\setminus}\MG\right)\right\}^{c}$. Pairs of K\"{a}hler cones -- corresponding to birationally equivalent pairs of Calabi-Yau spaces -- are glued along common faces in the interior of $\wh{\CK}(\MG)$. The boundaries of $\wh{\CK}(\MG)$ are of the following type  \cite{Witten:1996qb}:
\begin{itemize}
	\item Boundaries of K\"{a}hler cones of individual crepant resolutions: these are boundaries of $\CK(\wh{\MG}_{\ell}{\setminus}\MG)$, where the threefold $\wh{\MG}_{\ell}$ becomes singular. This happens when a 2-cycle in $\wh{\MG}_{\ell}$ shrinks to zero size and grows to negative volume in a birational K\"{a}hler cone, signaling a flop transition. This corresponds to a BPS particle becoming massless. At such points, the prepotential \eqref{eq:geometric prepot} becomes non-smooth.
	\item Exterior boundaries of $\wh{\cK}(\MG)$ where a 4-cycle $\bE_a$ can collapse to either (i) a 2-cycle, or (ii) a point.
\end{itemize}
The ``origin of moduli space'' is the origin of $\wh{\CK}(\MG)$, which is also the SCFT point. It corresponds to the singular geometry $\MG$, and is given by the connected union of 4-cycles (divisors) collapsing to a point.

\paragraph{Toric geometry and type IIA reduction.}\label{sec:m-theory-on-cy3-IIA-reduction}
In this paper, we will further assume that the isolated canonical singularity $\MG$ is also toric. This allows us to exploit the computational machinery of toric geometry (see \cite{alg-geom/9606016,Hofscheier,Cox2011,Fulton1993,Leung:1997tw,Hori:2003ic,Bouchard:2007ik,Closset:2009sv} for useful reviews). This restriction admittedly ignores many interesting cases by confining attention to a small subset of singularities. We leave a study of non-toric singularities for future work. 

In particular this implies that a resolution $\wh{\MG}$ of such a singularity is described by a two-dimensional toric diagram. This is specified as the convex hull of a set of lattice points $\bm{w}_i = (w_i^x, w_i^y) \in \Z^2$ (here $i = 1, \ldots, n$, where $n$ is the number of vertices), which contains a number $r \geq 0$ of internal points, denoting compact divisors. The $n_{E} \equiv n-r$ external points denote noncompact divisors. Edges in the toric diagram connecting two vertices denote curves in the geometry.

The toric variety $\wh{\MG}$ can be described using a gauged linear sigma model (GLSM) \cite{Witten:1993yc}. The key idea here is to realize the toric variety as the moduli space of vacua of a certain 2d $\mathcal{N}=(2,2)$ supersymmetric gauge theory. The defining data for this construction is (i) a set of $U(1)$ charges $Q_{i}^{\textbf{a}}$ (where $i = 1, \ldots, n$, and $a=1, \ldots, n-3$ labels a set of linearly independent compact curves -- the Mori cone generators), and (ii) a set of Fayet-Iliopololous (FI) parameters $\{\xi_{\textbf{a}}\}$ for the auxiliary gauge groups $U(1)_{\textbf{a}}$. Then, the toric CY$_3$ variety is defined as a K\"{a}hler quotient,
\begin{align}
\wh{\MG} &\cong \C^{n}/\!/_{\!\xi}U(1)^{n-3} = \bigg\{z_{i} \in \C^{n} \bigg{|}\sum_{i}Q_{i}^{\textbf{a}}|z_i|^2 = \xi_{\textbf{a}}\bigg\}\big/U(1)^{n-3}~, \qquad n \equiv n_{E} + r \,\, , \label{eq:kahler quotient}
\end{align}
which we recognize as the familiar quotienting of a set of ``D-term equations'' by some $U(1)$ actions. Different resolutions of the singularity -- which are related by flop transitions -- differ in their sets of $U(1)$ charges $Q_{i}^{\textbf{a}}$, which always obey the ``Calabi-Yau condition,'' namely, $\sum_{i=1}^{n} Q_{i}^{\textbf{a}} = 0$ $\forall \,\,\, \textbf{a} = 1, \ldots, n-3$. Arranged as a matrix of charges, the CY condition implies that the sum all charges in any row vanishes.
%\begin{align}
%\sum_{i=1}^{n} Q_{i}^{\textbf{a}} &= 0, \quad \forall \,\,\, \textbf{a} = 1, \ldots, n-3 \, \,.	
%\end{align}

%%%%%%%%%%%%%%%
 \begin{figure}[t]
\begin{center}
\subfigure[\small Allowed.]{
\includegraphics[height=4.5cm]{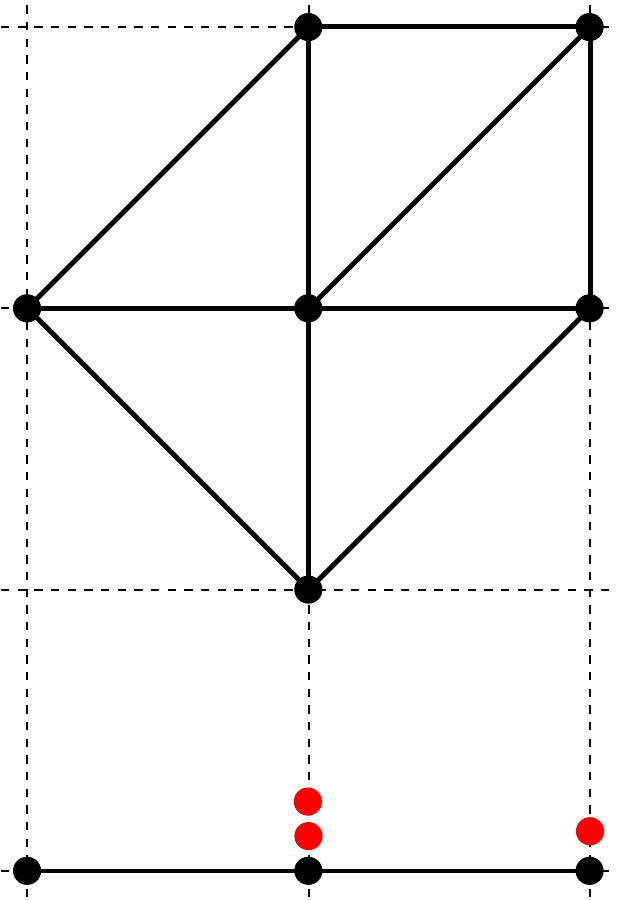}\label{fig:red exp1}}\qquad\qquad\qquad
\subfigure[\small Disallowed.]{
\includegraphics[height=4.5cm]{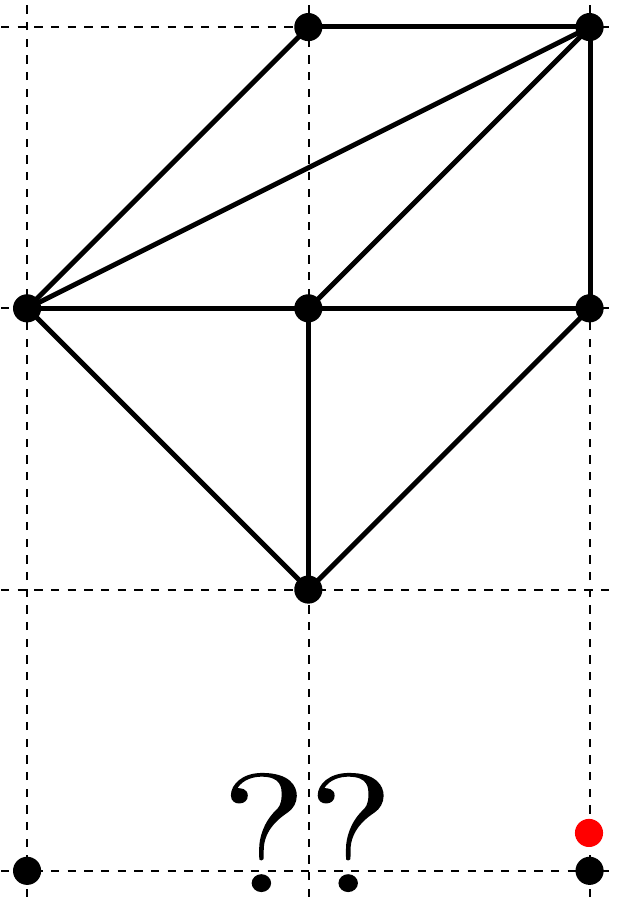}\label{fig:red exp2}}
\caption{{\bf Left:} An example of an allowed vertical reduction of a 2d toric diagram, which gives rise to a resolved $A_{1}$ singularity visualized by the 1d toric diagram below. {\bf Right:} An example of a disallowed vertical reduction, due to the presence of an edge that would collide with either one of the vertices along the vertical direction under such a reduction.  \label{fig:examples vert red}}
 \end{center}
 \end{figure} 
 %%%%%%%%%%

Assuming that the 2d toric diagram satisfies certain conditions (which we will revisit below), it is possible to collapse or project it down to a 1d toric diagram for the corresponding geometry in type IIA string theory. This is known as a ``vertical reduction,'' which was discussed in \cite{Closset:2018bjz}, which we refer the interested reader to.\footnote{This method relies on a technique that was originally introduced in \protect{\cite{Aganagic:2009zk}} and developed in \protect{\cite{Benini:2009qs,Jafferis:2009th,Benini:2011cma,Closset:2012ep,Closset:2012eq}} for M-theory on Calabi-Yau fourfold singularities.} The idea behind this method is to use a $U(1)_{M} \subset U(1)^3$ isometry of $\MG$ as an M-theory circle to view the $\wh{\MG}$ as a circle fibration over a five-dimensional base $\mathcal{M}_{5}$ (that is, $U(1)_{M} \hookrightarrow \wh{\MG} \longrightarrow \mathcal{M}_{5}$), such that the base itself is a fibration of an ALE space over the real line parametrized by $r_0$:
\begin{align}
\wh{\TY}(r_0) &\longrightarrow \mathcal{M}_5 \longrightarrow \R \cong \{r_0\} \, \, . \label{eq:fibration 2}
\end{align}
The complex two-dimensional space $\wh{\TY}(r_0)$, being toric, is the resolution of an A-type toric singularity, a hyperK\"{a}hler ALE space. The volumes of exceptional $\P^1$s in the resolution, denoted by $\chi_{s}(r_0)$, are piecewise-linear functions of $r_0$, the slopes of which jump at the locations of gauge- and flavor- D6-branes. The slope of $\chi_{s}(r_0)$ jumps by $2$ when we cross a gauge D6-brane, and by $1$ when we cross a flavor D6-brane. We refer to these functions as the ``IIA profiles''. From plots of these functions, one can infer various properties of the field theory and geometry. Let us briefly recall the dictionary developed in \cite{Closset:2018bjz}:\footnote{In addition, there are bifundamental hypermultiplets for quiver gauge theories realized by open strings stretched between two gauge D6-branes that wrap \textit{adjacent} exceptional curves, but we do not encounter them in this paper.}
\begin{itemize}
 \item \textbf{Effective Chern-Simons levels}: Due to the presence of a Wess-Zumino term on the worldvolume of gauge D6-branes \cite{Aganagic:2009zk}, there is an effective 5d Chern-Simons level $k_{s}$ for a probe D6-brane wrapping an exceptional $\P^{1}_{s}$. This can be computed directly from the slope of the IIA profile \cite{Benini:2011cma,Closset:2018bjz} $\chi_{s}(r_0)$ as follows. First, for every exceptional curve $\P^{1}_{s}$, define the asymptotic slopes,
\begin{align}
  \chi_{s,\pm}' &= \lim_{r_0 \rightarrow \pm \infty}\chi_{s}'(r_0) \,\,. \label{eq:asymptotic slope}
\end{align}
Then the effective Chern-Simons level $k_{s,\text{eff}}$ is given as the negative average of the asymptotic slopes:
\begin{align}
k_{s,\text{eff}} &= -\frac{1}{2}(\chi'_{s,-} + \chi'_{s,+}) \,\,. \label{eq:IIA effective CS level}
\end{align}
This ``effective CS level'' is in general half-integer, and equals the contact term $\kappa$ including half-integer contributions from matter fields, consistent with \eqref{eq:massive Dirac fermion collection U(1) half quantization}. 
\item \textbf{W-bosons of the $SU(n_s)$ gauge group}, given by open strings stretched between two gauge D6-branes at $r_0 = \xi_{s,(a_i)}$ and $r_0 = \xi_{s,(a_j)}$, have masses:
	\begin{align}
	    M(W_{s;i,j}) &= |\xi_{s,(a_i)} - \xi_{s,(a_j)}|	~.\label{eq:mass W boson IIA}
	\end{align}
\item \textbf{Fundamental hypermultiplets}, given by open strings stretched between a gauge D6-brane wrapping a compact 2-cycle at $r_0 = \xi_{s,(a_i)}$ and a flavor D6-brane wrapping a non-compact 2-cycle at $r_0 = \xi_{s,(\text{f})}$, have masses: 
\begin{align}
	    M(\cH_{s;i,\text{flavor}}) &= |\xi_{s,(a_i)} - \xi_{s,(\text{f})}| ~.\label{eq:mass fundamental hyper IIA}
	 \end{align}
\item \textbf{Tension of monopole strings}, given by the area under the IIA profile between the locations of two adjacent gauge D6-branes,
\begin{align}
T_{s,(a)} &= \int_{\xi_{s,(a)}}^{\xi_{s,(a+1)}}dr_0\,\chi(r_0) ~,	\label{eq:tension IIA}
\end{align}
which must match the first-derivatives of the gauge theory prepotential \eqref{eq:BPS monopole string tension} in the field-theory description.
\end{itemize}

\subsection{Graph-theoretic perspective}
The setting described in the previous subsection, especially the criterion in Figure \ref{fig:examples vert red}, strongly motivates the use of graph-theoretic techniques to study these geometries. In this work, we implement the idea of associating a graph to a toric diagram under study, with the aim of exploiting well-known notions and algorithms in the graph-theory literature. We introduce the relevant terminology briefly in this section, for readers unfamiliar with graph theory, but we focus only on the few features that are relevant to toric geometry. For comprehensive reviews and applications, we refer the reader to \cite{Tarjan:1983:DSN:3485,Chung1996,Bollobas1998Modern} and references therein.

A 2d toric diagram can be represented as an \textit{undirected graph} (with no loops) in $\Z^2$. A graph $G = (V,E)$ is specified by a set $V$ containing vertices and a set $E$ containing edges. An edge $e \in E$ connecting vertices $i, j \in V$ can be specified as a tuple $e = (i,j)$ of vertices. For an undirected graph, the set of tuples is unordered, i.e. the tuples $(i,j)$ and $(j,i)$ are considered to be equivalent. Therefore, for an undirected graph, the adjacency matrix, which is a map from $A_G: V\times V \rightarrow \{0,1\}$, defined by
\begin{align}
A_G(i,j) &= \left\{\begin{array}{ll} 1, & \text{if $\exists$ edge $e=(i,j) \in E$}\\ 
 0, & \text{otherwise} ~,	
\end{array}\right.
\end{align}
is symmetric. The no loops condition further implies that all diagonal entries are $0$, so it is sufficient to work with the upper (or lower) triangular part of the matrix, which is specified by $|V|(|V|-1)/2$ entries. The adjacency matrix is typically sparse. The spectral properties of the adjacency matrix contain useful information about the graph. One can show that the number of edges is given by,
\begin{align}
|E| &= \frac{1}{2}\text{tr}(A_{G}^2) ~.	
\end{align}
This counts all edges, including the non-compact curves that make up the toric skeleton (that is, the boundary of the convex hull of $V$). A cycle in a graph is defined as a non-empty path in which only the first and last path repeat. The number of triangles (3-cycles) in the toric diagram is given by \cite{Harary1971},
 \begin{align}
\mathcal{N}_{\Delta} &= \frac{1}{6}\text{tr}(A_{G}^3) ~.	
\end{align}
 Clearly, a simplex in a toric diagram is a cycle of length $3$, but not every cycle of length $3$ is a simplex. (Recall that a simplex in toric geometry must have a minimal simplical volume of $\frac{1}{2}$.)

In graph theory, the ``shape'' of a graph usually does not matter, only the connectivity does. However, in toric geometry, the ``shape'' (up to $SL(2,\Z)$ equivalence) does matter, since the locations of the divisors (vertices) critically dictate whether a given toric diagram corresponds to a crepant resolution, and also whether or not some curves can flop. In the previous section, we discussed the ``vertical reduction'' of the toric diagram. This has a natural interpretation in graph theory, where different ways of reducing the toric diagram can be viewed as different instances of an \textit{edge reduction}. This takes two vertices connected by an edge and eliminates the edge by mapping both vertices to a third vertex (which can be regarded as the fusion of the two vertices). Formally, if $A_{G}(u, v) = 1$ for a pair of vertices $u, v \in V$ (so that they are connected by an edge) and given a third vertex $w \in V$, we define a function $f: V \rightarrow V$ via its action on the vertices of $V$ by,
\begin{align}
  f_{u,v,w}(x) &= \left\{\begin{array}{ll}
                 x, &  x \in V\!\setminus\{u,v\},  \\
                 w, & \text{if } A_{G}(u,v) =1 \text{ and } x \in \{u, v\} ~.
       \end{array}
 \right.	
\end{align}
In this language, the allowed vertical reduction of Figure \ref{fig:examples vert red} corresponds to a sequence of (vertical) edge reductions such that at each step there is no obstruction due to an internal edge crossing an internal vertex. This systematizes the study of toric graphs and generalizes well to higher-rank examples.

A related motivation for viewing a toric diagram as a graph is the fact that different crepant resolutions (related by flops) differ only in their connectivities and so a combinatorial enumeration of crepant resolutions translates to a similar enumeration problem for graphs. The number of crepant resolutions grows very quickly with the rank, and although we restrict our attention in this paper -- for reasons of simplicity and brevity -- only to \textit{isolated} toric singularities at rank-two, a graph-based enumeration algorithm works even for nonisolated singularities at rank $> 2$. It might also be interesting to relate other ideas from spectral graph theory \cite{Chung1996} to toric geometry in the context of studying 5d SCFTs. These are possible avenues for future work. In the remainder of this paper, we focus on the isolated toric rank-two case.

\section{Rank-two isolated toric CY$_3$ singularities\label{sec:rank two section}}
\begin{figure}[ht]
%\vspace{-10pt}
\begin{center}
\subfigure[\small{$E_{0}{}^{2,\NL}$}]{
\includegraphics[width=2.8cm]{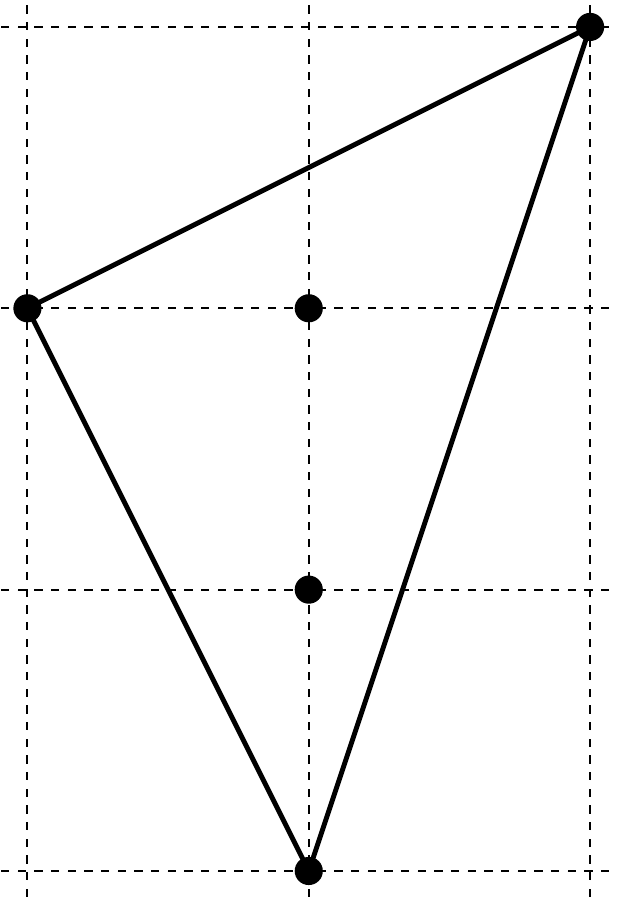}\label{fig:t2-sing}}\,
\subfigure[\small{$E_{1}{}^{2,2}$}]{
\includegraphics[width=2.8cm]{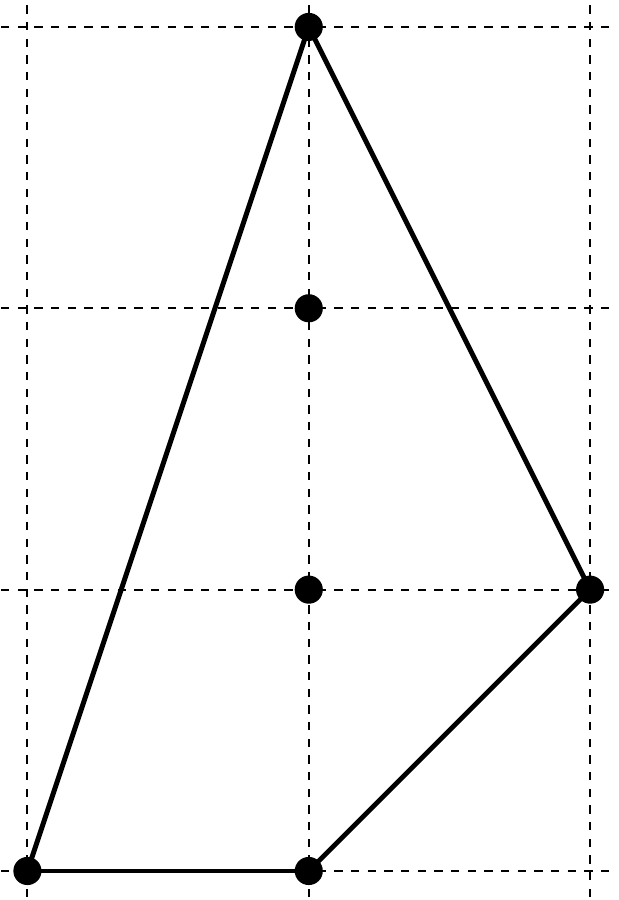}\label{fig:q1-sing}}\,
\subfigure[\small{$E_{1}{}^{2,1}$}]{
\includegraphics[width=2.8cm]{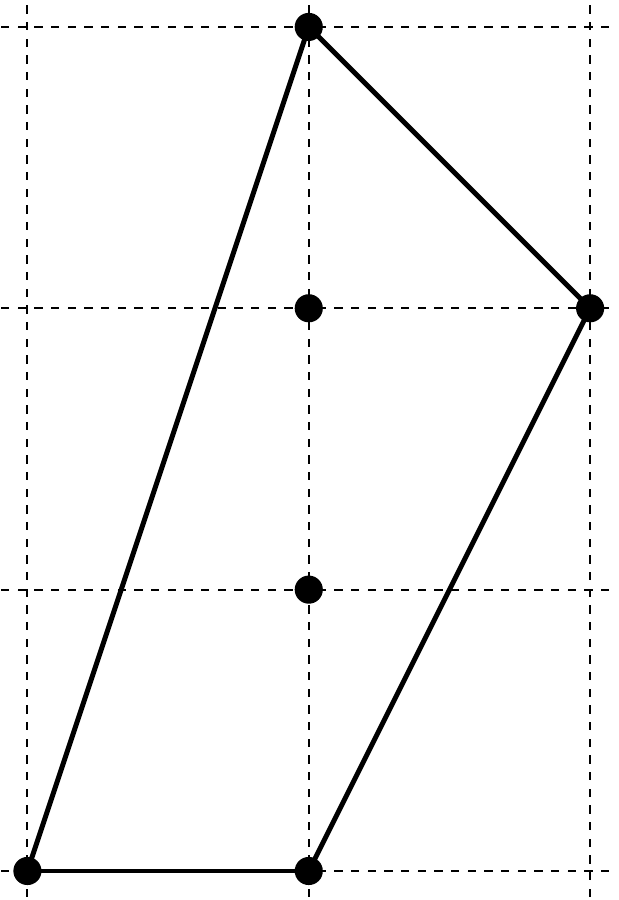}\label{fig:q2-sing}}\,
\subfigure[\small{$E_{1}{}^{2,0}$}]{
\includegraphics[width=2.8cm]{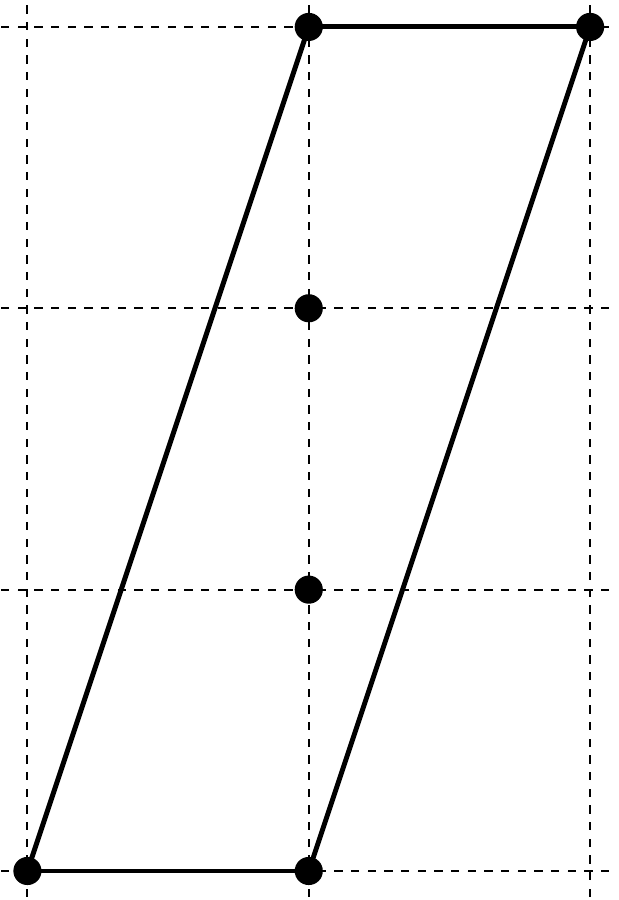}\label{fig:q3-sing}}\,
\subfigure[\small{$E_{1}{}^{2,\NL}$	}]{
\includegraphics[width=2.8cm]{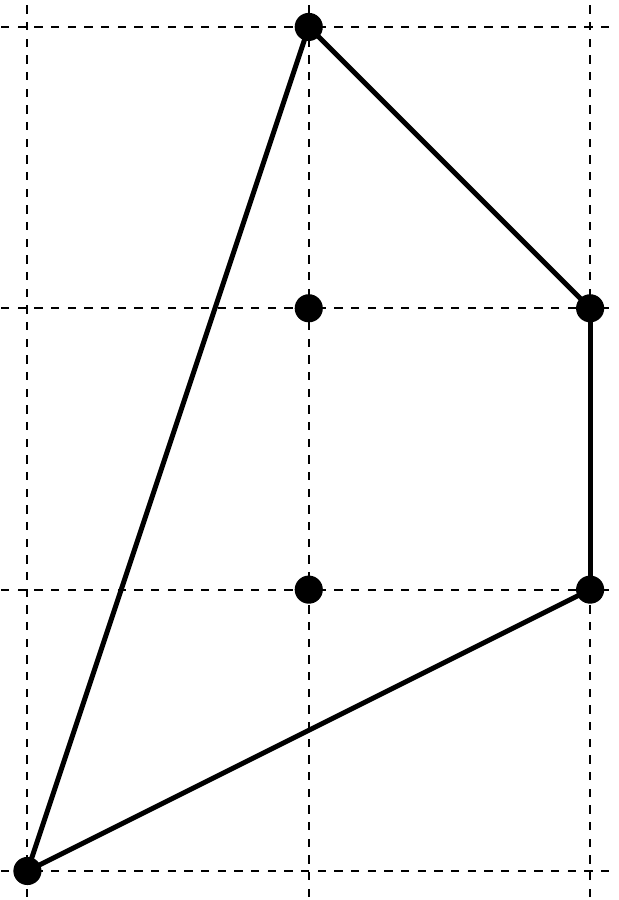}\label{fig:q13-sing}}\\
%%%%%
\subfigure[\small{$E_{2}{}^{2,\fthreehalf}$}]{
\includegraphics[width=2.8cm]{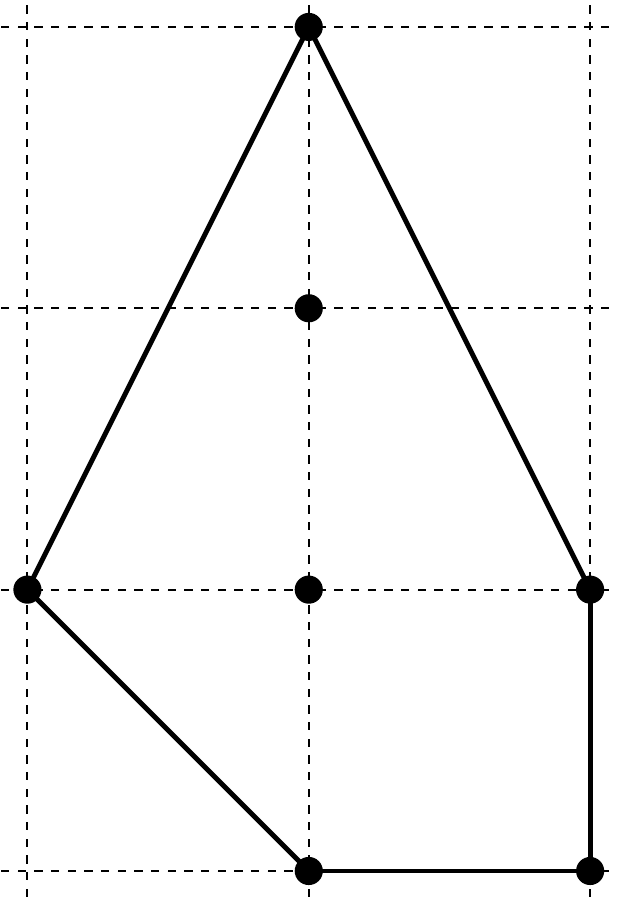}\label{fig:p1-sing}}\,
\subfigure[\small{$E_{2}{}^{2,\fhalf}$}]{
\includegraphics[width=2.8cm]{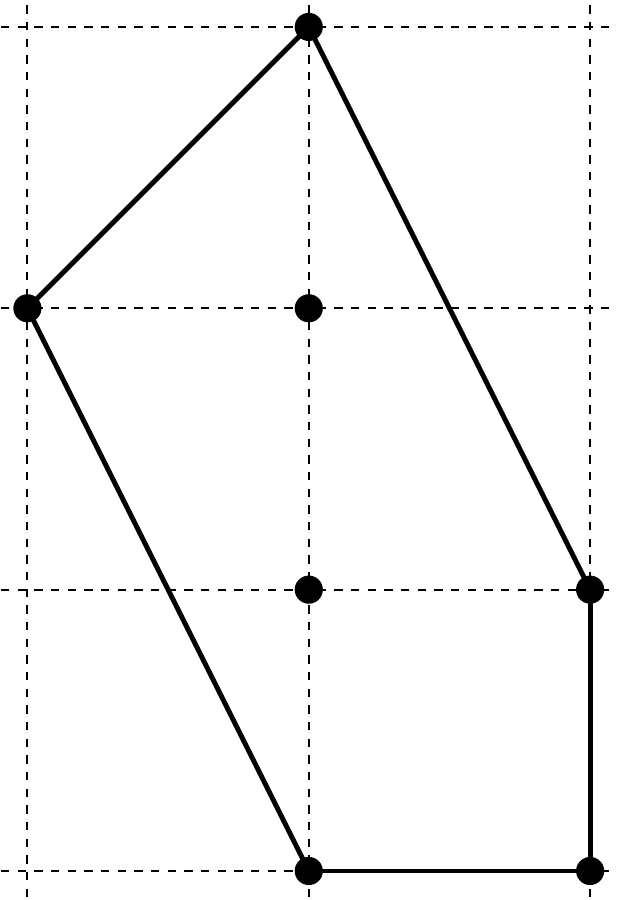}\label{fig:p2-sing}}\,
\subfigure[\small{$E_{2}{}^{2,\NL}$}]{
\includegraphics[width=2.8cm]{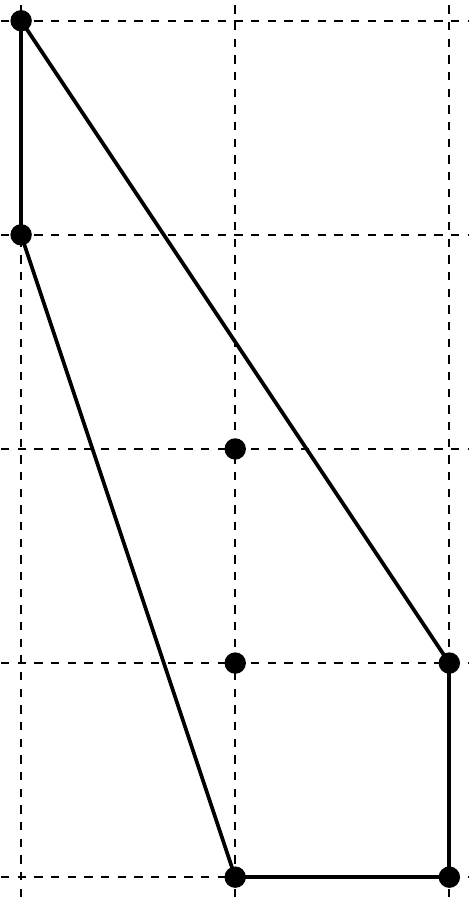}\label{fig:p15-sing}}\,
\subfigure[\small{$E_{3}{}^{2,1}$}]{
\includegraphics[width=2.8cm]{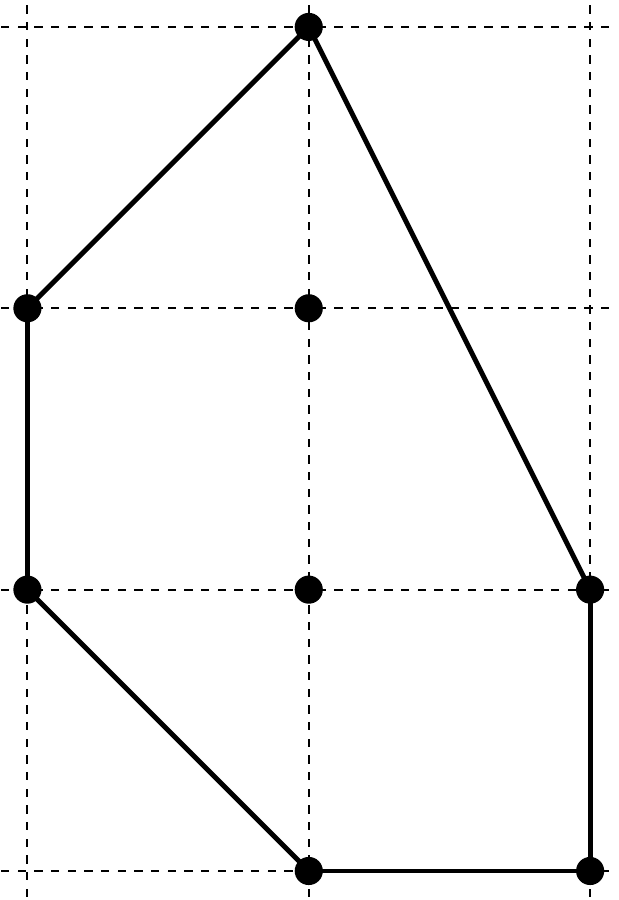}\label{fig:h1-sing}}\,
\subfigure[\small{$E_{3}{}^{2,0}$}]{
\includegraphics[width=2.8cm]{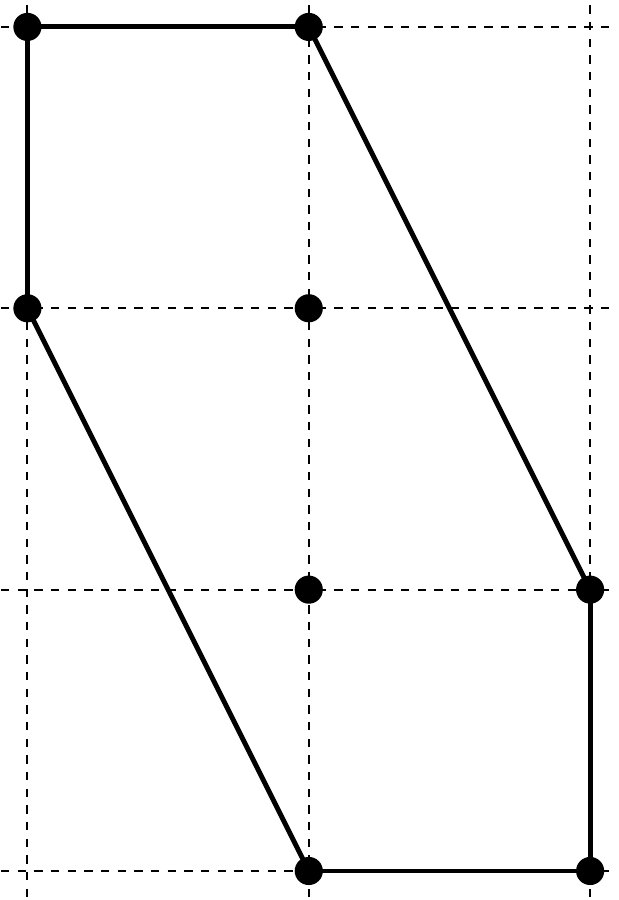}\label{fig:h4-sing}}
\caption{The 10 $SL(2,\Z)$-inequivalent isolated toric rank-2 singularities. \label{fig:isolated toric rank2}}
\end{center}
\end{figure}

An exhaustive list of rank-2 toric diagrams, i.e. toric diagrams with 2 interior points, was given by Xie and Yau in \cite{Xie:2017pfl}, based on earlier work by Wei and Ding \cite{Wei2012} which classified convex polygons with two interior points. Their list consists of 45 singularities, of which only 10 describe isolated toric singularities, i.e. toric diagrams with no lattice point on the boundary (except if it is a vertex). These 10 cases are listed in Figure \ref{fig:isolated toric rank2}. 
%%%%%%%%
%%%%%%%%

The rank-2 singularities $\MG$ are labeled as: 
\begin{align}
 E_{f}{}^{2,\kappa_{\text{eff}}} ~,	
\end{align}
where $f \equiv E-3 $ denotes the rank of the flavor symmetry\footnote{At the UV fixed point, the flavor symmetry $\bf{G}_{F}$ is sometimes enhanced to a larger global symmetry group. However, it is not obvious how the enhanced global symmetry group can be inferred from the singular geometry $\bf{X}$ in this approach. However, see footnote \protect{\ref{foot:new work}} for recent work in this direction.} of the theory $\FT$ (here $E$ is the number of external points in the toric diagram), the first superscript ($2$) denotes the rank of the singularity (the number of internal points), and the second superscript $\kappa_{\text{eff}}$ is the effective Chern-Simons level when a gauge-theory description exists or is `\NL' when the theory admits no Lagrangian interpretation.
%These geometries are labeled as T[$i$], Q[$i$], P[$i$] and H[$i$] (where $i$ ranges over an appropriate index set in each case) according to the shape of the outer skeleton, i.e. triangular (T), quadrilateral (Q), pentagonal (P) or hexagonal (H). It is worth mentioning that the diagrams in Figure \ref{fig:isolated toric rank2} are $SL(2,\Z)$ transformed relative to the ones that appear in \cite{Xie:2017pfl}, so that the two internal vertex points lie on a vertical line. %At the UV fixed point, the flavor symmetry $\bf{G}_{F}$ is sometimes enhanced to a larger global symmetry group. However, it is not obvious how the enhanced global symmetry group can be inferred from the singular geometry $\bf{X}$ in this approach.

In the remainder of this section, we consider each singularity that admits a gauge theory description, and examine its crepant resolutions, comparing the geometric description (using the M-theory and the type IIA interpretations) with the gauge-theory description. Along the way, we comment on various features of each model, including the BPS states and walls in moduli space, and also remark on resolutions that admit no gauge-theory interpretation. We use M-theory/type IIA duality to characterize the existence or non-existence of a ``gauge-theory phase'' of the resolved toric Calabi-Yau geometry based on whether or not the toric diagram admits a vertical reduction as explained in the previous section.\footnote{\protect{\label{foot:ruling}}Another approach \protect{\cite{Intriligator:1997pq,Jefferson:2018irk}} is to find a ruling of the exceptional set $\pi^{-1}(0) \subset \wh{\MG}$, that is, a set of surfaces $E$ which have the form of a fibration of $\P^1$ over a curve $\CC$ (i.e. $\P^1 \hookrightarrow E \rightarrow \CC$), such that M2-branes wrapping $\CC$ are identified with W-bosons.} 

\paragraph{Comparison with the literature.} As explained below, the resolutions of seven of the ten rank-2 isolated toric singularities of Figure \ref{fig:isolated toric rank2} describe the Coulomb branches of rank-2 five-dimensional gauge theories with gauge group $SU(3)$ and varying Chern-Simons levels and flavors. The models we consider here have appeared in the literature on five-dimensional dualities, notably in \cite{Jefferson:2018irk} (referred to as JKVZ below) and \cite{Hayashi:2018lyv} (referred to as HKLY below). Additionally,  \cite{Apruzzi:2019vpe,Apruzzi:2019opn,Bhardwaj:2019jtr,Bhardwaj:2019fzv,Apruzzi:2019enx} have focused on the \textit{classification} of 5d SCFTs treating the gauge theory phases as relevant deformations of the UV fixed point, with a view to connect 5d SCFTs to 6d SCFTs described by F-theory compactifications. Our goal here is not to classify SCFTs but to apply the methods developed in \cite{Closset:2018bjz} to study the mass deformations of a subset of models which are described by \textit{isolated} \textit{toric} CY$_3$ singularities. 

As a guide to the reader, Table \ref{tbl:ref geo table} translates between our terminology for the geometries in this paper and the terminology of JKVZ, and the corresponding five-brane web diagrams in HKLY.
\begin{table}[ht]
\centering
 \begin{tabular}{c|c|c|c}
 	Gauge-theory phase & JKVZ \protect{\cite{Jefferson:2018irk}} & HKLY \protect{\cite{Hayashi:2018lyv}} $(p,q)$-web & here \\ \hline
 	$SU(3)_{2}$ & $\mathbb{F}_{3} \cup dP_1$ & Fig. 119 & $E_{1}{}^{2,2}$ \\
 	$SU(3)_{1}$ & $\mathbb{F}_{2} \cup \mathbb{F}_{0}$ & Fig. 123 & $E_{1}{}^{2,1}$ \\
 	$SU(3)_{0}$ & $\mathbb{F}_{1} \cup dP_1$ & Fig. 124 & $E_{1}{}^{2,0}$ \\
  	$SU(3)_{\frac{3}{2}}$ N$_\text{f} = 1$ & $\mathbb{F}_{2} \cup dP_2$ & Fig. 118 & $E_{2}{}^{2,\frac{3}{2}}$ \\
  	$SU(3)_{\frac{1}{2}}$ N$_\text{f} = 1$ & $\mathbb{F}_{1} \mathop{\cup}\limits^{{X_1}} dP_2$ & Fig. 122 & $E_{2}{}^{2,\frac{1}{2}}$ \\
  	$SU(3)_{1}$ N$_\text{f} = 2$ & $\mathbb{F}_{1} \cup dP_3$ & Fig. 117 & $E_{3}{}^{2,1}$ \\
  	$SU(3)_{0}$ N$_\text{f} = 2$ & $\textrm{Bl}_1\mathbb{F}_{1} \mathop{\cup}\limits^{{X_1}} dP_2$ & Fig. 121 & $E_{3}{}^{2,0}$ \\ \hline
 \end{tabular}
 \caption{\label{tbl:ref geo table}Geometries considered in this paper and their descriptions in references  \protect{\cite{Jefferson:2018irk}} and \protect{\cite{Hayashi:2018lyv}}. Here $\mathbb{F}_{n}$ denotes the the $n^{th}$ Hirzebruch surface, a degree $(-n)$ fibration of $\mathbb{P}^{1}\times \mathbb{P}^1$, $dP_{n}$ denotes the $n^{th}$ del Pezzo surface, and $\textrm{Bl}_{k}$ denotes the blow-up in $k$ points. We refer the reader to Section 3 of JKVZ \cite{Jefferson:2018irk} and appendices therein, for explanations of the gluing terminology in the second column. The figure numbers in the third column are $(p,q)$-web diagrams in \protect{\cite{Hayashi:2018lyv}} corresponding to the geometries in Figure  \protect{\ref{fig:p1 crepant all}}, listed in the fourth column.}
\end{table}
Since the CY$_3$ geometries we consider are toric, we study the mass deformations in geometry by using toric diagrams for the resolved CY$_3$ singularities, rather than $(p,q)$-web diagrams. However, a large number of geometries considered in \cite{Jefferson:2018irk,Hayashi:2018lyv,Apruzzi:2019vpe,Apruzzi:2019opn,Bhardwaj:2019jtr,Bhardwaj:2019fzv,Apruzzi:2019enx} are non-toric. Whenever a Type IIB brane picture consisting of $(p,q)$ five-branes exists, $(p,q)$-webs are still good descriptions as used in HKLY \cite{Hayashi:2018lyv}, but it is not immediately obvious what the ``dual diagram'' of such a (non-toric) web might be. We leave this question for future work. Another caveat is that even within the toric realm, we focus on \textit{isolated} singularities because for such singularities, the number of crepant resolutions is equal to the number of gauge-theory chambers (whenever a gauge-theory interpretation exists). This is why Table \ref{tbl:ref geo table} has only seven entries. Recall from Section \ref{sec:m-theory-on-cy3-IIA-reduction} that the Type IIA interpretation of an M-theory Calabi-Yau geometry exists as long as the Calabi-Yau has (at least) a $U(1)$ isometry \cite{Closset:2018bjz}. So, we can indeed still apply the techniques of this paper to study \textit{non-isolated} toric singularities and obtain their Type IIA descriptions, but we will have to contend with relinquishing the one-to-one correspondence between gauge-theory chambers and crepant resolutions in that case. Some discussion of nonisolated singularities using these techniques already appeared in \cite{Closset:2018bjz}, so we will not revisit those issues here. But it is worth mentioning that \cite{Apruzzi:2019opn,Apruzzi:2019enx} do examine models that could be engineered using resolutions of nonisolated singularities. Their approach is based on an object called the Combined Fiber Diagram (also a graph, albeit a different kind than the toric graph of this paper), which among other things, also encodes the superconformal flavor symmetry. In their approach, transitions between such diagrams contain information about mass deformations that trigger flows between 5d SCFTs. 

The focus of the present work by contrast, is to analyze all mass deformations (which may or may not admit a gauge-theory interpretation) of a 5d SCFT engineered by a \textit{given} rank-2 isolated toric singularity, and discuss RG flows \text{between} different mass deformations (crepant resolutions) and also between crepant resolutions of \textit{different} singularities (i.e. between mass deformations of different parent UV SCFTs). The restriction to isolated toric singularities confines us generically to quiver gauge theories with $SU$ gauge groups, which is admittedly a limited class of examples. To this end, we are interested in the regime in which all mass deformations (dynamical K\"{a}hler moduli and non-dynamical K\"{a}hler deformations) are turned on, so that the prepotential is a function of not just the Coulomb vevs but of all mass deformations. In other words, we are rarely probing the conformal point and are mostly interested in physics away from it. In \cite{Closset:2018bjz} this motivated the need to slightly modify the IMS prepotential, as also discussed in Section \ref{sec:new prepotential} of this paper. As explained there, this modifies the parametrization between geometry and field theory. However, in probing the Coulomb branch of the SCFT from the perspective of geometry, one sets all mass deformations to zero, so the SCFT Coulomb branch prepotential -- which enters the analysis in \cite{Jefferson:2018irk,Bhardwaj:2018yhy,Bhardwaj:2018vuu,Bhardwaj:2019fzv,Bhardwaj:2019jtr} -- is unaffected, since terms in the cubic prepotential that are quadratic in Coulomb vevs necessarily involve linear powers of mass deformations, and such terms are killed on flowing to the SCFT Coulomb branch. Therefore, the classification program of 5d SCFTs as outlined in these papers is unaffected by such considerations. On the other hand, here we follow \cite{Closset:2018bjz} and work with the full cubic polynomial prepotential.\footnote{\protect{\label{foot:new work}}A few weeks after this paper appeared on the arXiv, HKLY uploaded their work \protect{\cite{Hayashi:2019jvx}}, where a ``complete'' prepotential for 5d $\mathcal{N}=1$ SCFTs is proposed, based on the modified prepotential introduced in \protect{\cite{Closset:2018bjz}}. In this approach, one can read off the enhanced global symmetry by writing the prepotential in terms of certain invariant Coulomb branch parameters. It will be interesting to extend their analysis to higher-rank theories.} %It may be interesting to study the implications of these modified quadratic and linear terms (which are relevant, as explained above, only for the gauge-theory Coulomb branch) for higher-rank theories, which is another avenue for future work.

\paragraph{Non-gauge-theoretic singularities.} Before proceeding, let us comment on the non-gauge-theoretic singularities $E_{\ell}{}^{2,\NL}$ for $\ell = 0, 1, 2$, which admit no vertical reduction. Let $\mathscr{C}(E_{\ell}{}^{2,\NL})$ denote the set of crepant resolutions of $E_{\ell}{}^{2,\NL}$, with a typical crepant resolution denoted by $\mathcal{R}_{\ell}{}^{\NL} \in \mathscr{C}(E_{\ell}{}^{2,\NL})$. Also, let $\mathscr{C}_{\text{vert}}^{(2)}$ be the space of all rank-2 crepant resolutions that admit a vertical reduction. There is a natural action of $g \in SL(2,\Z)$ action on every $\mathcal{R}_{\ell}$, denoted by $g\cdot \mathcal{R}_{\ell}$ (this simply applies an $SL(2,\Z)$ transformation given by $g$ on the toric vertices of $\mathcal{R}_{\ell}$). The fact that these singularities are non-gauge-theoretic is equivalent to saying that there exists no $SL(2,\Z)$ transformation that yields a resolution with a vertical reduction:
\begin{align}
  \forall\,\, g \in SL(2,\Z) \text{ and } \forall \,\,\mathcal{R}_{\ell}{}^{\NL} \in \mathscr{C}(E_{\ell}{}^{2,\NL}) \,\, ,~ \,\, g \cdot \mathcal{R}_{\ell}{}^{\NL} \notin \mathscr{C}_{\text{vert}}^{(2)}, \quad \text{ for } \ell = 0, 1, 2 ~.
\end{align}
The $E_{2}{}^{2,\NL}$ and $E_{3}{}^{3,\NL}$ singularities each admit an interpretation as a gauge theory coupled to a ``non-Lagrangian sector,'' due to the existence of a ruling (see footnote \ref{foot:ruling}). But $E_{0}{}^{2,\NL}$ does not admit such a ruling. (In this case, a ruling is equivalent to having a line with three points.) In fact, resolutions of these non-gauge-theoretic singularities can be arrived at by starting from resolutions of gauge-theoretic singularities $E_{\ell}{}^{2,\kappa_{\text{eff}}}$ (that is, the singularities whose crepant resolutions do admit a gauge theory interpretation) by a combination of flop transitions followed by decoupling divisors in the geometry by sending the volumes of certain compact curves to infinity. We interpret this as a ``generalized renormalization group (RG) flow,'' in the extended parameter space of the Calabi-Yau geometry.

\paragraph{Parity.} Parity $\bm{\mathcal{P}}$ acts on the toric diagram by the application of the central element $C_{0} = S^2 \in SL(2,\Z)$. If the effective Chern-Simons level $k_{\text{eff}}$ of a gauge-theory phase vanishes, the toric diagram is $\bm{\mathcal{P}}$-invariant. In Figure \ref{fig:RG flow chain}, this action of parity on the gauge-theory phases is indicated by arrows relating various geometries. Note that parity flips the sign of the effective Chern-Simons level $k_{\text{eff}}$.

\paragraph{RG flows at rank-two.} As we remark in more detail in various examples below and as also mentioned in the introduction, there are many RG flows that relate different geometries and field theories. The crucial point to note here is that stating from the two singularities labeled $E_{3}{}^{2,1}$ and $E_{3}{}^{2,0}$ one can recover all isolated toric singularities of rank-two shown in Figure \ref{fig:isolated toric rank2}, by a combination of flops and divisor decouplings, which we refer to as RG flow. For example, Figure \ref{fig:RG flow chain} already shows how the various rank-2 gauge-theory phases arise starting from these bigger geometries.

\paragraph{Vertical reductions.} For every toric rank-two singularity of Figure \ref{fig:isolated toric rank2} that admits a crepant resolution with a vertical reduction (that is, whenever a gauge-theory phase exists), the type IIA geometry takes the form of a resolved $A_{1}$ singularity fibered over the $x^{9} = r_{0}$ direction, with a set of D6-branes wrapping the exceptional $\P^1$s in the resolution. The fibration, as discussed in Section \ref{sec:m-theory-on-cy3}, is characterized by a piecewise linear function $\chi(r_0)$, the precise form of which depends on the specific details of the resolution. We refer to this function loosely as the ``IIA profile''. For a review of the vertical reduction method, see \cite{Closset:2018bjz}. Recall that the vertical reduction is defined by the choice of an auxiliary ``$U(1)_{M}$ line'' in the GLSM charge matrix, which specifies a redundant parametrization of the GLSM. The integer charges $Q_{i}^{M}$ ($i = 1, \ldots, n$, where $n$ is the number of toric vertices) of this line are required to satisfy $\sum_{i=1}^{n}Q_{i}^{M}=0$ and $\sum_{i=1}^{n}w_{i}^{y}Q_{i}^{M}=1$ where $\bm{w}=(w_i^x,w_i^y)\in \Z^2$ are the coordinates of the toric vertices. In all the geometries considered in this paper, the nonzero $U(1)_{M}$ charges satisfying these conditions are given by $Q_{\bE_1}^{M}=-1$ and $Q_{\bE_2}^{M}=1$ (and $Q_{i}^{M}=0$ for $i \neq \bE_1, \bE_2$), where $\bE_{1,2}$ denote the two compact divisors in any rank-two toric diagram (which are the two interior points). In every case, we begin by briefly outlining the toric geometry, listing the linear relations among divisors and curve classes, the GLSM charge matrix, the intersection numbers, and the geometric prepotential, followed by an analysis of the IIA profile leading to a map between geometry and field theory parameters. To keep the discussion brief, we spell out only the relevant details.  %We spell out all the details in the first example of each type, but to keep the discussion short, we only list the  

\subsection{The $E_{1}{}^{2,2}$ singularity and $SU(3)_{2}$ gauge theory}
\begin{figure}[t]
\centering
\subfigure[\small{}]{
\includegraphics[width=5cm]{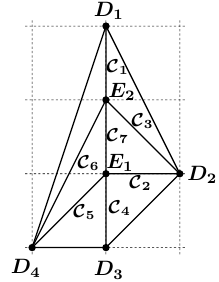}\label{fig:Q[1] res a labeled}}\,
\subfigure[\small{}]{
\includegraphics[width=5cm]{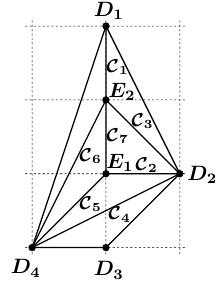}\label{fig:Q[1] res b labeled}}
\caption{The two crepant resolutions of the $E_{1}{}^{2,2}$ singularity. Resolution (a) admits a vertical reduction.\label{fig:q1 all res}}
\end{figure}

In this section, we consider the $E_{1}{}^{2,2}$ singularity of Figure \ref{fig:q1-sing}. There are two crepant resolutions, shown in Figure \ref{fig:q1 all res}, related by a flop of the curve $\CC_4$. Let us focus on resolution (a).
%\begin{figure}[H]
%  \centering
%  %\includegraphics[width=3in]{images/xieyau-q1}	
%  \includegraphics[width=2in]{images/xieyau-q1-transformed}
%  \caption{The toric resolution Q[1] after an $S$ transformation, with the curves and divisors labeled.}
%  \label{fig:q1}
%\end{figure}
There are four non-compact toric divisors $D_i$ ($i = 1, \ldots, 4$), and two compact toric diviors $\bE_1$ and $\bE_2$ with the following linear relations:
\begin{align}
    D_2 \simeq D_4~, \qquad \bE_1 \simeq D_1 - 3 D_2 - 2 D_3~, \qquad \bE_2 \simeq -2D_1 + D_2 + D_3 ~.
\end{align}
The curves $\CC$ are given as intersections of pairs of divisors according to:
\begin{align}\begin{split}
	\CC_1 &= \bE_2 \cdot D_1~, \quad \CC_2 = \bE_1 \cdot D_2~, \quad \CC_3 = \bE_2 \cdot D_2~, \quad \CC_4 = \bE_1 \cdot D_3 ~,\\
	\CC_5 &= \bE_1 \cdot D_4~, \quad \CC_6 = \bE_2 \cdot D_4~, \quad \CC_7 = \bE_2 \cdot \bE_1  ~.
	\end{split}
\end{align}
The linear relations among curve classes are
\begin{align}
  \CC_1 &\simeq \CC_2 + 3 \CC_3 + \CC_4~, \quad \CC_5 \simeq \CC_2~, \quad \CC_6 \simeq \CC_3~, \quad \CC_7 \simeq \CC_2 + \CC_4~.
\end{align}
We may take $\{\CC_2, \CC_3, \CC_4\}$ as generators of the Mori cone. Thus, the GLSM charge matrix is:
\be\label{intersect q1}
\begin{tabular}{l|cccccc|c}
 & $D_1$ &$D_2$& $D_3$ & $D_4$ & $\bE_1$& $\bE_2$ & vol($\CC$) \\
 \hline
$\CC_2$    &$0$&$0$ &$1$ &  $0$&$-2$& $1$ & $\xi_2$\\
$\CC_3$    &$1$&$0$ &$0$ &  $0$&$1$& $-2$ & $\xi_3$\\
$\CC_4$    &$0$&$1$ &$-1$ & $1$&$-1$& $0$ &$\xi_4$\\
\hline\hline
$U(1)_M$ & $0$& $0$ &$0$ & $0$ & $-1$ & $1$ & $r_0$
 \end{tabular}
\ee 
The FI terms $\xi_2 \geq 0$, $\xi_3 \geq 0$, and $\xi_4 \geq 0$ are, respectively, the volumes of the compact curves $\CC_2$, $\CC_3$ and $\CC_4$. In \eqref{intersect q1} we have shown also the last line (``$U(1)_M$ line'') which defines the GLSM of the vertical reduction, which we shall describe shortly.

\paragraph{Geometric prepotential.} The geometric prepotential can be computed from M-theory as follows. The K\"{a}hler cone can be parametrized by
\begin{align}
S &= \mu_4 D_4 + \nu_1 \bE_1 + \nu \bE_2	~.
\end{align}
By \eqref{eq:nef} the parameters $(\mu_1, \nu_1, \nu_2)$ are related to the FI parameters as:
\begin{align}
\xi_2 &= -2\nu_1 + \nu_2 \geq 0~, \qquad \xi_3 = \nu_1 - 2 \nu_2 \geq 0~, \qquad \xi_4 = \mu_4 -\nu_1 \geq 0~.	 \label{eq:q1 FI}
\end{align}
Using the charge matrix (the entries of which immediately give the intersection numbers between divisors and curves) and the linear equivalences among divisors, it is straightforward to compute the relevant triple-intersection numbers:
\begin{align}\label{eq:int nos q1}
\begin{array}{c@{~,\quad}c@{~,\quad}c@{~,\quad}c@{~\quad}c}
D_4 \bE_1 \bE_2 = 1 &  D_4^2 \bE_1 = 0 & D_4^2 \bE_2 = 0 & D_4 \bE_1^2 = -2 ~, & D_4\bE_2^2 = -2 ~, \\
 \bE_1^2 \bE_2 = -3 & \bE_1\bE_2^2 = 1 & \bE_1^3 = 8 & \bE_2^3 = 8 ~.
\end{array}
\end{align}
The value of $D_4^3$, the triple-intersection number for the noncompact divisor is ambiguous and regulator-dependent. Its coefficient, $\mu_4$, does not depend on Coulomb moduli (and thus its value does not affect subsequent analysis of BPS states), so we may as well drop this term.\footnote{The regulator dependence was explored and discussed in some detail in \protect{\cite{Closset:2018bjz}}, where a method based on the Jeffrey-Kirwan residue was proposed to compute it.} The compact part of the prepotential (i.e. $D_4^3$-independent part) is determined to be:
\begin{align}\label{eq:q1 geo prepot}
\hspace{-0.2in}\mathcal{F}(\nu_1, \nu_2; \mu_4) &= -\frac{1}{6}S^3 = -\frac{4}{3}(\nu_1^3+\nu_2^3) + \frac{3}{2}\nu_1^2\nu_2 - \frac{1}{2}\nu_1\nu_2^2 - \mu_4 \nu_1 \nu_2  + \mu_4 (\nu_1^2 + \nu_2^2) ~.
\end{align}
To relate to the non-abelian gauge theory description, we need to discuss the type-IIA string theory reduction of this geometry. 

\begin{figure}[ht]
\centering
\subfigure[\small{}]{
\includegraphics[width=4.5cm]{images/q1-res-a-new-labeled.pdf}}\,
\subfigure[\small{}]{
\includegraphics[width=7.5cm]{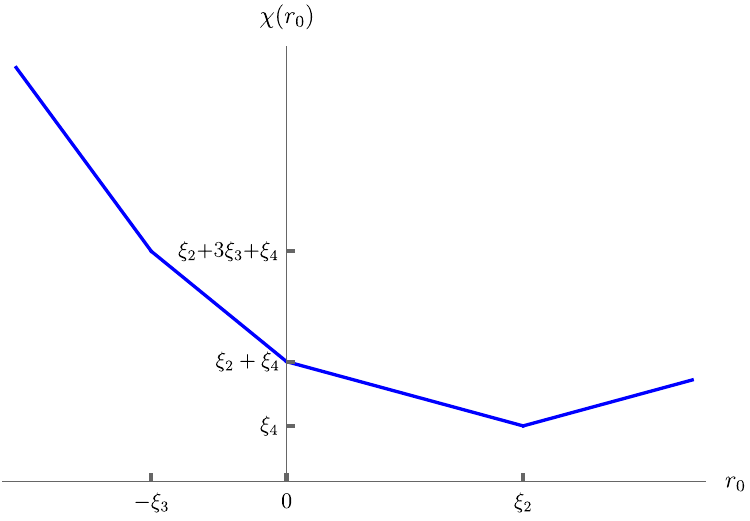}\label{fig:q1 chi final}}
\caption{Resolution (a) of the $E_{1}{}^{2,2}$ singularity and its vertical reduction.\label{fig:q1 chi}}
\end{figure}

\paragraph{Type IIA reduction and gauge theory description.} The vertical reduction of the toric diagram of Figure \ref{fig:Q[1] res a labeled} is represented in the GLSM approach via the $U(1)_M$ charges in the last line of \eqref{intersect q1}. The type IIA string background is a resolved $A_1$ singularity fibered over the $x^9 = r_0$ direction. The four vertical points in the toric diagram give rise to three D6-branes wrapping the exceptional $\P^1$ in the resolved $A_1$ singularity. This yields an $SU(3)$ gauge theory, as we explain below.

The volume of the exceptional $\P^1$ varies as a function of $r_0$, and is denoted by $\chi(r_0)$. This is a piecewise-linear function, which is determined to be:
\begin{align}\label{eq:q1 chi}
\chi(r_0) &= \left\{ 
                 \begin{array}{ll}
                 	-5r_0 + \xi_2 - 2 \xi_3 + \xi_4, & \text{for } r_0 \leq -\xi_3\\
                 	-3r_0 + \xi_2 + \xi_4, & \text{for } -\xi_3 \leq r_0 \leq 0\\
                 	-r_0 + \xi_2 + \xi_4, & \text{for } 0 \leq r_0 \leq \xi_2\\
                 	+r_0 - \xi_2 + \xi_4, & \text{for } r_0 \geq \xi_2 ~.
                 \end{array}
             \right.	
\end{align}
%%%%%%%%%%%%%%%%%%%%%%%%%%%%%
% \begin{figure}[t]
%\begin{center}
%\subfigure[\small{Resolution (a) of Q[1].}]{
%\includegraphics[width=4.5cm]{images/q1-res-a-new-labeled.pdf}\label{fig:Q[1]}}%
%\subfigure[\small Profile of $\chi(r_0)$ in IIA.]{
%\includegraphics[height=5cm]{images/q1-chi.pdf}\label{fig:Q[1] IIA}}
%\caption{The vertical reduction of the Q[1] geometry. \label{fig:q1 chi}}
% \end{center}
% \end{figure} 
% %%%%%%%%%%%%%%%
From a sketch of this function, shown in Figure \ref{fig:q1 chi final}, we can infer several features of the geometry. First of all, at each of the three kinks of the function where the slope changes by $2$, namely, at $r_0 = -\xi_3$, $r_0 = 0$ and $r_0 =+\xi_2$, there is a gauge D6-brane. When $\xi_2$ and $\xi_3$ are zero, the three wrapped D6-branes realize a 5d $SU(3)$ gauge group at $r_0 = 0$. The inverse coupling of the $SU(3)$ gauge group is given by the size of the $\P^1$ at $r_0 = 0$, which is $h_0 = \xi_4$ when $\xi_2 = \xi_3 = 0$. The effective Chern-Simons level is given by \eqref{eq:IIA effective CS level}, which yields:
\begin{align}
	\kappa_{s,\text{eff}} &= -\frac{1}{2}(1 - 5) = +2 ~.\label{eq:CS eff Q[1]}
\end{align}
Using \eqref{eq:u3 k prepot}, the gauge theory prepotential for $SU(3)_{k=2}$ gauge theory is given by:
\begin{align}
\mathcal{F}_{SU(3)_{k=2}} &= h_0(\varphi_1^2 + \varphi_2^2 - \varphi_1 \varphi_2) + \frac{1}{2}\varphi_1^2 \varphi_2 - \frac{3}{2}\varphi_1\varphi_2^2 + \frac{4}{3}(\varphi_1^3 + \varphi_2^3) ~.	\label{eq:su3 k=2 prepot field theory}
\end{align}
Finite FI parameters (i.e. finite volumes of the compact curves in the toric diagram) correspond to separating the D6-branes along the $r_0$ direction, which is equivalent to flowing onto the Coulomb branch. Open strings stretched between the gauge D6-branes yield W-bosons and their superpartners. In this case, the simple root W-bosons have masses given by \eqref{eq:mass W boson IIA}, which yields,
\begin{align}\label{eq:q1 wboson}
 M(W_1) &= 2\varphi_2 - \varphi_1 = \xi_2~, \qquad M(W_2) = 2\varphi_1 - \varphi_2 = \xi_3 ~.
\end{align}
We note the appearance of the Cartan matrix of $\mathfrak{su}(3)$ in the field-theoretic expressions for the W-boson masses, consistent with \eqref{eq:mass W boson}.

Instantons are engineered by D2-branes wrapping the gauge D6-branes, i.e. D2-branes wrapping the D6-branes at $r_0 = -\xi_3$, $r_0 = 0$ and $r_0 = \xi_2$. These states have masses given by the volumes of the exceptional $\P^1$'s at these values of $r_0$:
\begin{align}\begin{split}
M(\cI_1) &= h_0 + \varphi_2 = \xi_4 ~, \\
M(\cI_2) &= h_0 - \varphi_1 + 3 \varphi_2 = \xi_2 + \xi_4 ~,\\
M(\cI_3) &= h_0 + 5\varphi_1 = \xi_2 + 3 \xi_3 + \xi_4 	~.
\end{split}
\end{align}
%The field-theoretic expressions for the instanton masses can be obtained from the second-derivatives of the field theory prepotential of the $U(3)_{k}$ theory (with $k = +2)$  with respect to the $U(3)$ vevs, that is, $M(\cI_k) = \left.\frac{\partial^2 \cF_{U(3)_{k}}}{\partial \phi_k^2}\right|_{\mathfrak{su}(3)}$ for $k = 1, 2, 3$, where the bar denotes the substitution of $\phi_k$'s in terms of $\varphi_{k}$'s, the Coulomb moduli for the corresponding $SU(3)$ theory (i.e. imposing the ``traceless condition''). This method was employed in \cite{Closset:2018bjz} and was originally proposed in \cite{Assel:2018rcw}. 
The field-theoretic expressions for the instanton masses can be obtained using \eqref{eq:BPS instanton mass}. We note that $\cI_1$ is the instanton state of lowest mass, whereas the other instanton states can be viewed as bound states of this ``elementary instanton'' with other perturbative particles, e.g. $M(\cI_2) = M(\cI_1) + M(W_2)$, $M(\cI_3) = M(\cI_1) + M(W_1) + M(W_2)$, etc. 

Using the expressions for the K\"{a}hler volumes of the curves in terms of the FI terms \eqref{eq:q1 FI} and the expressions for the W-boson and instanton masses, we can complete the map between geometric quantities and field theory quantities. Specifically, we find:
\begin{align}
\mu_1 &= -3 h_0~, \qquad 	\nu_1 = -h_0-\varphi_2~, \qquad \nu_2 = -2h_0-\varphi_1 ~.\label{eq:q1 mu nu} 
\end{align}
Plugging \eqref{eq:q1 mu nu} into \eqref{eq:q1 geo prepot}, we find that the geometric prepotential \eqref{eq:q1 geo prepot} indeed matches the field theory prepotential \eqref{eq:su3 k=2 prepot field theory}.

As a final consistency check, we can compute the monopole string tensions from field theory via the first derivatives of the prepotential with respect to the Coulomb moduli (cf. \eqref{eq:BPS monopole string tension}). Using \eqref{eq:su3 k=2 prepot field theory}, these are:
\begin{align}
T_{1,\text{ft}} &= \frac{\partial \cF_{SU(3)_{2}}}{\partial \varphi_1} = 2 h_0 \varphi_1 + 4 \varphi_1^2 - h_0 \varphi_2 + \varphi_1 \varphi_2 - \frac{3\varphi_2^2}{2}~, \\
T_{2,\text{ft}} &= 	\frac{\partial \cF_{SU(3)_{2}}}{\partial \varphi_2} = -h_0 \varphi_1 + \frac{1}{2}\varphi_1^2 + 2 h_0\varphi_2 - 3 \varphi_1\varphi_2 + 4\varphi_2^2 ~.
\end{align}
whereas from geometry, these are given by the area under the $\chi(r_0)$ curve between the locations of gauge D6-branes (cf. \eqref{eq:tension IIA}):
\begin{align}\hspace{-0.05in}
T_{1,\text{geo}} &= \int_{-\xi_3}^{0}\chi(r_0)dr_0 = \xi_3\left(\frac{3}{2}\xi_3 + \xi_2 + \xi_4\right) ~,~%\\
T_{2,\text{geo}} = \int_{0}^{\xi_2}\chi(r_0)dr_0  = \xi_2\left(\frac{1}{2}\xi_2 + \xi_4\right)	~.
\end{align}
Using the map $\xi_2 = 2\varphi_2 - \varphi_1$, $\xi_3 = 2\varphi_1-\varphi_2$ and $\xi_4 = h_0 + \varphi_2$, we find that $T_{i,\text{geo}} = T_{i,\text{ft}}$ for $i = 1, 2$.
\paragraph{Magnetic walls.} The tensions vanish at the loci defined by:
\begin{align}\hspace{-0.15in}
(I) &: \{\xi_3 = 0 \} \cup \left\{\frac{3}{2}\xi_3 + \xi_2 + \xi_4  = 0	\right\}~, \text{ and}, (II) : \{\xi_2 = 0 \} \cup \left\{\frac{1}{2}\xi_2 + \xi_4 = 0 \right\} ~.
\end{align}
The loci $\{\xi_3 = 0\} \subset (I)$ and $\{\xi_2 = 0\} \subset (II)$, respectively correspond to hard walls along which the W-bosons $W_2$ and $W_1$ become massles. The loci $\{\tfrac{3}{2}\xi_3 + \xi_2 + \xi_4 = 0\} \subset (I)$ and $\{\tfrac{1}{2}\xi_2 + \xi_4 = 0 \}\subset (II)$ are not part of the K\"{a}hler chamber of this resolution. So there are no magnetic walls. But away from hard walls, the BPS instanton $\cI_1$ can become massless at $\xi_4 = 0$ (corresponding to a flop of the curve $\CC_4$), resulting in a traversable instantonic wall. In this case, the theory flows to a chamber (resolution (b)) that does not have a gauge theory interpretation.
%
%Along the submanifold $\{\xi_3 = 0\} \subset (I)$, the W-boson $W_2$ becomes massless. This is an impenetrable wall, being the boundary of the Weyl chamber. This is consistent with geometry, because the corresponding curve $\CC_3$ cannot flop in this K\"{a}hler chamber. The second subregion, namely $\{\tfrac{3}{2}\xi_3 + \xi_2 + \xi_4 = 0\} \subset (I)$ is not part of the K\"{a}hler chamber of resolution (a).
%
%Along the submanifold $\{\xi_2 =0 \} \subset (II)$, the W-boson $W_1$ becomes massless. This is also an impenetrable wall for the same reason as above. This is also consistent with geometry, because the corresponding curve $\CC_2$ cannot flop. And again, the second subregion $\{\tfrac{1}{2}\xi_2 + \xi_4 = 0 \}\subset (II)$ is also not part of the K\"{a}hler chamber of resolution (a). We remark that away from any hard wall, the BPS instanton $\cI_1$ can become massless at $\xi_4 = 0$ (corresponding to a flop of the curve $\CC_4$), resulting in a traversable instantonic wall. In this case, the theory flows to a chamber (resolution (b)) that does not have a gauge theory interpretation.

%In conclusion, the K\"{a}hler chamber of resolution (a) of $E_{1}{}^{2,2}$ has no magnetic walls. 

\paragraph{Parity.} Since the effective Chern-Simons level \eqref{eq:CS eff Q[1]} is nonvanishing, the theory breaks parity. This is consistent with the fact that the toric diagram of $E_{1}{}^{2,2}$ is not invariant under $C_{0}=S^2$, the central element of $SL(2,\Z)$.

%%%%%%%%%%%%%%%%%%%%%%%%%%%%%%%%%%%%%%%%%%%%%%%%%%%%%%%%%%%%%%%%%%%%%%%%%%%%%%%%%%%%%%%%
%%%%%%%%%%%%%%%%%%%%%%%%%%%%%%%%%%%%%%%%%%%%%%%%%%%%%%%%%%%%%%%%%%%%%%%%%%%%%%%%%%%%%%%%
% Maybe add something about the non-gauge theoretic phase of $E_{1}{}^{2,2}$, i.e. resolution (b)?
\paragraph{Resolution (b) and RG flow.} Resolution (b) of the $E_{1}{}^{2,2}$ singularity, shown in Figure \ref{fig:Q[1] res b labeled}, can be obtained by a flop of the instantonic curve $\CC_4$ in resolution (a). It does not admit a vertical reduction. This is consistent with the fact that the $SU(3)_{2}$ gauge theory has only one chamber that is geometrically engineered by resolution (a). However, note that in this resolution, one can decouple the divisor $D_3$ by sending the volume of the curve $\CC_4$ to infinity as shown in Figure \ref{fig:Q[1] res b RG flow}. This leads to the $SL(2,\Z)$-transformed version of the unique crepant singularity of the $E_{0}{}^{2,\NL}$ singularity (see Figure \ref{fig:t2-sing}).
%%%%%%%%%%%%%%%%%%%%%%%%%%%%%%%%%%%%%%%%%%%%%%%%%%%%%%%%%%%%%%%%%%%%%%%%%%%%%%%%%%%%%%%%
\begin{figure}[ht]
\centering
\begin{minipage}{0.35\textwidth}
	\includegraphics[width=5cm]{images/q1-res-b-new-labeled.pdf}
\end{minipage}%
\begin{minipage}{0.20\textwidth}
	{\Large $\xrightarrow[\vol(\CC_4)\rightarrow \infty]{\text{RG flow}}$ }
\end{minipage}%
\begin{minipage}{0.30\textwidth}
	\includegraphics[width=5cm]{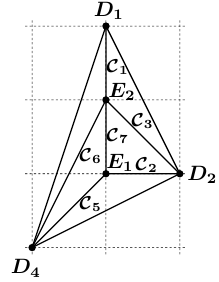}
\end{minipage}
\caption{Decoupling the divisor $D_3$ leads to the unique crepant resolution of the $E_{0}{}^{2,\NL}$ singularity.\label{fig:Q[1] res b RG flow}}
\end{figure}
We interpret this decoupling as a generalized renormalization group (RG) flow in the extended parameter space of the geometry. Physically, this amounts to sending the mass of the instanton particle $\cI_1$ in the gauge theory description of resolution (a) to zero (signaling a flop of $\CC_4$) and then blowing it up (in the opposite direction) in the K\"{a}hler cone of resolution (b), by sending the coupling to infinity. As we will see in subsequent examples, such a generalized RG flow, which involves some combination of flops (which are reversible operations) and decouplings (which are not reversible), frequently relates theories obtained by resolutions of distinct isolated toric singularities. In terms of geometry, one can ``flow'' to a toric diagram with fewer external points. Since the rank of the flavor symmetry is $f = E -3$ where $E$ is the number of external points, such a flow reduces the flavor symmetry of the theory. This parallels the field-theoretic operation of integrating out massive degrees of freedom.
%%%%%%%%%%%%%%%%%%%%%%%%%%%%%%%%%%%%%%%%%%%%%%%%%%%%%%%%%%%%%%%%%%%%%%%%%%%%%%%%%%%%%%%%
%%%%%%%%%%%%%%%%%%%%%%%%%%%%%%%%%%%%%%%%%%%%%%%%%%%%%%%%%%%%%%%%%%%%%%%%%%%%%%%%%%%%%%%%

\subsection{The $E_{1}{}^{2,1}$ singularity and $SU(3)_{1}$ gauge theory}
\begin{figure}[ht]
\centering 
\includegraphics[width=2in]{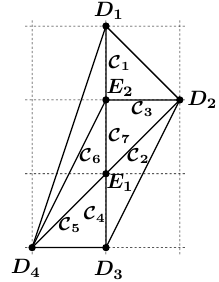}
\caption{The unique crepant resolution of the $E_{1}{}^{2,1}$ singularity. This admits a vertical reduction.\label{fig:fig q2 all res}}
\end{figure}

 This geometry has exactly one crepant resolution, shown in Figure \ref{fig:fig q2 all res}.
%%%%%%%%%%%%%%%%%%%%%%
%%%%%%%% 
The linear relations among divisors are:
\begin{align}
D_4 &\simeq D_2~, \qquad \bE_1 \simeq D_1 - 2 D_2 - 2 D_3~, \qquad \bE_2 \simeq -2D_1 + D_3	~.
\end{align} 
The compact curves $\CC$ are given by the intersection pairings of the divisors they connect in the toric diagram (for example, $\CC_1 = \bE_2 \cdot D_1$, $\CC_7 = \bE_2 \cdot \bE_1$ etc.), and can be read off from the toric diagram.
%\begin{align}\begin{split}
%    \CC_1 &= \bE_2 \cdot D_1, \quad \CC_2 = \bE_1 \cdot D_2, \quad \CC_3 = \bE_2 \cdot D_2, \qquad \CC_4 = \bE_1 \cdot D_3 ~,\\
%    \CC_5 &= \bE_1 \cdot D_4, \quad \CC_6 = \bE_2 \cdot D_4, \quad \CC_7 = \bE_2 \cdot \bE_1  ~.
%\end{split}	
%\end{align}
The linear relations among curve classes are:
\begin{align}
\CC_1 &\simeq 2 \CC_3 + \CC_4~, \quad \CC_5 \simeq \CC_2~, \quad \CC_6 \simeq \CC_3~, \quad \CC_7 \simeq \CC_4 ~.
\end{align}
We take $\{\CC_2, \CC_3, \CC_4\}$ as generators of the Mori cone. The GLSM charge matrix is:
\be\label{intersect q2}
\begin{tabular}{l|cccccc|c}
 & $D_1$ &$D_2$& $D_3$ & $D_4$ & $\bE_1$& $\bE_2$ & vol($\CC$) \\
 \hline
$\CC_2$    &$0$&$0$ &$1$ &  $0$&$-2$& $1$ & $\xi_2$\\
$\CC_3$    &$1$&$0$ &$0$ &  $0$&$1$& $-2$ & $\xi_3$\\
$\CC_4$    &$0$&$1$ &$0$ & $1$&$-2$& $0$ &$\xi_4$\\
\hline\hline
$U(1)_M$ & $0$& $0$ &$0$ & $0$ & $-1$ & $1$ & $r_0$
 \end{tabular} 
\ee 
%where we have also displayed the $U(1)_M$ line for the vertical reduction of the toric diagram. The FI terms $\xi_2 \geq 0$, $\xi_3 \geq 0$, and $\xi_4 \geq 0$ are, respectively, the volumes of the compact curves $\CC_2$, $\CC_3$ and $\CC_4$.
\paragraph{Geometric prepotential.} We parametrize the K\"{a}hler cone by $S = \mu_4 D_4 + \nu_1 \bE_1 + \nu_2 \bE_2$. The parameters $(\mu_4, \nu_1, \nu_2)$ are related to the FI parameters as:
\begin{align}
\xi_2 &= -2\nu_1 + \nu_2 \geq 0~, \quad \xi_3 = \nu_1 - 2\nu_2 \geq 0~, \quad \xi_4 = \mu_4 - 2\nu_1 \geq 0 ~. \label{eq:q2 FI}
\end{align}
The relevant triple-intersection numbers are:
\begin{align}
\begin{array}{c@{~,\quad}c@{~,\quad}c@{~,\quad}c@{~\quad}c}
	D_4 \bE_1 \bE_2 = 1 & D_4^2\bE_1 = 0 & D_4^2\bE_2 = 0 & D_4 \bE_1^2 = -2 ~,& D_4 \bE_2^2 = -2 ~,\\
	\bE_1^2 \bE_2 = -2 & \bE_1 \bE_2^2 = 0 & \bE_1^3 = 8 & \bE_2^3 = 8 ~.
\end{array}
\end{align}
So the compact part of the prepotential is:
\begin{align}
\cF(\nu_1, \nu_2; \mu_4) &= -\frac{1}{6}S^3 = -\frac{4}{3}\nu_1^3 - \frac{4}{3}\nu_2^3 + \nu_1^2\nu_2 + \mu_4(\nu_1^2+\nu_2^2-\nu_1\nu_2)  ~.	 \label{eq:prepot q2 geo}
\end{align}

\begin{figure}[ht]
\centering
\subfigure[\small{}]{
\includegraphics[width=4.5cm]{images/q2-res-new-labeled.pdf}}\,
\subfigure[\small{}]{
\includegraphics[width=7.5cm]{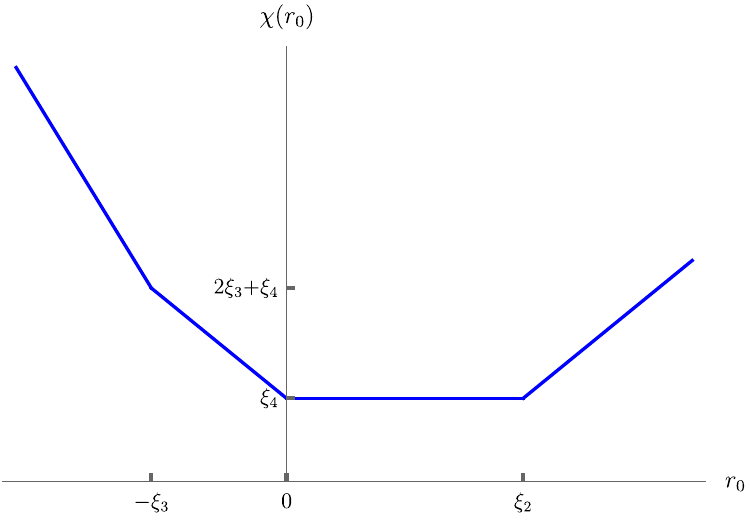}\label{fig:q2 chi final}}
\caption{Resolution (a) of the $E_{1}{}^{2,1}$ singularity and its vertical reduction.\label{fig:q2 chi}}
\end{figure}

\paragraph{Type IIA reduction and gauge theory description.} The type IIA profile is:
\begin{align}\label{eq:q2 chi}
\chi(r_0) &= \left\{ 
                 \begin{array}{ll}
                 	-4r_0 - 2 \xi_3 + \xi_4, & \text{for } r_0 \leq -\xi_3\\
                 	-2r_0 + \xi_4, & \text{for } -\xi_3 \leq r_0 \leq 0\\
                 	 \xi_4, & \text{for } 0 \leq r_0 \leq \xi_2\\
                 	+2r_0 - 2\xi_2 + \xi_4, & \text{for } r_0 \geq \xi_2 ~.
                 \end{array}
             \right.	
\end{align}
%%%%%%%%%%%%%%%%%%%%%%%%%%%%
% \begin{figure}[t]
%\begin{center}
%\subfigure[\small{The Q[2] geometry.}]{
%\includegraphics[width=4.5cm]{images/q2-res-new-labeled.pdf}\label{fig:Q[1]}}%
%\subfigure[\small Profile of $\chi(r_0)$ in IIA.]{
%\includegraphics[height=5cm]{images/q2-chi.pdf}\label{fig:Q[2] IIA}}
%\caption{The vertical reduction of the Q[2] geometry. \label{fig:q2 chi}}
% \end{center}
% \end{figure} 
 %%%%%%%%%%%%%%%
 This function is sketched in Figure \ref{fig:q2 chi final}. At the points $r_0 = -\xi_3$, $r_0 = 0$ and $r_0 = \xi_2$, there are gauge D6-branes wrapping exceptional $\P^1$'s. When $\xi_2 = \xi_3 = 0$, the three gauged D6-branes wrapping the exceptional $\P^1$ at $r_0 = 0$ engineer an $SU(3)$ gauge theory with gauge coupling $h_0 = \xi_4$. The effective CS level is given by $\kappa_{s,\text{eff}} = -\frac{1}{2}(-4 + 2) = +1$. Using \eqref{eq:u3 k prepot}, the prepotential for the $SU(3)_{k=1}$ gauge theory is given by:
 \begin{align}
 \mathcal{F}_{SU(3)_{1}} &= h_0(\varphi_1^2 + \varphi_2^2 - \varphi_1\varphi_2) - \varphi_1\varphi_2^2 + \frac{4}{3}(\varphi_1^3 + \varphi_2^3) ~. \label{eq:prepot ft q2}
 \end{align}
 The simple-root W-bosons have masses given by:
 \begin{align}
 M(W_1) &= 2\varphi_2 - \varphi_1 = \xi_2~, \qquad M(W_2) = 2\varphi_1 - \varphi_2 = \xi_3 ~.
 \end{align}
whereas the instantons have masses given by:
 \begin{align}%\begin{split}
    M(\cI_1) &= h_0 + 4 \varphi_1 = 	2\xi_3 + \xi_4 ~,\quad
    M(\cI_2) = M(\cI_3) = h_0 + 2 \varphi_2 = \xi_4 ~.
    %\end{split}
 \end{align}
From the K\"{a}hler volumes \eqref{eq:q2 FI} and the masses of W-bosons and instantons, we find:
\begin{align}	
\mu_4 &= h_0~, \quad \nu_1 = -\varphi_2~, \quad \nu_2 = -\varphi_1 ~. \label{eq:map q2}
\end{align}
Plugging \eqref{eq:map q2} into \eqref{eq:prepot q2 geo}, we recover the field theory prepotential \eqref{eq:prepot ft q2}.

The monopole string tensions from field theory are given by:
\begin{align}
T_{1,\text{ft}} &= \frac{\partial \cF_{SU(3)_{1}}}{\partial \varphi_1} = 4 \varphi_1^2 + h_0 (2\varphi_1 - \varphi_2) - 2\varphi_1\varphi_2 ~,\\
T_{2,\text{ft}} &= 	\frac{\partial \cF_{SU(3)_{1}}}{\partial \varphi_2} = 4\varphi_2^2 + h_0(2\varphi_2-\varphi_1) - \varphi_1^2 ~.
\end{align}
whereas from geometry, they are given by:
\begin{align}
T_{1,\text{geo}} &= \int_{-\xi_3}^{0}\chi(r_0)\,dr_0 = \xi_3(\xi_3 + \xi_4) ~,~
T_{2,\text{geo}} = \int_{0}^{\xi_2}\chi(r_0)\,dr_0 = \xi_2 \xi_4 ~.
\end{align}
Using the map $\xi_2 = 2\varphi_2 - \varphi_1$, $\xi_3 = 2\varphi_1 - \varphi_2$ and $\xi_4 = h_0 + 2\varphi_2$, we find that indeed $T_{i,\text{geo}} = T_{i,\text{ft}}$ for $i = 1, 2$. The tensions vanish along hard-walls where the W-bosons $W_1$ or $W_2$ become massless, or along the hard instanton wall $\xi_4=0$ where the instanton $\cI_2$ would become massless, which is not possible in this K\"{a}hler chamber (the corresponding curve $\CC_4$ cannot flop in this geometry). There are no walls in this geometry, except the hard walls along the boundary of the K\"{a}hler cone.

%The tensions vanish at the following loci:
%\begin{align}
%	(I)&: \{\xi_3 = 0\} \cup \{\xi_3 + \xi_4 = 0\}~, \text{ and }\,\,
%	(II): \{\xi_2 = 0\} \cup \{\xi_4 = 0\}~.
%\end{align}
%Along the submanifold $\{\xi_3 = 0\} \subset (I)$, the W-boson $W_2$ becomes massless, so this component of $(I)$ coincides with the hard wall that is the boundary of the Weyl chamber. Also the component $\{\xi_3 + \xi_4 = 0\} \subset (I)$ is not part of the K\"{a}hler chamber of the resolution of $E_{1}{}^{2,1}$. Along the submanifold $\{\xi_2 = 0\} \subset (I)$, the W-boson $W_1$ becomes massless, so this component also coincides with a hard wall. The component $\{\xi_4 = 0\} \subset (II)$ is the loci along with the instanton $\cI_2$ becomes massless -- this is not part of the K\"{a}hler chamber of this resolution, because the corresponding curve $\CC_4$ cannot flop. %Therefore, the K\"{a}hler chamber of the resolution of $E_{1}{}^{2,1}$ has no magnetic walls.

\subsection{The $E_{1}{}^{2,0}$ singularity and $SU(3)_{0}$ gauge theory}
\begin{figure}[t]
\centering
\subfigure[\small{}]{
\includegraphics[width=4.5cm]{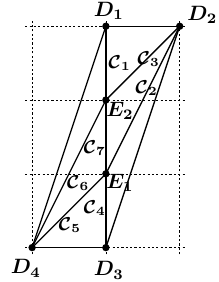}\label{fig:Q[3] res a labeled}}\,
\subfigure[\small{}]{
\includegraphics[width=4.5cm]{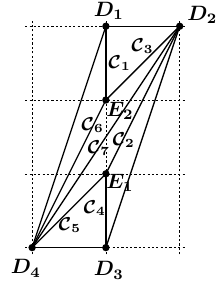}\label{fig:Q[3] res b labeled}}
\caption{The two resolutions of the $E_{1}{}^{2,0}$ singularity. Resolution (a) admits a vertical reduction.\label{fig:q3 all res}}
\end{figure}

This geometry has two crepant resolutions, shown in Figure \ref{fig:q3 all res}, related by a flop of the curve $\CC_7$. Only resolution (a) admits a vertical reduction, so we consider it first.
The linear relations among divisors are:
\begin{align}
D_2 &\simeq D_4~, \quad \bE_1 \simeq D_1 - D_2 - 2 D_3~, \quad \bE_2 \simeq -2D_1 - D_2 + D_3 ~.
\end{align}
%The compact curves $\CC$ can be read off from the toric diagram.
%\begin{align}
%\begin{split}
%	\CC_1 &= \bE_2 \cdot D_1,\quad \CC_2 = \bE_1 \cdot D_2,\quad \CC_3 = \bE_2 \cdot D_2,\quad \CC_4 = \bE_1 \cdot D_3 ~,\\
%	\CC_5 &= \bE_1 \cdot D_4, \quad \CC_6 = \bE_2 \cdot D_4, \quad \CC_7 = \bE_2 \cdot \bE_1 ~.
%\end{split}	
%\end{align}
The linear relations among curve classes are:
\begin{align}
 \CC_1 &\simeq \CC_3 + \CC_7~, \quad \CC_4 \simeq \CC_2 + \CC_7~, \quad \CC_5 \simeq \CC_2~, \quad \CC_6 \simeq \CC_3 ~.
\end{align}
We take $\{\CC_2, \CC_3, \CC_7\}$ as generators of the Mori cone. The GLSM charge matrix is:
\be\label{intersect q3}
\begin{tabular}{l|cccccc|c}
 & $D_1$ &$D_2$& $D_3$ & $D_4$ & $\bE_1$& $\bE_2$ & vol($\CC$) \\
 \hline
$\CC_2$    &$0$&$0$ &$1$ &  $0$&$-2$& $1$ & $\xi_2$\\
$\CC_3$    &$1$&$0$ &$0$ &  $0$&$1$& $-2$ & $\xi_3$\\
$\CC_7$    &$0$&$1$ &$0$ & $1$&$-1$& $-1$ &$\xi_7$\\
\hline\hline
$U(1)_M$ & $0$& $0$ &$0$ & $0$ & $-1$ & $1$ & $r_0$
 \end{tabular}
\ee 
%where the last line defines the vertical reduction of the 2d GLSM. The nonnegative FI terms $\xi_2 \geq 0$, $\xi_3 \geq 0$, and $\xi_7 \geq 0$ are respectively, the volumes of compact curves $\CC_2$, $\CC_3$ and $\CC_7$.
%%%%%%%%%%%%%%%%%%%%%%%%%%%%
% \begin{figure}[t]
%\begin{center}
%\subfigure[\small{The Q[3] geometry.}]{
%\includegraphics[width=4.5cm]{images/q3-res-a-new-labeled.pdf}\label{fig:Q[1]}}%
%\subfigure[\small Profile of $\chi(r_0)$ in IIA.]{
%\includegraphics[height=5cm]{images/q3-chi.pdf}\label{fig:Q[3] IIA}}
%\caption{The vertical reduction of the Q[3] geometry. \label{fig:q3 chi}}
% \end{center}
% \end{figure} 
 %%%%%%%%%%%%%%%
\paragraph{Geometric prepotential.} We parameterize the K\"{a}hler cone by $S = \mu_4 D_4 + \nu_1 \bE_1 + \nu_2 \bE_2$. The parameters $(\mu_4, \nu_1, \nu_2)$ are related to the FI parameters by
\begin{align}
\xi_2 &= -2\nu_1 + \nu_2 \geq 0~, \quad \xi_3 = \nu_1 - 2\nu_2 \geq 0~, \quad \xi_7 = \mu_4 - \nu_1 - \nu_2 \geq 0 ~. \label{eq:FI q3}	
\end{align}
The relevant triple-intersection numbers are:
\begin{align}
\begin{array}{c@{~,\quad}c@{~,\quad}c@{~,\quad}c@{~\quad}c}
	D_4 \bE_1 \bE_2 = 1 & D_4^2\bE_1 = 0 & D_4^2\bE_2 = 0 & D_4 \bE_1^2 = -2 ~, & D_4 \bE_2^2 = -2~,\\
	\bE_1^2 \bE_2 = -1 & \bE_1 \bE_2^2 = -1 & \bE_1^3 = 8 & \bE_2^3 = 8 ~.
\end{array}
\end{align}
Therefore, the compact part of the prepotential is determined to be:
\begin{align}
\cF(\nu_1, \nu_2; \mu_4) &= -\frac{1}{6}S^3 = -\frac{4}{3}(\nu_1^3 + \nu_2^3) + \frac{1}{2}(\nu_1^2\nu_2 + \nu_1\nu_2^2) + \mu_4(\nu_1^2 + \nu_2^2 - \nu_1\nu_2)	 ~.\label{eq:prepot q3 geo}
\end{align}

\begin{figure}[ht]
\centering
\subfigure[\small{}]{
\includegraphics[width=4.5cm]{images/q3-res-a-new-labeled.pdf}}\,
\subfigure[\small{}]{
\includegraphics[width=7.5cm]{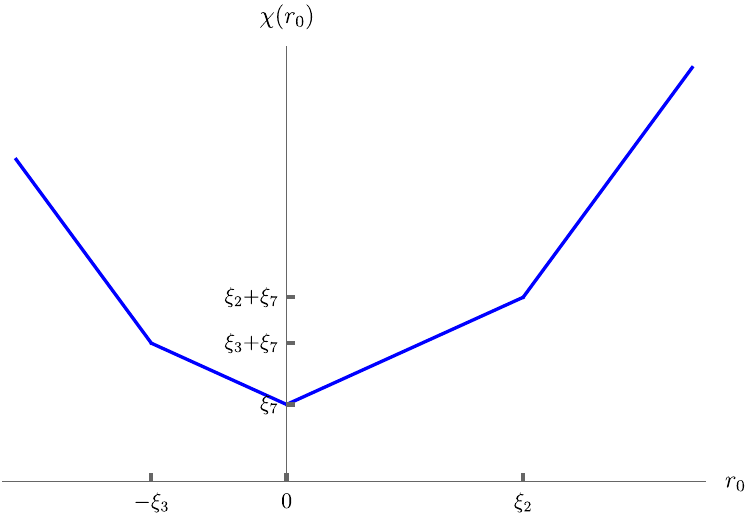}\label{fig:q3 chi final}}
\caption{Resolution (a) of the $E_{1}{}^{2,0}$ singularity and its vertical reduction.\label{fig:q3 chi final}}
\end{figure}

\paragraph{Type IIA reduction and gauge theory description.} The IIA profile function is:
\begin{align}\label{eq:q3 chi}
\chi(r_0) &= \left\{ 
                 \begin{array}{ll}
                 	-3r_0 - 2 \xi_3 + \xi_7, & \text{for } r_0 \leq -\xi_3\\
                 	-r_0 + \xi_7, & \text{for } -\xi_3 \leq r_0 \leq 0\\
                 	+r_0 + \xi_7, & \text{for } 0 \leq r_0 \leq \xi_2\\
                 	+3 r_0 - 2\xi_2 + \xi_7, & \text{for } r_0 \geq \xi_2 ~.
                 \end{array}
             \right.	
\end{align}
This function is sketched in Figure \ref{fig:q3 chi final}, where we have chosen $\xi_2 > \xi_3$ without loss of generality to plot the function. At the points $r_0 = -\xi_3$, $r_0 = 0$ and $r_0 = \xi_2$, there are gauge D6-branes wrapping exceptional $\P^1$'s in the resolution of the singularity. When $\xi_2 = \xi_3 = 0$, an $SU(3)$ gauge theory is realized with gauge coupling $h_0 = \xi_7$. The effective Chern-Simons level now vanishes: $\kappa_{s,\text{eff}} = -\frac{1}{2}(-3 + 3) = 0	$. Using \eqref{eq:u3 k prepot}, the prepotential for the $SU(3)_{k=0}$ gauge theory is given by:
\begin{align}
\cF_{SU(3)_{0}} &= \frac{4}{3}(\varphi_1^3 + \varphi_2^3) - \frac{1}{2}(\varphi_1^2\varphi_2 + \varphi_1\varphi_2^2) + h_0(\varphi_1^2 + \varphi_2^2 - \varphi_1\varphi_2)   ~.\label{eq:prepot q3 ft}
\end{align}
The simple-root W-bosons have massses given by:% the separation between adjacent gauge D6-branes:
\begin{align}
M(W_1) &= 2\varphi_2-\varphi_1 = \xi_2~, \quad M(W_2) = 2\varphi_1 - \varphi_2 = \xi_3	~.
\end{align}
whereas the instantons have masses given by:
\begin{align}
M(\cI_1) &= h_0 + \varphi_1 + \varphi_2 = \xi_7 ~,%\\
M(\cI_2) = h_0 + 3\varphi_2 = \xi_2 + \xi_7 ~,
M(\cI_3) = h_0 + 3\varphi_1 = \xi_3 + \xi_7	 ~.
\end{align}
From the K\"{a}hler volumes \eqref{eq:FI q3} of the compact curves and the masses of W-bosons and instantons, the map between geometry and field theory is determined to be:
\begin{align}
\mu_4 &= h_0, \quad \nu_1 = -\varphi_2, \quad \nu_2 = -\varphi_1	 ~. \label{eq:map q3}
\end{align}
Plugging \eqref{eq:map q3} into \eqref{eq:prepot q3 geo}, we recover the field theory prepotential \eqref{eq:prepot q3 ft}, up to $\varphi$-independent terms.

The monopole string tensions in field theory are given by:
\begin{align}
T_{1,\text{ft}} &= \frac{\partial \cF_{SU(3)_{0}}}{\partial \varphi_1} = 4 \varphi_1^2 + h_0 (2\varphi_1 - \varphi_2) - \varphi_1\varphi_2 - \frac{1}{2}\varphi_2^2 ~, \\
T_{2,\text{ft}} &= \frac{\partial \cF_{SU(3)_{0}}}{\partial \varphi_2} = 4\varphi_2^2 + h_0(2\varphi_2-\varphi_1) - \varphi_1\varphi_2 ~.
\end{align}
whereas from geometry, they are given by:
\begin{align}
\hspace{-0.15in}T_{1,\text{geo}} &= \int_{-\xi_3}^{0}\chi(r_0)\,dr_0 = \frac{1}{2}\xi_3(\xi_3 + 2\xi_7)~,~%\\
T_{2,\text{geo}} = \int_{0}^{\xi_2}\chi(r_0)\,dr_0 = \frac{1}{2}\xi_2(\xi_2 + 2\xi_7) ~.
\end{align}
Using the map $\xi_2 = 2\varphi_2 - \varphi_1$, $\xi_3 = 2\varphi_1 - \varphi_2$ and $\xi_7 = h_0 + \varphi_1 + \varphi_2$, we find that $T_{i,\text{geo}} = T_{i,\text{ft}}$ for $i = 1, 2$. The vanishing tension loci $\{xi_3 = 0\}$ and $\{xi_2 = 0\}$ respectively correspond to hard walls where the W-bosons $W_2$ and $W_1$ become massless, whereas the loci $\{\xi_3 + 2\xi_7=0\}$ and $\{\xi_2 + 2\xi_7=0\}$ do not belong to the K\"{a}hler chamber of this resolution. Away from any hard wall, the instanton particle $\cI_1$ can become massless at $\xi_7 = 0$ (signaling a flop of the curve $\CC_7$). This is a traversable instantonic wall, crossing which leads to a non-gauge-theoretic chamber (resolution (b)).
%Therefore, the K\"{a}hler chamber of resolution (a) of $E_{1}{}^{2,0}$ has no magnetic walls. 

\paragraph{Parity.} The effective Chern-Simons level vanishes, as observed above, and so the theory conserves parity. This is reflected by the symmetry of the toric diagram under the central element $C_{0} = S^2 \subset SL(2,\Z)$.

\paragraph{Resolution (b).} We remark that resolution (b) of the $E_{1}{}^{2,0}$, upon an $SL(2,Z)$ transformation, is seen to represent a coupling of two rank-1 $E_{0}$ non-Lagrangian singularities \cite{Seiberg:1996bd,Closset:2018bjz} (cf. the discussion around Figure \ref{fig:P[2] res f RG flow}).

% no comment made about resolution (b) of $E_{1}{}^{2,0}$
%
%%%%%%%%%%
% \begin{figure}[ht!]
%\begin{center}
%\subfigure[\small{Resolution (a) of the $E_{1}{}^{2,2}$ geometry and its vertical reduction: $SU(3)_{2}$.}]{
%\includegraphics[height=4.5cm]{images/q1-res-a-new-labeled.pdf}\label{fig:q1 chi final}\qquad\qquad\qquad
%\includegraphics[height=4.5cm]{images/q1-chi.pdf}}\\
%\subfigure[\small{The $E_{1}{}^{2,1}$ geometry and its vertical reduction: $SU(3)_{1}$.}]{
%\includegraphics[height=4.5cm]{images/q2-res-new-labeled.pdf}\label{fig:q2 chi final}\qquad\qquad\qquad
%\includegraphics[height=4.5cm]{images/q2-chi.pdf}}\\
%\subfigure[\small{Resolution (a) of the $E_{1}{}^{2,0}$ geometry and its vertical reduction: $SU(3)_{0}$.}]{
%\includegraphics[height=4.5cm]{images/q3-res-a-new-labeled.pdf}\label{fig:q3 chi final}\qquad\qquad\qquad
%\includegraphics[height=4.5cm]{images/q3-chi.pdf}}
%\caption{The three $SU(3)_{k}$ geometries and their corresponding IIA profiles for $\chi(r_0)$. \label{fig:pure SU(3)}}
% \end{center}
% \end{figure} 
% %%%%%%%%%%%

%\subsection{P[1]: $SU(3)_{2}$ $N_{\text{f}} =1$}
\subsection{The $E_{2}{}^{2,\fthreehalf}$ singularity and $SU(3)_{\fthreehalf}$ $N_{\text{f}} = 1$ gauge theory}
\begin{figure}[ht]
%\vspace{-10pt}
\begin{center}
\subfigure[\small{}]{
\includegraphics[width=2.5cm]{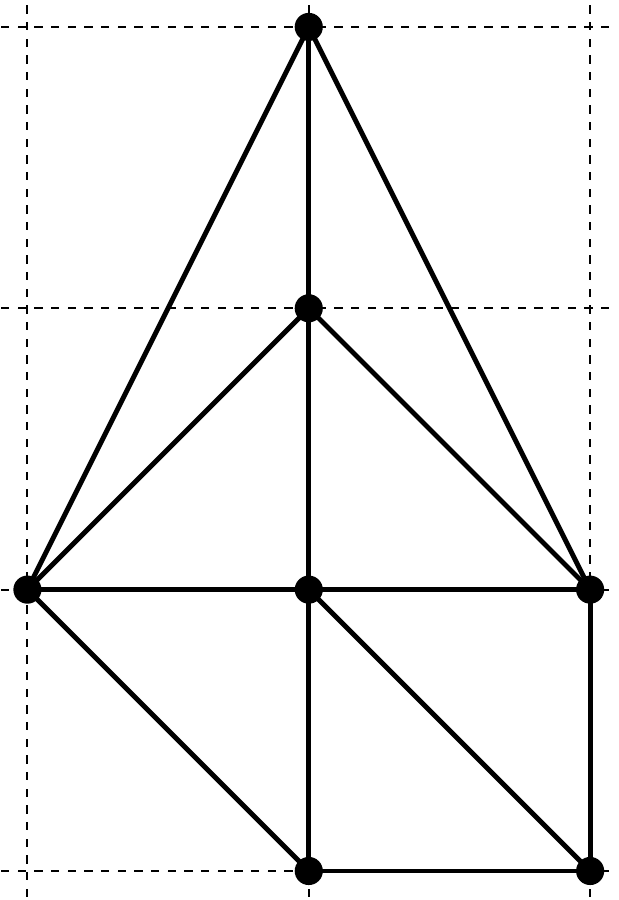}\label{fig:p1-res-a}}\,
\subfigure[\small{}]{
\includegraphics[width=2.5cm]{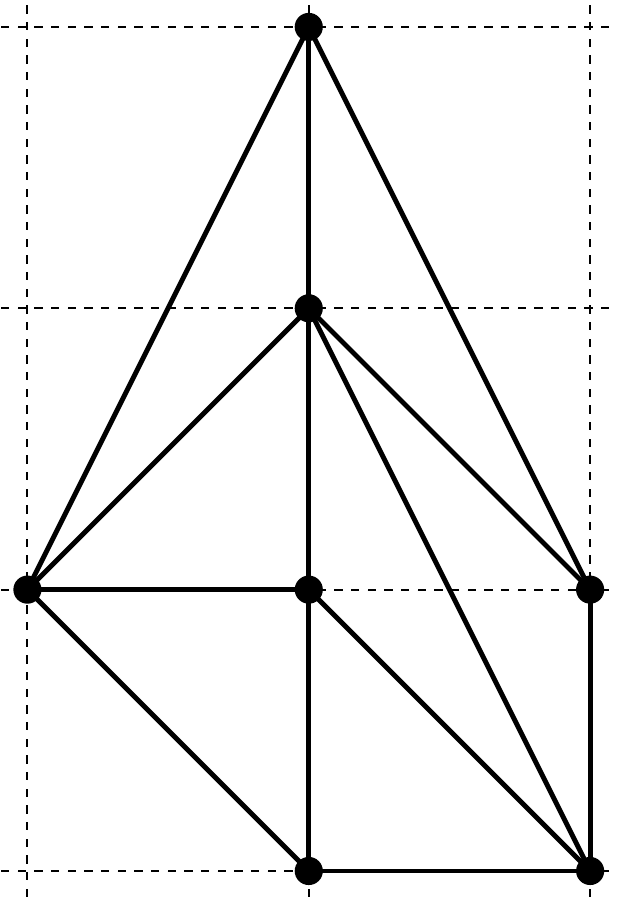}\label{fig:p1-res-b}}\,
\subfigure[\small{}]{
\includegraphics[width=2.5cm]{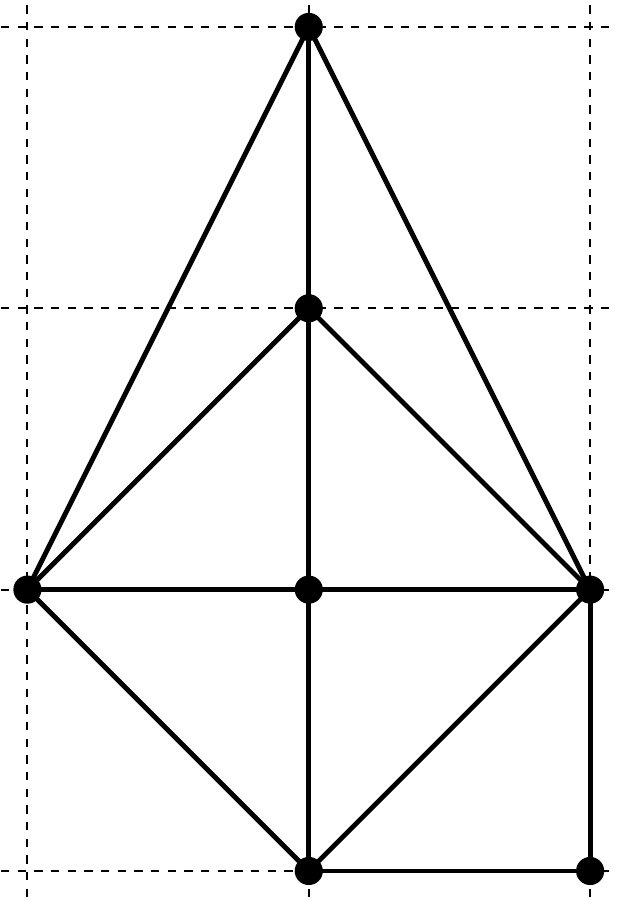}\label{fig:p1-res-c}}\,
\subfigure[\small{}]{
\includegraphics[width=2.5cm]{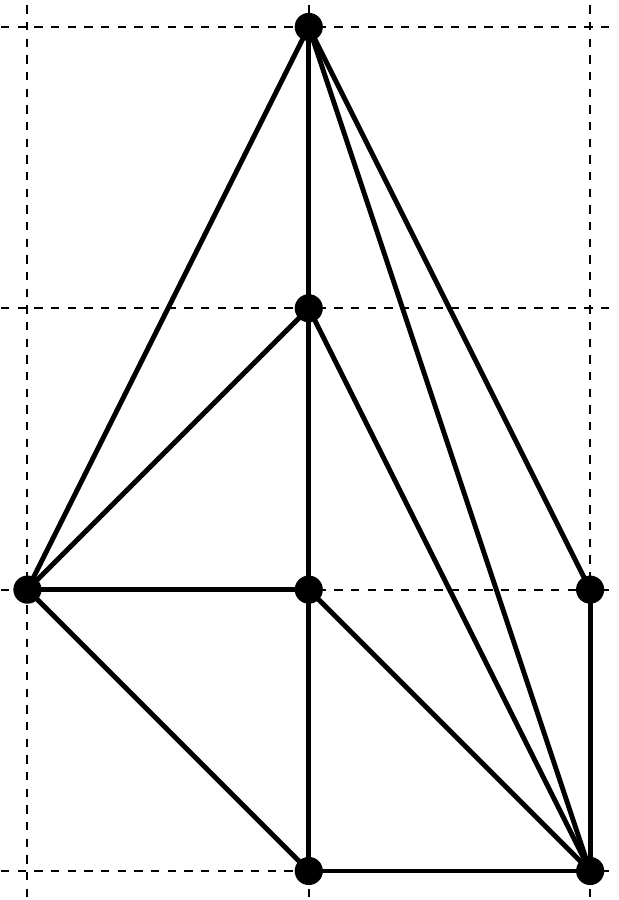}\label{fig:p1-res-d}}\\
%%%%%%%%
\subfigure[\small{}]{
\includegraphics[width=2.5cm]{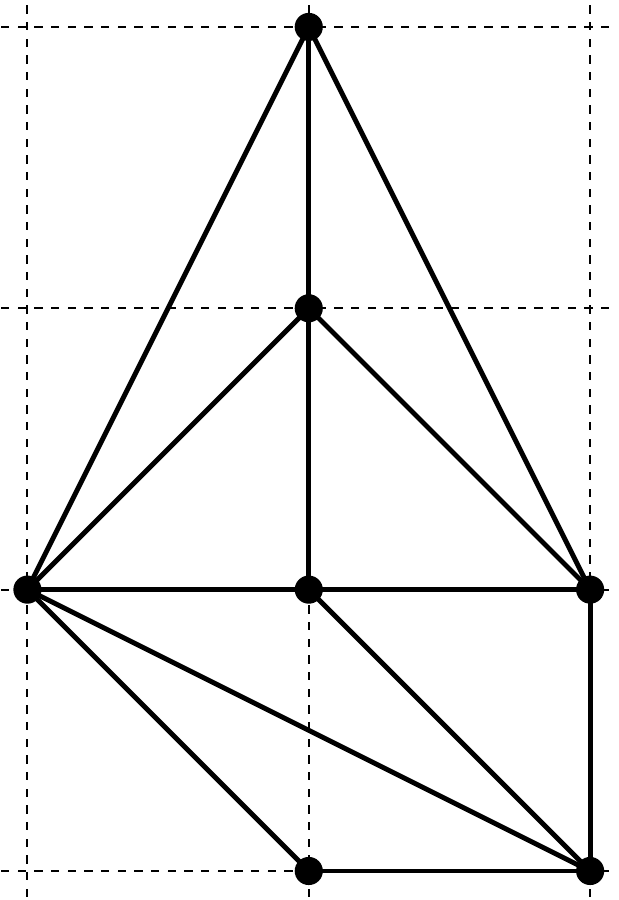}\label{fig:p1-res-e}}\,
\subfigure[\small{}]{
\includegraphics[width=2.4cm]{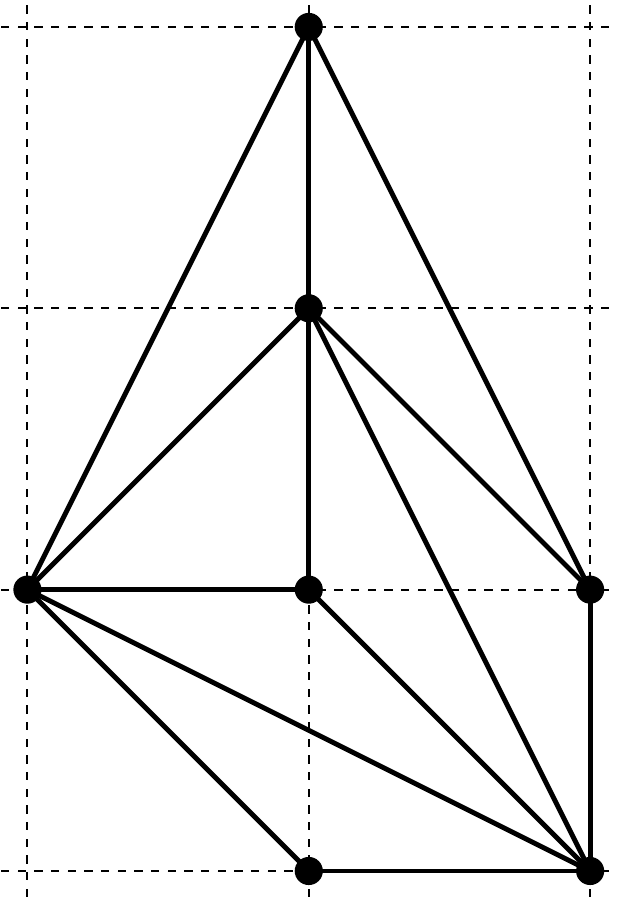}\label{fig:p1-res-f}}\,
\subfigure[\small{}]{
\includegraphics[width=2.4cm]{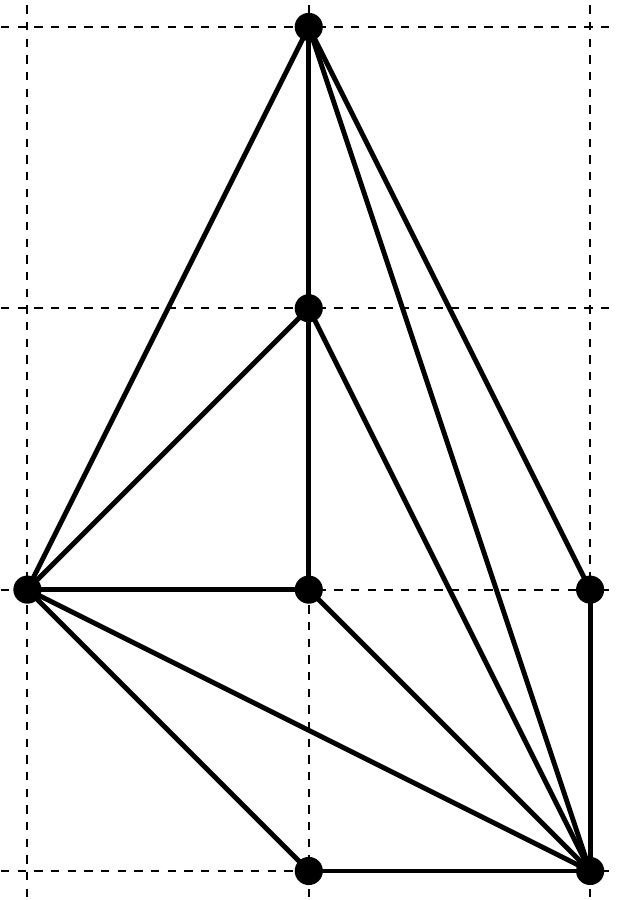}\label{fig:p1-res-g}}
\caption{The 7 crepant singularities of the $E_{2}{}^{2,\fthreehalf}$ singularity. The first four, (a)-(d) admit a vertical reduction, corresponding to chambers of the $SU(3)_{\fthreehalf}$ $N_{\text{f}}=1$ gauge theory. \label{fig:p1 crepant all}}
\end{center}
\end{figure}

The $E_{2}{}^{2,\fthreehalf}$ singularity (Figure \ref{fig:p1-sing}) admits 7 crepant resolutions, shown in Figures \ref{fig:p1 crepant all}. The first four resolutions, Figure \ref{fig:p1-res-a}-\ref{fig:p1-res-d}, admit vertical reductions to type IIA, which correspond to chambers of the $SU(3)_{\fthreehalf}$ $N_{\text{f}}=1$ gauge theory, as we illustrate below. Resolutions (e), (f), and (g) do not admit a gauge-theory interpretation. 
%%%%%%%%

\begin{figure}[ht]
\centering
\subfigure[\small{}]{
\includegraphics[width=4.5cm]{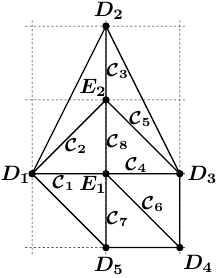}\label{fig:P[1] res a labeled}}\,
\subfigure[\small{}]{
\includegraphics[width=7.5cm]{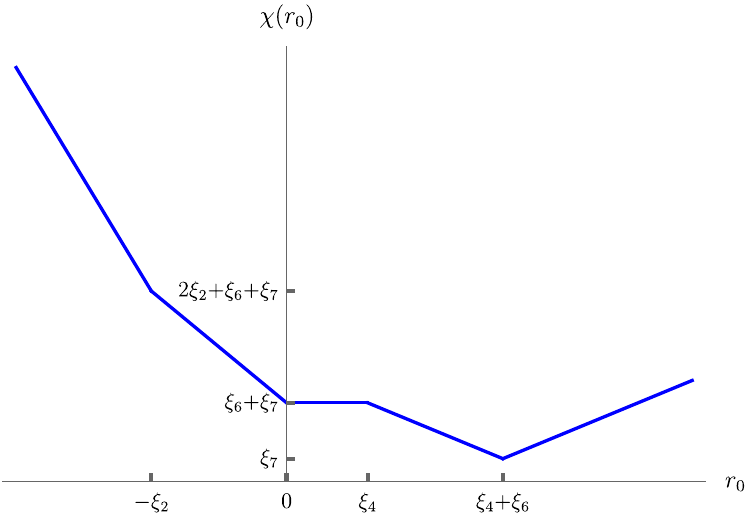}\label{fig:P[1] res a chi}}
\caption{Resolution (a) of the $E_{2}{}^{2,\fthreehalf}$ singularity and its vertical reduction.\label{fig:p1 res a final}}
\end{figure}

\paragraph{Resolution (a).} Consider the crepant resolution of Figure \ref{fig:p1-res-a}, with curves and divisors shown in Figure \ref{fig:P[1] res a labeled}.  
%%%%%%%%%%%%%%%
%%%%%%%%%%%%%%%
There are five non-compact toric divisors $D_{i}$ ($i = 1, \ldots, 5$), and two compact toric divisors $\bE_1$ and $\bE_2$ with the following linear relations:
\begin{align}
D_1 &\simeq D_3 + D_4~, \quad \bE_1 \simeq D_2 - 2D_3 - 3 D_4 - 2 D_5~, \quad 	\bE_2 \simeq -2D_2 +D_4 + D_5 ~. \label{eq:p1 res a divisor linear equivalences}
\end{align}
%The compact curves $\CC$ are given by:
%\begin{align}\begin{split}
%	\CC_1 &= \bE_1 \cdot D_1, \quad \CC_2 = \bE_2 \cdot D_1, \quad \CC_3 = \bE_2 \cdot D_2, \quad \CC_4 = \bE_1 \cdot D_3 ~,\\
%	\CC_5 &= \bE_2 \cdot D_3, \quad \CC_6 = \bE_1 \cdot D_4, \quad \CC_7 = \bE_1 \cdot D_5, \quad \CC_8 = \bE_2 \cdot \bE_1 ~.
%	\end{split}
%\end{align}
The linear relations among curve classes are:
\begin{align}
\CC_1 &\simeq \CC_4 + \CC_6~, \quad \CC_3 \simeq 2\CC_2 + \CC_6 + \CC_7~, \quad \CC_5 \simeq \CC_2 ~.
\end{align}
\noindent We take $\{\CC_2, \CC_4, \CC_6, \CC_7\}$ as generators of the Mori cone.
\newpage
\mbox{~}
\clearpage
\noindent The GLSM charge matrix is:
\be\label{intersect p1 res a}
\begin{tabular}{l|ccccccc|c}
 & $D_1$ &$D_2$& $D_3$ & $D_4$ & $D_5$ & $\bE_1$& $\bE_2$ & vol($\CC$) \\
 \hline
$\CC_2$  & $0$ & $1$ & $0$ &  $0$& $0$ & $1$ & $-2$ & $\xi_2$ \\ \hline
$\CC_4$  & $0$ & $0$ & $-1$ &  $1$& $0$ & $-1$ & $1$ & $\xi_4$ \\ \hline
$\CC_6$  & $0$ & $0$ & $1$ & $-1$ & $1$ & $-1$ & $0$ & $\xi_6$ \\ \hline
$\CC_7$  & $1$ & $0$ & $0$ & $1$ & $-1$ & $-1$ & $0$ & $\xi_7$ \\ \hline\hline
$U(1)_M$ & $0$& $0$ &$0$ & $0$ & $0$ & $-1$ & $1$ & $r_0$
 \end{tabular}
\ee 
%where the last line defines the vertical reduction of the 2d GLSM. The nonnegative FI terms $\xi_2 \geq 0$, $\xi_4 \geq 0$, $\xi_6 \geq 0$ and $\xi_7 \geq 0$ are, respectively, the volumes of compact curves $\CC_2$, $\CC_4$, $\CC_6$ and $\CC_7$.
\paragraph{Geometric prepotential.} We parametrize the K\"{a}hler cone by:
\begin{align}
 S &= \mu_1 D_1 + \mu_2 D_2 + \nu_1 \bE_1 + \nu_2 \bE_2 ~. \label{eq:kah cone p1}
\end{align}
The parameters $(\mu_1, \mu_2, \nu_1, \nu_2)$ are related to the FI parameters by:
\begin{align}
\xi_2 &= \mu_2 + \nu_1 - 2\nu_2 \geq 0, \quad \xi_4 = -\nu_1 + \nu_2 \geq 0, \quad \xi_6 = -\nu_1 \geq 0, \quad \xi_7 = \mu_1 - \nu_1 \geq 0 ~.	 \label{eq:p1 res a FI}
\end{align}
The relevant triple-intersection numbers are:
\begin{align} 
\begin{array}{c@{~,\quad}c@{~,\quad}c@{~,\quad}c@{~\quad}c}
  D_1 \bE_1 \bE_2 = 1 & D_2 \bE_1 \bE_2 = 0 & D_1 D_2 \bE_1 = 0 & D_1 D_2 \bE_2 = 1 ~, \\
  D_1\bE_1^2 = -2 & D_2 \bE_1^2 = 0 &  D_1\bE_2^2 = -2 & D_2\bE_2^2 = -4 ~,\\
  D_1^2 \bE_1 = 0 & D_1^2 \bE_2 = 0  & D_2^2 \bE_1 = 0 & D_2^2 \bE_2 = 2 ~,\\ 
  \bE_1^2 \bE_2 = -2 & \bE_1 \bE_2^2 = 0 &  \bE_1^3 = 7 & \bE_2^3 = 8 ~.
\end{array}
\end{align}
Therefore, the compact part of the prepotential is:
\begin{align}
\mathcal{F}_{(a)}(\nu_1, \nu_2; \mu_1, \mu_2) = -\frac{1}{6}S^3 &= -\frac{7}{6}\nu_1^3 - \frac{4}{3}\nu_2^3 + \nu_1^2 \nu_2 	+ \mu_1 \nu_1^2 - \mu_1 \nu_1 \nu_2 + (\mu_1 + 2\mu_2)\nu_2^2 \nonumber\\
& \quad - \mu_2^2 \nu_2 - \mu_1 \mu_2 \nu_2 ~. \label{eq:p1 res a geo prepot}
\end{align}
\paragraph{Type IIA reduction and gauge theory description.} The type IIA background is a resolved $A_1$ singularity fibered over the $x^9 = r_0$ direction. There are three D6-branes wrapping the exceptional $\P^1$ in the resolved $A_1$ singularity, resulting in an $SU(3)$ gauge theory. There is also a D6-brane wrapping a noncompact divisor in the resolved ALE space, which corresponds to one fundamental flavor. The volume of the exceptional $\P^1$ is given by the following piecewise linear function:
\begin{align}\label{eq:p1 res a chi}
\chi(r_0) &= \left\{ 
                 \begin{array}{ll}
                 	-4r_0 - 2 \xi_2 + \xi_6 + \xi_7, & \text{for } r_0 \leq -\xi_2\\
                 	-2r_0 + \xi_6 + \xi_7, & \text{for } -\xi_2 \leq r_0 \leq 0\\
                 	\xi_6 + \xi_7, & \text{for } 0 \leq r_0 \leq \xi_4\\
                 	-r_0 + \xi_4 + \xi_6 + \xi_7, & \text{for } \xi_4 \leq r_0 \leq \xi_4 + \xi_6 \\
                 	+r_0 - \xi_4 - \xi_6 + \xi_7, & \text{for } r_0 \geq \xi_4 + \xi_6 ~.
                 \end{array} 
             \right.	
\end{align}
This function is sketched in Figure \ref{fig:P[1] res a chi}. At the points $r_0 = -\xi_2$, $r_0 = 0$ and $r_0 = \xi_4 + \xi_6$, there are gauge D6-branes wrapping $\P^1$'s in the resolution of the singularity. When $\xi_2 = \xi_4 = \xi_6 =0$, an $SU(3)$ gauge theory is realized with coupling $h_0 = \xi_7$. There is a flavor D6-brane at $r_0 = \xi_4$. The effective Chern-Simons level is given by $\kappa_{s,\text{eff}} = -\frac{1}{2}(-4 + 1) = \frac{3}{2}$, which is interpreted as a bare CS level of $2$ plus the contribution $-\frac{1}{2}$ due to the single hypermultiplet (cf. \eqref{eq:massive Dirac fermion collection U(1) half quantization}). The simple-root W-bosons have masses given by:% the separation between adjacent gauge D6-branes:
\begin{align}
M(W_1) &= \xi_2 = 2\varphi_1 - \varphi_2~, \quad M(W_2) = \xi_4 + \xi_6 = 2\varphi_2 - \varphi_1 ~,
\end{align}
This resolution corresponds to gauge theory chamber 3 (cf. Table \ref{tbl:u3 nf1 instantons} and \eqref{eq:ft prepot SU(3) k Nf1 chamber 3}), with instanton masses given by:
\begin{align}
\begin{split}
	M(\cI_1) &= \chi(r_0 = -\xi_2) = 2\xi_2 + \xi_6 + \xi_7 = h_{0} + 4\varphi_1 -m  ~,\\
	M(\cI_2) &= \chi(r_0 = 0) = \xi_6 + \xi_7 = h_0 + 2\varphi_2 - m ~,\\
	M(\cI_3) &= \chi(r_0 = \xi_4 + \xi_6) = \xi_7 = h_0 + \varphi_2 ~.
\end{split}	
\end{align}
The masses of hypermultiplets (due to open strings stretched between gauge and flavor branes) are:
\begin{align}
	M(\cH_1) &= \xi_6 = \varphi_2 - m ~,~
	M(\cH_2) = \xi_4 = -\varphi_1 + \varphi_2 + m ~,~
	M(\cH_3) = \xi_2 + \xi_4  = \varphi_1 + m ~.
\end{align}
From the K\"{a}hler volumes \eqref{eq:p1 res a FI} of the compact curves and masses of W-bosons and instantons, the map between geometry and field theory variables is determined to be:
\begin{align}
\mu_1 &= h_{0} + m~, \quad \mu_2 = 3 m~, \quad \nu_1 = -\varphi_2 + m~, \quad \nu_2 = -\varphi_1 + 2m	~.\label{eq:p1 res a mu nu}
\end{align}
Plugging \eqref{eq:p1 res a mu nu} into \eqref{eq:p1 res a geo prepot}, we recover the field theory prepotential,
\begin{align}
\mathcal{F}_{SU(3)_{2},N_{\text{f}}=1}^{\text{chamber 3}} &= \frac{4}{3}\varphi_1^3 + \frac{7}{6}\varphi_2^3 - \varphi_1\varphi_2^2 + \left(h_{0}-\frac{m}{2}\right)\varphi_1^2 + h_{0}\varphi_2^2 - h_{0}\varphi_1\varphi_2 - \frac{m^2}{2}\varphi_1 ~,
\end{align}
up to $\varphi$-independent terms (i.e. terms independent of $\varphi_1$ and $\varphi_2$, as discussed in previous examples). From field theory, the monopole string tensions are given by:
\begin{align}
T_{1,\text{ft}} &= \frac{\partial \mathcal{F}_{SU(3)_{2},N_{\text{f}}=1}^{\text{chamber 3}}}{\partial \varphi_1} = 4\varphi_1^2 + 2(h_{0}-m)\varphi_{1} + (m-h_{0})\varphi_{2} - \varphi_2^2 ~,\\
T_{2,\text{ft}} &= \frac{\partial \mathcal{F}_{SU(3)_{2},N_{\text{f}}=1}^{\text{chamber 3}}}{\partial \varphi_2} = \frac{7}{2}\varphi_2^2 + (m-h_{0})\varphi_1 + (2h_{0}-m)\varphi_2 - 2\varphi_1\varphi_2 {-}\frac{m^2}{2} ~,
\end{align}
whereas from geometry, they are given by:
\begin{align}
T_{1,\text{geo}} &= \int_{-\xi_2}^{0}\chi(r_0)\,dr_{0} = \xi_2(\xi_2 + \xi_6 + \xi_7) ~,\\
T_{2,\text{geo}} &= \int_{0}^{\xi_4+\xi_6}\chi(r_0)\,dr_{0} = \frac{\xi _6^2}{2}+\xi _4 \xi _6+\xi _6 \xi _7+\xi _4 \xi _7 ~.
\end{align}
Using the map $\xi_2 = 2\varphi_1-\varphi_2$, $\xi_4 = -\varphi_1 + \varphi_2 +m$, $\xi_6 = \varphi_2-m$ and $\xi_7 = h_{0} + \varphi_2$, we find that $T_{i,\text{ft}} = T_{i,\text{geo}}$ for $i = 1, 2$. The tensions vanish at loci given by:
\begin{align}\hspace{-0.15in}
 (I)&: \{\xi_2 = 0\} \cup \{\xi_2 + \xi_6 + \xi_7 = 0\}, ~ \text{ and } ~
 (II): \left\{ \frac{\xi_6^2}{2}+\xi_4 \xi_6+\xi_6 \xi_7+\xi_4 \xi_7 =0 \right\} ~.
\end{align}
The loci $\{\xi_2 = 0\}\subset (I)$ coincides with the boundary of the Weyl chamber where the W-boson $W_1$ becomes massless, indicating a hard wall. The component $\{\xi_2 + \xi_6 + \xi_7 = 0\} \subset (I)$ is not part of the K\"{a}hler chamber of resolution (a). %In the degenerate case where $\xi_2 = 0$ (i.e. along the hard wall), this second component collapses to $\{\xi_6 + \xi_7 =0\}$, which is not part of this this K\"{a}hler chamber. %but this can be satisfied only if the curves $\CC_7$ and $\CC_6$ \textit{both} flop, which is not possible: the flop of either curve blocks the flop of the other. %Consequently, even in this degenerate case, this component does not contribute a magnetic wall.
As for the second component $(II)$, the solutions of the quadratic equation for $\xi_6$, defining the vanishing locus are:
\begin{align}
\xi_6 &\stackrel{(II)}{=} -\xi_4 - \xi_7 \pm \sqrt{\xi_4^2 + \xi_7^2}	 ~,\label{eq:quad for wall of P[1] res a}
\end{align}
which always lead to negative values of $\xi_6$ in resolution (a) (for both sign choices), which is unphysical in this K\"{a}hler chamber, and are hence rejected. Note that away from any hard wall, the BPS perturbative hypermultiplets $\cH_1$ or $\cH_2$ can become massless at $\xi_6 = 0$ or $\xi_4 = 0$ respectively (signaling flops of the curves $\CC_6$ or $\CC_4$). These are traversable walls that lead, respectively to gauge theory resolutions (c) and (b) respectively. Also away from any hard wall, the BPS instanton $\cI_3$ can become massless at $\xi_7 = 0$ (signaling a flop of $\CC_7$), which corresponds to a traversable instantonic wall that leads to a non-gauge-theoretic chamber (resolution (e)).
  
\paragraph{Parity.} Since the effective Chern-Simons level is nonvanishing, this theory breaks parity. In geometry, this is reflected by the non-invariance of the toric diagram under the central element $C_0 = S^2$ of $SL(2,\Z)$. This is true, of course, of all the crepant resolutions of $E_{2}{}^{2,\fthreehalf}$ as the CS level does not change under flops.

\begin{figure}[ht]
\centering
\subfigure[\small{}]{
\includegraphics[width=4.5cm]{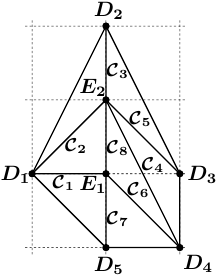}\label{fig:P[1] res b labeled}}\,
\subfigure[\small{}]{
\includegraphics[width=7.5cm]{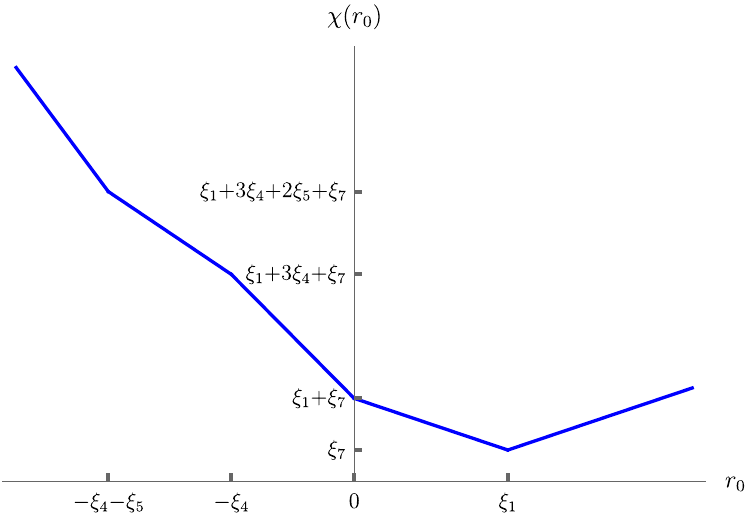}\label{fig:P[1] res b chi}}
\caption{Resolution (b) of the $E_{2}{}^{2,\fthreehalf}$ singularity and its vertical reduction.\label{fig:p1 res b final}}
\end{figure}

\paragraph{Resolution (b).} Consider the crepant resolution of Figure \ref{fig:p1-res-b}, with curves and divisors shown in Figure \ref{fig:P[1] res b labeled}. 
%%%%%%%%%%%%%%%
%%%%%%%%%%%%%%%
The linear equivalances among divisors remain unchanged, as in \eqref{eq:p1 res a divisor linear equivalences}. %The compact curves $\CC$ are given by:
%\begin{align}\begin{split}
%	\CC_1 &= \bE_1 \cdot D_1, \quad \CC_2 = \bE_2 \cdot D_1, \quad \CC_3 = \bE_2 \cdot D_2, \quad \CC_4 = \bE_2 \cdot D_4 ~,\\
%	\CC_5 &= \bE_2 \cdot D_3, \quad \CC_6 = \bE_1 \cdot D_4, \quad \CC_7 = \bE_1 \cdot D_5, \quad \CC_8 = \bE_2 \cdot \bE_1 ~.
%	\end{split}
%\end{align}
The linear relations among curve classes are:
\begin{align}
\CC_2 &\simeq \CC_4 + \CC_5~, \quad \CC_3 \simeq \CC_1 + 3\CC_4 + 2\CC_5 + \CC_7~, \quad \CC_6 \simeq \CC_1~, \quad \CC_8 \simeq \CC_1 + \CC_7 ~.
\end{align}
We take $\{\CC_1, \CC_4, \CC_5, \CC_7\}$ as generators of the Mori cone. The GLSM charge matrix is
\be\label{intersect p1 res b}
\begin{tabular}{l|ccccccc|c}
 & $D_1$ &$D_2$& $D_3$ & $D_4$ & $D_5$ & $\bE_1$& $\bE_2$ & vol($\CC$) \\
 \hline
$\CC_1$  & $0$ & $0$ & $0$ &  $0$& $1$ & $-2$ & $1$ & $\xi_1$ \\ \hline
$\CC_4$  & $0$ & $0$ & $1$ &  $-1$ & $0$ & $1$ & $-1$ & $\xi_4$ \\ \hline
$\CC_5$  & $0$ & $1$ & $-1$ & $1$& $0$ & $0$ & $-1$ & $\xi_5$ \\ \hline
$\CC_7$  & $1$ & $0$ & $0$ &  $1$ & $-1$ & $-1$ & $0$ & $\xi_7$ \\ \hline\hline
$U(1)_M$ & $0$& $0$ &$0$ & $0$ & $0$ & $-1$ & $1$ & $r_0$
 \end{tabular}
\ee 
The K\"{a}hler cone is parametrized by \eqref{eq:kah cone p1}. The parameters $(\mu_1, \mu_2, \nu_1, \nu_2)$ are now related to the FI parameters by:
\begin{align}\hspace{-0.15in}
\xi_1 &= -2\nu_1 + \nu_2 \geq 0~, \quad \xi_4 = \nu_1 - \nu_2 \geq 0~, \quad \xi_5 = \mu_2 - \nu_2 \geq 0~, \quad \xi_7 = \mu_1 - \nu_1 \geq 0 ~.	 \label{eq:p1 res b FI}
\end{align}
The relevant triple-intersection numbers are:
\begin{align}
\begin{array}{c@{~,\quad}c@{~,\quad}c@{~,\quad}c@{~\quad}}
  D_1 \bE_1 \bE_2 = 1 &  D_2 \bE_1 \bE_2 = 0 & D_1 D_2 \bE_1 = 0 & D_1 D_2 \bE_2 = 1 ~, \\
  D_1\bE_1^2 = -2 & D_2 \bE_1^2 = 0 & D_1\bE_2^2 = -2 & D_2\bE_2^2 = -4 ~,\\
  D_1^2 \bE_1 = 0 &  D_1^2 \bE_2 = 0 & D_2^2 \bE_1 = 0 & D_2^2 \bE_2 = 2 ~,\\
  \bE_1^2 \bE_2 = -3 & \bE_1 \bE_2^2 = 1 & \bE_1^3 = 8 & \bE_2^3 = 7 ~.
\end{array}
\end{align}
Therefore, the compact part of the prepotential is:
\begin{align}
\mathcal{F}_{(b)}(\nu_1, \nu_2; \mu_1, \mu_2) = -\frac{1}{6}S^3 &= -\frac{4}{3}\nu_1^3 - \frac{7}{6}\nu_2^3 + \frac{3}{2}\nu_1^2\nu_2 -\frac{1}{2}\nu_1\nu_2^2 + \mu_1 \nu_1^2 + (\mu_1 + 2\mu_2)\nu_2^2 ~, \nonumber\\
& \quad - \mu_1 \nu_1 \nu_2 - \mu_2^2 \nu_2 - \mu_1 \mu_2 \nu_2 ~. \label{eq:p1 res b geo prepot}
\end{align}
The type IIA profile is:
\begin{align}\label{eq:p1 res b chi}
\chi(r_0) &= \left\{ 
                 \begin{array}{ll}
                 	-4r_0 + \xi_1 - \xi_4 - 2\xi_5 + \xi_7, & \text{for } r_0 \leq -\xi_4-\xi_5\\
                 	-2r_0 + \xi_1 + \xi_4 + \xi_7, & \text{for } -\xi_4-\xi_5 \leq r_0 \leq -\xi_4\\
                 	-3r_0 + \xi_1 + \xi_7, & \text{for } -\xi_4 \leq r_0 \leq 0\\
                 	-r_0 + \xi_1 + \xi_7, & \text{for } 0 \leq r_0 \leq \xi_1  \\
                 	+r_0 - \xi_1 + \xi_7, & \text{for } r_0 \geq \xi_1 ~.
                 \end{array}
             \right.	
\end{align}
This function is sketched in Figure \ref{fig:P[1] res b chi}. At the points $r_0 = -\xi_4-\xi_5$, $r_0 = 0$ and $r_0 = \xi_1$, there are gauge D6-branes wrapping $\P^1$'s in the resolution of the singularity. There is a flavor D6-brane at $r_0 = -\xi_4$.  The simple-root W-bosons have masses given by: 
\begin{align}
M(W_1) &= \xi_4 + \xi_5 = 2\varphi_1 - \varphi_2~, \quad M(W_2) = \xi_1 = 2\varphi_2 - \varphi_1 ~.
\end{align}
This resolution corresponds to gauge theory chamber 2 (cf. Table \ref{tbl:u3 nf1 instantons} and \eqref{eq:ft prepot SU(3) k Nf1 chamber 2}) with instanton masses given by:
\begin{align}
\begin{split}
	M(\cI_1) &= \chi(r_0 = -\xi_4-\xi_6) = \xi_1 + 3 \xi_4 + 2 \xi_5 + \xi_7 = h_0 + 4\varphi_1 - m  ~,\\
	M(\cI_2) &= \chi(r_0 = 0) = \xi_1 + \xi_7 = h_0 - \varphi_1 + 3 \varphi_2 ~,\\
	M(\cI_3) &= \chi(r_0 = \xi_1) = \xi_7 = h_0 + \varphi_2 ~.
\end{split}	
\end{align}
The masses of hypermultiplets are:
\begin{align}
\begin{split}
	M(\cH_1) &= \xi_4 = \varphi_1 - \varphi_2 - m ~,~
	M(\cH_2) = \xi_5 = \varphi_1 + m ~,~
	M(\cH_3) = \xi_1 + \xi_4 = \varphi_2 - m ~.
\end{split}	
\end{align}
One can verify that the map \eqref{eq:p1 res a chi} still holds, and plugging it into \eqref{eq:p1 res b geo prepot}, we recover the field theory prepotential,
\begin{align}
\mathcal{F}_{SU(3)_{2},N_{\text{f}}=1}^{\text{chamber 2}} &= \frac{7}{6}\varphi_1^3 + \frac{4}{3}\varphi_2^3 + \frac{1}{2}\varphi_1^2 \varphi_2 - \frac{3}{2} \varphi_1\varphi_2^2 + \left(h_{0}-\frac{m}{2}\right)\varphi_1^2 + h_{0}\varphi_2^2\nonumber\\& - h_{0}\varphi_1\varphi_2 - \frac{m^2}{2}\varphi_1 ~,
\end{align}
up to $\varphi$-independent terms. %From field theory, the monopole string tensions are given by
%\begin{align}
%T_{1,\text{ft}} &= \frac{\partial \mathcal{F}_{SU(3)_{2},N_{\text{f}}=1}^{\text{chamber 2}}}{\partial \varphi_1} = \frac{7}{2}\varphi_1^2 - \frac{3}{2}\varphi_2^2 + \varphi_1\varphi_2 + (2h_0{-}m)\varphi_1 - h_0 \varphi_2 - \frac{m^2}{2} ~,\\
%T_{2,\text{ft}} &= \frac{\partial \mathcal{F}_{SU(3)_{2},N_{\text{f}}=1}^{\text{chamber 2}}}{\partial \varphi_2} = \frac{1}{2}\varphi_1^2 + 4\varphi_2^2 -3 \varphi_1\varphi_2 - h_{0}\varphi_1 + 2 h_{0}\varphi_2 ~,
%\end{align}
The monopole string tensions are given from $\chi(r_0)$ by:
\begin{align}
T_{1,\text{geo}} &= \int_{-\xi_4-\xi_5}^{0}\chi(r_0)\,dr_{0} = \frac{3 \xi _4^2}{2}+ \xi _4\left(3 \xi _5+\xi _7\right) +\xi _1 \left(\xi _4+\xi _5\right)+\xi _5 \left(\xi _5+\xi _7\right) ~,\\
T_{2,\text{geo}} &= \int_{0}^{\xi_1}\chi(r_0)\,dr_{0} = \frac{\xi _1^2}{2}+\xi _1\xi _7 ~.
\end{align}
One can verify, using the map $\xi_1 = 2\varphi_2-\varphi_1$, $\xi_4 = \varphi_1 - \varphi_2 -m$, $\xi_5 = \varphi_1+m$ and $\xi_7 = h_0 + \varphi_2$, that $T_{i,\text{ft}} = T_{i,\text{geo}}$ for $i = 1, 2$. The tensions vanish at loci given by:
\begin{align}\begin{split}
 &(I): \{\xi_1 = 0\} \cup \{\xi_1 + 2\xi_7 = 0\}~, \text{ and } ~,\\
 &(II): \left\{ \frac{3}{2} \xi _4^2 + \xi _4\left(3 \xi _5+\xi _7\right) +\xi _1 \left(\xi _4+\xi _5\right)+\xi _5 \left(\xi _5+\xi _7\right) =0 \right\} ~.
 \end{split}
\end{align}
Along the submanifold $\{\xi_1 = 0\} \subset (I)$, the W-boson $W_2$ becomes massless, signaling a hard wall. Also $\{\xi_1 + 2\xi_7=0\}$ is not part of the K\"{a}hler chamber of resolution (b). As for the condition $(II)$, the solutions to the quadratic equation for $\xi_4$ are:
\begin{align}
\xi_4 &= -\xi_5 - \frac{1}{3}\left(\xi_1 + \xi_7 \pm \sqrt{(\xi_1+\xi_7)^2 + 3 \xi_5^2}\right) ~.
\end{align}
Both sign choices lead to a negative value of $\xi_4$, which is inconsistent in this K\"{a}hler chamber. Also note that the curve $\CC_5$ cannot flop in this chamber, so $\xi_5$ cannot vanish. 

Away from any hard wall, the perturbative BPS hypermultiplet $\cH_1$ can become massless at $\xi_4 = 0$ (signaling a flop of $\CC_4$), indicating a traversable wall that leads back to gauge theory resolution (a). Alternatively, the BPS instanton $\cI_3$ can become massless at $\xi_7 = 0$ (signaling a flop of $\CC_7$), corresponding to a traversable instantonic wall that leads to non-gauge-theoretic resolution (f).
%Consequently there are no magnetic walls in the K\"{a}hler chamber of resolution (b).

\begin{figure}[ht]
\centering
\subfigure[\small{}]{
\includegraphics[width=4.5cm]{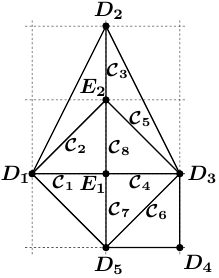}\label{fig:P[1] res c labeled}}\,
\subfigure[\small{}]{
\includegraphics[width=7.5cm]{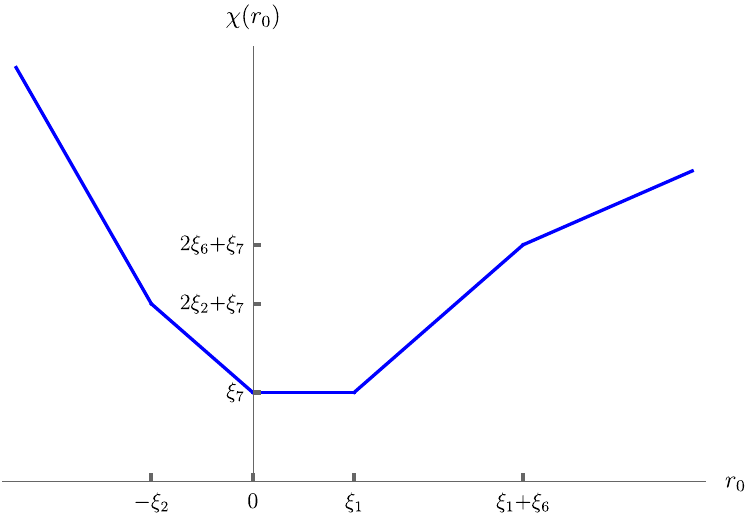}\label{fig:P[1] res c chi}}
\caption{Resolution (c) of the $E_{2}{}^{2,\fthreehalf}$ singularity and its vertical reduction.\label{fig:p1 res c final}}
\end{figure}

\paragraph{Resolution (c).} Consider the crepant resolution of Figure \ref{fig:p1-res-c}, with curves and divisors shown in Figure \ref{fig:P[1] res c labeled}.  
%%%%%%%%%%%%%%%
%%%%%%%%%%%%%%%
%The linear relations among divisors are given by \eqref{eq:p1 res a divisor linear equivalences}. The compact curves $\CC$ are given by:
%\begin{align}\begin{split}
%	\CC_1 &= \bE_1 \cdot D_1, \quad \CC_2 = \bE_2 \cdot D_1, \quad \CC_3 = \bE_2 \cdot D_2, \quad \CC_4 = \bE_1 \cdot D_3 ~,\\
%	\CC_5 &= \bE_2 \cdot D_3, \quad \CC_6 = D_3 \cdot D_5, \quad \CC_7 = \bE_1 \cdot D_5, \quad \CC_8 = \bE_2 \cdot \bE_1 ~.
%	\end{split}
%\end{align}
The linear relations among curve classes are:
\begin{align}
\CC_3 &\simeq 2\CC_2 + \CC_7~, \quad \CC_4 \simeq \CC_1~, \quad \CC_5 \simeq \CC_2~, \quad \CC_8 \simeq \CC_7 ~.
\end{align}
We take $\{\CC_1, \CC_2, \CC_6, \CC_7\}$ as generators of the Mori cone. The GLSM charge matrix is:
\be\label{intersect p1 res c}
\begin{tabular}{l|ccccccc|c}
 & $D_1$ &$D_2$& $D_3$ & $D_4$ & $D_5$ & $\bE_1$& $\bE_2$ & vol($\CC$) \\
 \hline
$\CC_1$  & $0$ & $0$ & $0$ &  $0$& $1$ & $-2$ & $1$         & $\xi_1$ \\ \hline
$\CC_2$  & $0$ & $1$ & $0$ &  $0$ & $0$ & $1$ & $-2$       & $\xi_2$ \\ \hline
$\CC_6$  & $0$ & $0$ & $-1$ & $1$& $-1$ & $1$ & $0$         & $\xi_6$ \\ \hline
$\CC_7$  & $1$ & $0$ & $1$ &  $0$ & $0$ & $-2$ & $0$       & $\xi_7$ \\ \hline\hline
$U(1)_M$ & $0$& $0$ &$0$ & $0$ & $0$ & $-1$ & $1$ & $r_0$
 \end{tabular}
\ee 
The parameters $(\mu_1, \mu_2, \nu_1, \nu_2)$ are related to the FI parameters by:
\begin{align}%\hspace{-0.35in}
\xi_1 &= -2\nu_1 + \nu_2 \geq 0~, \quad \xi_2 = \mu_2 + \nu_1 - 2\nu_2 \geq 0~, \quad \xi_6 = \nu_1 \geq 0~, \quad \xi_7 = \mu_1 - 2\nu_1 \geq 0 ~.	 \label{eq:p1 res c FI}
\end{align}
The relevant triple-intersection numbers are:
\begin{align}
\begin{array}{c@{~,\quad}c@{~,\quad}c@{~,\quad}c@{~\quad}}
  D_1 \bE_1 \bE_2 = 1 & D_2 \bE_1 \bE_2 = 0 & D_1 D_2 \bE_1 = 0 & D_1 D_2 \bE_2 = 1 ~, \\
  D_1\bE_1^2 = -2 & D_2 \bE_1^2 = 0 & D_1\bE_2^2 = -2 & D_2\bE_2^2 = -4~,\\
  D_1^2 \bE_1 = 0 & D_1^2 \bE_2 = 0 & D_2^2 \bE_1 = 0 & D_2^2 \bE_2 = 2 ~,\\
  \bE_1^2 \bE_2 = -2 & \bE_1 \bE_2^2 = 0 &  \bE_1^3 = 8 & \bE_2^3 = 8 ~.
\end{array}
\end{align}
Therefore, the compact part of the prepotential is:
\begin{align}
\mathcal{F}_{(c)}(\nu_1, \nu_2; \mu_1, \mu_2) = -\frac{1}{6}S^3 &= -\frac{4}{3}(\nu_1^3 + \nu_2^3)  + \nu_1^2\nu_2 +  \mu_1\nu_1^2 + (\mu_1+2\mu_2)\nu_2^2\nonumber\\
& - \mu_1\nu_1\nu_2 - \mu_2(\mu_1+\mu_2)\nu_2 ~. \label{eq:p1 res c geo prepot}
\end{align}
The type IIA profile is:
\begin{align}\label{eq:p1 res c chi}
\chi(r_0) &= \left\{ 
                 \begin{array}{ll}
                 	-4r_0 - 2\xi_2 + \xi_7, & \text{for } r_0 \leq -\xi_2\\
                 	-2r_0 + \xi_7, & \text{for } -\xi_2 \leq r_0 \leq 0\\
                 	\xi_7, & \text{for } 0 \leq r_0 \leq \xi_1\\
                 	+2r_0-2\xi_1+\xi_7, & \text{for } \xi_1 \leq r_0 \leq \xi_1+\xi_6  \\
                 	+r_0 - \xi_1 + \xi_6 + \xi_7, & \text{for } r_0 \geq \xi_1 + \xi_6 ~.
                 \end{array}
             \right.	
\end{align}
This function is sketched in Figure \ref{fig:P[1] res c chi}. At the points $r_0 = -\xi_2$, $r_0 = 0$ and $r_0 = \xi_1$, there are gauge D6-branes wrapping $\P^1$'s in the resolution of the singularity. There is a flavor D6-brane at $r_0 = \xi_1+\xi_6$. The simple-root W-bosons have masses given by:
\begin{align}
M(W_1) &= \xi_2  = 2\varphi_1 - \varphi_2~, \quad M(W_2) = \xi_1 = 2\varphi_2 - \varphi_1 ~.
\end{align}
This resolution corresponds to gauge theory chamber 4 (cf. Table \ref{tbl:u3 nf1 instantons} and \eqref{eq:ft prepot SU(3) k Nf1 chamber 4}), with instanton masses given by:
\begin{align}
\begin{split}
	M(\cI_1) &= \chi(r_0 = -\xi_2) = 2\xi_2+\xi_7 = h_0 + 4\varphi_1 - m ~, \\
	M(\cI_2) &= \chi(r_0 = 0) = \xi_7 = h_0 + 2\varphi_2 - m ~,\\
	M(\cI_3) &= \chi(r_0 = \xi_1) = \xi_7 = h_0 + 2\varphi_2 - m ~.
\end{split}	
\end{align}
The masses of hypermultiplets are:
\begin{align}
\begin{split}
	M(\cH_1) &= \xi_6 = -\varphi_2 + m ~,\\
	M(\cH_2) &= \xi_1+\xi_6 = -\varphi_1 + \varphi_2 + m ~,\\
	M(\cH_3) &= \xi_1 + \xi_2 + \xi_6 = \varphi_1 + m ~.
\end{split}	
\end{align}
Plugging the map between $(\bm{\nu}, \bm{\mu})$ parameters and field-theory parameters given by \eqref{eq:p1 res a mu nu}, into \eqref{eq:p1 res c geo prepot}, we recover the field theory prepotential,
\begin{align}
\mathcal{F}_{SU(3)_{2},N_{\text{f}}=1}^{\text{chamber 4}} &= \frac{4}{3}(\varphi_1^3 + \varphi_2^3) - \varphi_1 \varphi_2^2 + (h_0-m)\varphi_1^2 + (h_0-m)\varphi_2^2 + (m-h_0)\varphi_1\varphi_2 ~,
\end{align}
up to $\varphi$-independent terms. %From field theory, the monopole string tensions are given by:
%\begin{align}
%T_{1,\text{ft}} &= \frac{\partial \mathcal{F}_{SU(3)_{2},N_{\text{f}}=1}^{\text{chamber 4}}}{\partial \varphi_1} =4\varphi_1^2 - \varphi_2^2 + 2(h_0-m)\varphi_1 + (m-h_0)\varphi_2 ~,\\
%T_{2,\text{ft}} &= \frac{\partial \mathcal{F}_{SU(3)_{2},N_{\text{f}}=1}^{\text{chamber 4}}}{\partial \varphi_2} = 4\varphi_2^2 - 2 \varphi_1\varphi_2 + (m-h_0)\varphi_1 + 2(h_0-m)\varphi_2 ~.
%\end{align}
The monopole string tensions from $\chi(r_0)$ are given by:
\begin{align}
T_{1,\text{geo}} &= \int_{-\xi_2}^{0}\chi(r_0)\,dr_{0} = \xi_2(\xi_2 + \xi_7) ~,~%\\
T_{2,\text{geo}} = \int_{0}^{\xi_1}\chi(r_0)\,dr_{0} = \xi_1 \xi_7 ~,
\end{align}
Using the map $\xi_1 = 2\varphi_2-\varphi_1$, $\xi_2 = 2\varphi_1 - \varphi_2$, $\xi_6 = -\varphi_2+m$ and $\xi_7 = h_0 + 2\varphi_2-m$, one can verify that $T_{i,\text{ft}} = T_{i,\text{geo}}$ for $i = 1, 2$. It is easy to see that subloci of vanishing tension lie along hard walls where either W-boson becomes massless, or along hard walls that are not in this K\"{a}hler chamber. Away from a hard wall, $\cH_1$ can become massless signaling a flop of $\CC_6$ leading back to resolution (a).

%The tensions vanish at loci given by:
%\begin{align}%\begin{split}
% (I)&: \{\xi_2 = 0\} \cup \{\xi_2 + \xi_7 = 0\},~ \text{ and } ~ (II) : \{ \xi_1 = 0 \} \cup \{\xi_7 = 0\} ~.
% %\end{split}
%\end{align}
%Along $\{\xi_2 = 0\} \subset (I)$, the W-boson $W_1$ becomes massless, signaling a hard wall. The component $\{\xi_2 + \xi_7 = 0\}$ is not part of the K\"{a}hler chamber of resolution (c). Along $\{\xi_1 =0\} \subset (II)$, the W-boson $W_2$ becomes massless, again indicating a hard wall. The component $\{\xi_7 = 0\} \subset (II)$ is also not part of the K\"{a}hler chamber of resolution (c), as the corresponding curve $\CC_7$ cannot flop. Away from any hard wall, the perturbative hypermultiplet $\cH_1$ can become massless at $\xi_6 = 0$ (signaling a flop of $\CC_6$), leading to gauge theory resolution (a). %So there are no magnetic walls in the K\"{a}hler chamber of resolution (c).

\begin{figure}[ht]
\centering
\subfigure[\small{}]{
\includegraphics[width=4.5cm]{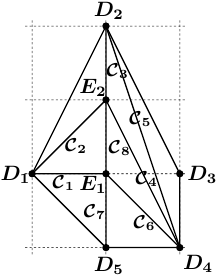}\label{fig:P[1] res d labeled}}\,
\subfigure[\small{}]{
\includegraphics[width=7.5cm]{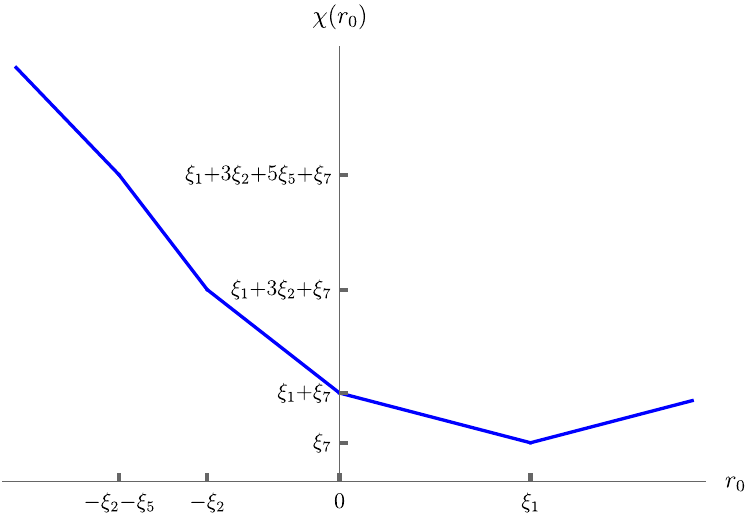}\label{fig:P[1] res d chi}}
\caption{Resolution (d) of the $E_{2}{}^{2,\fthreehalf}$ singularity and its vertical reduction.\label{fig:p1 res d final}}
\end{figure}

\paragraph{Resolution (d).} Consider the crepant resolution of Figure \ref{fig:p1-res-d}, with curves and divisors shown in Figure \ref{fig:P[1] res d labeled}. 
%%%%%%%%%%%%%%%
%%%%%%%%%%%%%%%
%The linear equivalances among divisors are given by \eqref{eq:p1 res a divisor linear equivalences}. The compact curves $\CC$ are given by:
%\begin{align}\begin{split}
%	\CC_1 &= \bE_1 \cdot D_1, \quad \CC_2 = \bE_2 \cdot D_1, \quad \CC_3 = \bE_2 \cdot D_2, \quad \CC_4 = \bE_2 \cdot D_4 ~,\\
%	\CC_5 &= D_4 \cdot D_2, \quad \CC_6 = \bE_1 \cdot D_4, \quad \CC_7 = \bE_1 \cdot D_5, \quad \CC_8 = \bE_2 \cdot \bE_1 ~.
%	\end{split}
%\end{align}
The linear relations among curve classes are:
\begin{align}
\CC_3 &\simeq \CC_1 + 3\CC_2 + \CC_7~, \quad \CC_4 \simeq \CC_2~, \quad \CC_6 \simeq \CC_1~, \quad \CC_8 \simeq \CC_1 + \CC_7 ~.
\end{align}
We take $\{\CC_1, \CC_2, \CC_5, \CC_7\}$ as generators of the Mori cone. The GLSM charge matrix is:
\be\label{intersect p1 res c}
\begin{tabular}{l|ccccccc|c}
 & $D_1$ &$D_2$& $D_3$ & $D_4$ & $D_5$ & $\bE_1$& $\bE_2$ & vol($\CC$) \\
 \hline
$\CC_1$  & $0$ & $0$ & $0$ &  $0$& $1$ & $-2$ & $1$         & $\xi_1$ \\ \hline
$\CC_2$  & $0$ & $1$ & $0$ &  $0$ & $0$ & $1$ & $-2$       & $\xi_2$ \\ \hline
$\CC_5$  & $0$ & $-1$ & $1$ & $-1$& $0$ & $0$ & $1$         & $\xi_5$ \\ \hline
$\CC_7$  & $1$ & $0$ & $0$ &  $1$ & $-1$ & $-1$ & $0$       & $\xi_7$ \\ \hline\hline
$U(1)_M$ & $0$& $0$ &$0$ & $0$ & $0$ & $-1$ & $1$ & $r_0$
 \end{tabular}
\ee 
The parameters $(\mu_1, \mu_2, \nu_1, \nu_2)$ are related to the FI parameters by:
\begin{align}
\xi_1 &= -2\nu_1 + \nu_2 \geq 0~, \quad \xi_2 = \mu_2 + \nu_1 - 2\nu_2 \geq 0~, \quad \xi_5 = -\mu_2 + \nu_2 \geq 0~, \quad \xi_7 = \mu_1 - \nu_1 \geq 0 ~.	 \label{eq:p1 res d FI}
\end{align}
The relevant triple-intersection numbers are:
\begin{align}
\begin{array}{c@{~,\quad}c@{~,\quad}c@{~,\quad}c@{~\quad}}
  D_1 \bE_1 \bE_2 = 1 & D_2 \bE_1 \bE_2 = 0 & D_1 D_2 \bE_1 = 0 & D_1 D_2 \bE_2 = 1 ~,\\
  D_1\bE_1^2 = -2 & D_2 \bE_1^2 = 0 & D_1\bE_2^2 = -2 & D_2\bE_2^2 = -5 ~,\\ 
  D_1^2 \bE_1 = 0 & D_1^2 \bE_2 = 0 & D_2^2 \bE_1 = 0 & D_2^2 \bE_2 = 3 ~,\\
  \bE_1^2 \bE_2 = -3 & \bE_1 \bE_2^2 = 1 &  \bE_1^3 = 8 & \bE_2^3 = 8 ~.
\end{array}
\end{align}
Therefore, the compact part of the prepotential is:
\begin{align}
\mathcal{F}_{(d)}(\nu_1, \nu_2; \mu_1, \mu_2) = -\frac{1}{6}S^3 &= -\frac{4}{3}(\nu_1^3 + \nu_2^3)  + \frac{3}{2}\nu_1^2\nu_2 - \frac{1}{2}\nu_1\nu_2^2 + \mu_1 \nu_1^2  - \mu_1\nu_1\nu_2\nonumber\\
&+ \left(\mu_1 + \frac{5}{2}\mu_2\right)\nu_2^2 + \left(\frac{3}{2}\mu_2^2 -\mu_1 \mu_2\right)\nu_2 ~. \label{eq:p1 res d geo prepot}
\end{align}
The IIA profile is:
\begin{align}\label{eq:p1 res c chi}
\chi(r_0) &= \left\{ 
                 \begin{array}{ll}
                 	-4r_0 + \xi_1-\xi_2 + \xi_5 + \xi_7, & \text{for } r_0 \leq -\xi_2-\xi_5\\
                 	-5r_0 + \xi_1 - 2\xi_2 + \xi_7, & \text{for } -\xi_2-\xi_5 \leq r_0 \leq -\xi_2\\
                 	-3r_0 + \xi_1 + \xi_7, & \text{for } -\xi_2 \leq r_0 \leq 0\\
                 	-r_0+\xi_1+\xi_7, & \text{for } \xi_1 \leq r_0 \leq \xi_1  \\
                 	+r_0 - \xi_1 + \xi_7, & \text{for } r_0 \geq \xi_1 ~.
                 \end{array}
             \right.	
\end{align}
This function is sketched in Figure \ref{fig:P[1] res d chi}. At the points $r_0 = -\xi_2$, $r_0 = 0$ and $r_0 = \xi_1$, there are gauge D6-branes wrapping $\P^1$'s in the resolution of the singularity. There is a flavor D6-brane at $r_0 = -\xi_2-\xi_5$. The simple-root W-bosons have masses given by:
\begin{align}
M(W_1) &= \xi_2  = 2\varphi_1 - \varphi_2~, \quad M(W_2) = \xi_1 = 2\varphi_2 - \varphi_1 ~.
\end{align}
This resolution corresponds to gauge theory chamber 1 (cf. Table \ref{tbl:u3 nf1 instantons} and \eqref{eq:ft prepot SU(3) k Nf1 chamber 1}), with instanton masses given by:
\begin{align}
\begin{split}
	M(\cI_1) &= \chi(r_0 = -\xi_2) = \xi_1 + 3\xi_2 + \xi_7 = h_0 + 5\varphi_1 ~,  \\
	M(\cI_2) &= \chi(r_0 = 0) = \xi_1 + \xi_7 = h_0 - \varphi_1 + 3\varphi_2 ~,\\
	M(\cI_3) &= \chi(r_0 = \xi_1) = \xi_7 = h_0 + \varphi_2 ~.
\end{split}	
\end{align}
The masses of hypermultiplets are:
\begin{align}
\begin{split}
	M(\cH_1) &= \xi_5 = -\varphi_1 - m ~,\\
	M(\cH_2) &= \xi_2+\xi_5 = \varphi_1 - \varphi_2 - m ~,\\
	M(\cH_3) &= \xi_1 + \xi_2 + \xi_5 = \varphi_2 - m ~.
\end{split}	
\end{align}
Plugging \eqref{eq:p1 res a mu nu} into \eqref{eq:p1 res d geo prepot}, we recover the field theory prepotential,
\begin{align}\hspace{-0.25in}
\mathcal{F}_{SU(3)_{2},N_{\text{f}}=2}^{\text{chamber 1}} &= \frac{4}{3}(\varphi_1^3 + \varphi_2^3) - \varphi_1 \varphi_2^2 + (h_0-m)\varphi_1^2 + (h_0-m)\varphi_2^2 + (m-h_0)\varphi_1\varphi_2 ~,
\end{align}
up to $\varphi$-independent terms. %From field theory, the monopole string tensions are given by:
%\begin{align}
%T_{1,\text{ft}} &= \frac{\partial \mathcal{F}_{SU(3)_{2},N_{\text{f}}=1}^{\text{chamber 1}}}{\partial \varphi_1} = 4\varphi_1^2 -\frac{3}{2}\varphi_2^2 + \varphi_1\varphi_2 + h_{0}(2\varphi_1-\varphi_2) ~,\\
%T_{2,\text{ft}} &= \frac{\partial \mathcal{F}_{SU(3)_{2},N_{\text{f}}=1}^{\text{chamber 1}}}{\partial \varphi_2} = \frac{1}{2}\varphi_1^2 + 4\varphi_2^2 - 3\varphi_1\varphi_2 + h_{0}(2\varphi_2-\varphi_1) ~,
%\end{align}
The tensions from $\chi(r_0)$ are given by:
\begin{align}
T_{1,\text{geo}} &= \int_{-\xi_2}^{0}\chi(r_0)\,dr_{0} = \xi_2\left(\frac{3}{2} \xi_2 + \xi_1+\xi_7\right) ~,~%\\
T_{2,\text{geo}} = \int_{0}^{\xi_1}\chi(r_0)\,dr_{0} = \frac{1}{2}\xi_1(\xi_1 + 2\xi_7) ~.
\end{align}
Using the map $\xi_1 = 2\varphi_2-\varphi_1$, $\xi_2 = 2\varphi_1 - \varphi_2$, $\xi_5 = -\varphi_1-m$ and $\xi_7 = h_0 + \varphi_2$, we find that $T_{i,\text{ft}} = T_{i,\text{geo}}$ for $i = 1, 2$. It is easy to check that in this case too, loci of vanishing tension are either hard walls where W-bosons become massless, or walls that do not lie in this K\"{a}hler chamber. There is a perturbative wall corresponding to a flop of $\CC_5$ (when $\cH_1$ becomes massless), but also a traversible instantonic wall at leading to non-gauge theory resolution (g).
%\paragraph{Magnetic walls.} The tensions vanish at loci given by:
%\begin{align}%\begin{split}
% &(I): \{\xi_2 = 0\} \cup \left\{\frac{3}{2}\xi_2 + \xi_1 + \xi_7 = 0\right\}, \text{ and }(II): \{ \xi_1 = 0 \} \cup \{\xi_1 + 2\xi_7 = 0\} ~.
% %\end{split}
%\end{align}
%Along the submanifold $\{\xi_2 = 0\} \subset (I)$, the W-boson $W_1$ becomes massless, indicating a hard wall. Also the locus $\{\tfrac{3}{2}\xi_2 + \xi_1 + \xi_7 = 0\}$ is not part of the K\"{a}hler chamber of resolution (d). Along the submanifold $\{\xi_1 = 0\} \subset (II)$, the W-boson $W_2$ becomes massless, indicating again a hard wall. The locus $\{\xi_2 + 2\xi_7 = 0\} \subset (II)$ is likewise not part of the K\"{a}hler chamber of resolution (d). Away from any hard wall, the perturbative BPS hypermultiplet $\cH_1$ can be come massless at $\xi_5 = 0$ (signaling a flop of $\CC_5$), which leads back to gauge theory resolution (b). Alternatively, away from the hard wall, the instantonic particle $\cI_3$ can become massless at $\xi_7 = 0$ (signaling a flop of $\CC_7$), indicating a traversable instantonic wall that leads to non-gauge-theoretic resolution (g).  %So there are no magnetic walls in the K\"{a}hler chamber of resolution (d).

\begin{figure}[ht]
\centering
\begin{minipage}{0.35\textwidth}
	%\subfigure[]{\includegraphics[width=4.5cm]{images/p1-res-e-new-labeled.pdf}\label{fig:p1-res-rg-flow-1}}
	\includegraphics[width=4cm]{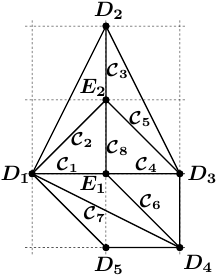}
\end{minipage}%
\begin{minipage}{0.20\textwidth}
	{\Large $\xrightarrow[\vol(\CC_7)\rightarrow \infty]{\text{RG flow}}$ }
\end{minipage}%
\begin{minipage}{0.30\textwidth}
	%\subfigure[]{\includegraphics[width=4.5cm]{images/p1-res-e-labeled-decouple-d5.pdf}\label{fig:p1-res-rg-flow-2}}
	\includegraphics[width=4cm]{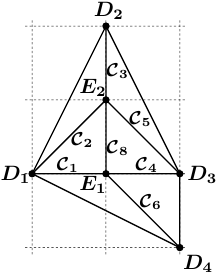}
\end{minipage}\label{fig:Q[1] res b RG flow}
\caption{Decoupling a divisor from $E_{2}{}^{2,\fthreehalf}$ resolution (e) yields an $SL(2,\Z)$-transformed version of a crepant resolution of the $E_{1}{}^{2,\NL}$ singularity.\label{fig:p1 non lagrangian rgflow}}
\end{figure}

\begin{figure}[ht]
\centering
\begin{minipage}{0.35\textwidth}
	%\subfigure[]{\includegraphics[width=4.5cm]{images/p1-res-f-new-labeled.pdf}\label{fig:p1-res-rg-flow-3}}
	\includegraphics[width=4cm]{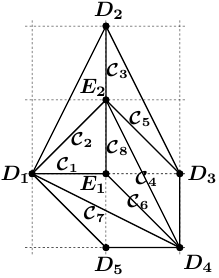}
\end{minipage}%
\begin{minipage}{0.20\textwidth}
	{\Large $\xrightarrow[\vol(\CC_7)\rightarrow \infty]{\text{RG flow}}$ }
\end{minipage}%
\begin{minipage}{0.30\textwidth}
	%\subfigure[]{\includegraphics[width=4.5cm]{images/p1-res-f-labeled-decouple-d5.pdf}\label{fig:p1-res-rg-flow-4}}
	\includegraphics[width=4cm]{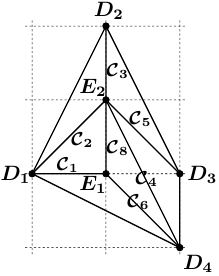}
\end{minipage}
\caption{Decoupling a divisor from $E_{2}{}^{2,\fthreehalf}$ resolution (f) yields an $SL(2,\Z)$-transformed version of a crepant resolution of the $E_{1}{}^{2,\NL}$ singularity.\label{fig:p1 non lagrangian rgflow 2}}
\end{figure}

\begin{figure}[ht]
\centering
\begin{minipage}{0.35\textwidth}
	%\subfigure[]{\includegraphics[width=4.5cm]{images/p1-res-g-new-labeled.pdf}\label{fig:p1-res-rg-flow-5}}
	\includegraphics[width=4cm]{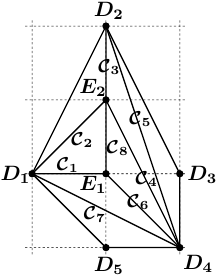}
\end{minipage}%
\begin{minipage}{0.20\textwidth}
	{\Large $\xrightarrow[\substack{\vol(\CC_5)\rightarrow \infty\\\vol(\CC_7)\rightarrow \infty}]{\text{RG flow}}$ }
\end{minipage}%
\begin{minipage}{0.30\textwidth}
	%\subfigure[]{\includegraphics[width=4.5cm]{images/p1-res-g-labeled-decouple-d3d5.pdf}\label{fig:p1-res-rg-flow-6}}
	\includegraphics[width=4cm]{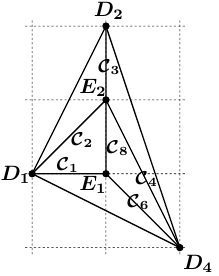}
\end{minipage}
\caption{Decoupling two divisors from $E_{2}{}^{2,\fthreehalf}$ resolution (g) yields an $SL(2,\Z)$-transformed version of the unique crepant resolution of the $E_{0}{}^{2,\NL}$ singularity.\label{fig:p1 non lagrangian rgflow 3}}
\end{figure}

\paragraph*{Resolutions (e), (f), (g) and RG flow.} As noted above, the crepant resolutions in Figures \ref{fig:p1-res-e}-\ref{fig:p1-res-g} do not admit vertical reductions. Nevertheless, they have interesting roles to play in the K\"{a}hler moduli space of the $E_{2}{}^{2,\fthreehalf}$ singularity. %They are shown more explicitly in Figures \ref{fig:p1-res-e-labeled}-\ref{fig:p1-res-g-labeled}.
%%%%%%%%%%%%%%%%
%\begin{figure}[ht]
%%\vspace{-10pt}
%\begin{center}
%\setcounter{subfigure}{4}
%\subfigure[]{
%\includegraphics[width=4.5cm]{images/p1-res-e-new-labeled}\label{fig:p1-res-e-labeled}}\,
%\subfigure[]{
%\includegraphics[width=4.5cm]{images/p1-res-f-new-labeled}\label{fig:p1-res-f-labeled}}\,
%\subfigure[]{
%\includegraphics[width=4.5cm]{images/p1-res-g-new-labeled}\label{fig:p1-res-g-labeled}}
%\caption{The three crepant resolutions of the $E_{2}{}^{2,\fthreehalf}$ singularity that do not admit a vertical reduction. These are ``non-gauge-theoretic phases.''\label{fig:p1 non lagrangian}}
%\end{center}
%\end{figure}
%%%%%%%%
In resolution (e), one can send the volume of the curve $\CC_7$ to infinity, thereby decoupling the divisor $D_5$. This leads to a crepant resolution of the non-Lagrangian $E_{1}{}^{2,\NL}$ singularity (see Figure \ref{fig:p1 non lagrangian rgflow}).\footnote{More precisely, an $SL(2,\Z)$ transformation (using, for instance an $S^2 T S T S^{-2}$ transformation) of the toric diagram on the right in Figure \protect{\ref{fig:p1 non lagrangian rgflow}} brings it into crepant resolution of $E_{1}{}^{2,\NL}$ of Figure \protect{\ref{fig:q13-sing}}.} Similarly in resolution (f), one can decouple $D_5$ by sending $\text{vol}(\CC_7)$ to infinity. This results in yet another crepant resolution of the $E_{1}{}^{2,\NL}$ singularity, as shown in Figure \ref{fig:p1 non lagrangian rgflow 2}. Finally in resolution (g), one can decouple $D_3$ and $D_5$ by sending both $\text{vol}(\CC_5)$ and $\text{vol}(\CC_7)$ to infinity as shown in Figure \ref{fig:p1 non lagrangian rgflow 3}. This leads to the unique crepant resolution of the $SL(2,\Z)$-transformed version of the $E_{0}{}^{2,\NL}$ singularity, which as we stated above, is also non-Lagrangian. 
%%%%%%%%%%%%%
%%%%%%%%%%%%
%%%%%%%%%%%%%%%%
%%%%%%%%%%%%
%%%%%%%%%%%%%%%%
%%%%%%%%%%%%

In summary, starting from the non-Lagrangian deformations of the $E_{2}{}^{2,\fthreehalf}$ singularity, one can obtain the non-Lagrangian deformations of $E_{0}{}^{2,\NL}$ and $E_{1}{}^{2,\NL}$ via RG flow in parameter space. A careful analysis of the phase boundaries -- carried out in the next section -- reveals that resolutions (e), (f) and (g) do not survive the limit in which the mass deformations are set to zero (that is, the Coulomb branch of the SCFT). 

\subsubsection{Sample slicings of the $E_{2}{}^{2,\fthreehalf}$ moduli space}
\begin{figure}[ht]
\begin{center}
\begin{tabular}{ccc}
\includegraphics[width=2.8in]{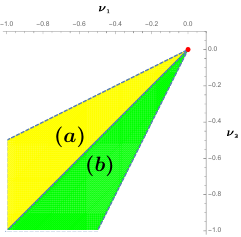}&&\includegraphics[width=2.8in]{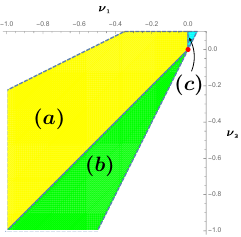}\\
{\small (i): phases: (a), (b)}& & {\small (ii): phases: (a), (b), (c)}\\
{\small  $\bm{\mu} = (0,0)$}&&{\small $\bm{\mu} = (-0.15,0.55)$}\\
&&\\
\includegraphics[width=2.8in]{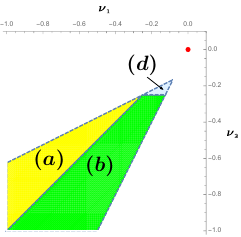}& &\includegraphics[width=2.8in]{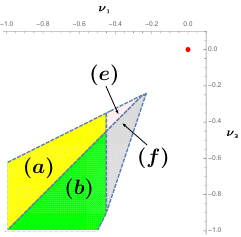}\\
{\small (iii): phases: (a), (b), (d)} & &{\small (iv): phases: (a), (b), (e), (f)}\\
{\small  $\bm{\mu} = (-0.25,0.25)$}&&{\small $\bm{\mu} = (-0.45,-0.25)$}\\
\end{tabular}
\end{center}
\caption{Sample slices of the moduli space of the $E_{2}{}^{2,\fthreehalf}$ geometry. Turning on different mass deformations reveals more $SU(3)$ $N_{\text{f}}=1$ phases such as (c) in (ii) and (d) in (iii), but also non-gauge theoretic phases such as (e) and (f) in (iv).\label{fig:P[1] slices}}
\end{figure}

In Figure \ref{fig:P[1] slices}, we show some sample slices of the moduli space of the $E_{2}{}^{2,\fthreehalf}$ geometry. The phase diagram of this geometry, parametrized by $(\bm{\nu}; \bm{\mu}) \equiv  (\nu_1, \nu_2; \mu_1, \mu_2)$ is a four-dimensional region, given by the disjoint union of the regions described by the defining inequalities of $7$ K\"{a}hler chambers, which are listed in Table \ref{tbl:P[1] Nef conditions}.
\begin{table}[ht]
\centering
\begin{tabular}{c|l}\hline
 \text{(a)} & $\{\mu _2+\nu _1-2 \nu _2 \geq 0\}$ $\cap$ $\{\nu _2-\nu _1 \geq 0\}$ $\cap$ $\{-\nu _1 \geq 0\}$ $\cap$ $\{\mu _1-\nu _1 \geq 0\}$ \\ \hline
 \text{(b)} & $\{\nu _2-2 \nu _1 \geq 0\}$ $\cap$ $\{\nu _1-\nu _2 \geq 0\}$ $\cap$ $\{\mu _2-\nu _2 \geq 0\}$ $\cap$ $\{\mu _1-\nu _1 \geq 0\}$ \\ \hline
 \text{(c)} & $\{\nu _2-2 \nu _1 \geq 0\}$ $\cap$ $\{\mu _2+\nu _1-2 \nu _2 \geq 0\}$ $\cap$ $\{\nu _1 \geq 0\}$ $\cap$ $\{\mu _1-2 \nu _1 \geq 0\}$ \\ \hline
 \text{(d)} & $\{\nu _2-2 \nu _1 \geq 0\}$ $\cap$ $\{\mu _2+\nu _1-2 \nu _2\geq 0\}$ $\cap$ $\{\nu _2-\mu _2\geq 0\}$ $\cap$ $\{\mu _1-\nu _1 \geq 0\}$ \\ \hline
 \text{(e)} & $\{\nu _2-\nu _1 \geq 0\}$ $\cap$ $\{\mu _2+\nu _1-2 \nu _2\geq 0\}$ $\cap$ $\{\mu _1-2 \nu _1\geq 0\}$ $\cap$ $\{\nu _1-\mu _1 \geq 0\}$ \\ \hline
 \text{(f)} & $\{\mu _1-3 \nu _1+\nu _2 \geq 0\}$ $\cap$ $\{\nu _1-\nu _2 \geq 0\}$ $\cap$ $\{\mu _2-\nu _2 \geq 0\}$ $\cap$ $\{\nu _1-\mu _1 \geq 0\}$ \\ \hline
 \text{(g)} & $\{\mu _1-3 \nu _1+\nu _2 \geq 0\}$ $\cap$ $\{\mu _2+\nu _1-2 \nu _2 \geq 0\}$ $\cap$ $\{\nu _2-\mu _2 \geq 0\}$ $\cap$ $\{\nu _1-\mu _1 \geq 0\}$ \\ \hline
\end{tabular}
\caption{Geometric inequalities (``Nef conditions'') defining the K\"{a}hler chambers of the 7 resolutions of the $E_{2}{}^{2,\fthreehalf}$ geometry. \label{tbl:P[1] Nef conditions}}
\end{table}

The phase diagram can be visualized by taking slices at different values of $(\mu_1, \mu_2)$, which reveal different chambers. For some values of $(\bm{\nu}; \bm{\mu})$ some regions vanish altogether while other regions collapse to real codimension-one walls in this parameter space (along which flops may occur). The origin $(\nu_1, \nu_2) = (0, 0)$ is denoted by a red dot on the top right of each plot. To make the plots readable, we only highlight chambers that have a finite area in parameter space in the slices that are considered. When $\bm{\mu} \neq 0$, the origin $\nu_1 = \nu_2 = 0$ is generally \textit{not} the origin of the Coulomb branch of the gauge theory (when such a description exists), since the map \eqref{eq:p1 res a mu nu} between $\nu_1, \nu_2$ and $\varphi_1, \varphi_2$ for $SU(3)$ $N_{\text{f}} = 2$ involves a contribution from the real mass $m$.
%%%%%
%%%%%%%
The slicings of Figure \ref{fig:P[1] slices} can also be used to highlight some geometric features. For instance, resolution (b) is obtained from resolution (a) by flopping curve $\CC_4$. This corresponds to the volume $\text{vol}(\CC_4) = \xi_4 = -\nu_1+\nu_2$ shrinking to zero size in the K\"{a}hler chamber defining resolution (a), before it grows in the birational K\"{a}hler chamber of resolution (b). The real codimension-1 wall separating phases (a) and (b) is clearly visible in Figure \ref{fig:P[1] slices}(i). In order to reach resolution (c), one just needs to flop curve $\CC_6$ which has volume $\text{vol}(\CC_6) = \xi_6 = -\nu_1$. This vanishes along the vertical line $\nu_1 = 0$ indicating a wall separating regions (a) and (c) in Figure \ref{fig:P[1] slices}(ii). On the other hand, to reach chamber (c) from chamber (b), one needs to perform \textit{two} flops, which necessitates going through the origin, as is also clear from the figure. 

Finally, turning on generic mass deformations reveals non-gauge theoretic phases, and, as is clear from Figure \ref{fig:P[1] slices}(iii) and Figure \ref{fig:P[1] slices}(iv), these phases -- which also admit no type IIA reduction -- are not compatible with the SCFT Coulomb branch. This is consistent with the results of \cite{Closset:2018bjz}. 

\subsubsection{Probing the Coulomb branch of the 5d SCFT}
To probe the Coulomb branch of the 5d SCFT, we set the mass parameters to zero. From \eqref{eq:p1 res a mu nu}, this implies that $\nu_1 = -\varphi_2$ and $\nu_2 = -\varphi_1$. On the field theory side, we observe that only chambers $2$ and $3$ of the $SU(3)_{\fthreehalf}$ $N_{\text{f}}=1$ theory (see Appendix \ref{sec:field theory prepotentials}) survive in this limit, and they are given by:
\begin{align}
\begin{split}
\text{chamber 2} &: \left\{\begin{array}{l}	\varphi_1 \geq 0 ~,\\
                                           	-\varphi_2 + \varphi_2 < 0 ~,\\
                                           	-\varphi_2 < 0 ~,
                           \end{array}\right. \qquad 
\text{chamber 3} : \left\{\begin{array}{l}	\varphi_1 \geq 0 ~,\\
                                           	-\varphi_2 + \varphi_2 \geq 0 ~,\\
                                           	-\varphi_2 < 0 ~.
                           \end{array}\right.
\end{split}
\end{align}
One can also verify from the Nef conditions in Table \ref{tbl:P[1] Nef conditions} that only resolutions (a) and (b) survive in this limit. The Coulomb branch of the SCFT is sketched in Figure \ref{fig:P[1] slices}(i). The red dot on the right in the figure is the conformal point.

\subsection{The $E_{2}{}^{2,\fhalf}$ singularity and $SU(3)_{\fhalf}$ $N_{\text{f}} =1$ gauge theory}
\begin{figure}[ht]
%\vspace{-10pt}
\begin{center}
\subfigure[\small{}]{
\includegraphics[width=2.5cm]{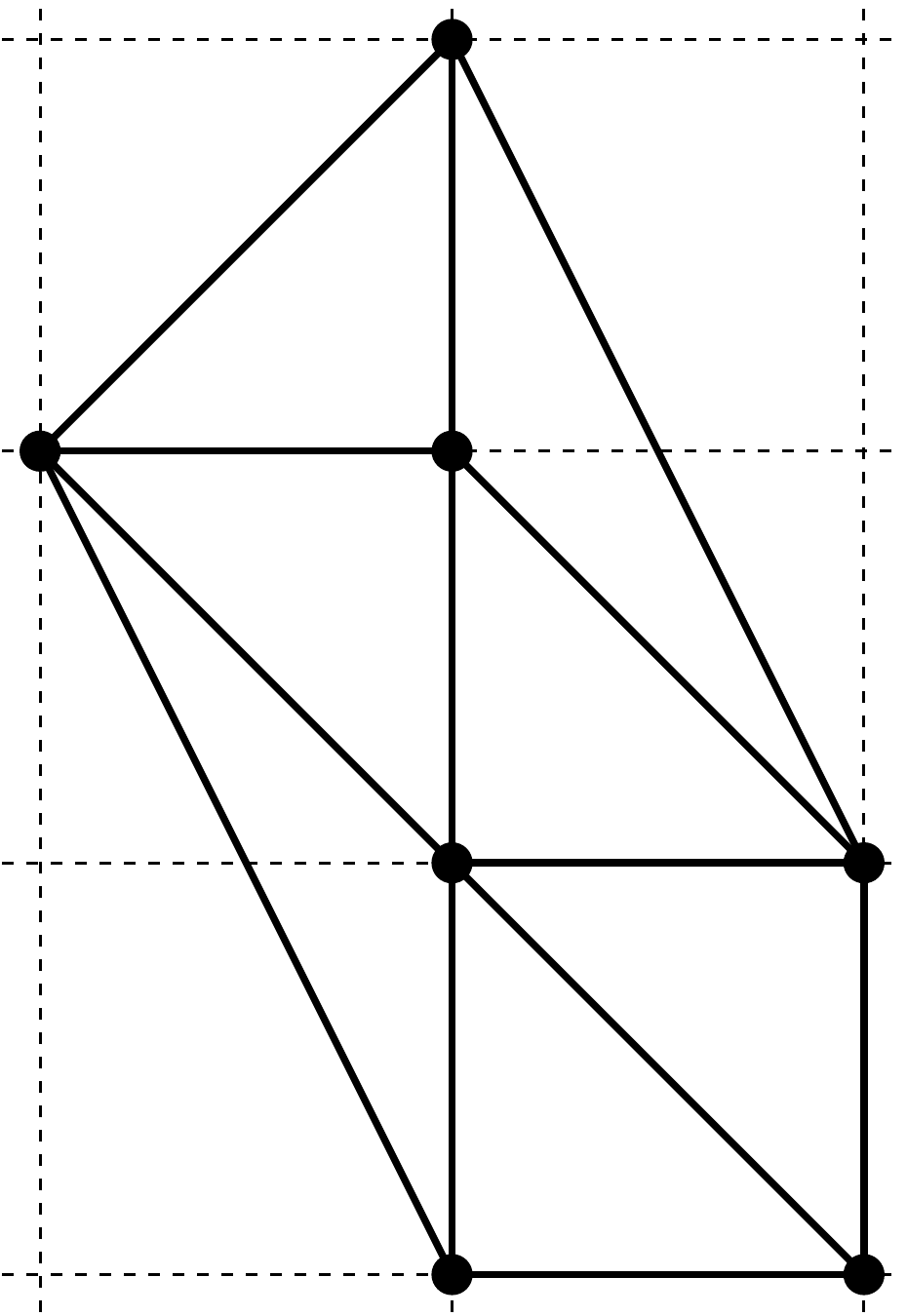}\label{fig:p2-res-a}}\,
\subfigure[\small{}]{
\includegraphics[width=2.5cm]{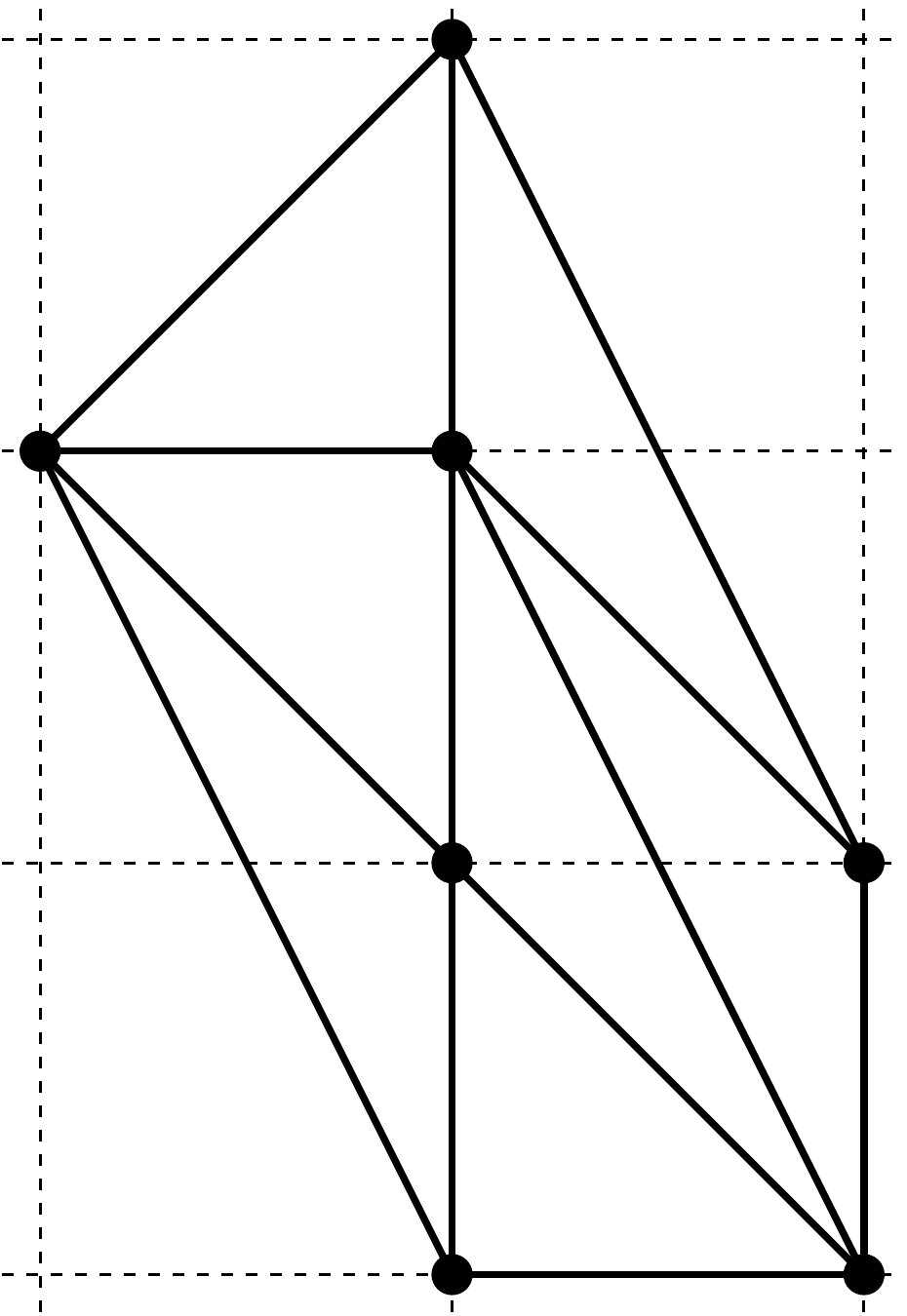}\label{fig:p2-res-b}}\,
\subfigure[\small{}]{
\includegraphics[width=2.5cm]{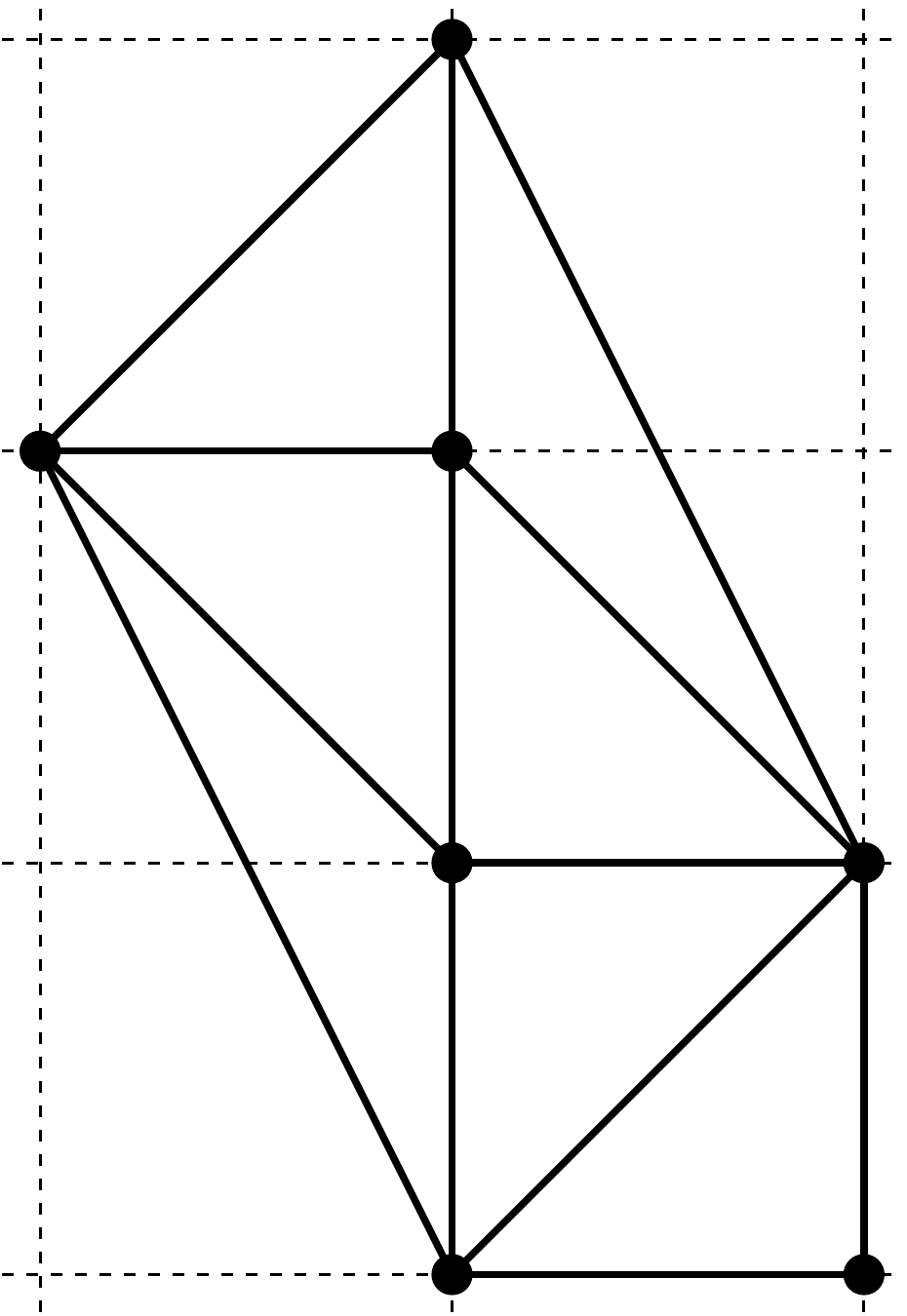}\label{fig:p2-res-c}}\,
\subfigure[\small{}]{
\includegraphics[width=2.5cm]{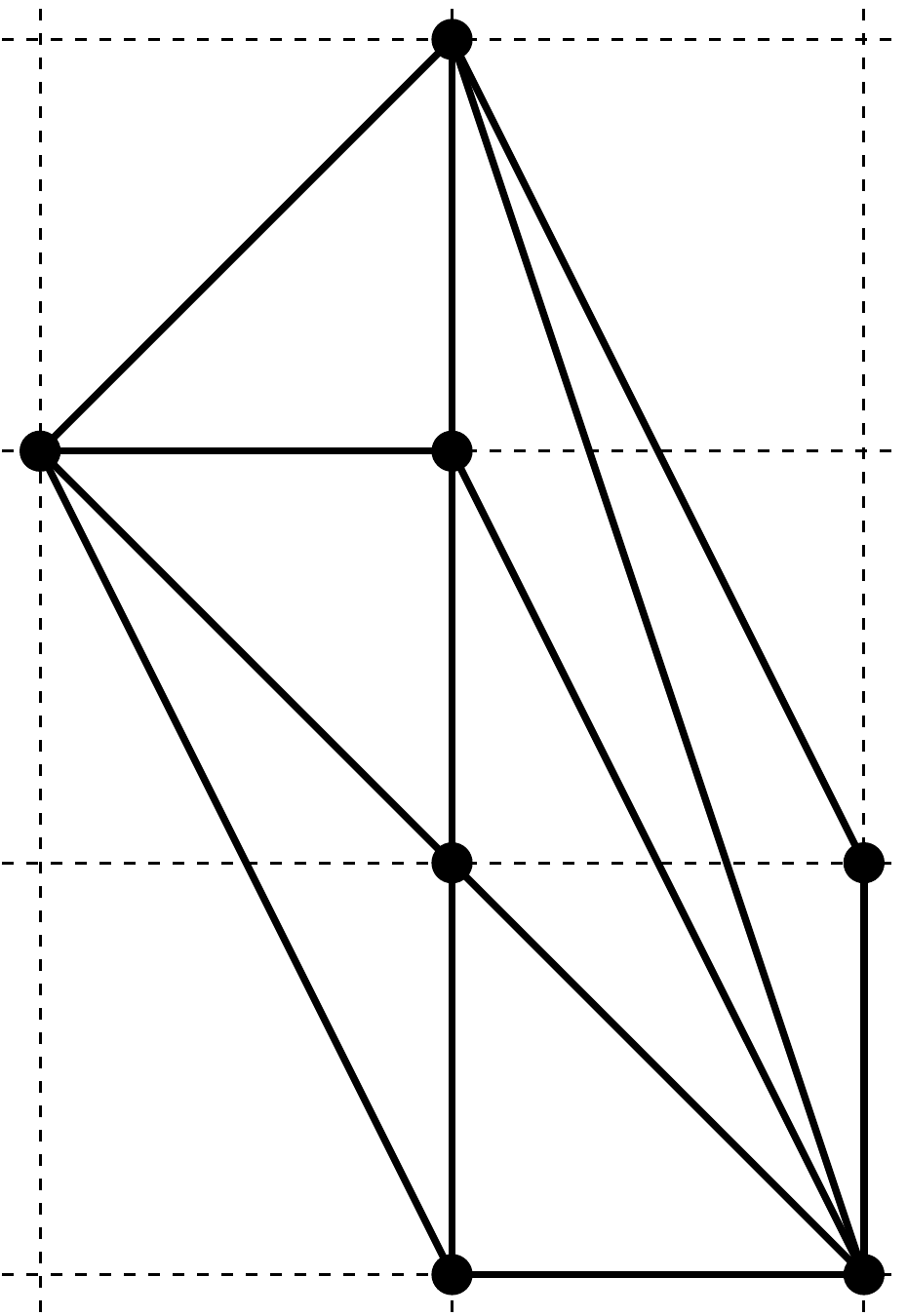}\label{fig:p2-res-d}}\\
%%%%%%%%
\subfigure[\small{}]{
\includegraphics[width=2.5cm]{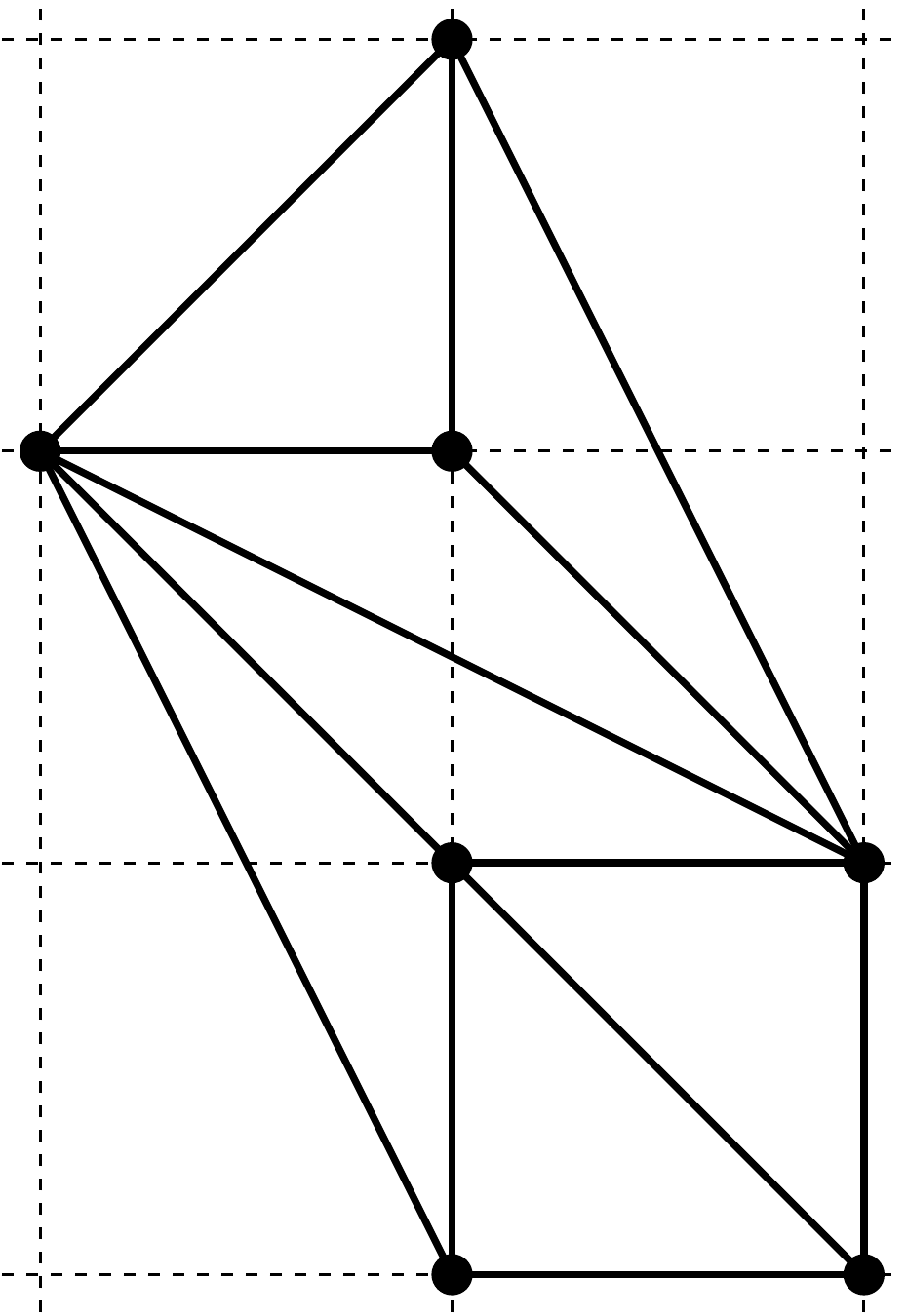}\label{fig:p2-res-e}}\,
\subfigure[\small{}]{
\includegraphics[width=2.4cm]{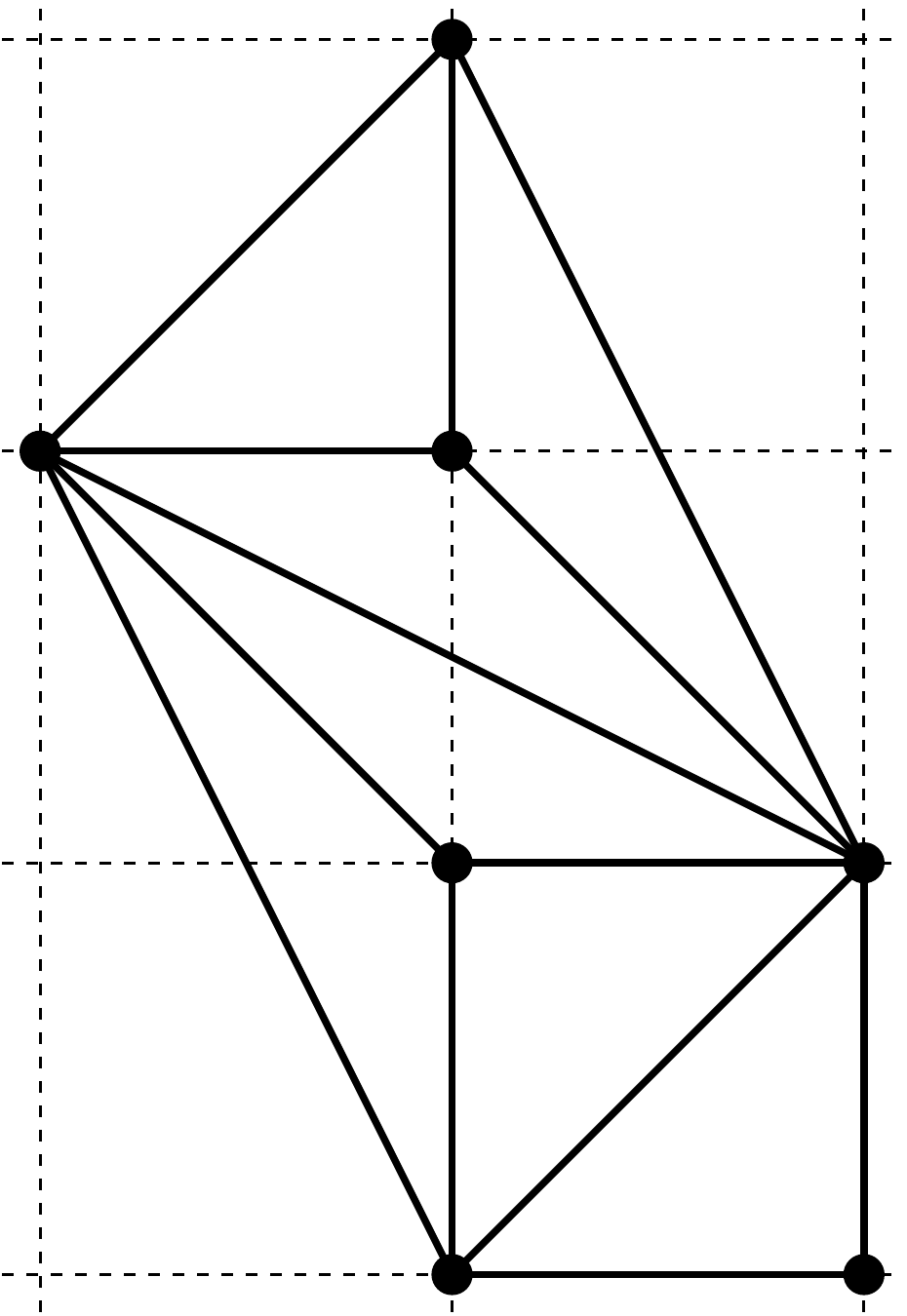}\label{fig:p2-res-f}}
\caption{The 6 crepant singularities of the $E_{2}{}^{2,\fhalf}$ singularity. The first four, (a)-(d) admit a vertical reduction, corresponding to chambers of the $SU(3)_{1}$ $N_{\text{f}}=1$ gauge theory. \label{fig:p2 crepant all}}
\end{center}
\end{figure}

\begin{figure}[ht]
\centering
\subfigure[\small{}]{
\includegraphics[width=4.5cm]{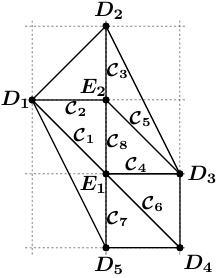}\label{fig:P[2] res a labeled}}\,
\subfigure[\small{}]{
\includegraphics[width=7.5cm]{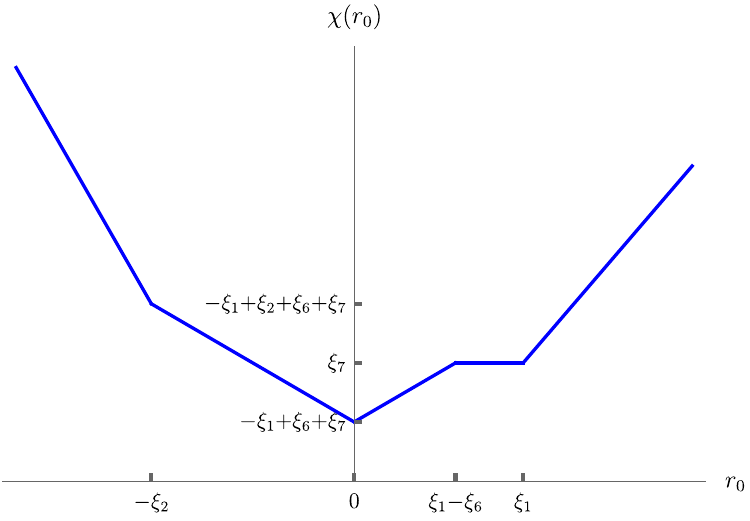}\label{fig:P[2] res a chi}}
\caption{Resolution (a) of the $E_{2}{}^{2,\fhalf}$ singularity and its vertical reduction.\label{fig:p2 res a final}}
\end{figure}

The $E_{2}{}^{2,\fhalf}$ singularity (Figure \ref{fig:p2-sing}) admits $6$ crepant resolutions shown in Figure \ref{fig:p2 crepant all}. The first four resolutions, Figures \ref{fig:p2-res-a}-\ref{fig:p2-res-d}, admit vertical reductions to type IIA, which correspond to chambers of the $SU(3)_{1}$ $N_{\text{f}}=1$ gauge thoery, as we illustrate below. Phases (e) and (f) do not admit a Lagrangian description. 
%%%%%%%%
%\vspace{-0.25in}
\paragraph{Resolution (a).} Consider the crepant resolution of Figure \ref{fig:p2-res-a}, with curves and divisors shown in Figure \ref{fig:P[2] res a labeled}.  
%%%%%%%%%%%%%%%
%%%%%%%%%%%%%%%
There are five non-compact toric divisors $D_{i}$ ($i = 1, \ldots, 5$), and two compact toric divisors $\bE_1$ and $\bE_2$ with the following linear relations:
\begin{align}\hspace{-0.25in}
D_1 &\simeq D_3 + D_4~, \quad \bE_1 \simeq D_2-D_3-2D_4-2D_5~, \quad \bE_2 \simeq -D_1-2D_2+D_4+D_5 ~. \label{eq:p2 res a divisor linear equivalences}
\end{align}
The compact curves $\CC$ are:
\begin{align}\begin{split}
	\CC_1 &= \bE_1 \cdot D_1~, \quad \CC_2 = \bE_2 \cdot D_1~, \quad \CC_3 = \bE_2 \cdot D_2~, \quad \CC_4 = \bE_1 \cdot D_3 ~,\\
	\CC_5 &= \bE_2 \cdot D_3~, \quad \CC_6 = \bE_1 \cdot D_4~, \quad \CC_7 = \bE_1 \cdot D_5~, \quad \CC_8 = \bE_2 \cdot \bE_1 ~.
	\end{split}
\end{align}
The linear relations among curve classes are:
\begin{align}
\CC_3 &\simeq -\CC_1 + \CC_2 + \CC_6 + \CC_7~, \quad \CC_4 \simeq \CC_1 - \CC_6, \quad \CC_5 \simeq \CC_2, \quad \CC_8 \simeq -\CC_1 + \CC_6 + \CC_7~.
\end{align}
We take $\{\CC_1, \CC_2, \CC_6, \CC_7\}$ as generators of the Mori cone. The requirement the compact curves $\CC_3$, $\CC_4$ and $\CC_8$ have non-negative volume imposes the following additional conditions on the FI parameters in this chamber:
\begin{align}	
-\xi_1 + \xi_2 + \xi_6 + \xi_7 &\geq 0~, \quad \xi_1 -\xi_6 \geq 0~, \quad -\xi_1 + \xi_6 + \xi_7 \geq 0 ~. \label{eq:additional FI conditions P[2] res a}
\end{align}
The GLSM charge matrix is:
\be\label{intersect p2 res a}
\begin{tabular}{l|ccccccc|c}
 & $D_1$ &$D_2$& $D_3$ & $D_4$ & $D_5$ & $\bE_1$& $\bE_2$ & vol($\CC$) \\
 \hline
$\CC_1$  & $0$ & $0$ & $0$ &  $0$& $1$ & $-2$ & $1$ & $\xi_1$ \\ \hline
$\CC_2$  & $0$ & $1$ & $0$ &  $0$& $0$ & $1$ & $-2$ & $\xi_2$ \\ \hline
$\CC_6$  & $0$ & $0$ & $1$ & $-1$ & $1$ & $-1$ & $0$ & $\xi_6$ \\ \hline
$\CC_7$  & $1$ & $0$ & $0$ & $1$ & $0$ & $-2$ & $0$ & $\xi_7$ \\ \hline\hline
$U(1)_M$ & $0$& $0$ &$0$ & $0$ & $0$ & $-1$ & $1$ & $r_0$
 \end{tabular}
\ee 
%where the last line defines the vertical reduction of the 2d GLSM. The nonnegative FI terms $\xi_1 \geq 0$, $\xi_2 \geq 0$, $\xi_6 \geq 0$ and $\xi_7 \geq 0$ are, respectively, the volumes of compact curves $\CC_1$, $\CC_2$, $\CC_6$ and $\CC_7$. 
\paragraph{Geometric prepotential.} We parametrize the K\"{a}hler cone by:
\begin{align}
 S &= \mu_1 D_1 + \mu_2 D_2 + \nu_1 \bE_1 + \nu_2 \bE_2 ~.	\label{eq:kah cone p2}
\end{align}
The parameters $(\mu_1, \mu_2, \nu_1, \nu_2)$ are related to the FI parameters by:
\begin{align}
\xi_1 &= -2\nu_1+\nu_2 \geq 0~, \quad \xi_2 = \mu_2 + \nu_1 - 2\nu_2 \geq 0~, \quad \xi_6 = -\nu_1 \geq 0~, \quad \xi_7 = \mu_1 - 2\nu_1 \geq 0 ~.	 \label{eq:p2 res a FI}
\end{align}
The relevant triple-intersection numbers are:
\begin{align}
\begin{array}{c@{~,\quad}c@{~,\quad}c@{~,\quad}c@{~\quad}}
  D_1 \bE_1 \bE_2 = 1 & D_2 \bE_1 \bE_2 = 0 & D_1 D_2 \bE_1 = 0 & D_1 D_2 \bE_2 = 1 ~,\\
  D_1\bE_1^2 = -2 & D_2 \bE_1^2 = 0 & D_1\bE_2^2 = -2 & D_2\bE_2^2 = -3 ~,\\
  D_1^2 \bE_1 = 0 & D_1^2 \bE_2 = 0 & D_2^2 \bE_1 = 0 & D_2^2 \bE_2 = 1 ~,\\
  \bE_1^2 \bE_2 = -1 & \bE_1 \bE_2^2 = -1 & \bE_1^3 = 7 & \bE_2^3 = 8 ~.
\end{array}
\end{align}
Therefore, the compact part of the prepotential is:
\begin{align}
\mathcal{F}_{(a)}(\nu_1, \nu_2; \mu_1, \mu_2) = -\frac{1}{6}S^3 &= -\frac{7}{6}\nu_1^3 - \frac{4}{3}\nu_2^3 + \frac{1}{2}(\nu_1^2\nu_2 + \nu_1\nu_2^2)+ \mu_1 \nu_1^2-  \mu_1 \nu_1 \nu_2\nonumber\\
& \quad  + \left(\mu_1 + \frac{3}{2}\mu_2\right)\nu_2^2 - \left(\mu_1\mu_2 + \frac{1}{2}\mu_2^2\right)\nu_2 ~. \label{eq:p2 res a geo prepot}
\end{align}
\paragraph{Type IIA reduction and gauge theory description.} The type IIA background is again resolved $A_1$ singularity fibered over the $x^9 = r_0$ direction. There are three D6-branes wrapping the exceptional $\P^1$ in the resolved $A_1$ singularity, resulting in an $SU(3)$ gauge theory. There is also a D6-brane wrapping a noncompact divisor in the resolved ALE space, which corresponds to one fundamental flavor. The volume of the exceptional $\P^1$ is given by the following piecewise linear function:
\begin{align}\label{eq:p2 res a chi}
\chi(r_0) &= \left\{ 
                 \begin{array}{ll}
                 	-3r_0 - \xi_1 - 2 \xi_2 + \xi_6 + \xi_7, & \text{for } r_0 \leq -\xi_2\\
                 	-r_0 - \xi_1 + \xi_6 + \xi_7, & \text{for } -\xi_2 \leq r_0 \leq 0\\
                 	+r_0 - \xi_1 + \xi_6 + \xi_7, & \text{for } 0 \leq r_0 \leq \xi_1-\xi_6\\
                 	\xi_7, & \text{for } \xi_1-\xi_6 \leq r_0 \leq \xi_1 \\
                 	+2r_0 - 2\xi_1 + \xi_7, & \text{for } r_0 \geq \xi_1 ~.
                 \end{array}
             \right.
\end{align}
This function is sketched in Figure \ref{fig:P[2] res a chi}. At the points $r_0 = -\xi_2$, $r_0 = 0$ and $r_0 = \xi_1$, there are gauge D6-branes wrapping $\P^1$'s in the resolution of the singularity. When $\xi_1 = \xi_2 =0$, an $SU(3)$ gauge theory is realized with coupling $h_0 = \xi_6 + \xi_7$. There is a flavor D6-brane at $r_0 = \xi_1-\xi_6$. The effective Chern-Simons level is given by $\kappa_{s,\text{eff}} = -\frac{1}{2}(-3 + 2) = \frac{1}{2}$, which is interpreted as a bare CS level of $1$ plus the contribution $-\frac{1}{2}$ due to the single hypermultiplet (cf. \eqref{eq:massive Dirac fermion collection U(1) half quantization}). The simple-root W-bosons have masses:
\begin{align}
M(W_1) &= \xi_2 = 2\varphi_1 - \varphi_2~, \quad M(W_2) = \xi_1 = 2\varphi_2 - \varphi_1  ~. 
\end{align}
From the instanton masses, one can identify that this resolution corresponds to gauge theory chamber 3 (cf. Table \ref{tbl:u3 nf1 instantons} and \eqref{eq:ft prepot SU(3) k Nf1 chamber 3}):
\begin{align}
\begin{split}
	M(\cI_1) &= \chi(r_0 = -\xi_2) = -\xi_1+\xi_2+\xi_6+\xi_7 = h_0 + 3\varphi_1 - m ~,  \\
	M(\cI_2) &= \chi(r_0 = 0) = -\xi_1+\xi_6+\xi_7 = h_0 + \varphi_1+\varphi_2 - m ~,\\
	M(\cI_3) &= \chi(r_0 = \xi_1) = \xi_7 = h_0 + 2\varphi_2 ~.
\end{split}	
\end{align}
The masses of hypermultiplets are:
\begin{align}
\begin{split}
	M(\cH_1) &= \xi_6 = \varphi_2 - m ~,\\
	M(\cH_2) &= \xi_1-\xi_6 = -\varphi_1 + \varphi_2 + m ~,\\
	M(\cH_3) &= \xi_1 + \xi_2-\xi_6  = \varphi_1 + m ~.
\end{split}	
\end{align}
From the K\"{a}hler volumes \eqref{eq:p2 res a FI} of the compact curves and masses of W-bosons and instantons, the map between geometry and field theory variables is determined to be:
\begin{align}
\mu_1 &= h_{0} + 2m~, \quad \mu_2 = 3 m~, \quad \nu_1 = -\varphi_2 + m~, \quad \nu_2 = -\varphi_1 + 2m ~.	\label{eq:p2 res a mu nu}
\end{align} 
Plugging \eqref{eq:p2 res a mu nu} into \eqref{eq:p2 res a geo prepot}, we recover the field theory prepotential,
\begin{align}
\mathcal{F}_{SU(3)_{\fhalf},N_{\text{f}}=1}^{\text{chamber 3}} &= \frac{4}{3}\varphi_1^3 + \frac{7}{6}\varphi_2^3 -\frac{1}{2}(\varphi_1^2\varphi_2 + \varphi_1\varphi_2^2) + (h_0-m)\varphi_1^2 + \left(h_{0}-\tfrac{m}{2}\right)\varphi_2^2\nonumber\\
&\quad + (m-h_{0})\varphi_1\varphi_2 - \tfrac{m^2}{2}\varphi_2 ~,
\end{align}
up to $\varphi$-independent terms. %From field theory, the monopole string tensions are given by:
%\begin{align} 
%T_{1,\text{ft}} &= \frac{\partial \mathcal{F}_{SU(3)_{3/2},N_{\text{f}}=1}^{\text{chamber 3}}}{\partial \varphi_1} = 4\varphi_1^2 - \varphi_1 \varphi_2 - \frac{1}{2}\varphi_2^2 + 2(h_{0}{-}m)\varphi_{1} + (m{-}h_{0})\varphi_{2} ~,\\
%T_{2,\text{ft}} &= \frac{\partial \mathcal{F}_{SU(3)_{3/2},N_{\text{f}}=1}^{\text{chamber 3}}}{\partial \varphi_2} = \frac{7}{2}\varphi_2^2 -\varphi_1\varphi_2 - \frac{1}{2}\varphi_1^2 + (m{-}h_{0})\varphi_1 + (2h_{0}{-}m)\varphi_2 -\frac{m^2}{2} ~,
%\end{align}
The monopole string tensions from $\chi(r_0)$ are given by:
\begin{align} 
T_{1,\text{geo}} &= \int_{-\xi_2}^{0}\chi(r_0)\,dr_{0} = \xi_2\left(-\xi_1 + \frac{1}{2}\xi_2 + \xi_6 + \xi_7\right) ~,\\
T_{2,\text{geo}} &= \int_{0}^{\xi_1}\chi(r_0)\,dr_{0} = -\frac{\xi _1^2}{2}+\xi _6 \xi _1+\xi _7 \xi _1-\frac{\xi _6^2}{2} ~.
\end{align} 
Using the map $\xi_1 = 2\varphi_2-\varphi_1$, $\xi_2 = 2\varphi_1 - \varphi_2$, $\xi_6 = \varphi_2-m$ and $\xi_7 = h_{0} + 2\varphi_2$, one can verify that indeed $T_{i,\text{ft}} = T_{i,\text{geo}}$ for $i = 1, 2$. The tensions vanishes at loci given by:
\begin{align}
\begin{split}
 (I)&: \{\xi_2 = 0\} \cup \left\{-\xi_1 + \frac{1}{2}\xi_2 + \xi_6 + \xi_7 = 0\right\}~, ~ \text{ and },\\
 (II)&: \left\{ -\frac{\xi _1^2}{2}+\xi _6 \xi _1+\xi _7 \xi _1-\frac{\xi _6^2}{2} =0 \right\} ~.
 \end{split}
\end{align}
Along the submanifold $\{\xi_2 = 0\} \subset (I)$, the W-boson $W_1$ becomes massless, signaling a hard wall. The submanifold $\{-\xi_1 + \tfrac{1}{2}\xi_2 + \xi_6 + \xi_7 = 0\}$ is not part of the K\"{a}hler chamber of resolution (a). Solving the quadratic equation in $(II)$, we get two solutions: $\xi_6 = \xi_1 \pm \sqrt{2\xi_1 \xi_7}$. Both sign choices are inconsistent with \eqref{eq:additional FI conditions P[2] res a}, and are hence rejected. Note that away from any hard wall, the perturbative hypermultiplet $\cH_1$ can become massless at $\xi_6 = 0$ (signaling a flop of $\CC_6$), leading to gauge theory resolution (c), or the hypermultiplet $\cH_2$ can become massless along the locus $\xi_4 = \xi_1{-}\xi_6 = 0$ (signaling a flop of $\CC_4$), leading to gauge theory resolution (b). Note that the intersection of the loci $(II)$ above with the loci $\{\xi_1 = \xi_6\}$ is $\xi_1 \xi_7 = 0$, is inconsistent in this K\"{a}hler chamber, as neither the W-boson $W_2$ (with mass $\xi_1$) can become massless (except at the hard wall) nor can the curve $\CC_7$ flop in this chamber. This is a reassuring consistency check. % there is no magnetic wall in resolution (a) of $E_{2}{}^{2,\fhalf}$.

%%%%%%
% add phases (b), (c), (d) for $E_{2}{}^{2,\fhalf}$, comment on RG flow for phases (e), (f) too
% P[2] res C chi profile is drawn for xi2 > xi1 maybe make a note of this somewhere
%%%%%%

\begin{figure}[ht]
\centering
\subfigure[\small{}]{
\includegraphics[width=4.5cm]{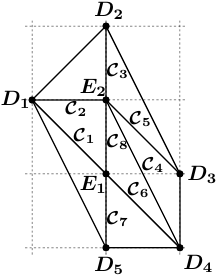}\label{fig:P[2] res b labeled}}\,
\subfigure[\small{}]{
\includegraphics[width=7.5cm]{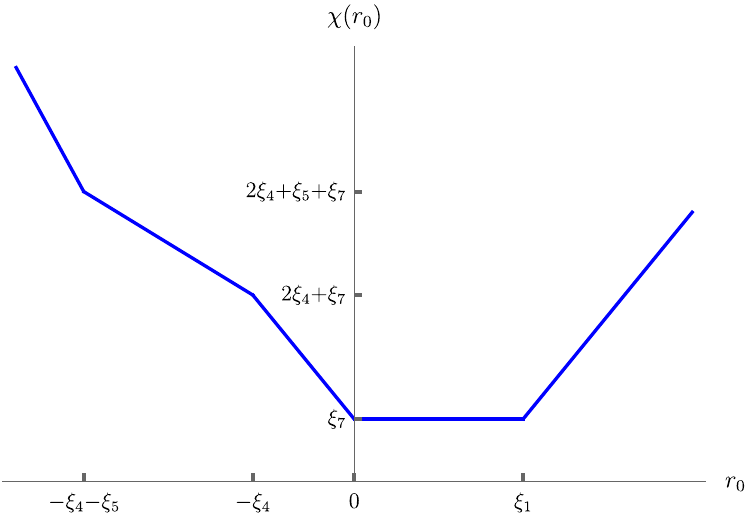}\label{fig:P[2] res b chi}}
\caption{Resolution (b) of the $E_{2}{}^{2,\fhalf}$ singularity and its vertical reduction.\label{fig:p2 res b final}}
\end{figure}

\paragraph{Resolution (b).} Consider the crepant resolution of Figure \ref{fig:p2-res-b}, with curves and divisors shown in Figure \ref{fig:P[2] res b labeled}.  
%%%%%%%%%%%%%%%
%%%%%%%%%%%%%%%
The linear relations among the toric divisors are still given by \eqref{eq:p2 res a divisor linear equivalences}.
%\begin{align}
%D_1 &\simeq D_3 + D_4, \quad \bE_1 \simeq D_2-D_3-2D_4-2D_5 \bE_1, \quad \bE_2 \simeq -D_1-2D_2+D_4+D_5 ~. \label{eq:p2 res a divisor linear equivalences}
%\end{align}
The compact curves $\CC$ can be read off the toric diagram.
%\begin{align}\begin{split}
%	\CC_1 &= \bE_1 \cdot D_1, \quad \CC_2 = \bE_2 \cdot D_1, \quad \CC_3 = \bE_2 \cdot D_2, \quad \CC_4 = \bE_2 \cdot D_4 ~,\\
%	\CC_5 &= \bE_2 \cdot D_3, \quad \CC_6 = \bE_1 \cdot D_4, \quad \CC_7 = \bE_1 \cdot D_5, \quad \CC_8 = \bE_2 \cdot \bE_1 ~.
%	\end{split}
%\end{align}
The linear relations among curve classes are:
\begin{align}
\CC_2 &\simeq \CC_4 + \CC_5~, \quad \CC_3 \simeq 2\CC_4 + \CC_5 + \CC_7~, \quad \CC_6 \simeq \CC_1~, \quad \CC_8 \simeq  \CC_7 ~.
\end{align}
We take $\{\CC_1, \CC_4, \CC_5, \CC_7\}$ as generators of the Mori cone. The GLSM charge matrix is:
\be\label{intersect p2 res b}
\begin{tabular}{l|ccccccc|c}
 & $D_1$ &$D_2$& $D_3$ & $D_4$ & $D_5$ & $\bE_1$& $\bE_2$ & vol($\CC$) \\
 \hline
$\CC_1$  & $0$ & $0$ & $0$ &  $0$& $1$ & $-2$ & $1$ & $\xi_1$ \\ \hline
$\CC_4$  & $0$ & $0$ & $1$ &  $-1$& $0$ & $1$ & $-1$ & $\xi_4$ \\ \hline
$\CC_5$  & $0$ & $1$ & $-1$ & $1$ & $0$ & $0$ & $-1$ & $\xi_5$ \\ \hline
$\CC_7$  & $1$ & $0$ & $0$ & $1$ & $0$ & $-2$ & $0$ & $\xi_7$ \\ \hline\hline
$U(1)_M$ & $0$& $0$ &$0$ & $0$ & $0$ & $-1$ & $1$ & $r_0$
 \end{tabular}
\ee 
%where the last line defines the vertical reduction of the 2d GLSM. The nonnegative FI terms $\xi_1 \geq 0$, $\xi_4 \geq 0$, $\xi_5 \geq 0$ and $\xi_7 \geq 0$ are, respectively, the volumes of compact curves $\CC_1$, $\CC_4$, $\CC_5$ and $\CC_7$. 
The K\"{a}hler cone is parametrized by \eqref{eq:kah cone p2}. The parameters $(\mu_1, \mu_2, \nu_1, \nu_2)$ are related to the FI parameters by: 
\begin{align}
\xi_1 &= -2\nu_1 + \nu_2 \geq 0~, \quad \xi_4 =  \nu_1 - \nu_2 \geq 0~, \quad \xi_5 = \mu_2 - \nu_2 \geq 0~, \quad \xi_7 = \mu_1 - 2\nu_1 \geq 0  ~.	 \label{eq:p2 res b FI}
\end{align}
The relevant triple-intersection numbers are:
\begin{align}
\begin{array}{c@{~,\quad}c@{~,\quad}c@{~,\quad}c@{~\quad}}
  D_1 \bE_1 \bE_2 = 1 & D_2 \bE_1 \bE_2 = 0 & D_1 D_2 \bE_1 = 0 & D_1 D_2 \bE_2 = 1 ~,\\
  D_1\bE_1^2 = -2 & D_2 \bE_1^2 = 0 & D_1\bE_2^2 = -2 & D_2\bE_2^2 = -3 ~,\\
  D_1^2 \bE_1 = 0 & D_1^2 \bE_2 = 0 & D_2^2 \bE_1 = 0 & D_2^2 \bE_2 = 1 ~,\\
  \bE_1^2 \bE_2 = -2 & \bE_1 \bE_2^2 = 0 & \bE_1^3 = 8 & \bE_2^3 = 7 ~.
\end{array}
\end{align}
Therefore, the compact part of the prepotential is:
\begin{align}
\mathcal{F}_{(b)}(\nu_1, \nu_2; \mu_1, \mu_2) = -\frac{1}{6}S^3 &= -\frac{4}{3}\nu_1^3 - \frac{7}{6}\nu_2^3 + \nu_1^2 \nu_2 + \mu_1 \nu_1^2 + \left(\mu_1 + \tfrac{3}{2}\mu_2\right)\nu_2^2\nonumber\\
&\qquad - \mu_1 \nu_1 \nu_2 - \left(\mu_1 \mu_2 + \tfrac{1}{2}\mu_2^2\right)\nu_2 ~. \label{eq:p2 res b geo prepot}
\end{align}
The type IIA profile is:
\begin{align}\label{eq:p2 res b chi}
\chi(r_0) &= \left\{ 
                 \begin{array}{ll}
                 	-3r_0 -\xi_4-2\xi_5+\xi_7, & \text{for } r_0 \leq -\xi_4-\xi_5\\
                 	-r_0 + \xi_4 + \xi_7, & \text{for } -\xi_4-\xi_5 \leq r_0 \leq -\xi_4\\
                 	-2r_0 + \xi_7, & \text{for } -\xi_4 \leq r_0 \leq 0\\
                 	\xi_7, & \text{for } 0 \leq r_0 \leq \xi_1 \\
                 	+2r_0 - 2\xi_1 + \xi_7, & \text{for } r_0 \geq \xi_1 ~.
                 \end{array}
             \right.
\end{align}
This function is sketched in Figure \ref{fig:P[2] res b chi}. At the points $r_0 = -\xi_4-\xi_5$, $r_0 = 0$ and $r_0 = \xi_1$, there are gauge D6-branes wrapping $\P^1$'s in the resolution of the singularity. There is a flavor D6-brane at $r_0 = -\xi_4$. The effective Chern-Simons level is, of course, still $\frac{1}{2}$, as for resolution (a). The simple-root W-bosons have masses given by:
\begin{align}
M(W_1) &= \xi_4 + \xi_5 = 2\varphi_1 - \varphi_2~, \qquad M(W_2) = \xi_1 = 2\varphi_2 - \varphi_1  ~. 
\end{align}
This resolution corresponds to gauge theory chamber 2 (cf. Table \ref{tbl:u3 nf1 instantons} and \eqref{eq:ft prepot SU(3) k Nf1 chamber 2}), with instanton masses given by:
\begin{align}
\begin{split}
	M(\cI_1) &= \chi(r_0 = -\xi_4-\xi_5) = 2\xi_4 + \xi_5 + \xi_7 = h_0 + 3\varphi_1 - m ~,  \\
	M(\cI_2) &= \chi(r_0 = 0) = \xi_7 = h_0 + 2\varphi_2 ~,\\
	M(\cI_3) &= \chi(r_0 = \xi_1) = \xi_7 = h_0 + 2\varphi_2 ~.
\end{split}	
\end{align}
The masses of hypermultiplets are:
\begin{align}
\begin{split}
	M(\cH_1) &= \xi_5 = \varphi_1 + m ~,\\
	M(\cH_2) &= \xi_4 = \varphi_1 - \varphi_2 - m ~,\\
	M(\cH_3) &= \xi_1 + \xi_4  = \varphi_2 - m ~.
\end{split}	
\end{align}
The map between geometry and field theory variables is still given by \eqref{eq:p2 res a mu nu}, and plugging it into \eqref{eq:p2 res b geo prepot}, we recover the field theory prepotential,
\begin{align}\hspace{-0.25in}
\mathcal{F}_{SU(3)_{3/2},N_{\text{f}}=1}^{\text{chamber 2}} &= \frac{7}{6}\varphi_1^3 + \frac{4}{3}\varphi_2^3 - \varphi_1\varphi_2^2 + \left(h_0-\frac{m}{2}\right)\varphi_1^2 + h_{0}\varphi_2^2 - h_{0}\varphi_1\varphi_2 -\frac{m^2}{2}\varphi_2 ~,
\end{align}
up to $\varphi$-independent terms. %From field theory, the monopole string tensions are:
%\begin{align} 
%T_{1,\text{ft}} = \frac{\partial \mathcal{F}_{SU(3)_{3/2},N_{\text{f}}=1}^{\text{chamber 2}}}{\partial \varphi_1} &= -\frac{m^2}{2}+\left(-m+2 h_0\right) \varphi _1+\frac{7 \varphi _1^2}{2}-h_0 \varphi _2-\varphi _2^2 ~,\\
%T_{2,\text{ft}} = \frac{\partial \mathcal{F}_{SU(3)_{3/2},N_{\text{f}}=1}^{\text{chamber 2}}}{\partial \varphi_2} &= -h_0 \varphi _1+2 h_0 \varphi _2-2 \varphi _1 \varphi _2+4 \varphi _2^2 ~,
%\end{align}
The monopole string tensions from $\chi(r_0)$ are given by:
\begin{align} 
T_{1,\text{geo}} &= \int_{-\xi_4-\xi_5}^{0}\chi(r_0)\,dr_{0} = \xi _4^2+\left(2 \xi _5+\xi _7\right) \xi _4+\frac{1}{2} \xi _5 \left(\xi _5+2 \xi _7\right) ~,\\
T_{2,\text{geo}} &= \int_{0}^{\xi_1}\chi(r_0)\,dr_{0} =  \xi _1 \xi _7 ~.
\end{align} 
Using the map $\xi_1 = 2\varphi_2-\varphi_1$, $\xi_4 = \varphi_1 - \varphi_2-m$, $\xi_5 = \varphi_1+m$ and $\xi_7 = h_{0} + 2\varphi_2$, we find that $T_{i,\text{ft}} = T_{i,\text{geo}}$ for $i = 1, 2$. The tensions vanish at loci given by:
\begin{align}\hspace{-0.2in}
 (I)&: \left\{\xi _4^2+\left(2 \xi _5+\xi _7\right) \xi _4+\frac{1}{2} \xi _5 \left(\xi _5+2 \xi _7\right) = 0\right\}~, ~ \text{and}~%\\
 (II): \left\{ \xi_1 =0 \right\} \cup \{\xi_7 = 0\}~.
\end{align}
The solution to the quadratic equation from $(I)$ is $\xi_4 = \frac{1}{2}(-2\xi_5-\xi_7 \pm \sqrt{2\xi_5^2 + \xi_7^2})$. Both sign choices lead to a negative value for $\xi_4$ in this chamber, and are hence rejected. The loci $\{\xi_1=0\}\subset (II)$ and $\{\xi_7=0\}\subset (II)$ coincide with hard walls, which are, respectively, loci along which the W-boson $W_2$ becomes massless and the instanton particles $\cI_2$, $\cI_3$ become massless. These are both non-traversible walls. Away from the hard wall, either hypermultiplet $\cH_1$ can become massless at $\xi_5 = 0$ (signaling a flop of $\CC_5$), leading to gauge theory resolution (d), or hypermultiplet $\cH_2$ can become massless at $\xi_4 = 0$ (signaling a flop of $\CC_4$) leading to back to gauge theory resolution (a).

\begin{figure}[ht]
\centering
\subfigure[\small{}]{
\includegraphics[width=4.5cm]{images/p2-res-c-new-labeled}\label{fig:P[2] res c labeled}}\,
\subfigure[\small{}]{
\includegraphics[width=7.5cm]{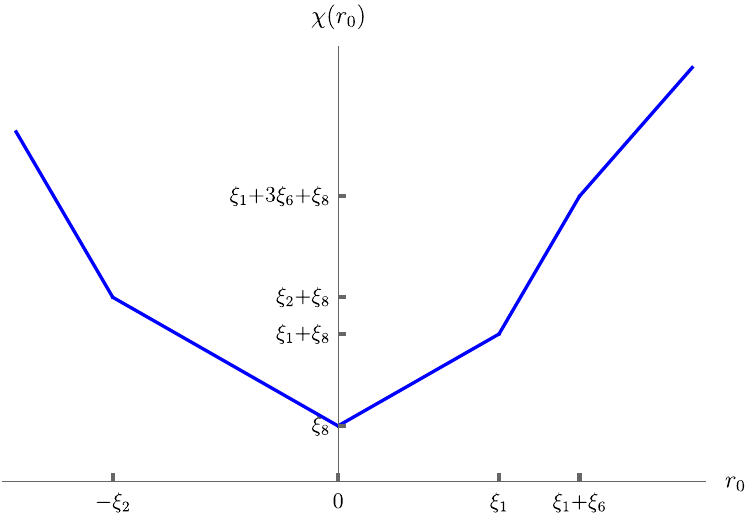}\label{fig:P[2] res c chi}}
\caption{Resolution (c) of the $E_{2}{}^{2,\fhalf}$ singularity and its vertical reduction.\label{fig:p2 res c final}}
\end{figure}

\paragraph{Resolution (c).} Consider the crepant resolution of Figure \ref{fig:p2-res-c}, with curves and divisors shown in Figure \ref{fig:P[2] res c labeled}.  
%%%%%%%%%%%%%%%
%%%%%%%%%%%%%%%
%The linear relations among the toric divisors given by \eqref{eq:p2 res a divisor linear equivalences}. The compact curves $\CC$ are given by:
%\begin{align}\begin{split}
%	\CC_1 &= \bE_1 \cdot D_1, \quad \CC_2 = \bE_2 \cdot D_1, \quad \CC_3 = \bE_2 \cdot D_2, \quad \CC_4 = \bE_1 \cdot D_3 ~,\\
%	\CC_5 &= \bE_2 \cdot D_3, \quad \CC_6 = D_3 \cdot D_5, \quad \CC_7 = \bE_1 \cdot D_5, \quad \CC_8 = \bE_2 \cdot \bE_1 ~.
%	\end{split}
%\end{align}

\noindent The linear relations among curve classes are:
\begin{align}
\CC_3 &\simeq \CC_5 + \CC_8~, \quad \CC_4 \simeq \CC_1~, \quad \CC_5 \simeq \CC_2~, \quad \CC_7 \simeq  \CC_1 + \CC_8 ~.
\end{align}
We take $\{\CC_1, \CC_2, \CC_6, \CC_8\}$ as generators of the Mori cone. The GLSM charge matrix is:
\be\label{intersect p2 res c}
\begin{tabular}{l|ccccccc|c}
 & $D_1$ &$D_2$& $D_3$ & $D_4$ & $D_5$ & $\bE_1$& $\bE_2$ & vol($\CC$) \\
 \hline
$\CC_1$  & $0$ & $0$ & $0$ &  $0$& $1$ & $-2$ & $1$ & $\xi_1$ \\ \hline
$\CC_2$  & $0$ & $1$ & $0$ &  $0$& $0$ & $1$ & $-2$ & $\xi_2$ \\ \hline
$\CC_6$  & $0$ & $0$ & $-1$ & $1$ & $-1$ & $1$ & $0$ & $\xi_6$ \\ \hline
$\CC_8$  & $1$ & $0$ & $1$ & $0$ & $0$ & $-1$ & $-1$ & $\xi_8$ \\ \hline\hline
$U(1)_M$ & $0$& $0$ &$0$ & $0$ & $0$ & $-1$ & $1$ & $r_0$
 \end{tabular}
\ee 
The K\"{a}hler cone is parametrized by \eqref{eq:kah cone p2}. The parameters $(\mu_1, \mu_2, \nu_1, \nu_2)$ are related to the FI parameters by
\begin{align}
\xi_1 &= -2\nu_1 + \nu_2 \geq 0~, \quad \xi_2 = \mu_2 + \nu_1 - 2\nu_2 \geq 0~, \quad \xi_6 = \nu_1 \geq 0~, \quad \xi_8 = \mu_1 - \nu_1 - \nu_2 \geq 0 ~.	 \label{eq:p2 res c FI}
\end{align}
The relevant triple-intersection numbers are:
\begin{align}
\begin{array}{c@{~,\quad}c@{~,\quad}c@{~,\quad}c@{~\quad}}
  D_1 \bE_1 \bE_2 = 1 & D_2 \bE_1 \bE_2 = 0 & D_1 D_2 \bE_1 = 0 & D_1 D_2 \bE_2 = 1 ~,\\
  D_1\bE_1^2 = -2 & D_2 \bE_1^2 = 0 & D_1\bE_2^2 = -2 & D_2\bE_2^2 = -3 ~,\\
  D_1^2 \bE_1 = 0 & D_1^2 \bE_2 = 0 & D_2^2 \bE_1 = 0 & D_2^2 \bE_2 = 1 ~,\\
  \bE_1^2 \bE_2 = -1 & \bE_1 \bE_2^2 = -1 & \bE_1^3 = 8 & \bE_2^3 = 8 ~.
\end{array}
\end{align}
Therefore, the compact part of the prepotential is:
\begin{align}
\mathcal{F}_{(c)}(\nu_1, \nu_2; \mu_1, \mu_2) = -\frac{1}{6}S^3 &= -\frac{4}{3}\nu_1^3 - \frac{4}{3}\nu_2^3 + \frac{1}{2}(\nu_1^2\nu_2 + \nu_1\nu_2^2) + \mu_1 \nu_1^2 + \left(\mu_1 + \frac{3}{2}\mu_2\right)\nu_2^2\nonumber\\
&\qquad - \mu_1 \nu_1 \nu_2 -\left(\mu_1\mu_2 + \frac{1}{2}\mu_2^2\right)\nu_2 ~. \label{eq:p2 res c geo prepot}
\end{align}
The IIA profile is:
\begin{align}\label{eq:p2 res c chi}
\chi(r_0) &= \left\{ 
                 \begin{array}{ll}
                 	-3r_0 -2\xi_2 + \xi_8, & \text{for } r_0 \leq -\xi_2\\
                 	-r_0 + \xi_8, & \text{for } -\xi_2 \leq r_0 \leq 0\\
                 	r_0 + \xi_8, & \text{for } 0 \leq r_0 \leq \xi_1\\
                 	3r_0 - 2\xi_1 + \xi_8, & \text{for } \xi_1 \leq r_0 \leq \xi_1 + \xi_6 \\
                 	+2r_0 -\xi_1 + \xi_6 + \xi_8, & \text{for } r_0 \geq \xi_1 + \xi_6 ~.
                 \end{array}
             \right.
\end{align}
This function is sketched in Figure \ref{fig:P[2] res c chi}. At the points $r_0 = -\xi_2$, $r_0 = 0$ and $r_0 = \xi_1$, there are gauge D6-branes wrapping $\P^1$'s in the resolution of the singularity. There is a flavor D6-brane at $r_0 = \xi_1 + \xi_6$.  The effective Chern-Simons level is still $\frac{1}{2}$. The simple-root W-bosons have masses given by:
\begin{align}
M(W_1) &= \xi_2 = 2\varphi_1 - \varphi_2~, \qquad M(W_2) = \xi_1 = 2\varphi_2 - \varphi_1  ~. 
\end{align}
This resolution corresponds to gauge theory chamber 4 (cf. Table \ref{tbl:u3 nf1 instantons} and \eqref{eq:ft prepot SU(3) k Nf1 chamber 4}), with instanton masses given by:
\begin{align}
\begin{split}
	M(\cI_1) &= \chi(r_0 = -\xi_2) = \xi_2 + \xi_8 = h_0 + 3\varphi_1 - m ~,  \\
	M(\cI_2) &= \chi(r_0 = 0) = \xi_8 = h_0 + \varphi_1 + \varphi_2 - m ~,\\
	M(\cI_3) &= \chi(r_0 = \xi_1) = \xi_1 + \xi_8 = h_0 + 3\varphi_2 - m ~.
\end{split}	
\end{align}
The masses of hypermultiplets are:
\begin{align}
\begin{split}
	M(\cH_1) &= \xi_6 = -\varphi_2 + m ~,\\
	M(\cH_2) &= \xi_1 + \xi_6 = -\varphi_1 + \varphi_2 + m ~,\\
	M(\cH_3) &= \xi_1 + \xi_2 + \xi_6  = \varphi_1 + m ~.
\end{split}	
\end{align}
The map between geometry and field theory variables is given by \eqref{eq:p2 res a mu nu}. Plugging \eqref{eq:p2 res a mu nu} into \eqref{eq:p2 res c geo prepot}, we recover the field theory prepotential,
\begin{align}
\mathcal{F}_{SU(3)_{3/2},N_{\text{f}}=1}^{\text{chamber 4}} &= \frac{4}{3}\varphi_1^3 + \frac{4}{3}\varphi_2^3 - \frac{1}{2}(\varphi_1^2\varphi_2 + \varphi_1\varphi_2^2) + (h_0-m)\varphi_1^2 + (h_{0}-m)\varphi_2^2\nonumber\\
&\qquad + (m-h_{0})\varphi_1\varphi_2 ~,
\end{align}
up to $\varphi$-independent terms. %From field theory, the monopole string tensions are:
%\begin{align} 
%T_{1,\text{ft}} = \frac{\partial \mathcal{F}_{SU(3)_{3/2},N_{\text{f}}=1}^{\text{chamber 4}}}{\partial \varphi_1} &= \left(-2 m+2 h_0\right) \varphi _1+4 \varphi _1^2+\left(m-h_0\right) \varphi _2-\varphi _1 \varphi
%   _2-\frac{\varphi _2^2}{2} ~,\\
%T_{2,\text{ft}} = \frac{\partial \mathcal{F}_{SU(3)_{3/2},N_{\text{f}}=1}^{\text{chamber 4}}}{\partial \varphi_2} &= \left(m-h_0\right) \varphi _1-\frac{\varphi _1^2}{2}+\left(-2 m+2 h_0\right) \varphi _2-\varphi _1
%   \varphi _2+4 \varphi _2^2 ~,
%\end{align}
%whereas from geometry, they are given by:
The monopole string tensions from $\chi(r_0)$ are:
\begin{align} \hspace{-0.30in}
T_{1,\text{geo}} &= \int_{-\xi_2}^{0}\chi(r_0)\,dr_{0} = \frac{1}{2} \xi _2 \left(\xi _2+2 \xi _8\right) ~,~
T_{2,\text{geo}} = \int_{0}^{\xi_1}\chi(r_0)\,dr_{0} =  \frac{1}{2}\xi_1\left(\xi_1 + 2\xi_8\right) ~.
\end{align} 
Using the map $\xi_1 = 2\varphi_2-\varphi_1$, $\xi_2 = 2\varphi_1 - \varphi_2$, $\xi_6 = -\varphi_2+m$ and $\xi_8 = h_{0} + \varphi_1+\varphi_2-m$, one can verify that $T_{i,\text{ft}} = T_{i,\text{geo}}$ for $i = 1, 2$. %The tensions vanish at loci given by:
%\begin{align}
%\begin{split}
% (I)&:  \{\xi_2 = 0\} \cup \{\xi_2 + 2\xi_8 =0 \}, ~ \text{ and },\\
% (II)&: \left\{ \xi_1 =0 \right\} \cup \{\xi_1 + 2\xi_8 = 0\}~.
% \end{split}
%\end{align}
The loci $\xi_1=$ and $\xi_2 = 0$ correspond, respectively, to hard walls along which the W-bosons $W_2$ and $W_1$ become massless, whereas the loci $\{\xi_2 + 2\xi_8 = 0\}$ and $\{\xi_1 + 2\xi_8=0\}$ are both not part of the K\"{a}hler chamber of resolution (c). Away from any hard wall, the BPS instanton particle $\cI_2$ can become massless at $\xi_8 = 0$ (signaling a flop of $\CC_8$), indicating a traversable instantonic wall which leads to a non-gauge-theoretic chamber (f). Alternatively, away from any hard wall, the perturbative hypermultiplet $\cH_1$ can become massless at $\xi_6 = 0$ (signaling a flop of $\CC_6$), leading back to gauge theory resolution (a).

\begin{figure}[ht]
\centering
\begin{minipage}{0.35\textwidth}
	\includegraphics[width=4cm]{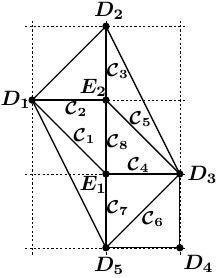}
\end{minipage}%
\begin{minipage}{0.20\textwidth}
	{\Large $\xrightarrow[\vol(\CC_6)\rightarrow \infty]{\text{RG flow}}$ }
\end{minipage}%
\begin{minipage}{0.30\textwidth}
	\includegraphics[width=4cm]{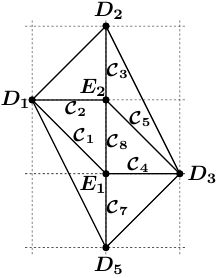}
\end{minipage}
\caption{Decoupling the divisor $D_4$ leads to a crepant resolution of the $E_{1}{}^{2,0}$ singularity.\label{fig:P[2] res c RG flow}}
\end{figure}

\paragraph{RG flow and decoupling limits.} In this resolution, we can decouple divisor $D_4$ by sending the volume of the curve $\CC_6$ to infinity. As $M(\cH_1) = \xi_6 = -\varphi_2 + m$, this is equivalent to taking the limit $m \rightarrow +\infty$, i.e. integrating out the massive fermion, which results in an $SU(3)_{0}$ pure gauge theory. From the perspective of geometry, this leads to an $SL(2,\Z)$-transformed version of a resolution of the $E_{1}{}^{2,0}$ singularity, as shown in Figure \ref{fig:P[2] res c RG flow}. For example, one can apply a $(TS)^2 T^{-1}S^{-1}$ transformation to the toric diagram on the right in Figure \protect{\ref{fig:P[2] res c RG flow}} to get to resolution (a) of the $E_{1}{}^{2,0}$ singularity.

\begin{figure}[ht]
\centering
\subfigure[\small{}]{
\includegraphics[width=4.5cm]{images/p2-res-d-new-labeled}\label{fig:P[2] res d labeled}}\,
\subfigure[\small{}]{
\includegraphics[width=7.5cm]{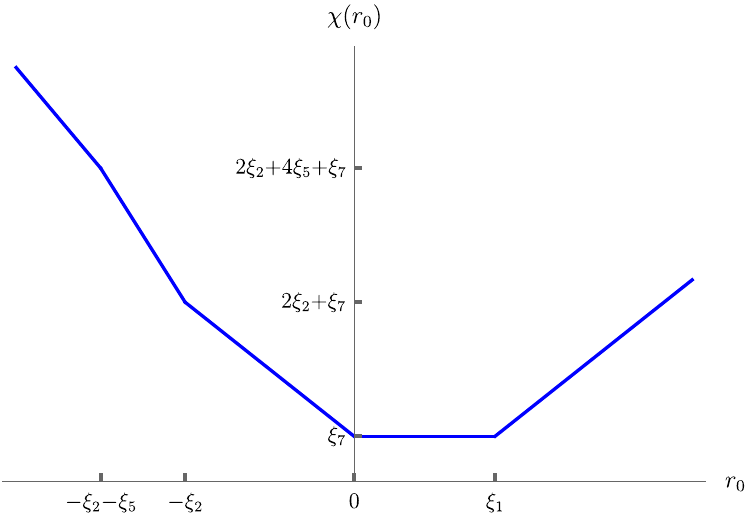}\label{fig:P[2] res d chi}}
\caption{Resolution (d) of the $E_{2}{}^{2,\fhalf}$ singularity and its vertical reduction.\label{fig:p2 res d final}}
\end{figure}

\paragraph{Resolution (d).} Consider the crepant resolution of Figure \ref{fig:p2-res-d}, with curves and divisors shown in Figure \ref{fig:P[2] res d labeled}.  
%%%%%%%%%%%%%%%
%%%%%%%%%%%%%%%
%The linear relations among the toric divisors are given by \eqref{eq:p2 res a divisor linear equivalences}. The compact curves $\CC$ are given by:
%\begin{align}\begin{split}
%	\CC_1 &= \bE_1 \cdot D_1, \quad \CC_2 = \bE_2 \cdot D_1, \quad \CC_3 = \bE_2 \cdot D_2, \quad \CC_4 = \bE_2 \cdot D_4 ~,\\
%	\CC_5 &= D_2 \cdot D_4, \quad \CC_6 = \bE_1 \cdot D_4, \quad \CC_7 = \bE_1 \cdot D_5, \quad \CC_8 = \bE_2 \cdot \bE_1 ~.
%	\end{split}
%\end{align}
The linear relations among curve classes are:
\begin{align}
\CC_3 &\simeq 2\CC_4 + \CC_7~, \quad \CC_4 \simeq \CC_2~, \quad \CC_6 \simeq \CC_1, \quad \CC_8 \simeq  \CC_7 ~.
\end{align}
We take $\{\CC_1, \CC_2, \CC_5, \CC_7\}$ as generators of the Mori cone. The GLSM charge matrix is:
\be\label{intersect p2 res d}
\begin{tabular}{l|ccccccc|c}
 & $D_1$ &$D_2$& $D_3$ & $D_4$ & $D_5$ & $\bE_1$& $\bE_2$ & vol($\CC$) \\
 \hline
$\CC_1$  & $0$ & $0$ & $0$ &  $0$& $1$ & $-2$ & $1$ & $\xi_1$ \\ \hline
$\CC_2$  & $0$ & $1$ & $0$ &  $0$& $0$ & $1$ & $-2$ & $\xi_2$ \\ \hline
$\CC_5$  & $0$ & $-1$ & $1$ & $-1$ & $0$ & $0$ & $1$ & $\xi_5$ \\ \hline
$\CC_7$  & $1$ & $0$ & $0$ & $1$ & $0$ & $-2$ & $0$ & $\xi_7$ \\ \hline\hline
$U(1)_M$ & $0$& $0$ &$0$ & $0$ & $0$ & $-1$ & $1$ & $r_0$
 \end{tabular}
\ee 
%where the last line defines the vertical reduction of the 2d GLSM. The nonnegative FI terms $\xi_1 \geq 0$, $\xi_2 \geq 0$, $\xi_5 \geq 0$ and $\xi_7 \geq 0$ are, respectively, the volumes of compact curves $\CC_1$, $\CC_2$, $\CC_5$ and $\CC_7$.
The K\"{a}hler cone is parametrized by \eqref{eq:kah cone p2}. The parameters $(\mu_1, \mu_2, \nu_1, \nu_2)$ are related to the FI parameters by:
\begin{align}
\xi_1 &= -2\nu_1 + \nu_2 \geq 0~, \quad \xi_2 = \mu_2 + \nu_1 - 2\nu_2 \geq 0~, \quad \xi_5 = -\mu_2 + \nu_2 \geq 0~, \quad \xi_7 = \mu_1 - 2\nu_1 \geq 0 ~.	 \label{eq:p2 res d FI}
\end{align}
The relevant triple-intersection numbers are:
\begin{align}
\begin{array}{c@{~,\quad}c@{~,\quad}c@{~,\quad}c@{~\quad}}
  D_1 \bE_1 \bE_2 = 1 & D_2 \bE_1 \bE_2 = 0 & D_1 D_2 \bE_1 = 0 & D_1 D_2 \bE_2 = 1 ~,\\
  D_1\bE_1^2 = -2 & D_2 \bE_1^2 = 0 & D_1\bE_2^2 = -2 & D_2\bE_2^2 = -4 ~,\\
  D_1^2 \bE_1 = 0 & D_1^2 \bE_2 = 0 & D_2^2 \bE_1 = 0 & D_2^2 \bE_2 = 2 ~,\\
  \bE_1^2 \bE_2 = -2 & \bE_1 \bE_2^2 = 0 & \bE_1^3 = 8 & \bE_2^3 = 8 ~.
\end{array}
\end{align}
Therefore, the compact part of the prepotential is:
\begin{align}
\mathcal{F}_{(d)}(\nu_1, \nu_2; \mu_1, \mu_2) = -\frac{1}{6}S^3 &= -\frac{4}{3}\nu_1^3 - \frac{4}{3}\nu_2^3 + \nu_1^2\nu_2 + \mu_1 \nu_1^2 + \left(\mu_1 + 2\mu_2\right)\nu_2^2\nonumber\\
&\qquad - \mu_1 \nu_1 \nu_2 -(\mu_1\mu_2 + \mu_2^2)\nu_2 ~. \label{eq:p2 res d geo prepot}
\end{align}
The IIA profile is:
\begin{align}\label{eq:p2 res d chi}
\chi(r_0) &= \left\{ 
                 \begin{array}{ll}
                 	-3r_0 -2\xi_2 + \xi_5 + \xi_7, & \text{for } r_0 \leq -\xi_2-\xi_5\\
                 	-4r_0 -2\xi_2 + \xi_7, & \text{for } -\xi_3-\xi_5 \leq r_0 \leq -\xi_2\\
                 	-2r_0 + \xi_7, & \text{for } -\xi_2 \leq r_0 \leq 0\\
                 	\xi_7, & \text{for } 0 \leq r_0 \leq \xi_1  \\
                 	+2r_0 -2\xi_1 + \xi_7, & \text{for } r_0 \geq \xi_1 ~.
                 \end{array}
             \right.
\end{align}
This function is sketched in Figure \ref{fig:P[2] res d chi}. At the points $r_0 = -\xi_2$, $r_0 = 0$ and $r_0 = \xi_1$, there are gauge D6-branes wrapping $\P^1$'s in the resolution of the singularity. There is a flavor D6-brane at $r_0 = -\xi_2-\xi_5$. The effective Chern-Simons level is still $\frac{1}{2}$. The simple-root W-bosons have masses given by:
\begin{align}
M(W_1) &= \xi_2 = 2\varphi_1 - \varphi_2~, \qquad M(W_2) = \xi_1 = 2\varphi_2 - \varphi_1  ~. 
\end{align}
This resolution corresponds to gauge theory chamber 1 (cf. Table \ref{tbl:u3 nf1 instantons} and \eqref{eq:ft prepot SU(3) k Nf1 chamber 1}), with instanton masses given by:
\begin{align}
\begin{split}
	M(\cI_1) &= \chi(r_0 = -\xi_2) = 2\xi_2 + \xi_7 = h_0 + 4\varphi_1 ~,  \\
	M(\cI_2) &= \chi(r_0 = 0) = \xi_7 = h_0 + 2\varphi_2 ~,\\
	M(\cI_3) &= \chi(r_0 = \xi_1) = \xi_7 = h_0 + 2\varphi_2 ~.
\end{split}	
\end{align}
The masses of hypermultiplets are:
\begin{align}
\begin{split}
	M(\cH_1) &= \xi_5 = -\varphi_1 - m ~,\\
	M(\cH_2) &= \xi_2 + \xi_5 = \varphi_1 - \varphi_2 - m ~,\\
	M(\cH_3) &= \xi_1 + \xi_2 + \xi_5  = \varphi_2 - m ~.
\end{split}	
\end{align}
Plugging \eqref{eq:p2 res a mu nu} into \eqref{eq:p2 res d geo prepot}, we recover the field theory prepotential,
\begin{align}
\mathcal{F}_{SU(3)_{3/2},N_{\text{f}}=1}^{\text{chamber 1}} &= \frac{4}{3}\varphi_1^3 + \frac{4}{3}\varphi_2^3 - \varphi_1^2\varphi_2 + h_{0}(\varphi_1^2 + \varphi_2^2 - \varphi_1\varphi_2)  ~,
\end{align}
up to $\varphi$-independent terms. %From field theory, the monopole string tensions are:
%\begin{align} 
%T_{1,\text{ft}} = \frac{\partial \mathcal{F}_{SU(3)_{3/2},N_{\text{f}}=1}^{\text{chamber 1}}}{\partial \varphi_1} &= 2 h_0 \varphi _1+4 \varphi _1^2-h_0 \varphi _2-\varphi _2^2 ~,\\
%T_{2,\text{ft}} = \frac{\partial \mathcal{F}_{SU(3)_{3/2},N_{\text{f}}=1}^{\text{chamber 1}}}{\partial \varphi_2} &= -h_0 \varphi _1+2 h_0 \varphi _2-2 \varphi _1 \varphi _2+4 \varphi _2^2 ~,
%\end{align}
The monopole string tensions from $\chi(r_0)$ are given by:
\begin{align} 
T_{1,\text{geo}} &= \int_{-\xi_2}^{0}\chi(r_0)\,dr_{0} = \xi _2 \left(\xi _2+\xi _7\right) ~,~
T_{2,\text{geo}} = \int_{0}^{\xi_1}\chi(r_0)\,dr_{0} =  \xi _1 \xi _7 ~.
\end{align} 
Using the map $\xi_1 = 2\varphi_2-\varphi_1$, $\xi_2 = 2\varphi_1 - \varphi_2$, $\xi_5 = -\varphi_1-m$ and $\xi_7 = h_{0} + 2\varphi_2$, one can verify that $T_{i,\text{ft}} = T_{i,\text{geo}}$ for $i = 1, 2$. 
%%\paragraph{Magnetic walls.} 
%The tensions vanish at loci given by:
%\begin{align}
%\begin{split}
% (I)&:  \{\xi_2 = 0\} \cup \{\xi_2 + \xi_7 =0 \}, ~ \text{ and },\\
% (II)&: \left\{ \xi_1 =0 \right\} \cup \{\xi_7 = 0\}~.
% \end{split}
%\end{align}
The loci $\xi_2 = 0$ and $\xi_1=0$ are both hard walls, being the boundaries of the Weyl chamber where either W-boson becomes massless. The loci $\{\xi_2 + \xi_7 = 0\} \subset (I)$ and $\{\xi_7= 0\} \subset (II)$ are both not part of the K\"{a}hler chamber of resolution (d) (the curve $\CC_7$ cannot flop). Away from any hard wall, the perturbative hypermultiplet $\cH_1$ can become massless at $\xi_5 = 0$ (signaling a flop of $\CC_5$), leading back to gauge theory resolution (a). %So there are no magnetic walls in resolution (d) of $E_{2}{}^{2,\fhalf}$. 

\begin{figure}[ht]
\centering
\begin{minipage}{0.35\textwidth}
	\includegraphics[width=5cm]{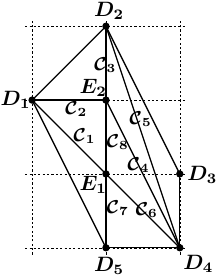}
\end{minipage}%
\begin{minipage}{0.20\textwidth}
	{\Large $\xrightarrow[\vol(\CC_5)\rightarrow \infty]{\text{RG flow}}$ }
\end{minipage}%
\begin{minipage}{0.30\textwidth}
	\includegraphics[width=5cm]{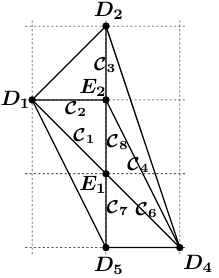}
\end{minipage}
\caption{Decoupling the divisor $D_3$ leads to a crepant resolution of the $E_{1}{}^{2,1}$ singularity.\label{fig:P[2] res d RG flow}}
\end{figure}

\paragraph{RG flow and decoupling limits.} In this resolution, we can decouple divisor $D_3$ by sending the volume of the curve $\CC_5$ to infinity. As $M(\cH_1) = \xi_5 = -\varphi_2 - m$, this is equivalent to taking the limit $m \rightarrow -\infty$,  which results in an $SU(3)_{1}$ pure gauge theory. On the geometry side, this leads to an $SL(2,\Z)$-transformed version of a resolution of the $E_{1}{}^{2,1}$ singularity, as shown in Figure \ref{fig:P[2] res d RG flow}. For instance, one can apply a $(TS)^2 T^{-1}S^{-1}$ transformation to the figure on the right in Figure \ref{fig:P[2] res d RG flow} to get to resolution (a) of the $E_{1}{}^{2,1}$ singularity.

\paragraph{Resolutions (e) and (f).} These are non-gauge-theoretic resolutions, related to each other by a flop of a single curve ($\CC_6$). In resolution (f), the divisor $D_4$ can be decoupled, which leads to an $SL(2,\Z)$-transformed version of resolution (b) of the $E_{1}{}^{2,0}$ singularity. For example, one such transformation is $(TS)^2 T^{-1}S^{-1}$, which leads precisely to resolution (b) of the $E_{1}{}^{2,0}$ singularity discussed above.

\begin{figure}[ht]
\centering
\begin{minipage}{0.35\textwidth}
	\includegraphics[width=5cm]{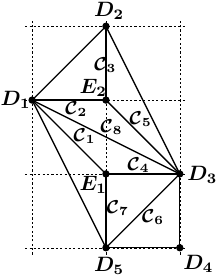}
\end{minipage}%
\begin{minipage}{0.20\textwidth}
	{\Large $\xrightarrow[\vol(\CC_6)\rightarrow \infty]{\text{RG flow}}$ }
\end{minipage}%
\begin{minipage}{0.30\textwidth}
	\includegraphics[width=5cm]{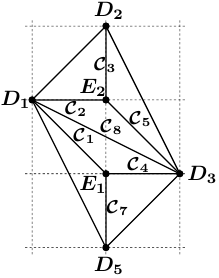}
\end{minipage}
\caption{Decoupling the divisor $D_4$ leads to a crepant resolution of the $E_{1}{}^{2,1}$ singularity.\label{fig:P[2] res f RG flow}}
\end{figure}

We remark that the resolutions (e) and (f) represent the coupling of a rank-1 $E_{1}$ singularity with a rank-1 non-Lagrangian $E_{0}$ singularity \cite{Seiberg:1996bd,Aharony:1997ju,Aharony:1997bh}. In the RG flow shown in Figure \ref{fig:P[2] res f RG flow}, we end up with a pair of coupled $E_{0}$ theories, as is evident from the shape of the final toric diagram.

\subsection{The $E_{3}{}^{2,1}$ singularity and $SU(3)_{1}$ $N_{\text{f}} =2$ gauge theory}
\begin{figure}[ht]
%\vspace{-10pt}
\begin{center}
%\renewcommand*{\thesubfigure}{(\arabic{subfigure})}
%%%%%%%%%%%%%%%%%%
\subfigure[\small{}]{
\includegraphics[width=1.85cm]{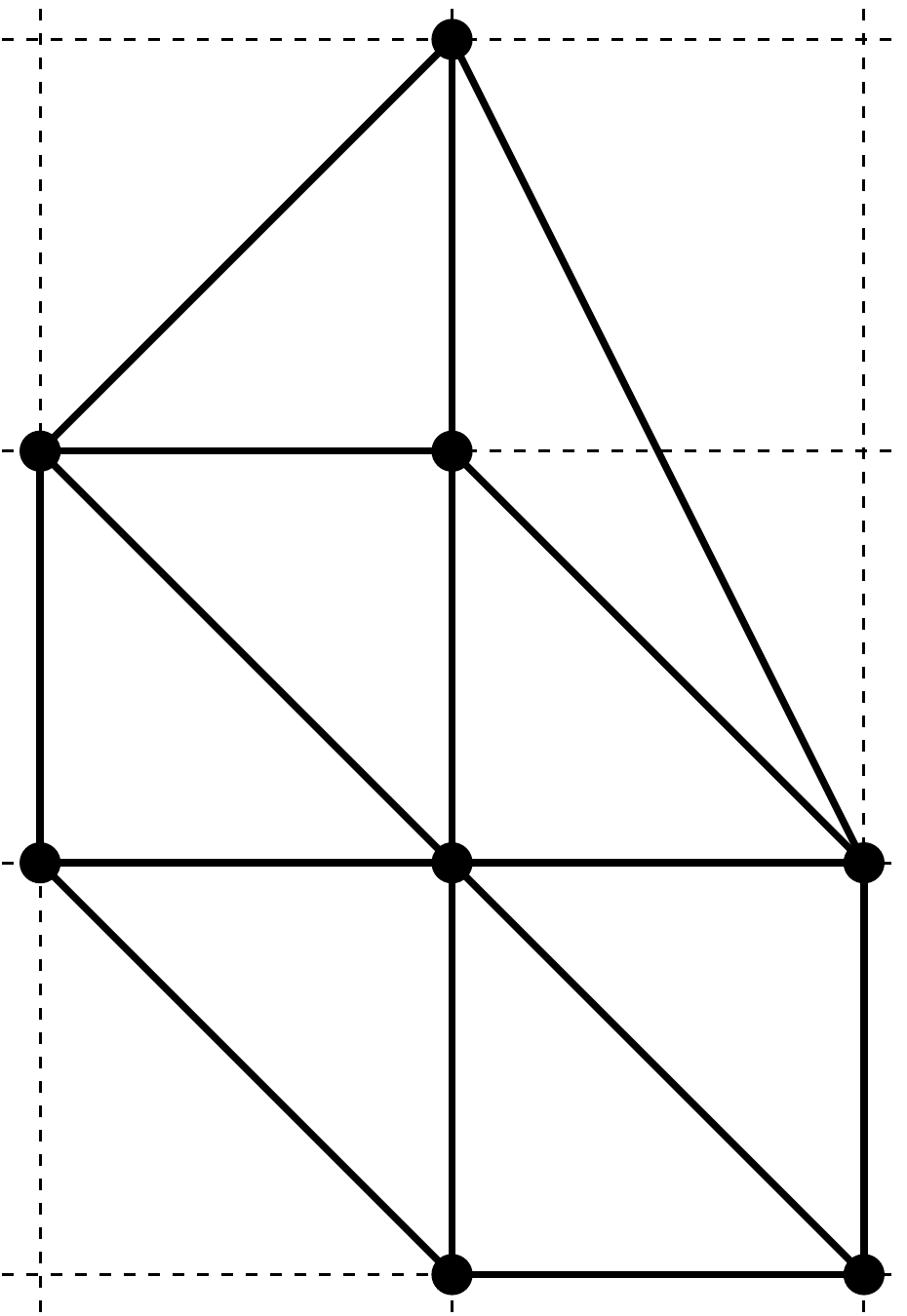}\label{fig:h1-res-a}}\,
\subfigure[\small{}]{
\includegraphics[width=1.85cm]{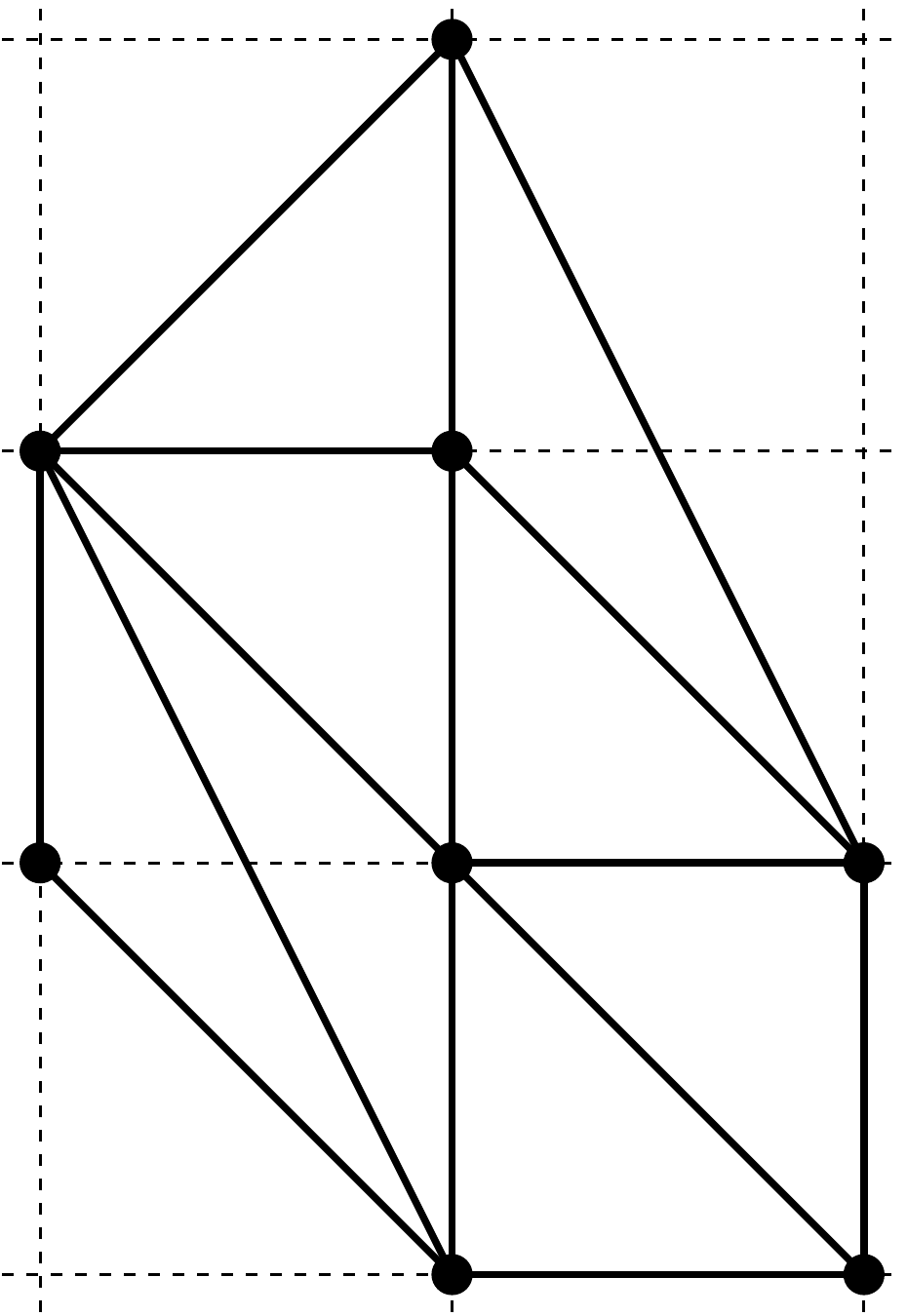}\label{fig:h1-res-b}}\,
\subfigure[\small{}]{
\includegraphics[width=1.85cm]{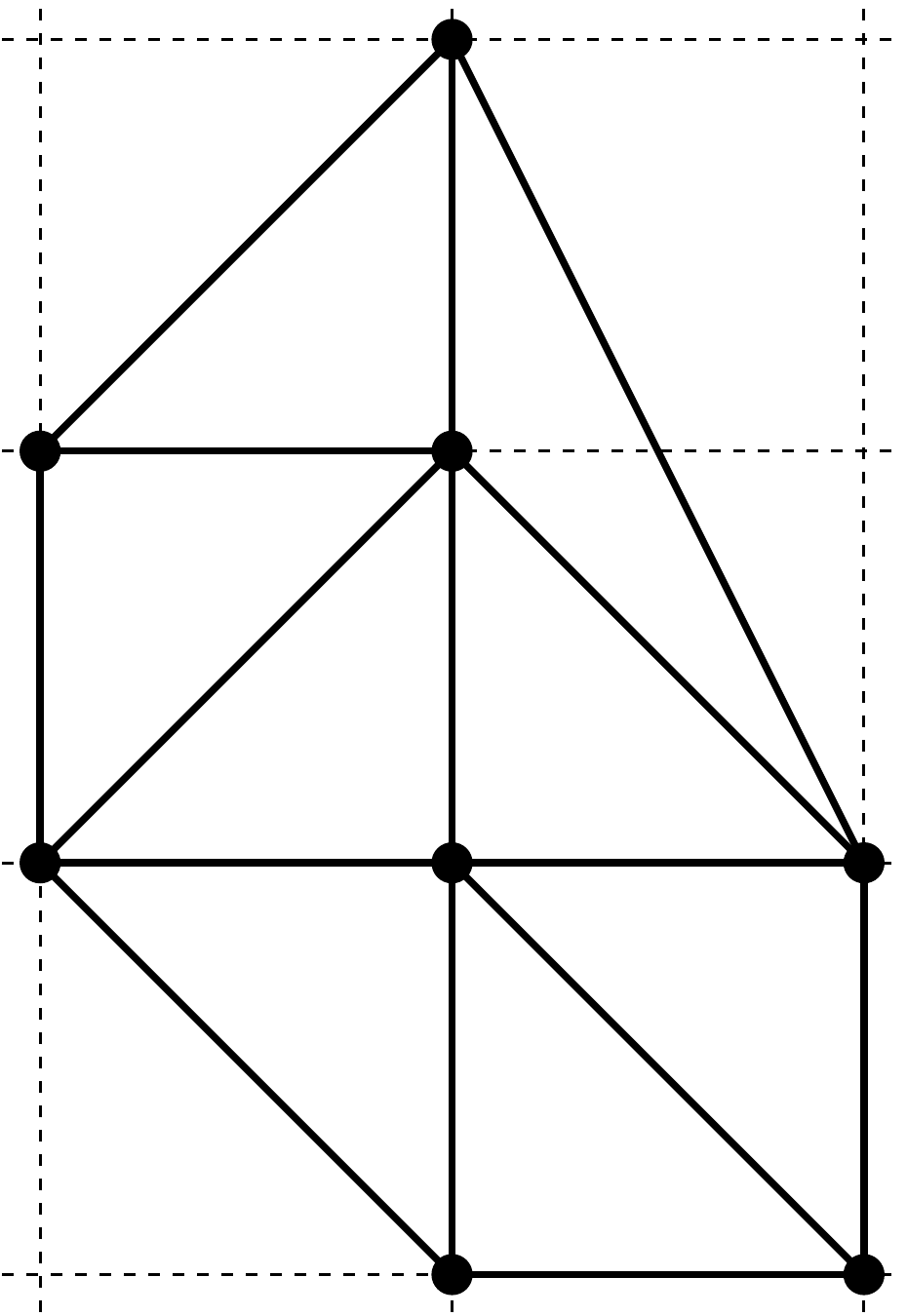}\label{fig:h1-res-c}}\,
\subfigure[\small{}]{
\includegraphics[width=1.85cm]{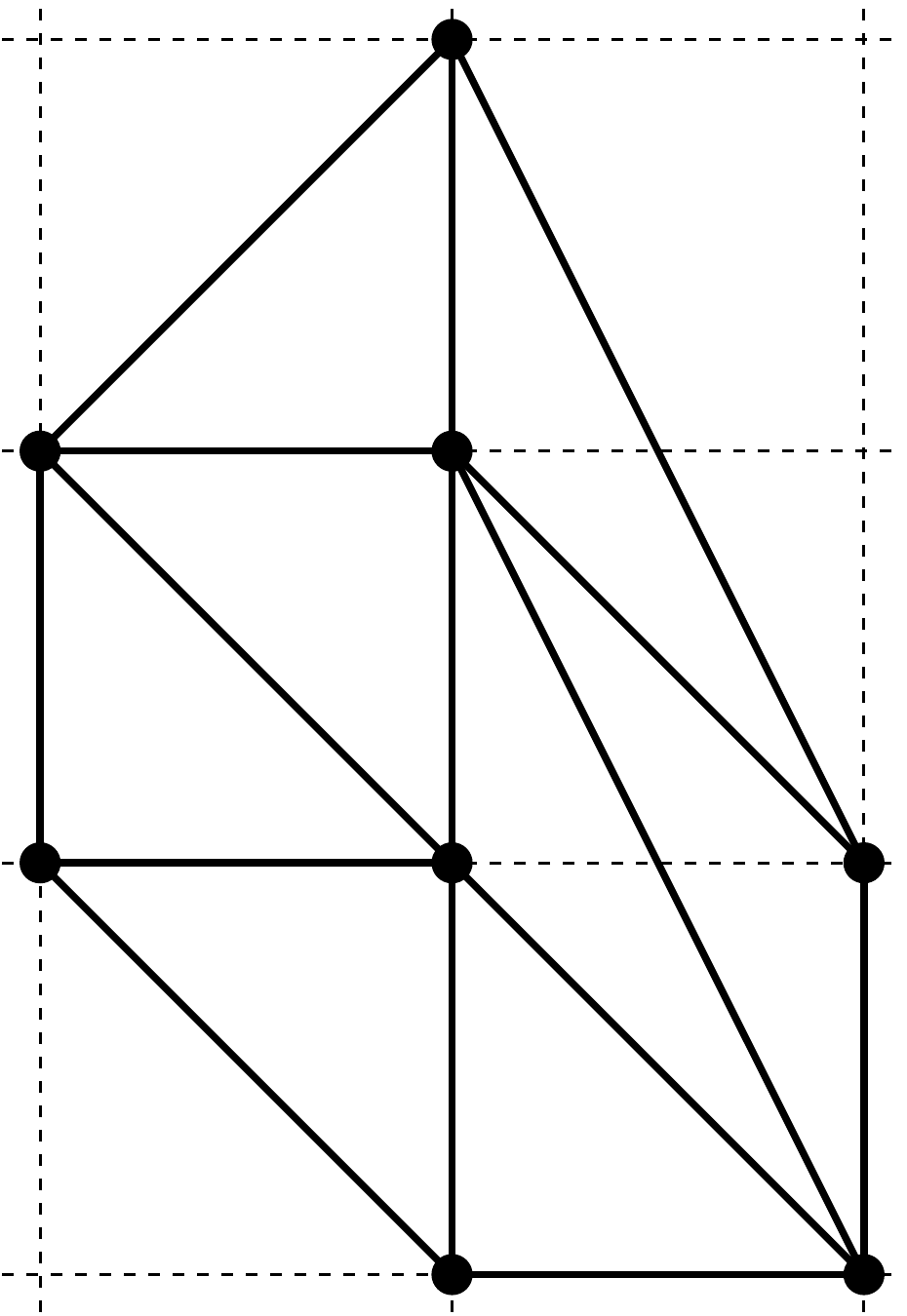}\label{fig:h1-res-d}}\,
\subfigure[\small{}]{
\includegraphics[width=1.85cm]{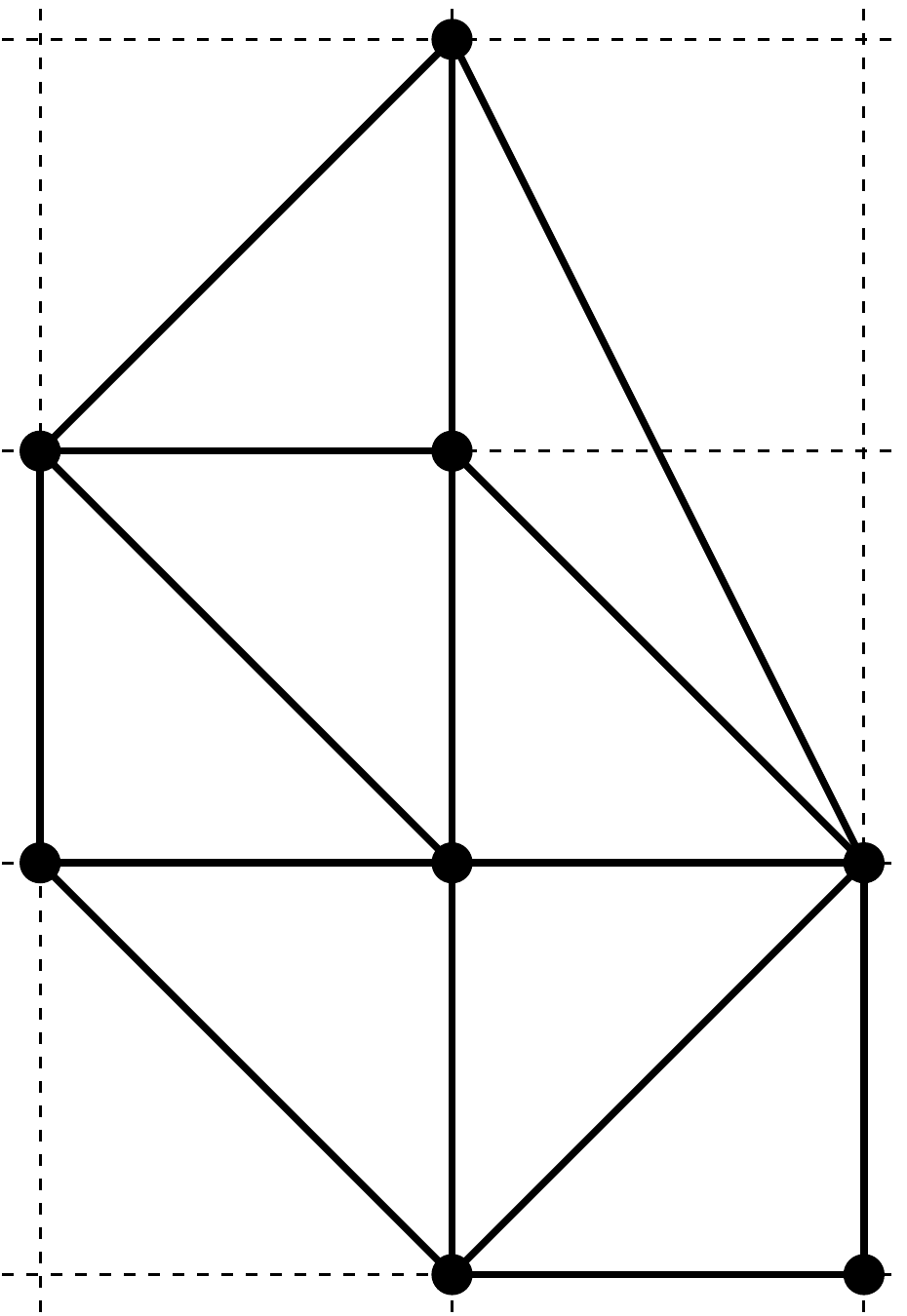}\label{fig:h1-res-e}}\,
\subfigure[\small{}]{
\includegraphics[width=1.85cm]{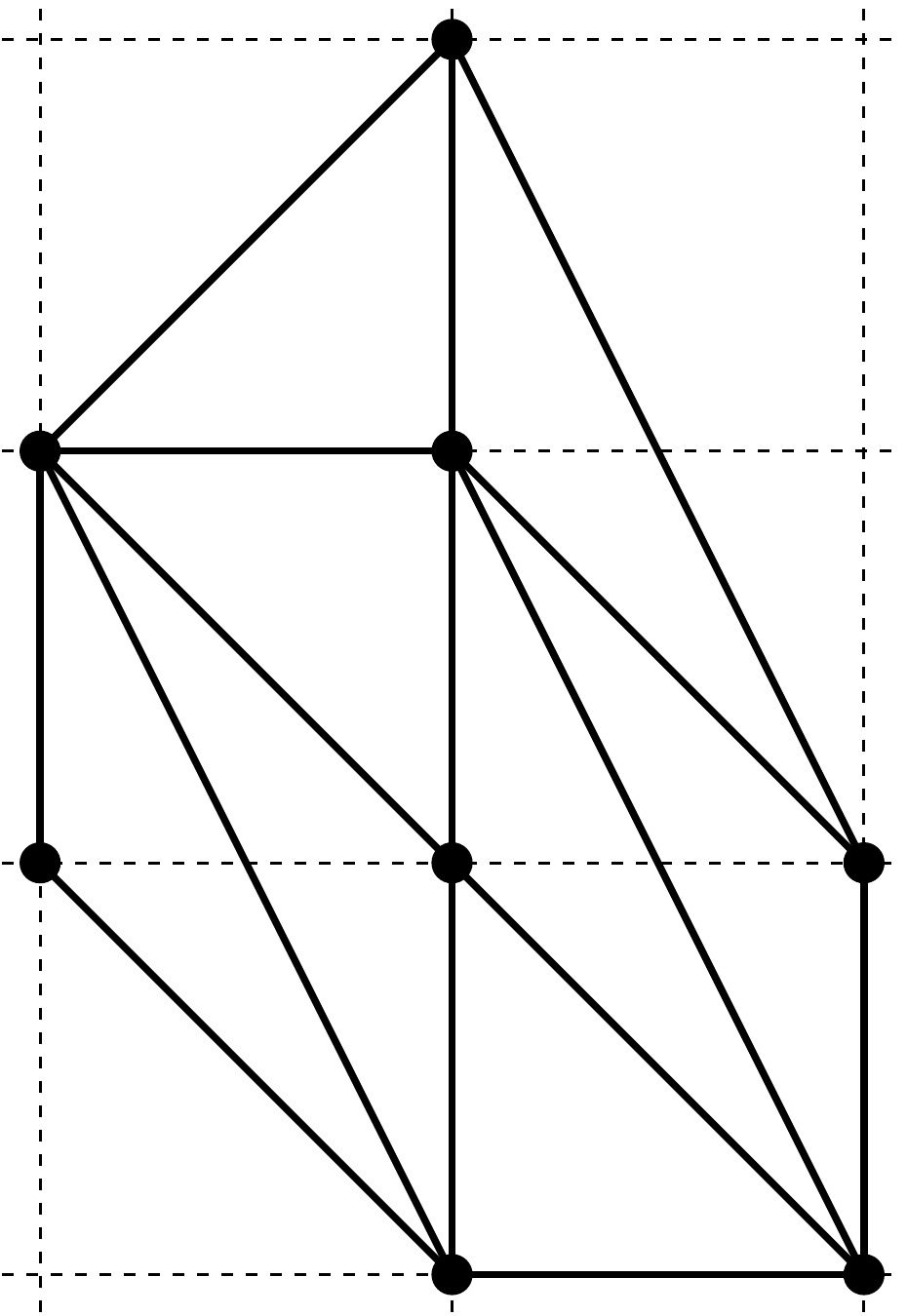}\label{fig:h1-res-f}}\,
\subfigure[\small{}]{
\includegraphics[width=1.85cm]{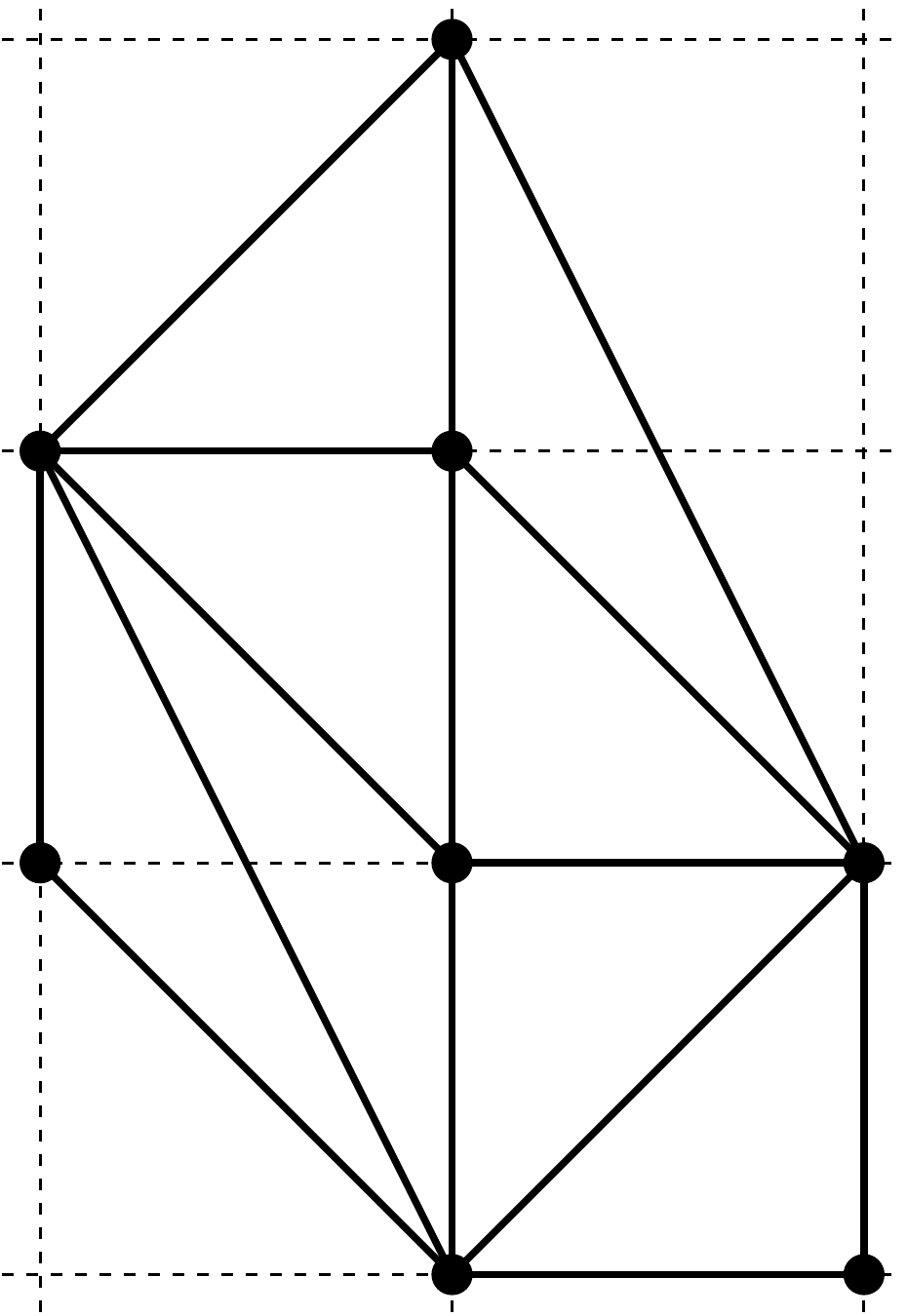}\label{fig:h1-res-g}}\,
\subfigure[\small{}]{
\includegraphics[width=1.85cm]{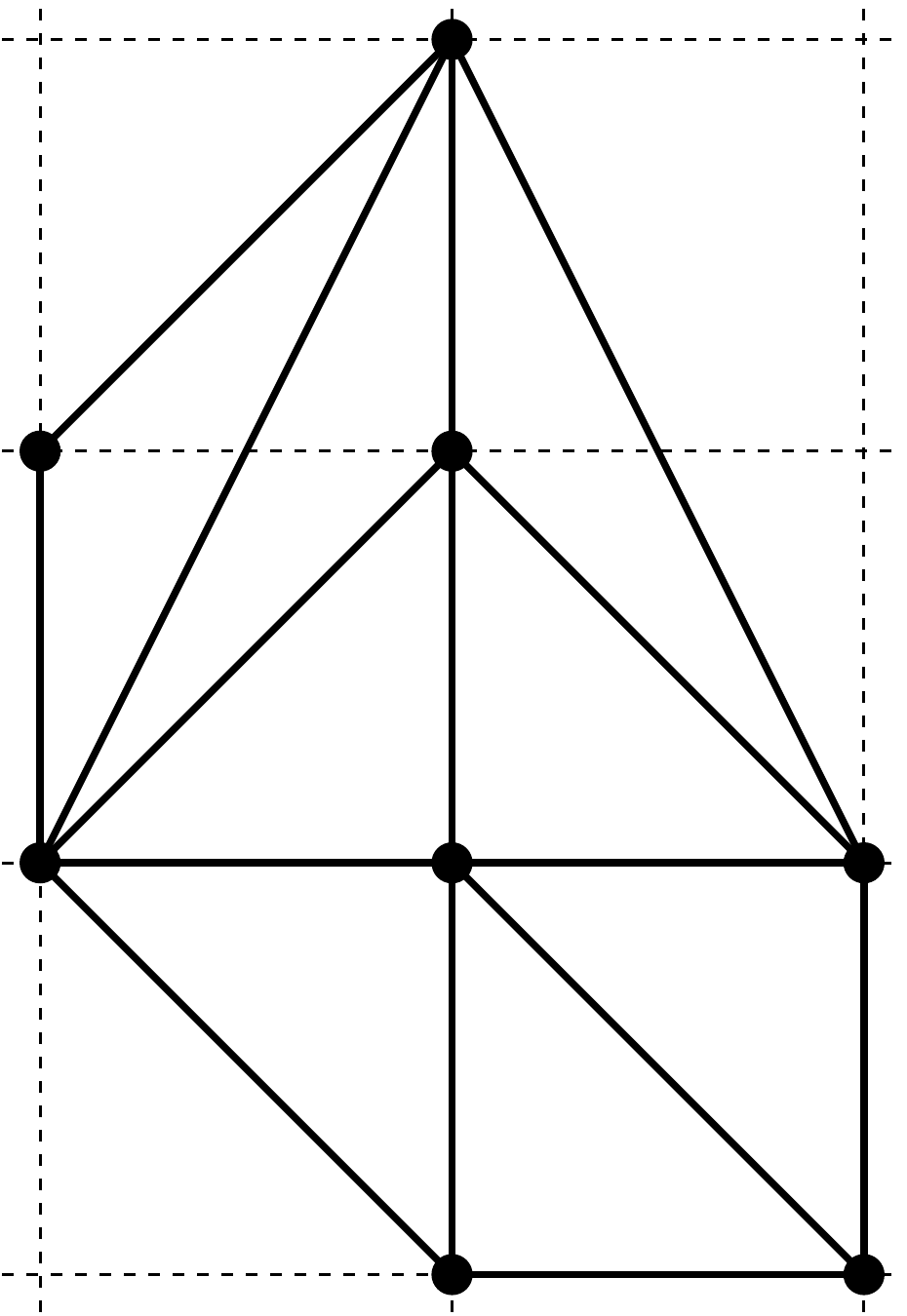}\label{fig:h1-res-h}}\,
\subfigure[\small{}]{
\includegraphics[width=1.85cm]{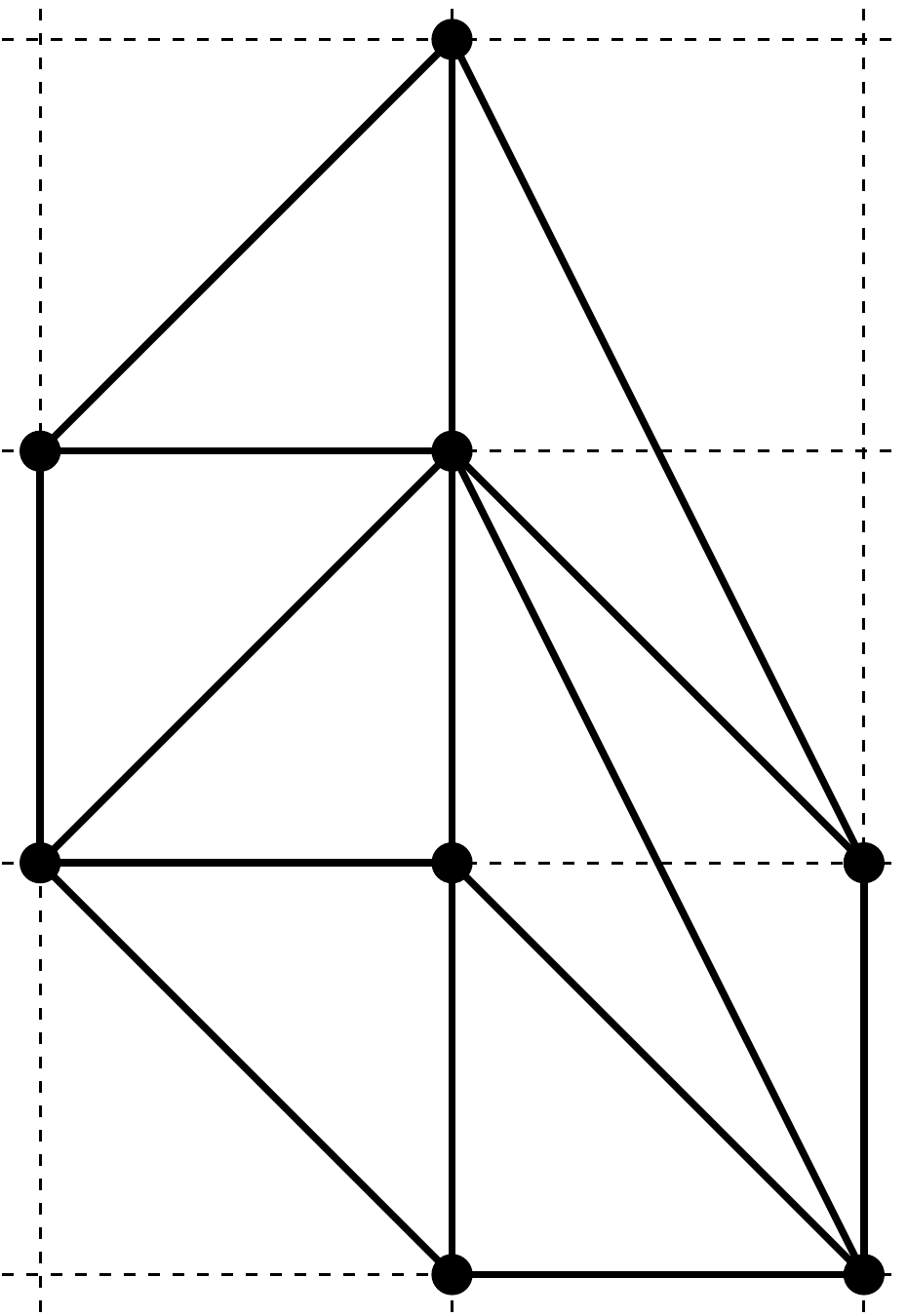}\label{fig:h1-res-i}}\,
\subfigure[\small{}]{
\includegraphics[width=1.85cm]{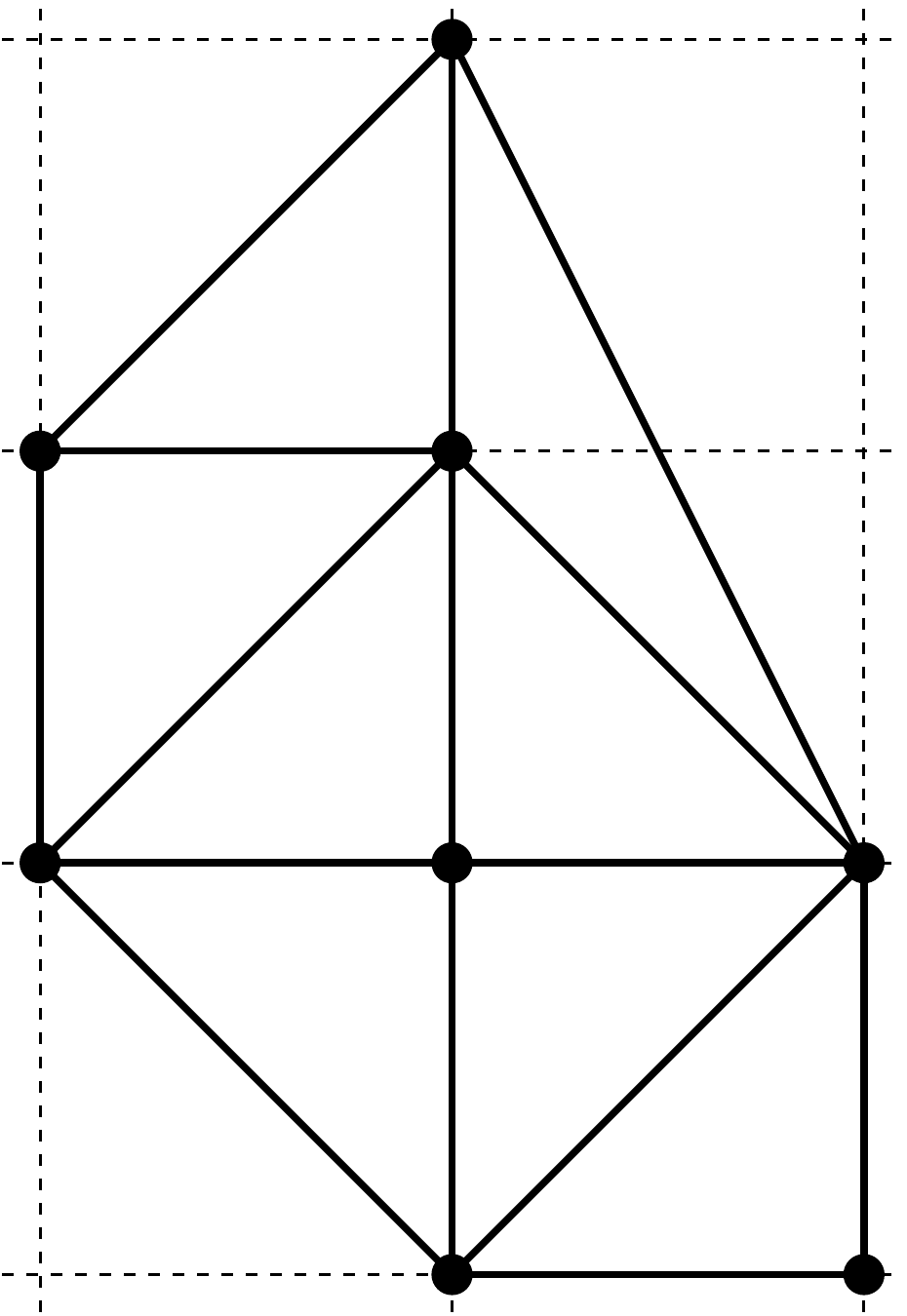}\label{fig:h1-res-j}}\,
\subfigure[\small{}]{
\includegraphics[width=1.85cm]{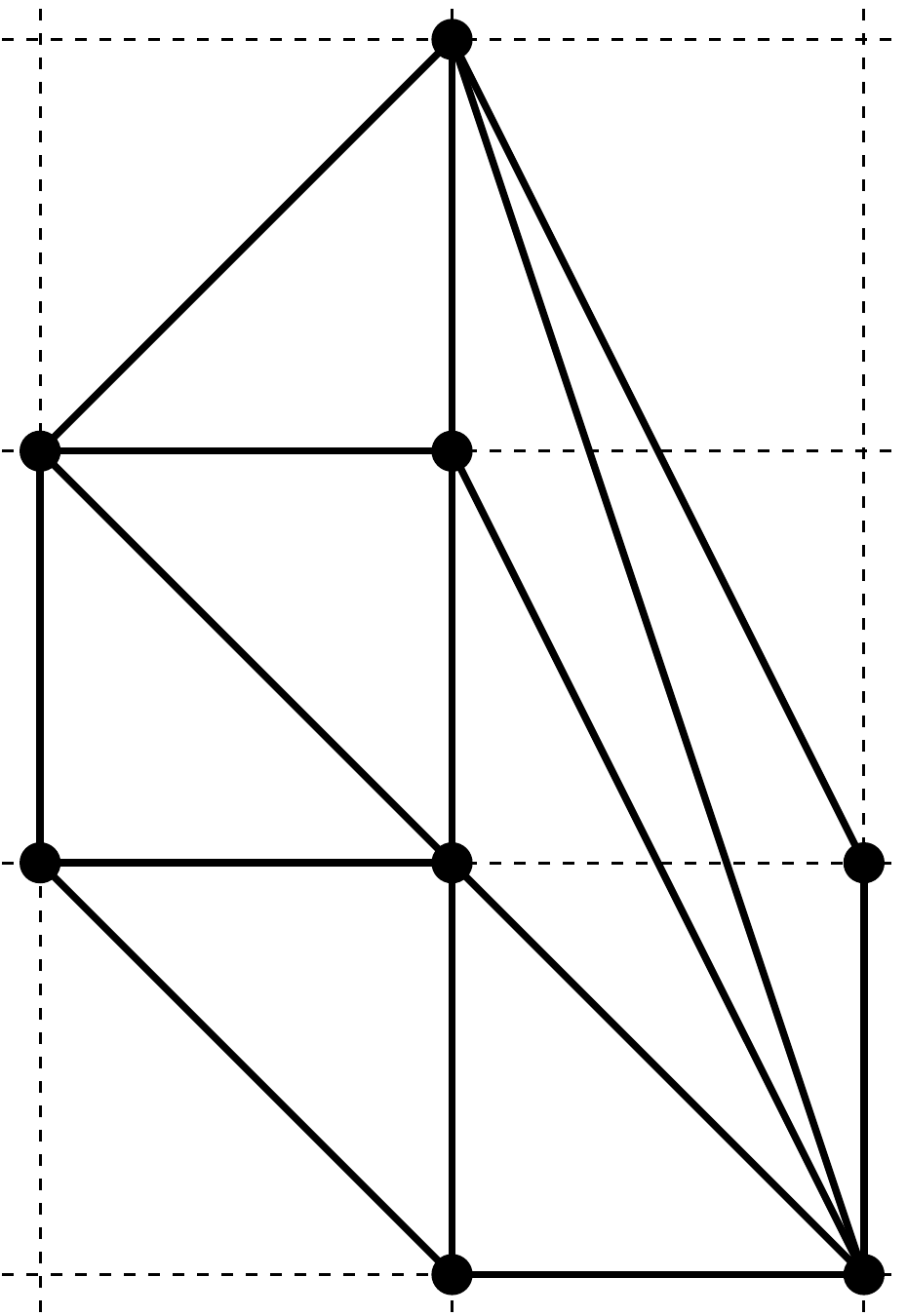}\label{fig:h1-res-k}}\,
\subfigure[\small{}]{
\includegraphics[width=1.85cm]{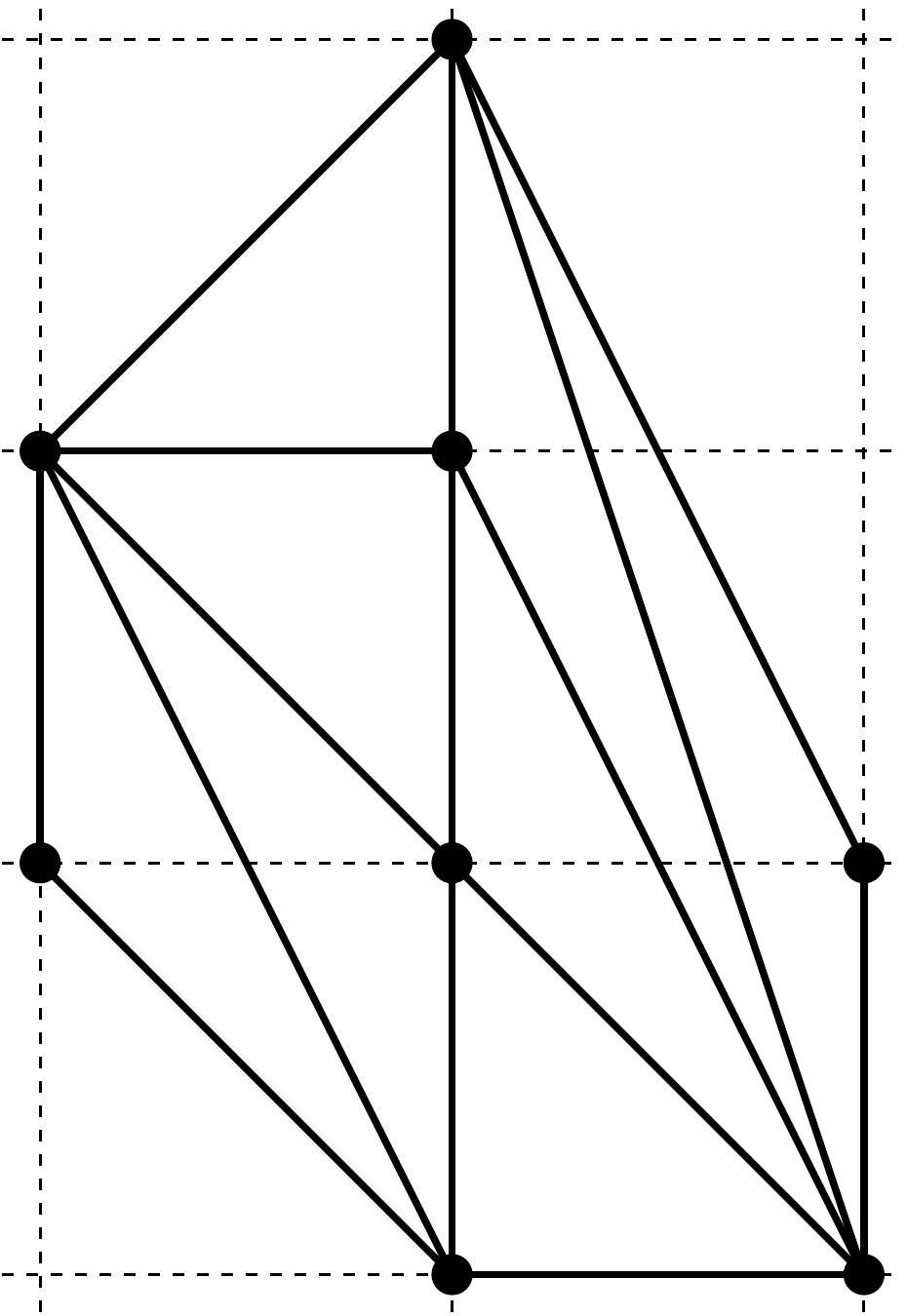}\label{fig:h1-res-l}}\,
\subfigure[\small{}]{
\includegraphics[width=1.85cm]{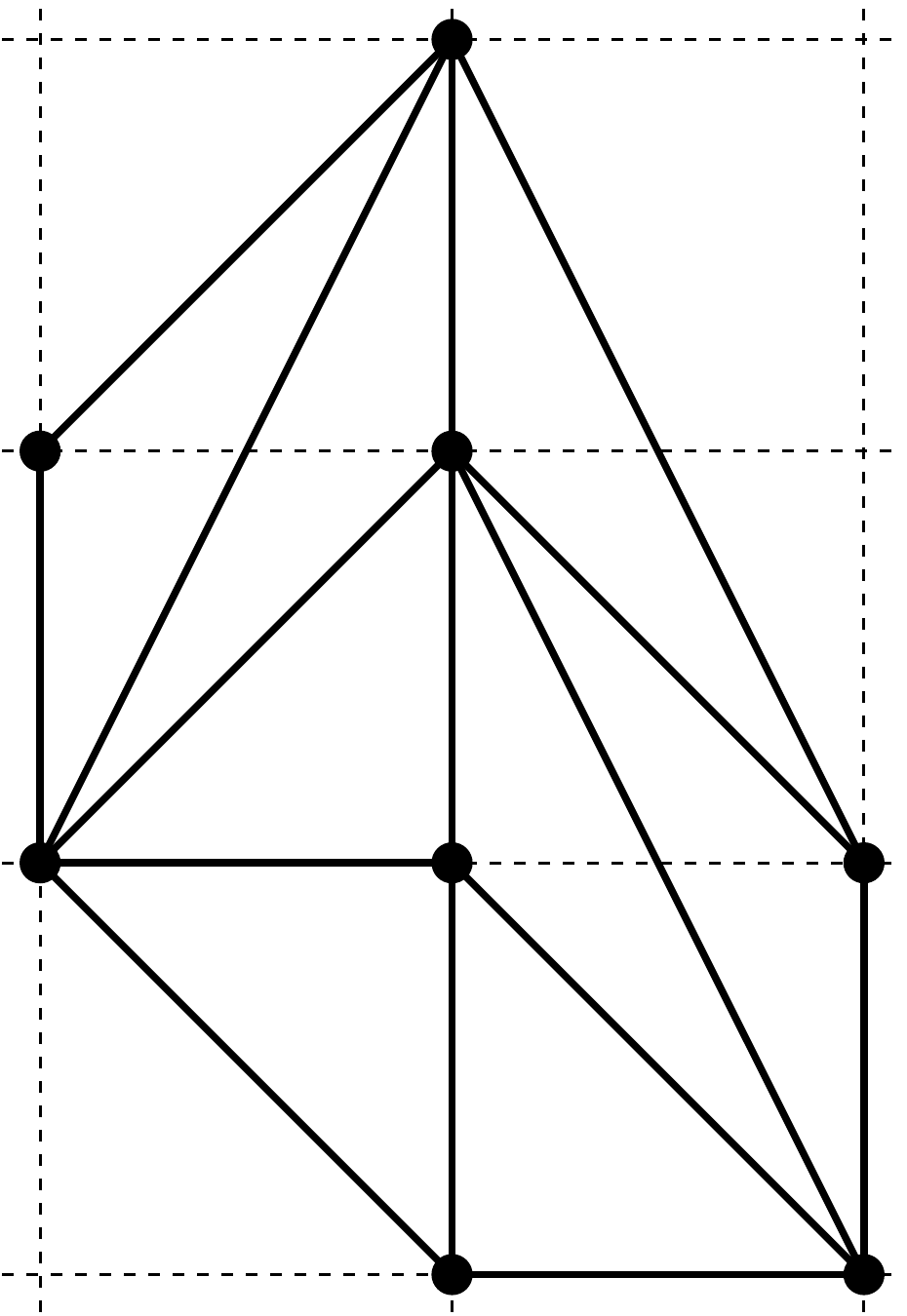}\label{fig:h1-res-m}}\,
\subfigure[\small{}]{
\includegraphics[width=1.85cm]{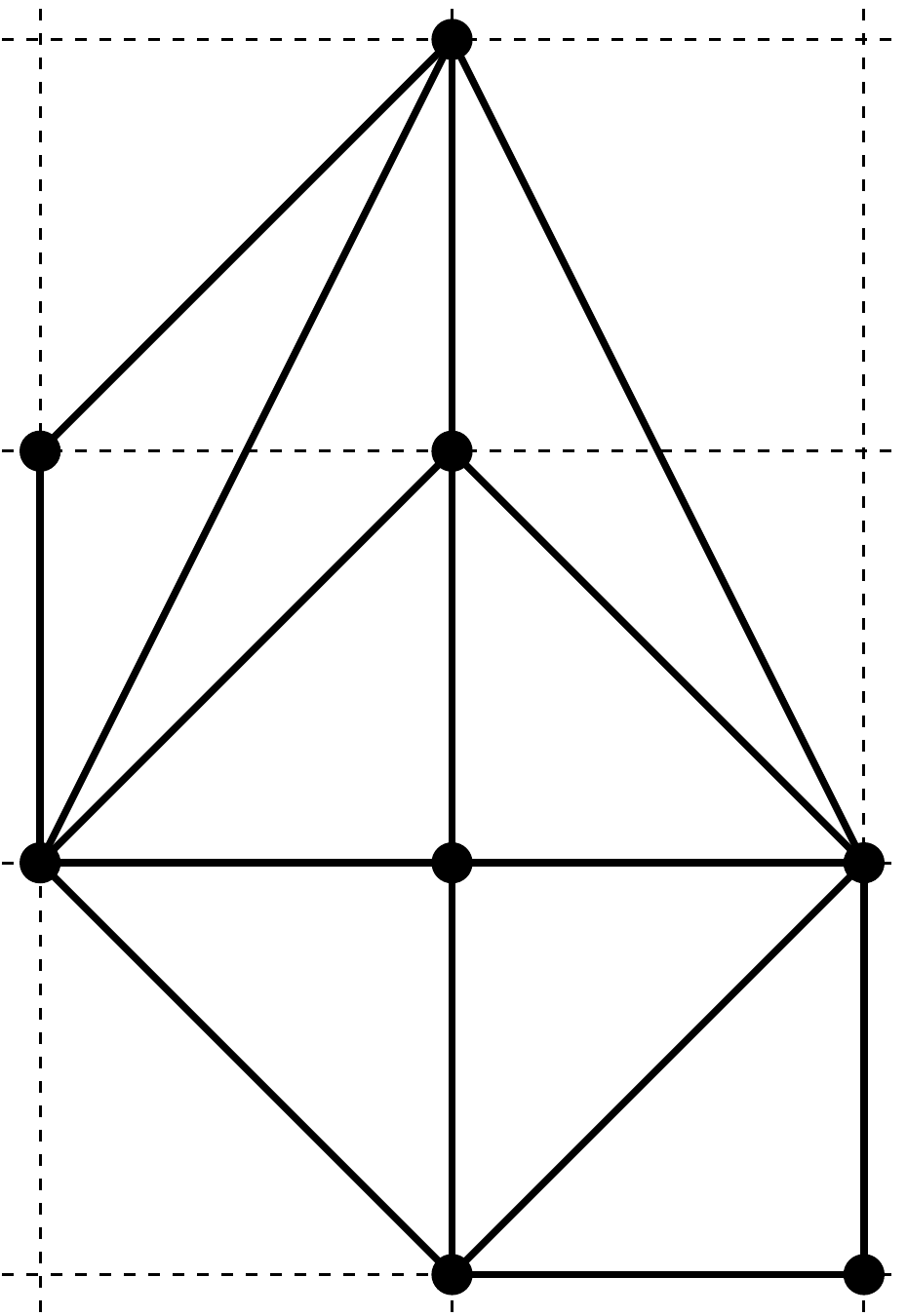}\label{fig:h1-res-n}}\,
\subfigure[\small{}]{
\includegraphics[width=1.85cm]{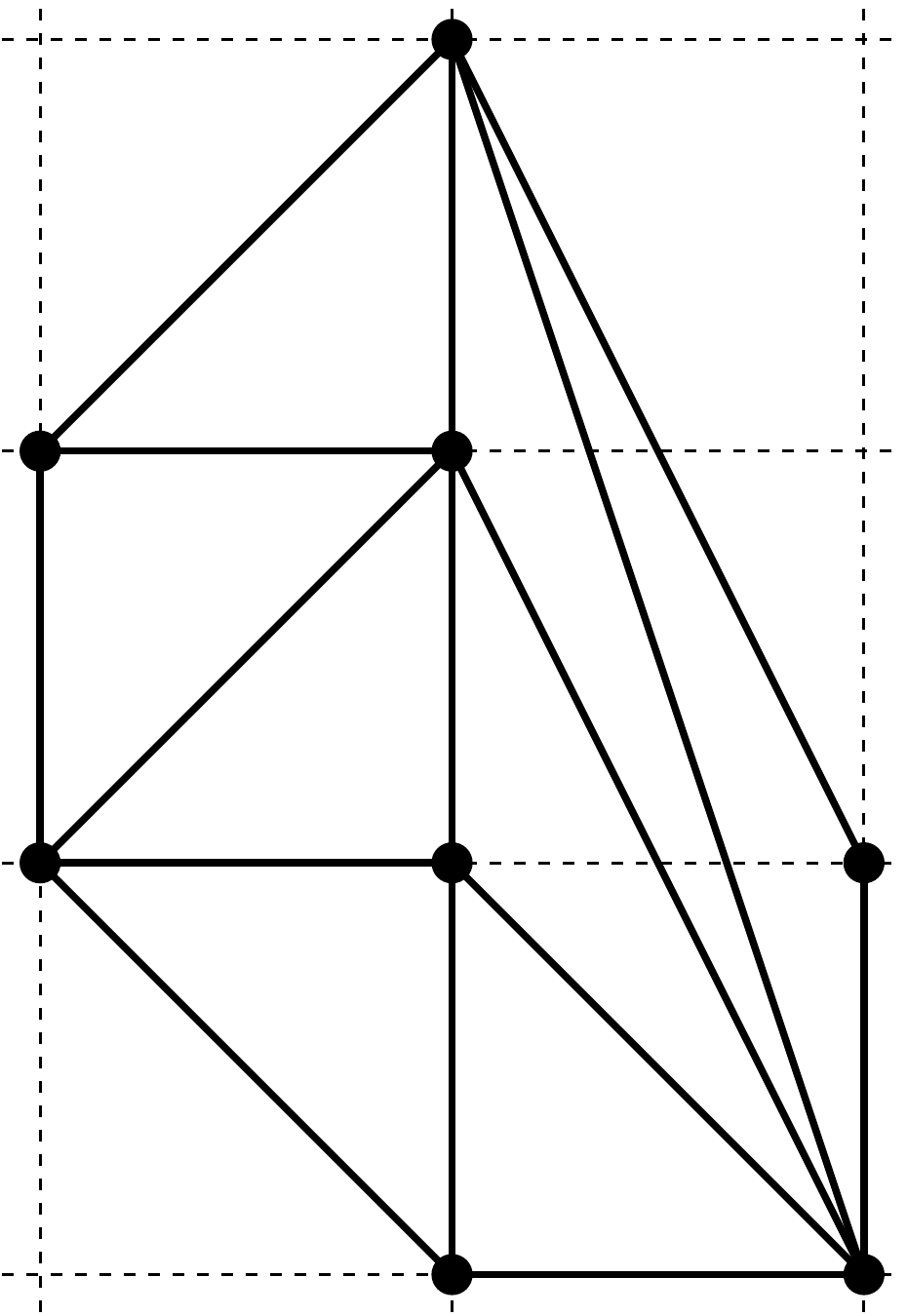}\label{fig:h1-res-o}}\,
\subfigure[\small{}]{
\includegraphics[width=1.85cm]{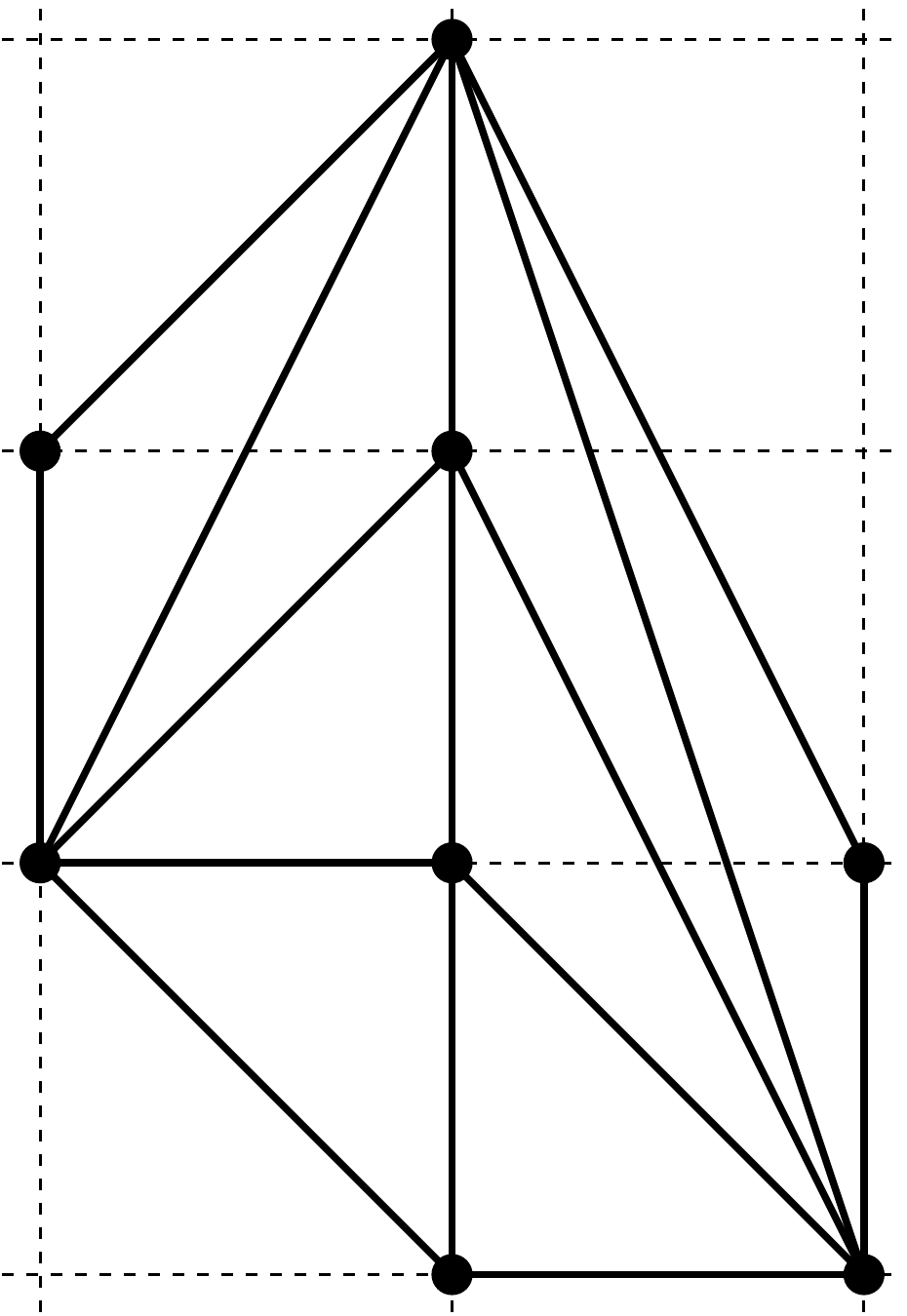}\label{fig:h1-res-p}}\,
%%%%%%%%%%%%%%%%%%NON LAGRANGIAN PHASES%%%%%%%%%%%%%%%%%%%%%%%%%%
\addtocounter{subfigure}{-16}
\renewcommand*{\thesubfigure}{($\text{q}_{\arabic{subfigure}}$)}
\subfigure[\small{}]{
\includegraphics[width=1.85cm]{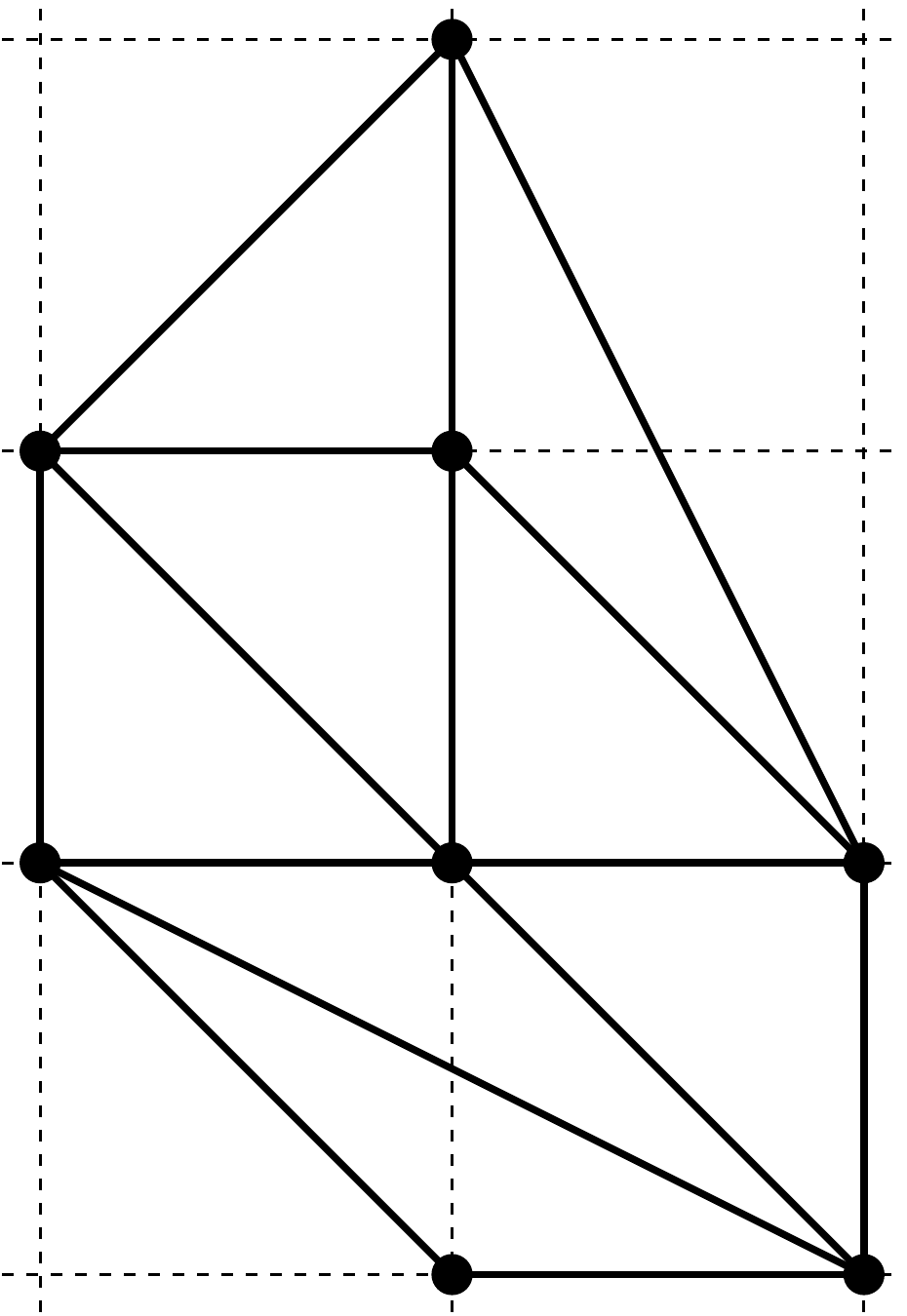}\label{fig:h1-res-n1}}\,
\subfigure[\small{}]{
\includegraphics[width=1.85cm]{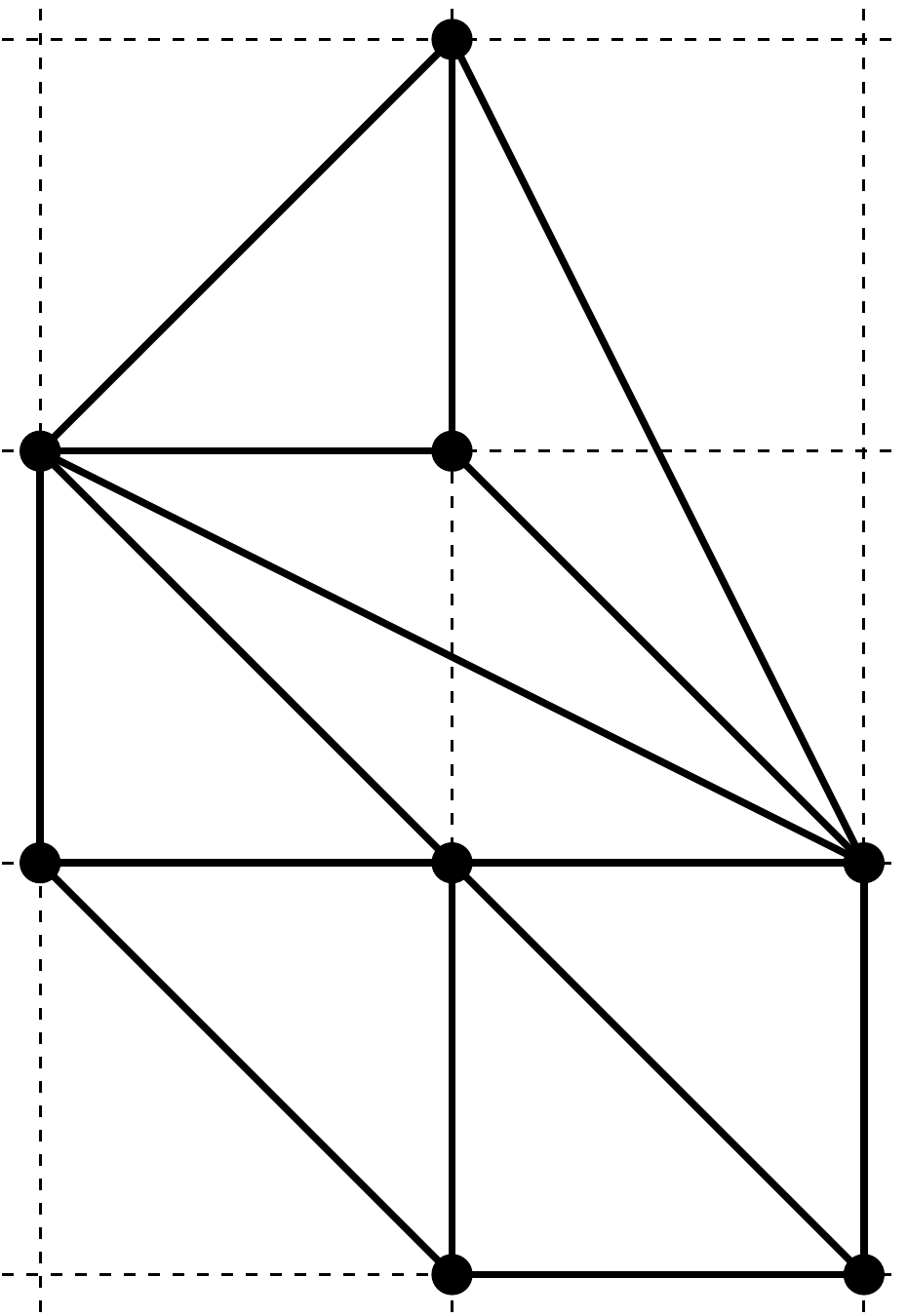}\label{fig:h1-res-n2}}\,
\subfigure[\small{}]{
\includegraphics[width=1.85cm]{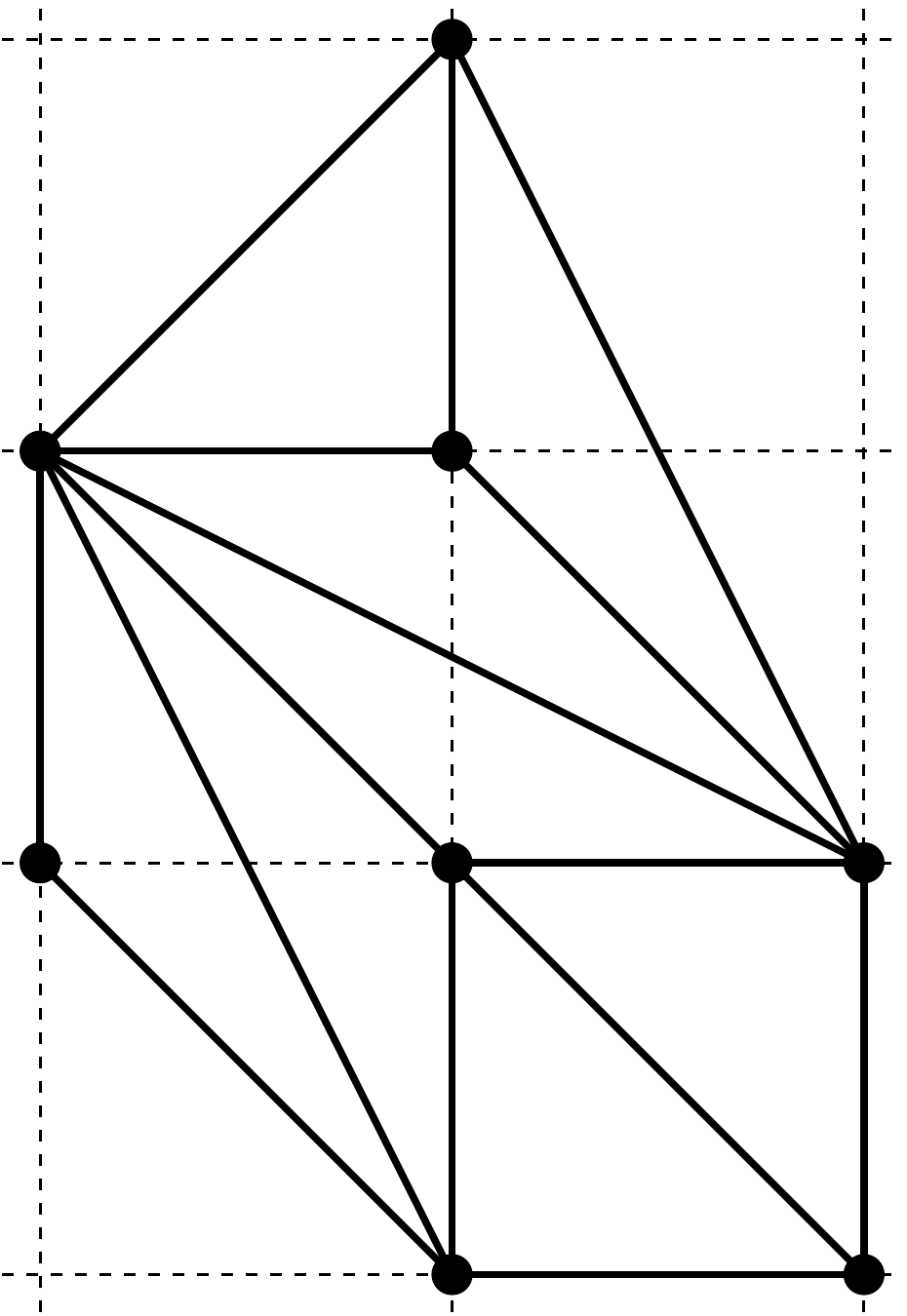}\label{fig:h1-res-n3}}\,
\subfigure[\small{}]{
\includegraphics[width=1.85cm]{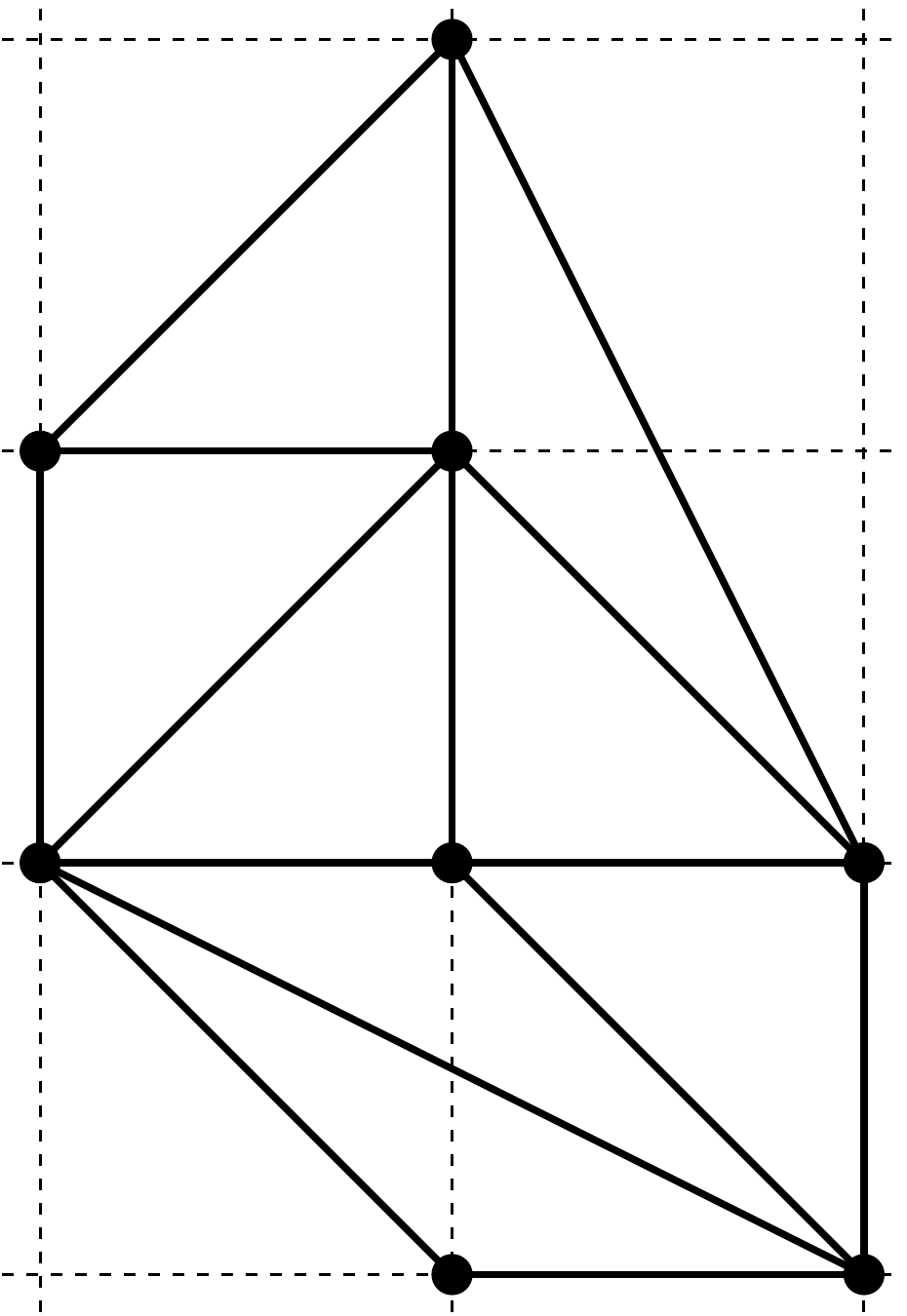}\label{fig:h1-res-n4}}\,
\subfigure[\small{}]{
\includegraphics[width=1.85cm]{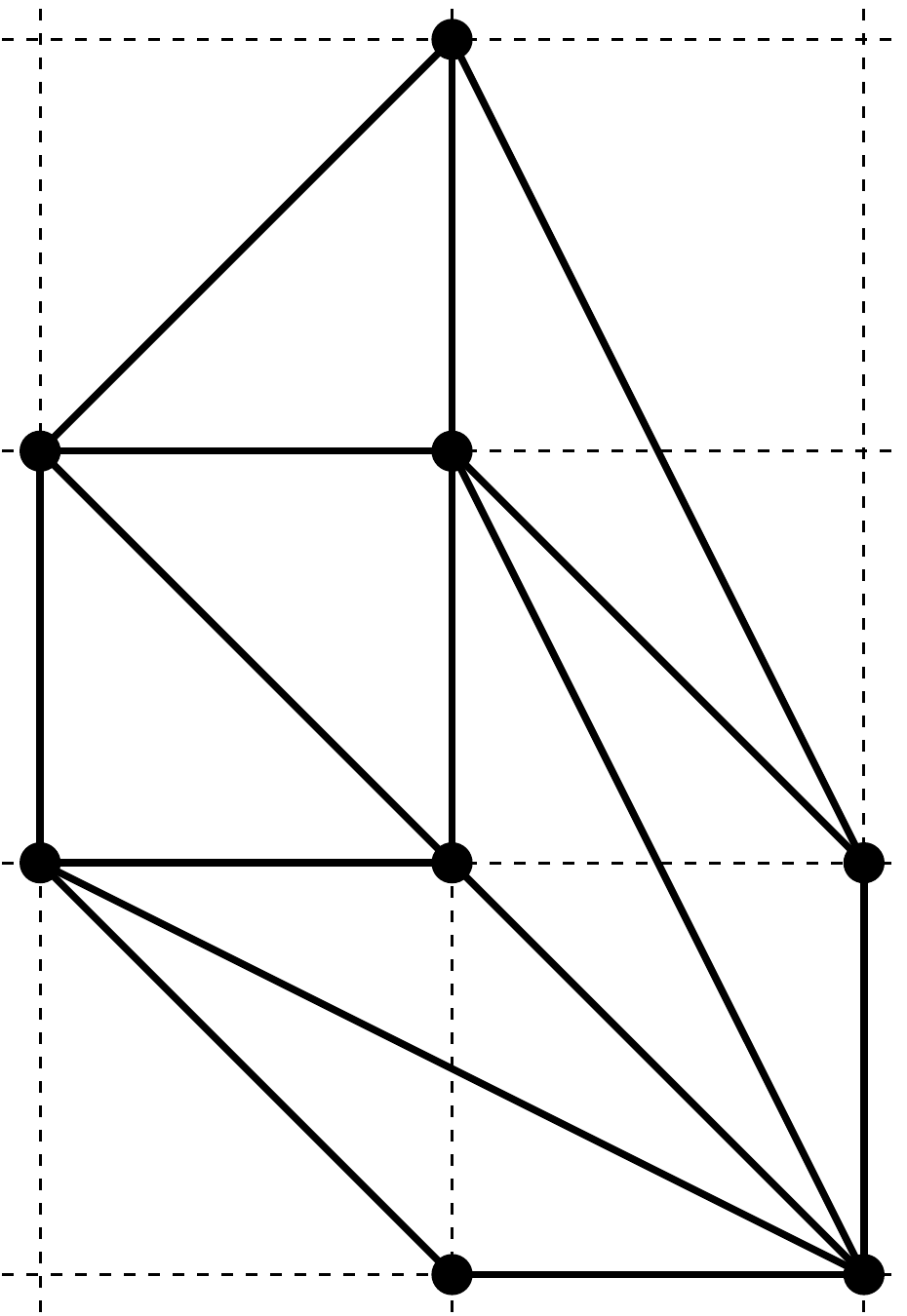}\label{fig:h1-res-n5}}\,
\subfigure[\small{}]{
\includegraphics[width=1.85cm]{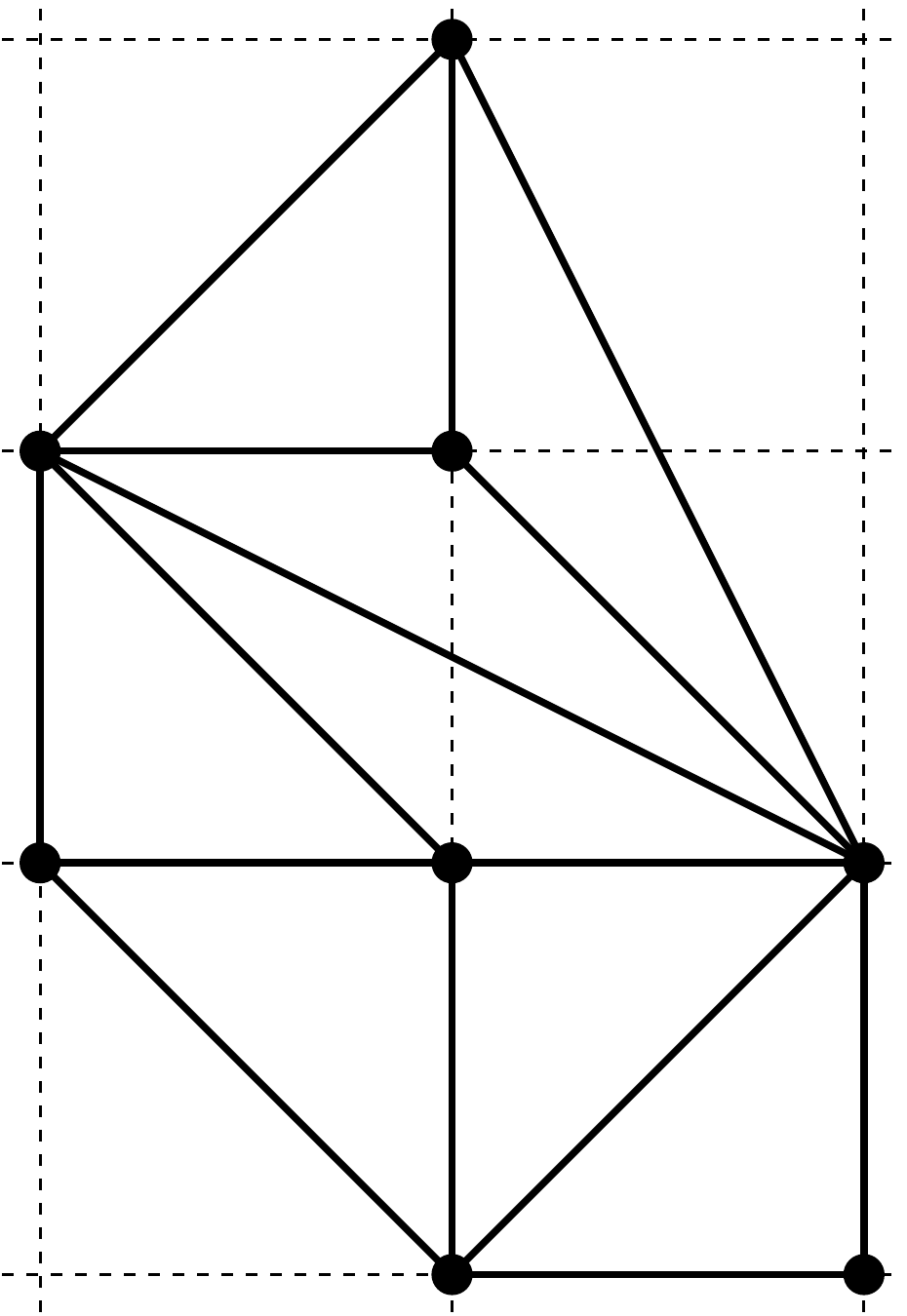}\label{fig:h1-res-n6}}\,
\subfigure[\small{}]{
\includegraphics[width=1.85cm]{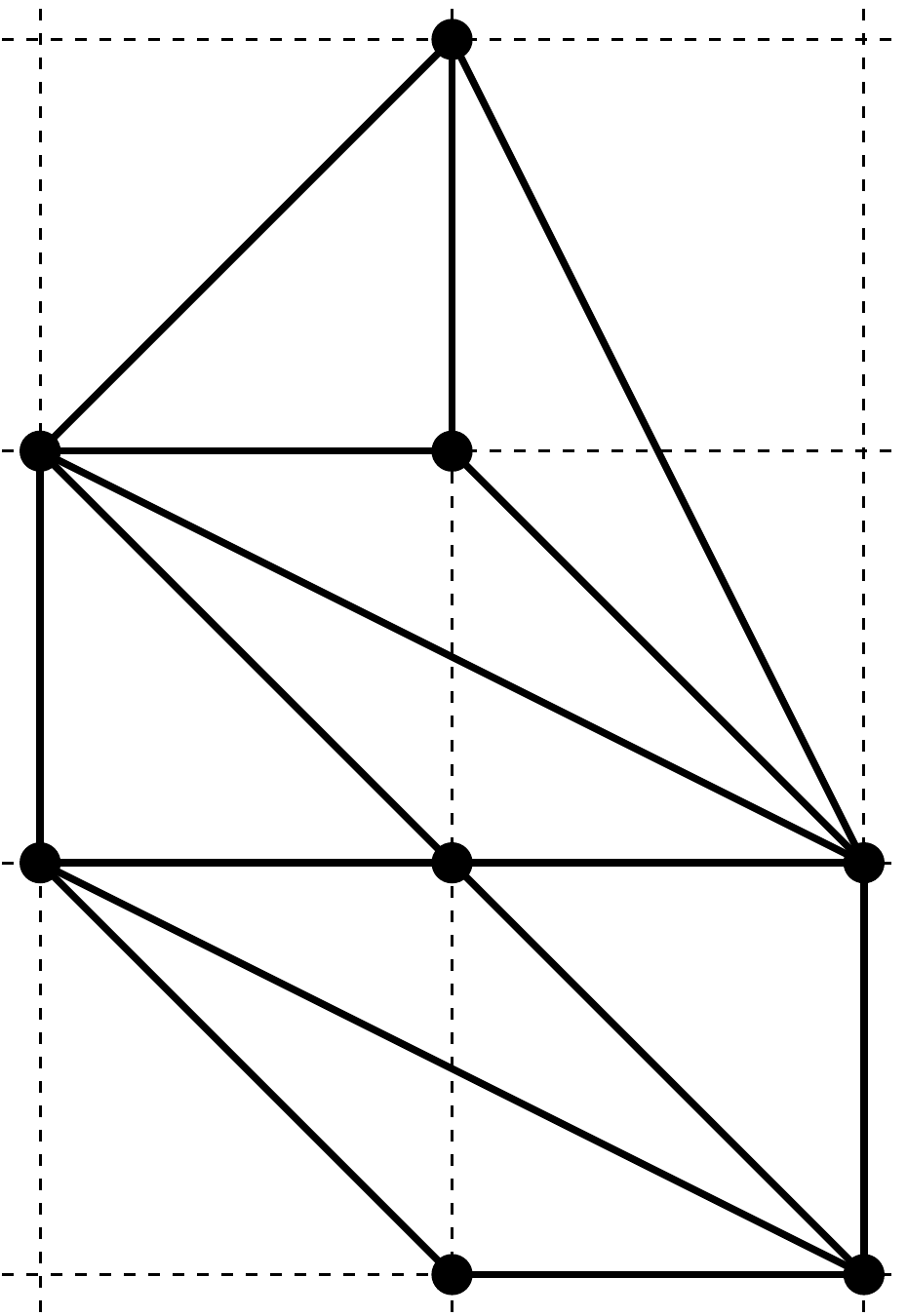}\label{fig:h1-res-n7}}\,
\subfigure[\small{}]{
\includegraphics[width=1.85cm]{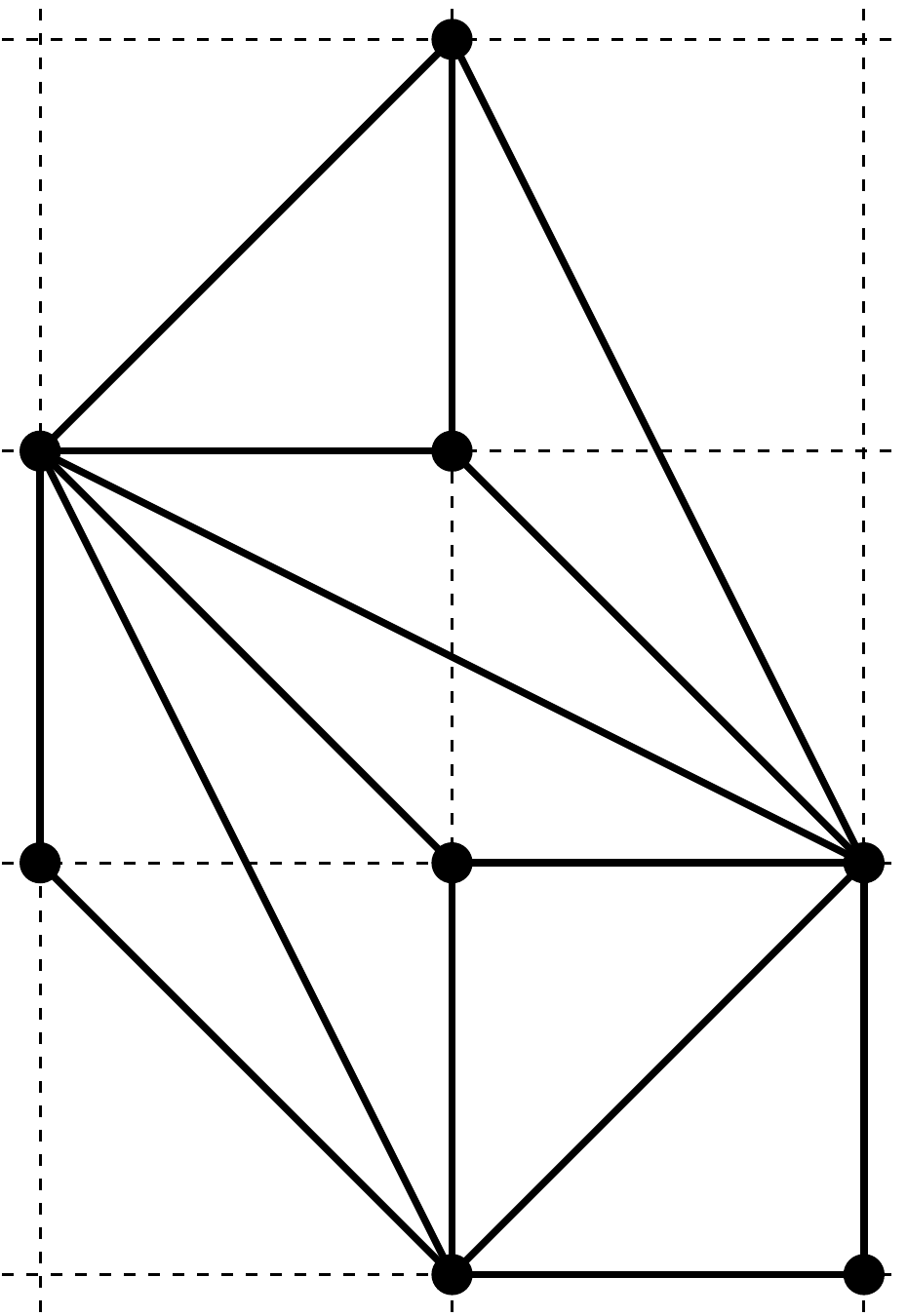}\label{fig:h1-res-n8}}\,
\subfigure[\small{}]{
\includegraphics[width=1.85cm]{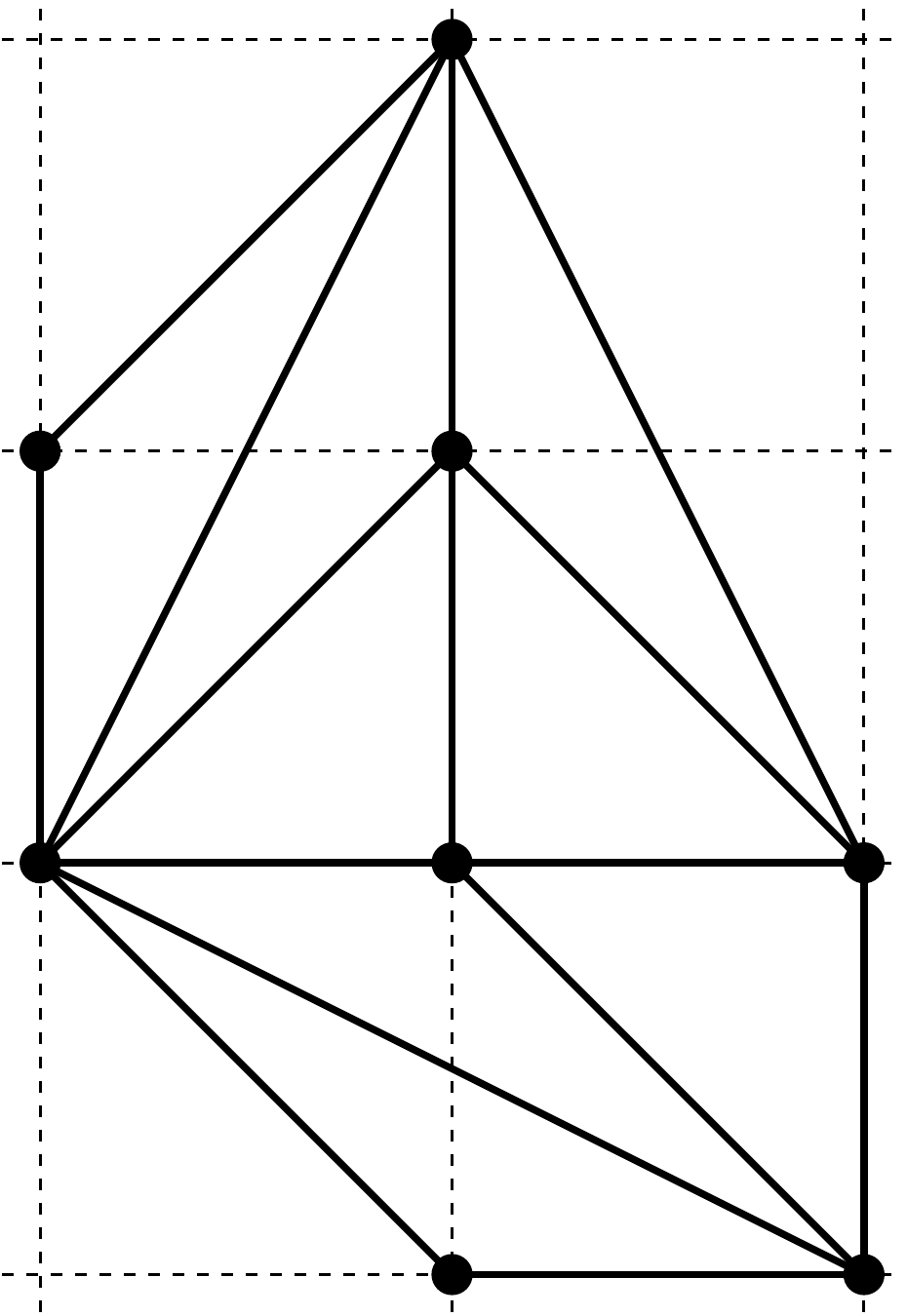}\label{fig:h1-res-n9}}\,
\subfigure[\small{}]{
\includegraphics[width=1.85cm]{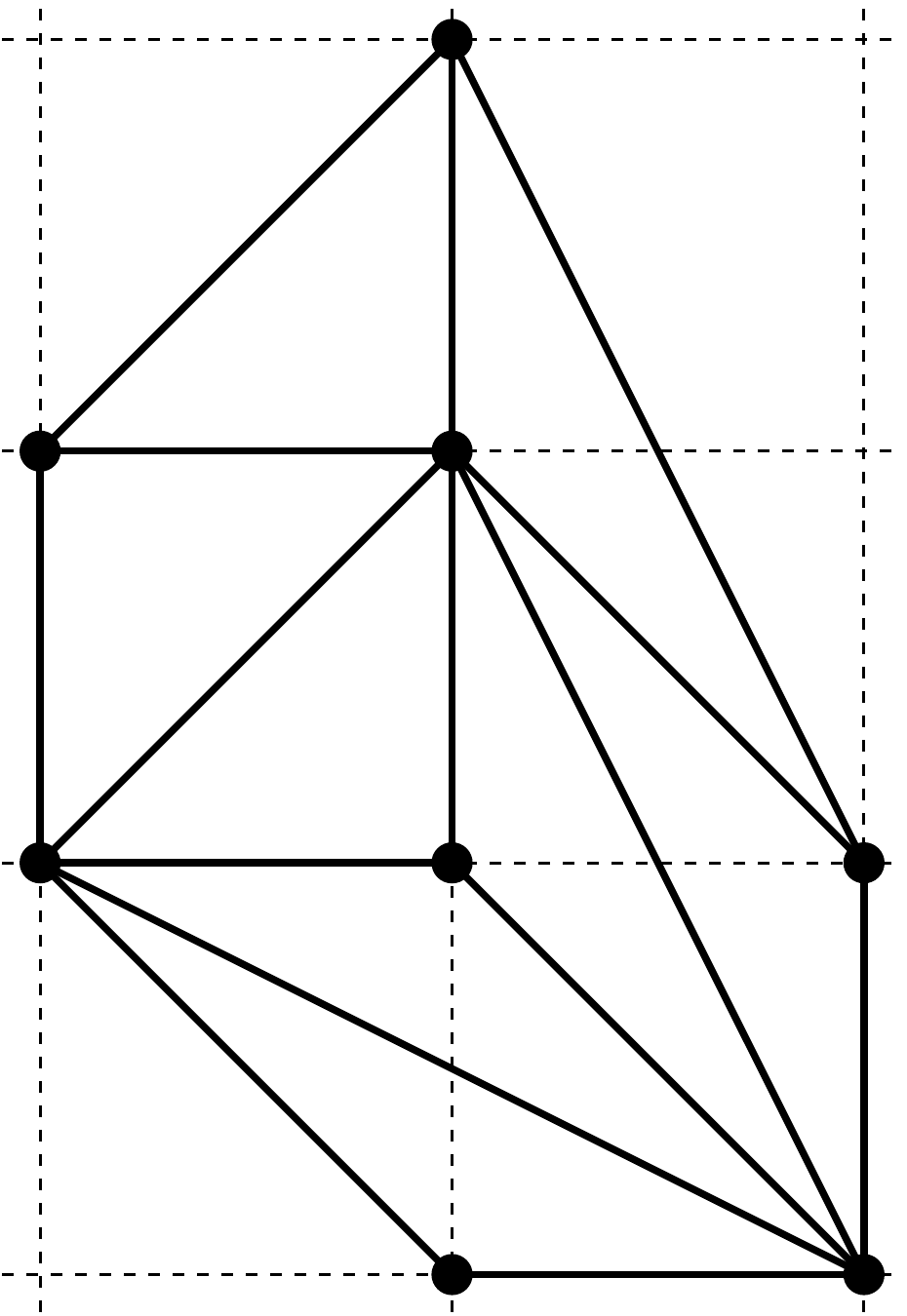}\label{fig:h1-res-n10}}\,
\subfigure[\small{}]{
\includegraphics[width=1.85cm]{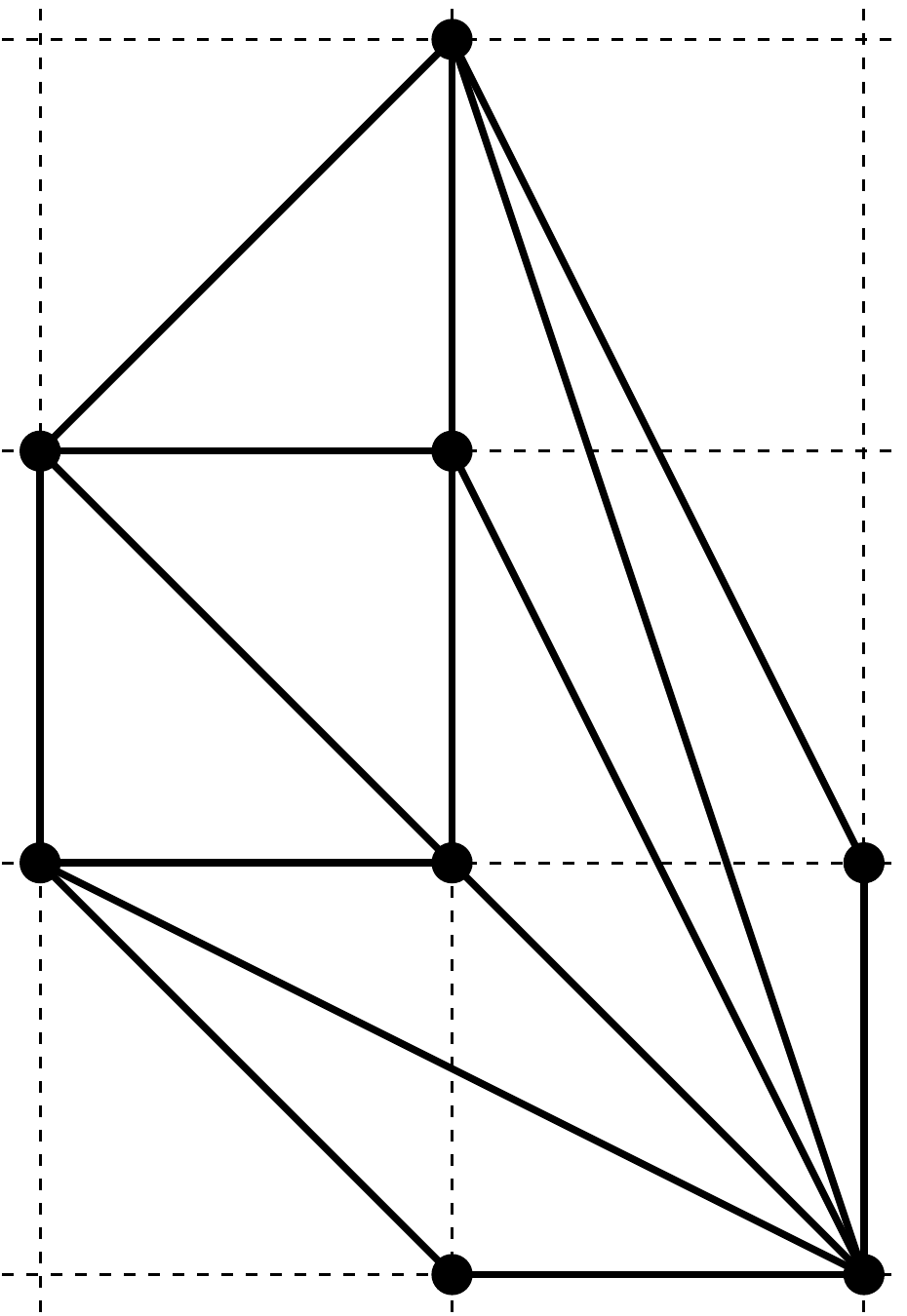}\label{fig:h1-res-n11}}\,
\subfigure[\small{}]{
\includegraphics[width=1.85cm]{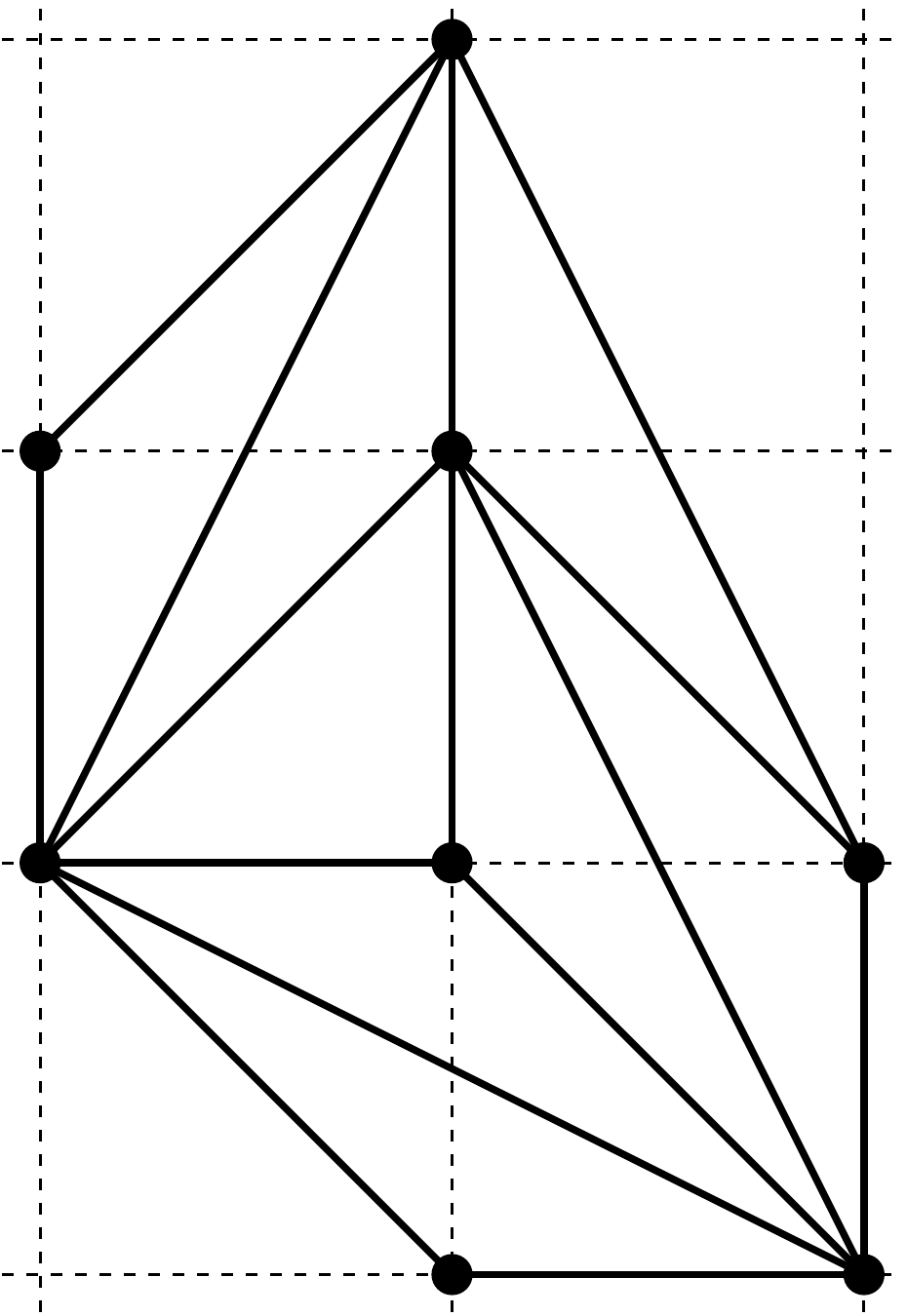}\label{fig:h1-res-n12}}\,
\subfigure[\small{}]{
\includegraphics[width=1.85cm]{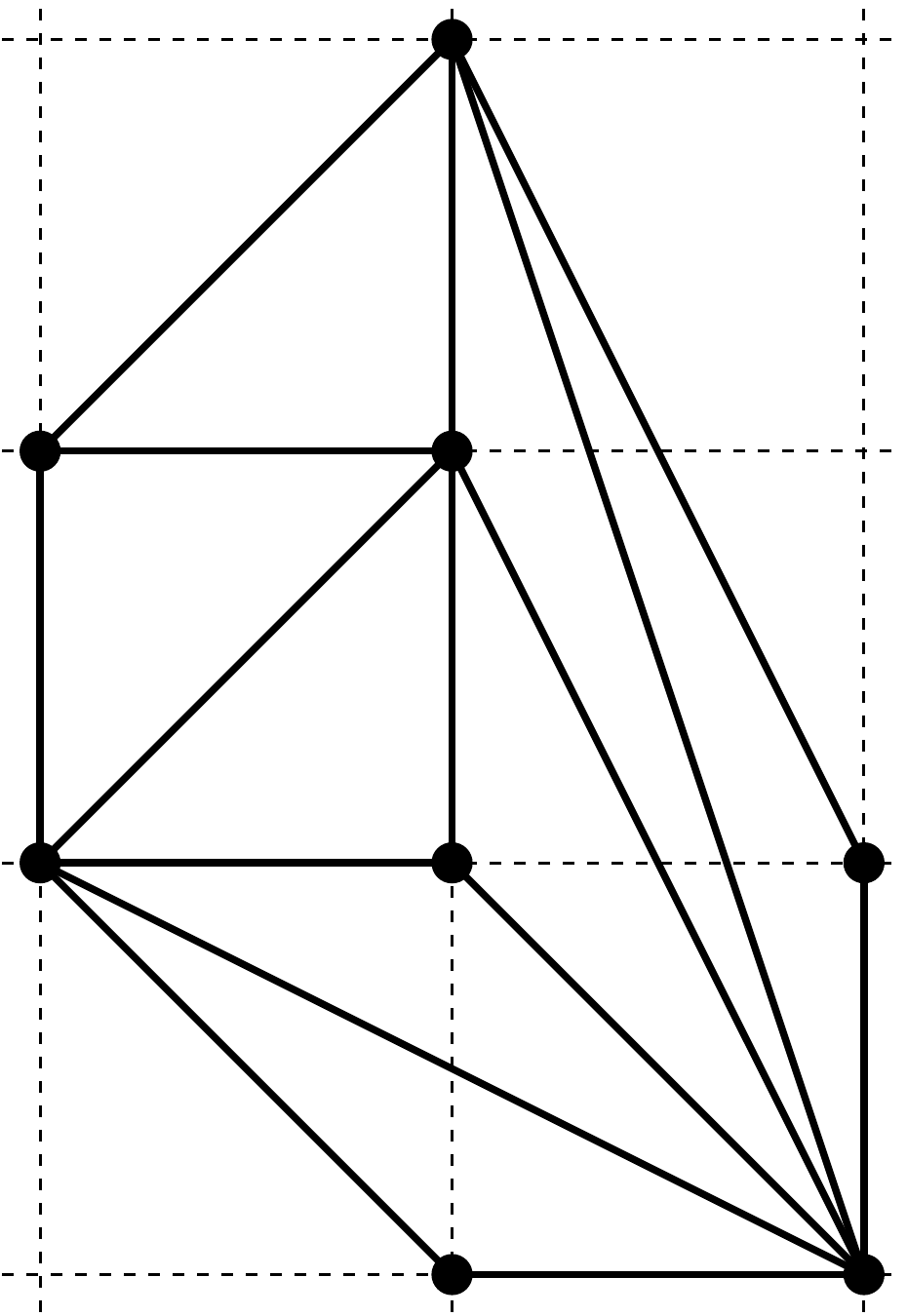}\label{fig:h1-res-n13}}\,
\subfigure[\small{}]{
\includegraphics[width=1.85cm]{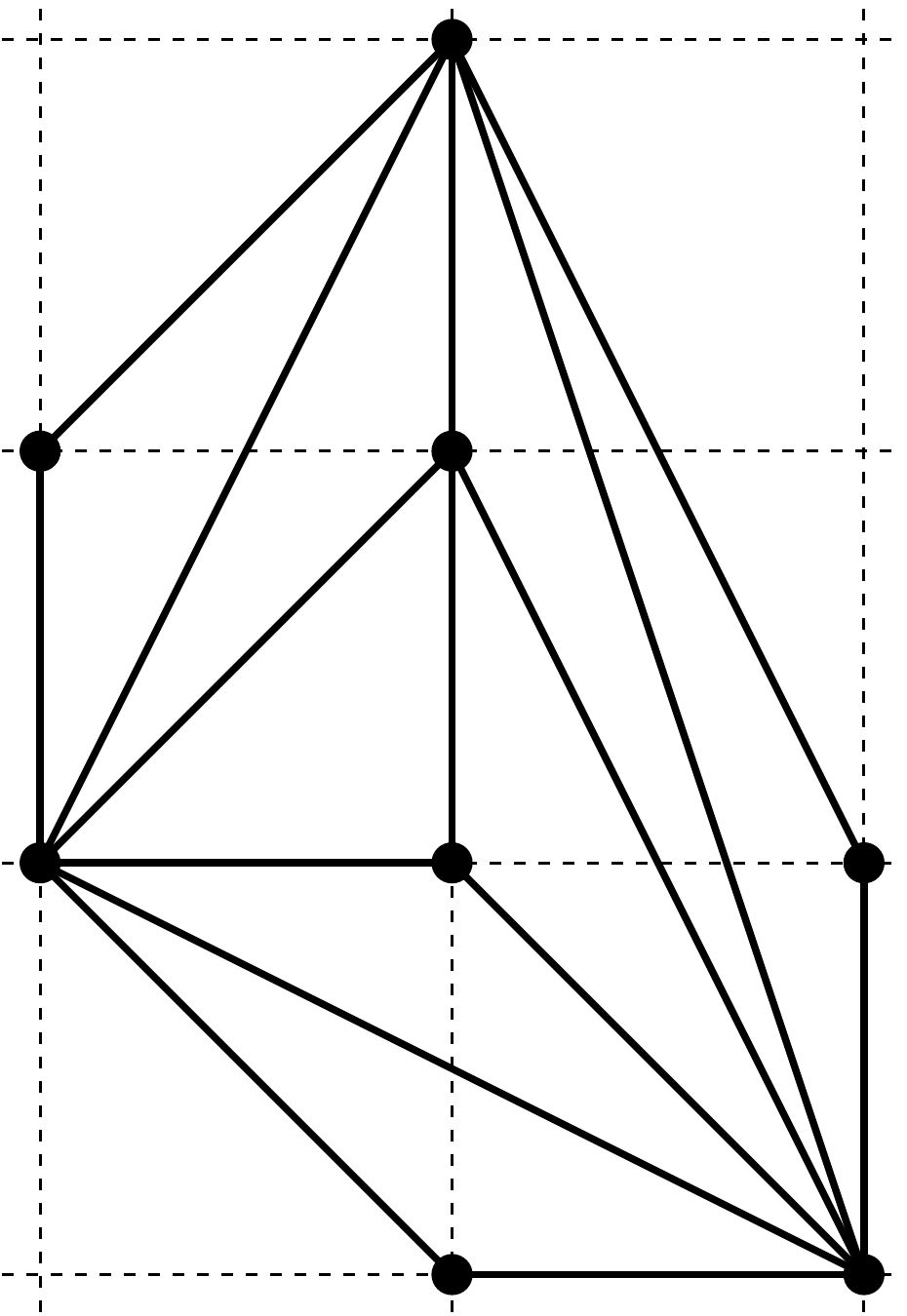}\label{fig:h1-res-n14}}\,
%%%%%%%%%%%%%%%%%%
\caption{The 30 crepant singularities of the $E_{3}{}^{2,1}$ singularity. The first 16, (a)-(p) admit a vertical reduction, corresponding to chambers of the $SU(3)_{2}$ $N_{\text{f}}=2$ gauge theory. \label{fig:h1 crepant all}}
\end{center}
\end{figure}

The $E_{3}{}^{2,1}$ singularity (Figure \ref{fig:h1-sing}) admits $30$ crepant resolutions shown in Figure \ref{fig:h1 crepant all}. The first 16 resolutions, Figures \ref{fig:h1-res-a}-\ref{fig:h1-res-p}, admit vertical reductions to type IIA, which correspond to chambers of the $SU(3)_{2}$ $N_{\text{f}}=2$ gauge theory, as we illustrate below. 
%%%%%%%%%%%%%%%%%%
%%%%%%%%

\begin{figure}[t]
\centering
\subfigure[\small{}]{
\includegraphics[width=4.5cm]{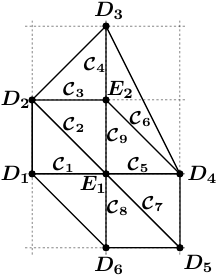}\label{fig:H1[1] res a labeled}}\,
\subfigure[\small{}]{
\includegraphics[width=7.5cm]{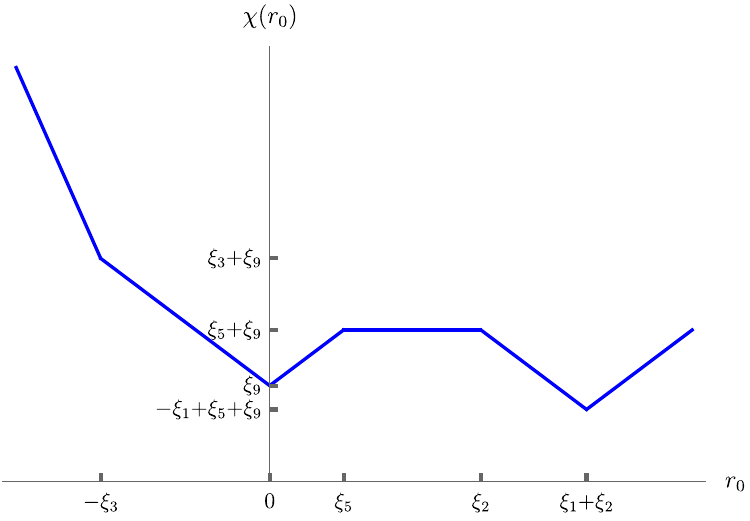}\label{fig:H[1] res a chi}}
\caption{Resolution (a) of the $E_{3}{}^{2,1}$ singularity and its vertical reduction.\label{fig:h1 res a final}}
\end{figure}

\paragraph{Resolution (a).} Consider the crepant resolution of Figure \ref{fig:h1-res-a}, with curves and divisors shown in Figure \ref{fig:H1[1] res a labeled}. 
%%%%%%%%%%%%%%%
%%%%%%%%%%%%%%%
There are six non-compact toric divisors $D_{i}$ ($i = 1, \ldots, 6$), and two compact toric divisors $\bE_1$ and $\bE_2$ with the following linear relations:
\begin{align}
D_5 &\simeq D_1 + D_2-D_4~, \quad \bE_1 \simeq -3D_1-2D_2+D_3+D_4-2D_6~, \quad \bE_2 \simeq D_1-2D_3-D_4+D_6 ~. \label{eq:h1 res a divisor linear equivalences}
\end{align}
The compact curves $\CC$ are given by:
\begin{align}\begin{split}
	\CC_1 &= \bE_1 \cdot D_1~, \quad \CC_2 = \bE_1 \cdot D_2~, \quad \CC_3 = \bE_2 \cdot D_2~, \quad \CC_4 = \bE_2 \cdot D_3 ~,\\
	\CC_5 &= \bE_1 \cdot D_4~, \quad \CC_6 = \bE_2 \cdot D_4~, \quad \CC_7 = \bE_1 \cdot D_5~, \quad \CC_8 = \bE_2 \cdot \bE_1 ~.
	\end{split}
\end{align}
The linear relations among curve classes are:
\begin{align}
\CC_4 &\simeq \CC_3 + \CC_9~, \quad \CC_6 \simeq \CC_3~, \quad \CC_7 \simeq \CC_1 + \CC_2 - \CC_5~, \quad \CC_8 \simeq -\CC_1 + \CC_5 + \CC_9 ~.
\end{align}
We take $\{\CC_1, \CC_2, \CC_3, \CC_5, \CC_9\}$ as Mori cone generators. The GLSM charge matrix is:
\be\label{intersect h1 res a}
\begin{tabular}{l|cccccccc|c}
 & $D_1$ &$D_2$& $D_3$ & $D_4$ & $D_5$ & $D_6$ & $\bE_1$& $\bE_2$ & vol($\CC$) \\
 \hline
$\CC_1$  & $-1$ & $1$ & $0$ &  $0$& $0$ & $1$ & $-1$ & $0$ & $\xi_1$ \\ \hline
$\CC_2$  & $1$ & $-1$ & $0$ &  $0$& $0$ & $0$ & $-1$ & $1$ & $\xi_2$ \\ \hline
$\CC_3$  & $0$ & $0$ & $1$ & $0$ & $0$ & $0$ & $1$ & $-2$ & $\xi_3$ \\ \hline
$\CC_5$  & $0$ & $0$ & $0$ & $-1$ & $1$ & $0$ & $-1$ & $1$ & $\xi_5$ \\ \hline
$\CC_9$  & $0$ & $1$ & $0$ & $1$ & $0$ & $0$ & $-1$ & $-1$ & $\xi_9$ \\ \hline\hline
$U(1)_M$ & $0$& $0$ &$0$ & $0$ & $0$ & $0$ & $-1$ & $1$ & $r_0$
 \end{tabular}
\ee 
where the last line defines the vertical reduction of the 2d GLSM. The non-negative FI terms $\xi_1 \geq 0$, $\xi_2 \geq 0$, $\xi_3 \geq 0$, $\xi_5 \geq 0$ and $\xi_9 \geq 0$ are, respectively, the volumes of compact curves $\CC_1$, $\CC_2$, $\CC_3$, $\CC_5$ and $\CC_9$. Note that the requirement that the curves $\CC_7$ and $\CC_9$ have non-negative volume translates to the following conditions on the FI terms in this chamber:
\begin{align}
 \xi_1 + \xi_2 - \xi_5 &\geq 0 ~,~ -\xi_1 + \xi_5 + \xi_9 \geq 0 ~.\label{eq:H[1] additional fi conditions}	
\end{align}
\paragraph{Geometric prepotential.} We parametrize the K\"{a}hler cone by:
\begin{align}
 S &= \mu_1 D_1 + \mu_2 D_2 + \mu_3 D_3 \nu_1 \bE_1 + \nu_2 \bE_2 ~.	 \label{eq:ess for H[1]}
\end{align}
The parameters $(\mu_1, \mu_2, \mu_3, \nu_1, \nu_2)$ are related to the FI parameters by
\begin{align}\begin{split}
\xi_1 &= -\mu_1 + \mu_2 - \nu_1 \geq 0~, \quad \xi_2 = \mu_1-\mu_2 -\nu_1 +\nu_2 \geq 0~, \quad \xi_3 = \mu_3 + \nu_1 - 2\nu_2 \geq 0 ~,\\
\xi_5 &= -\nu_1 + \nu_2 \geq 0~, \quad \xi_9 = \mu_2-\nu_1-\nu_2 \geq 0 ~. 
\end{split}\label{eq:h1 res a FI}
\end{align}
The relevant triple-intersection numbers are:
\begin{align}\hspace{-0.25in}\begin{array}{c@{~,\quad}c@{~,\quad}c@{~,\quad}c@{~,\quad}c@{~}}
 \bE_1^3=6 & \bE_2^3=8 & D_1 \bE_1 \bE_2=0 & D_2 \bE_1 \bE_2=1 & D_3 \bE_1 \bE_2=0 ~, \\
 \bE_1^2 \bE_2=-1 & \bE_1 \bE_2^2=-1 & D_1^2 \bE_1=-1 & D_1^2 \bE_2=0 & D_1 D_2 \bE_1=1 ~,\\
 D_1 D_2 \bE_2=0 & D_1 D_3 \bE_1=0 & D_1 D_3 \bE_2=0 & D_2^2 \bE_1=-1 & D_2^2 \bE_2=0 ~,\\
 D_2 D_3 \bE_1=0 & D_2 D_3 \bE_2=1 & D_3^2 \bE_1=0 & D_3^2 \bE_2=1 & D_1 \bE_1^2=-1 ~,\\
 D_1 \bE_2^2=0 & D_2 \bE_1^2=-1 & D_2 \bE_2^2=-2 & D_3 \bE_1^2=0 & D_3 \bE_2^2=-3 ~.
\end{array}
\end{align}
Therefore, the compact part of the prepotential is:
\begin{align}
\mathcal{F}_{(a)}(\nu_1, \nu_2; \mu_1, \mu_2, \mu_3) &= -\nu_1^3 - \frac{4}{3}\nu_2^3 + \frac{1}{2}(\nu_1^2\nu_2 + \nu_1\nu_2^2) + \frac{1}{2}(\mu_1+\mu_2)\nu_1^2 + \left(\mu_2 + \frac{3}{2}\mu_3\right)\nu_2^2\nonumber\\
& \quad -\mu_2 \nu_1\nu_2 + \frac{1}{2}(\mu_1-\mu_2)^2\nu_1 - \left(\mu_2\mu_3 + \frac{1}{2}\mu_3^2\right)\nu_2 ~.
 \label{eq:h1 res a geo prepot}
\end{align}
\paragraph{Type IIA reduction and gauge theory description.} The type IIA background is a resolved $A_1$ singularity fibered over the $x^9 = r_0$ direction. There are three D6-branes wrapping the exceptional $\P^1$ in the resolved $A_1$ singularity, resulting in an $SU(3)$ gauge theory. There are two D6-branes wrapping the two noncompact divisors in the resolved ALE space, which give rise to the two fundamental flavors. The volume of the exceptional $\P^1$ in type IIA is given by the following piecewise linear function,
\begin{align}\label{eq:h1 res a chi}
\chi(r_0) &= \left\{ 
                 \begin{array}{ll}
                 	-3r_0 - 2\xi_3 + \xi_9, & \text{for } r_0 \leq -\xi_3\\
                 	-r_0 + \xi_9, & \text{for } -\xi_3 \leq r_0 \leq 0\\
                 	+r_0 + \xi_9, & \text{for } 0 \leq r_0 \leq \xi_5\\
                 	\xi_5 + \xi_9, & \text{for } \xi_5 \leq r_0 \leq \xi_2 \\
                 	-r_0 + \xi_2 + \xi_5 + \xi_9, & \text{for } \xi_2 \leq r_0 \leq \xi_1 + \xi_2 \\
                 	+r_0 - 2\xi_1 - \xi_2 + \xi_5 + \xi_9, & \text{for } r_0 \geq \xi_1 + \xi_2 ~.
                 \end{array}
             \right.	
\end{align}
This function is sketched in Figure \ref{fig:H[1] res a chi} (where we have chosen $\xi_3 > \xi_5$ for convenience of plotting). At the points $r_0 = -\xi_3$, $r_0 = 0$ and $r_0 = \xi_1+\xi_2$, there are gauge D6-branes wrapping $\P^1$'s in the resolution of the singularity (we denote them by $\cG_1$, $\cG_2$ and $\cG_3$ respectively) . When $\xi_1 = \xi_2 = \xi_3 = 0$, an $SU(3)$ gauge theory is realized (at $r_0 = 0$) with inverse coupling $h_0 = \xi_9$. There are two flavor D6-branes at $r_0 = \xi_5$ and $r_0 = \xi_2$, denoted by $\cF_1$ and $\cF_2$ respectively. In the evaluation of $\chi(r_0)$, we have assumed without loss of generality that $\xi_2 \geq \xi_5$, which is of course consistent with \eqref{eq:H[1] additional fi conditions}.

The effective Chern-Simons level is given by:
\begin{align}
\kappa_{s,\text{eff}} &= -\frac{1}{2}(-3 + 1) = 1	~.
\end{align}
which is interpreted as a bare CS level of $2$ plus the contribution $-\frac{1}{2}-\frac{1}{2}=-1$ due to the two hypermultiplets (cf. \eqref{eq:massive Dirac fermion collection U(1) half quantization}). The simple-root W-bosons have masses given by the separation between adjacent gauge D6-branes:
\begin{align}
M(W_1) &= \xi_3 = 2\varphi_1 - \varphi_2~, \	qquad M(W_2) = \xi_1 + \xi_2 = 2\varphi_2 - \varphi_1 ~.
\end{align}
From the instanton masses, one can identify that this resolution corresponds to gauge theory chamber 11 (cf. Table \ref{tbl:u3 nf2 instantons} and \eqref{eq:ft prepot SU(3) k Nf2 chamber all}):
\begin{align}
\begin{split}
	M(\cI_1) &= \chi(r_0 = -\xi_3) = \xi_3 + \xi_9 = h_0 - m_1 - m_2 + 3 \varphi_1 ~,  \\
	M(\cI_2) &= \chi(r_0 = 0) = \xi_9 = h_0 - m_1 - m_2 + \varphi_1 + \varphi_2 ~,\\
	M(\cI_3) &= \chi(r_0 = \xi_1+\xi_2) = -\xi_1 + \xi_5 + \xi_9 = h_0 + \varphi_2 ~.
\end{split}	
\end{align}
The masses of hypermultiplets (due to open strings stretched between gauge and flavor branes) are:
\begin{align}
\begin{split}
	M(\cH_1) &= M(\cG_1 \cF_1) = \xi_3+\xi_5 = \varphi_1 + m_1 ~,\\
	M(\cH_2) &= M(\cG_2 \cF_1) = \xi_5 = -\varphi_1 + \varphi_2 + m_1 ~,\\
	M(\cH_3) &= M(\cG_3 \cF_1) = \xi_1 + \xi_2 - \xi_5 = \varphi_2 - m_1 ~,\\
	M(\cH_4) &= M(\cG_1 \cF_2) = \xi_2 + \xi_3 = \varphi_1 + m_2 ~,\\
	M(\cH_5) &= M(\cG_2 \cF_2) = \xi_2 = -\varphi_1 + \varphi_2 + m_2 ~,\\
	M(\cH_6) &= M(\cG_3 \cF_2) = \xi_1 = \varphi_2 - m_2 ~.
\end{split}	
\end{align}
Note that the choice $\xi_2 \geq \xi_5$ made above while computing $\chi(r_0)$ therefore implies that $m_2 \geq m_1$ in this chamber. From the K\"{a}hler volumes \eqref{eq:h1 res a FI} of the compact curves and masses of W-bosons and instantons, the map between geometry and field theory variables is determined to be:
\begin{align}
\text{$E_{3}{}^{2,1}$ geometry}&: \quad \left\{ 
\begin{array}{l}
\mu_1 = h_{0} + m_1, \\
\mu_2 = h_0 + 2m_1 - m_2, \\
\mu_3 = 3 m_1, \\
\nu_1 = -\varphi_2 + m_1, \\
\nu_2 = -\varphi_1 + 2m_1 ~.
\end{array}
\right.\label{eq:h1 res a mu nu}
\end{align} 
Plugging \eqref{eq:h1 res a mu nu} into \eqref{eq:h1 res a geo prepot}, we recover the field theory prepotential 
\begin{align}
\mathcal{F}_{SU(3)_{2},N_{\text{f}}=2}^{\text{chamber 11}} &= \frac{4}{3}\varphi_1^3 + \varphi_2^3 - \frac{1}{2}(\varphi_1^2\varphi_2 + \varphi_1\varphi_2^2) + (-h_0 + m_1 + m_2)\varphi_1\varphi_2 \nonumber\\
&+ (h_0-m_1-m_2)\varphi_1^2 + \left(h_{0}-\frac{m_1+m_2}{2}\right)\varphi_2^2 -\frac{1}{2}(m_1^2+m_2^2)\varphi_2 -\frac{1}{3}(m_1^3 + m_2^3)~.
\end{align}
up to $\varphi$-independent terms. From field theory, the monopole string tensions are given by:
\begin{align} 
T_{1,\text{ft}} = \frac{\partial \mathcal{F}_{SU(3)_{2},N_{\text{f}}=2}^{\text{chamber 11}}}{\partial \varphi_1} &= 4\varphi_1^2 -\varphi_1\varphi_2 - \frac{1}{2}\varphi_2^2 + 2(h_0-m_1-m_2)\varphi_1 \nonumber\\ &\quad + (-h_0+m_1+m_2)\varphi_2 ~,\\
T_{2,\text{ft}} = \frac{\partial \mathcal{F}_{SU(3)_{2},N_{\text{f}}=2}^{\text{chamber 11}}}{\partial \varphi_2} &= -\frac{1}{2}\varphi_1^2 - \varphi_1\varphi_2 + 3\varphi_2^2 + (-h_0 + m_1 + m_2)\varphi_1\nonumber\\
&\quad  + (2h_0-m_1-m_2)\varphi_2 - \frac{1}{2}(m_1^2+m_2^2) ~,
\end{align}
whereas from geometry, they are given by:
\begin{align} 
T_{1,\text{geo}} &= \int_{-\xi_3}^{0}\chi(r_0)\,dr_{0} = \frac{\xi _3^2}{2}+\xi _9 \xi _3~,\\
T_{2,\text{geo}} &= \int_{0}^{\xi_1+\xi_2}\chi(r_0)\,dr_{0} = -\frac{\xi _1^2}{2}+\xi _5 \xi _1+\xi _9 \xi _1+\xi _2 \xi _5+\xi _2 \xi _9-\frac{\xi _5^2}{2} ~.
\end{align} 
Using the map,
\begin{align}
\text{Resolution (a)}: \quad \left\{ 
   \begin{array}{l}
   	\xi_1 = \varphi_2-m_2~,\\
   	\xi_2 = -\varphi_1 + \varphi_2 + m_2~,\\
   	\xi_3 = 2\varphi_1-\varphi_2~,\\
   	\xi_5 = -\varphi_1 + \varphi_2 + m_1~,\\
   	\xi_9 = h_0 + \varphi_1 + \varphi_2 - m_1 - m_2~,
   \end{array}
\right.	
\end{align}
we find that $T_{i,\text{ft}} = T_{i,\text{geo}}$ for $i = 1, 2$. 
\paragraph{Magnetic walls.} The tensions vanish at loci given by:
\begin{align}\begin{split}
 (I)&: \{\xi_3 = 0\} \cup \left\{\frac{1}{2}\xi_3 + \xi_9 = 0\right\}~, \text{ and },\\
 (II)&: \left\{ \xi_1\xi_5 - \frac{1}{2}(\xi_1^2+\xi_5^2) + \xi_1\xi_9 + \xi_2\xi_5 + \xi_2\xi_9 = 0 \right\} ~.
 \end{split}
\end{align}
Along the submanifold $\{\xi_{3} = 0\} \subset (I)$, the W-boson $W_1$ in this chamber becomes massless. So this submanifold coincides with the hard wall which is the boundary of the Weyl chamber. The submanifold $\{\tfrac{1}{2}\xi_3 + \xi_9 \} = 0$ is not part of the K\"{a}hler chamber of resolution (1). So the locus $(I)$ contributes no magnetic walls.

The locus defined by $(II)$ is more intricate. The condition $(II)$ has two solutions:
\begin{align}
\xi _1 &\stackrel{(II)}{=} \xi _5+\xi _9 \pm \sqrt{(\xi_5+\xi_9)^2 +2 \xi_2(\xi_5+\xi_9)-\xi_5^2}\label{eq:magnetic wall H[1] condition II} ~.
\end{align}
Since \eqref{eq:H[1] additional fi conditions} requires that $\xi_8 = -\xi_1 + \xi_5 + \xi_9$ in this chamber, only the negative sign in \eqref{eq:magnetic wall H[1] condition II} is acceptable. However, in this chamber, recall that $\xi_2 \geq \xi_5$. So the quantity $2\xi_2(\xi_5 + \xi_9) - \xi_5^2 = 2\xi_2 \xi_9 + \xi_5(2\xi_2 - \xi_5)$ is always non-negative in this chamber. Therefore, the square root in \eqref{eq:magnetic wall H[1] condition II} is $\geq \xi_5 + \xi_9$, which implies that the right-hand-side of \eqref{eq:magnetic wall H[1] condition II} is negative, which is unphysical in the K\"{a}hler chamber of resolution (a). This implies that there are no magnetic walls in resolution (a) of the $E_{3}{}^{2,1}$ singularity.

Away from any hard wall, any one of a number of perturbative hypermultiplets, namely $\cH_2$, $\cH_3$, $\cH_5$ and $\cH_6$, can become massless, respectively at $\xi_5 = 0$, $\xi_7 = 0$, $\xi_2 = 0$ and $\xi_1 = 0$, signaling flops of the corresponding compact curves $\CC_5$, $\CC_7$, $\CC_2$ and $\CC_1$. These lead, respectively, to gauge theory phases described by resolutions (d), (e), (c) and (b).  Alternatively, away from the hard wall, the BPS instanton particles $\cI_2$ or $\cI_3$ can become massless, respectively, at $\xi_9 = 0$ (signaling a flop of $\CC_9)$ or $\xi_8 = -\xi_1+\xi_5+\xi_9 = 0$ (signaling a flop of $\CC_8$). These correspond to traversable instantonic wall, which lead, respectively to non-gauge-theoretic resolutions (q$_2$) or (q$_1$).

\begin{figure}[t]
\centering
\subfigure[\small{}]{
\includegraphics[width=4.5cm]{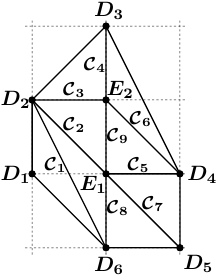}\label{fig:H1[1] res b labeled}}\,
\subfigure[\small{}]{
\includegraphics[width=7.5cm]{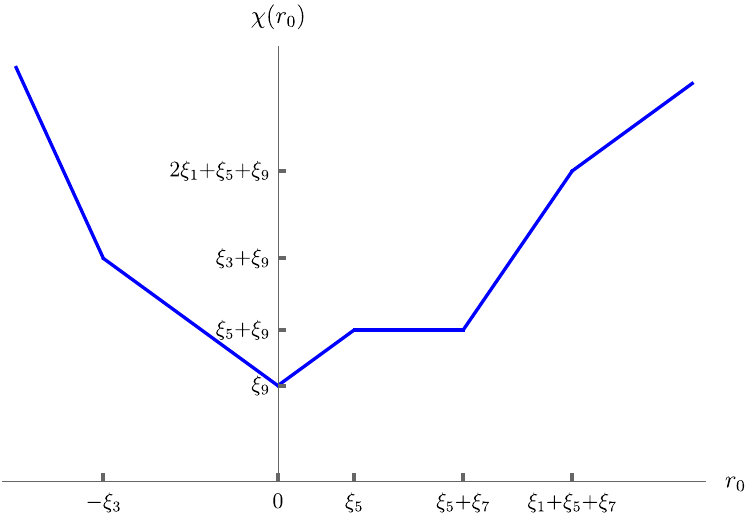}\label{fig:H[1] res b chi}}
\caption{Resolution (b) of the $E_{3}{}^{2,1}$ singularity and its vertical reduction.\label{fig:h1 res b final}}
\end{figure}

\paragraph {Resolution (b).} Consider the crepant resolution of Figure \ref{fig:h1-res-b}, with curves and divisors shown in Figure \ref{fig:H1[1] res b labeled}. This resolution can be obtained by a flop of the curve $\CC_1$ in resolution (a).
%%%%%%%%%%%%%%%
%%%%%%%%%%%%%%%
The linear equivalences among divisors are given by \eqref{eq:h1 res a divisor linear equivalences}. The compact curves $\CC$ can be read off the toric diagram.
%\begin{align}\begin{split}
%	\CC_1 &= D_2 \cdot D_6, \quad \CC_2 = \bE_1 \cdot D_2, \quad \CC_3 = \bE_2 \cdot D_2, \quad \CC_4 = \bE_2 \cdot D_3 ~,\\
%	\CC_5 &= \bE_1 \cdot D_4, \quad \CC_6 = \bE_2 \cdot D_4, \quad \CC_7 = \bE_1 \cdot D_5, \quad \CC_8 = \bE_2 \cdot \bE_1 ~.
%	\end{split}
%\end{align}
The linear relations among curve classes are:
\begin{align}
 \CC_2 &= \CC_5 + \CC_7~, \quad \CC_4 \simeq \CC_3 + \CC_9~, \quad \CC_6 \simeq \CC_3~, \quad \CC_8 = \CC_5 + \CC_9 ~.
\end{align}
We take $\{\CC_1, \CC_3, \CC_5, \CC_7, \CC_9\}$ as Mori cone generators. The GLSM charge matrix is:
\be\label{intersect h1 res b}
\begin{tabular}{l|cccccccc|c}
 & $D_1$ &$D_2$& $D_3$ & $D_4$ & $D_5$ & $D_6$ & $\bE_1$& $\bE_2$ & vol($\CC$) \\
 \hline
$\CC_1$  & $1$ & $-1$ & $0$ &  $0$& $0$ & $-1$ & $1$ & $0$ & $\xi_1$ \\ \hline
$\CC_3$  & $0$ & $0$ & $1$ &  $0$& $0$ & $0$ & $1$ & $-2$ & $\xi_3$ \\ \hline
$\CC_5$  & $0$ & $0$ & $0$ & $-1$ & $1$ & $0$ & $-1$ & $1$ & $\xi_5$ \\ \hline
$\CC_7$  & $0$ & $0$ & $0$ & $1$ & $-1$ & $1$ & $-1$ & $0$ & $\xi_7$ \\ \hline
$\CC_9$  & $0$ & $1$ & $0$ & $1$ & $0$ & $0$ & $-1$ & $-1$ & $\xi_9$ \\ \hline\hline
$U(1)_M$ & $0$& $0$ &$0$ & $0$ & $0$ & $0$ & $-1$ & $1$ & $r_0$
 \end{tabular}
\ee  
The K\"{a}hler cone is parametrized by \eqref{eq:ess for H[1]}. The parameters $(\mu_1, \mu_2, \mu_3, \nu_1, \nu_2)$ are related to the FI parameters by
\begin{align}\begin{split}
\xi_1 &= \mu_1-\mu_2 + \nu_1 \geq 0~, \quad \xi_3 =  \mu_3 + \nu_1 - 2\nu_2 \geq 0~, \quad \xi_5 =  -\nu_1 + \nu_2 \geq 0 ~,\\
\xi_7 &= -\nu_1 \geq 0~, \quad \xi_9 = \mu_2-\nu_1-\nu_2 \geq 0 ~. 
\end{split}\label{eq:h1 res b FI}
\end{align}
The relevant triple-intersection numbers are:
\begin{align}\hspace{-0.15in}\begin{array}{c@{~,\quad}c@{~,\quad}c@{~,\quad}c@{~,\quad}c@{~}}
 \bE_1^3=7 & \bE_2^3=8 & D_1 \bE_1 \bE_2=0 & D_2 \bE_1 \bE_2=1 & D_3 \bE_1 \bE_2=0 ~,\\
 \bE_1^2 \bE_2=-1 & \bE_1 \bE_2^2=-1 & D_1^2 \bE_1=0 & D_1^2 \bE_2=0 & D_1 D_2 \bE_1=0 ~,\\
 D_1 D_2 \bE_2=0 & D_1 D_3 \bE_1=0 & D_1 D_3 \bE_2=0 & D_2^2 \bE_1=0 & D_2^2 \bE_2=0 ~,\\
 D_2 D_3 \bE_1=0 & D_2 D_3 \bE_2=1 & D_3^2 \bE_1=0 & D_3^2 \bE_2=1 & D_1 \bE_1^2=0 ~,\\
 D_1 \bE_2^2=0 & D_2 \bE_1^2=-2 & D_2 \bE_2^2=-2 & D_3 \bE_1^2=0 & D_3 \bE_2^2=-3 ~.
\end{array}
\end{align}
Therefore, the compact part of the prepotential is:
\begin{align}
\mathcal{F}_{(b)}(\nu_1, \nu_2; \mu_1, \mu_2, \mu_3) &= -\frac{7}{6}\nu_1^3 - \frac{4}{3}\nu_2^3 + \frac{1}{2}(\nu_1^2\nu_2 + \nu_1\nu_2^2) + \mu_2^2 \nu_1^2 - \mu_2 \nu_1\nu_2 + \left(\mu_2 + \frac{3}{2}\mu_3\right)\nu_2^2\nonumber\\
&\qquad - \left(\mu_2 \mu_3 + \frac{1}{2}\mu_3^2\right)
 \label{eq:h1 res b geo prepot}
\end{align}
The IIA profile is:
\begin{align}\label{eq:h1 res b chi}
\chi(r_0) &= \left\{ 
                 \begin{array}{ll}
                 	-3r_0 - 2\xi_3 + \xi_9, & \text{for } r_0 \leq -\xi_3\\
                 	-r_0 + \xi_9, & \text{for } -\xi_3 \leq r_0 \leq 0\\
                 	+r_0 + \xi_9, & \text{for } 0 \leq r_0 \leq \xi_5\\
                 	\xi_5 + \xi_9, & \text{for } \xi_5 \leq r_0 \leq \xi_5+\xi_7 \\
                 	+2r_0 - \xi_5 - 2\xi_7 + \xi_9, & \text{for } \xi_5+\xi_7 \leq r_0 \leq \xi_1 + \xi_5 + \xi_7 \\
                 	+r_0 +\xi_1-\xi_7+\xi_9, & \text{for } r_0 \geq \xi_1 + \xi_5 + \xi_7 ~.
                 \end{array}
             \right.	
\end{align}
This function is sketched in Figure \ref{fig:H[1] res b chi} (where we have chosen $\xi_3 > \xi_5$ for convenience of plotting). At the points $r_0 = -\xi_3$, $r_0 = 0$ and $r_0 = \xi_5 + \xi_7$, there are gauge D6-branes wrapping $\P^1$'s in the resolution of the singularity (we denote them by $\cG_1$, $\cG_2$ and $\cG_3$ respectively). When $\xi_3 = \xi_5 = \xi_7 = 0$, an $SU(3)$ gauge theory is realized (at $r_0 = 0$) with inverse coupling $h_0 = \xi_9$. There are two flavor D6-branes at $r_0 = \xi_5$ and $r_0 = \xi_5+\xi_7$, denoted by $\cF_1$ and $\cF_2$ respectively.

The effective Chern-Simons level is, of course, still $1$. The simple-root W-bosons have masses given by:
\begin{align}
M(W_1) &= \xi_3 = 2\varphi_1 - \varphi_2~, \quad M(W_2) = \xi_5 + \xi_7 = 2\varphi_2 - \varphi_1 ~. 
\end{align}
This resolution can be identified with gauge theory chamber 12 (cf. Table \ref{tbl:u3 nf2 instantons} and \eqref{eq:ft prepot SU(3) k Nf2 chamber all}), with instanton masses given by:
\begin{align}
\begin{split}
	M(\cI_1) &= \chi(r_0 = -\xi_3) = \xi_3 + \xi_9 = h_0 - m_1 - m_2 + 3\varphi_1 ~,\\
	M(\cI_2) &= \chi(r_0 = 0) = \xi_9 = h_0 - m_1 - m_2 + \varphi_1 + \varphi_2 ~\\
	M(\cI_3) &= \chi(r_0 = \xi_5+\xi_7) = \xi_5 + \xi_9 = h_0 - m_2 + 2\varphi_2 ~.
\end{split}	
\end{align}
The masses of hypermultiplets are:
\begin{align}
\begin{split}
	M(\cH_1) &= M(\cG_1 \cF_1) = \xi_3 + \xi_5 = \varphi_1 + m_1 ~,\\
	M(\cH_2) &= M(\cG_2 \cF_1) = \xi_5 = -\varphi_1 + \varphi_2 + m_1 ~,\\
	M(\cH_3) &= M(\cG_3 \cF_1) = \xi_7 = \varphi_2 - m_1 ~,\\
	M(\cH_4) &= M(\cG_1 \cF_2) = \xi_1 + \xi_3 + \xi_5 + \xi_7 = \varphi_1 + m_2 ~,\\
	M(\cH_5) &= M(\cG_2 \cF_2) = \xi_1 + \xi_5 + \xi_7 = -\varphi_1 + \varphi_2 + m_2 ~,\\
	M(\cH_6) &= M(\cG_3 \cF_2) = \xi_1 = -\varphi_2 + m_2 ~.
\end{split}	
\end{align}
It is easy to verify that the map between geometry and field theory variables is the still \eqref{eq:h1 res a mu nu}, as it must be.
%From the K\"{a}hler volumes \eqref{eq:h1 res b FI} of the compact curves and masses of W-bosons and instantons, the map between geometry and field theory variables is determined to be
%\begin{align}
%\text{$E_{3}{}^{2,1}$ geometry}&: \quad \left\{ 
%\begin{array}{l}
%\mu_1 = h_{0} + m_1, \\
%\mu_2 = h_0 + 2m_1 - m_2, \\
%\mu_3 = 3 m_1, \\
%\nu_1 = -\varphi_2 + m_1, \\
%\nu_2 = -\varphi_1 + 2m_1 \, \,.
%\end{array}
%\right.\label{eq:h1 res a mu nu}
%\end{align} 
Plugging \eqref{eq:h1 res a mu nu} into \eqref{eq:h1 res b geo prepot}, we recover the field theory prepotential:
\begin{align}
\mathcal{F}_{SU(3)_{2},N_{\text{f}}=2}^{\text{chamber 12}} &= \frac{4}{3}\varphi_1^3 + \frac{7}{6}\varphi_2^3 + \frac{1}{2}(\varphi_1^2\varphi_2 + \varphi_1\varphi_2^2) + (h_0-m_1-m_2)\varphi_1^2 + (-h_0 + m_1 + m_2)\varphi_1\varphi_2 \nonumber\\
&\qquad + \left(h_0 -\frac{1}{2}m_1 - m_2\right)\varphi_2^2 - \frac{1}{2}m_1^2\varphi_2 ~,
\end{align}
up to $\varphi$-independent terms. %From field theory, the monopole string tensions are given by:
%\begin{align} 
%T_{1,\text{ft}} = \frac{\partial \mathcal{F}_{SU(3)_{2},N_{\text{f}}=2}^{\text{chamber 12}}}{\partial \varphi_1} &= 2 \left(h_0-m_1-m_2\right) \varphi _1+4 \varphi _1^2+\left(-h_0+m_1+m_2\right) \varphi _2\nonumber\\&\qquad -\varphi _1 \varphi
%   _2-\frac{\varphi _2^2}{2}~,\\
%T_{2,\text{ft}} \frac{\partial \mathcal{F}_{SU(3)_{2},N_{\text{f}}=2}^{\text{chamber 12}}}{\partial \varphi_2} &= -\frac{m_1^2}{2}+\left(-h_0+m_1+m_2\right) \varphi _1-\frac{\varphi _1^2}{2}+\left(2h_0-m_1-2m_2\right)
%   \varphi _2\nonumber\\&\qquad-\varphi _1 \varphi _2+\frac{7 \varphi _2^2}{2}~,
%\end{align}
The monopole string tensions from the IIA profile are:
\begin{align} 
T_{1,\text{geo}} &= \int_{-\xi_3}^{0}\chi(r_0)\,dr_{0} = \frac{\xi _3^2}{2}+\xi _3\xi _9 ~,\\
T_{2,\text{geo}} &= \int_{0}^{\xi_5+\xi_7}\chi(r_0)\,dr_{0} = \frac{\xi _5^2}{2}+\xi _5 \xi _9+\xi _7 \left(\xi _5+\xi _9\right) ~.
\end{align} 
Using the map,
\begin{align}
\text{Resolution (b)}: \quad \left\{ 
   \begin{array}{l}
   	\xi_1 = -\varphi_2+m_2~,\\
   	\xi_3 = 2\varphi_1-\varphi_2~,\\
   	\xi_5 = -\varphi_1+\varphi_2 + m_1~,\\
   	\xi_7 = -\varphi_2 + m_1~,\\
   	\xi_9 = h_0 - m_1 - m_2 + \varphi_1 + \varphi_2~,
   \end{array}
\right.	
\end{align}
we find that $T_{i,\text{ft}} = T_{i,\text{geo}}$ for $i = 1, 2$. The tensions vanish at loci given by:
\begin{align}\hspace{-0.1in}%\begin{split}
 (I)&: \{\xi_3 = 0\} \cup \left\{\frac{1}{2}\xi_3 + \xi_9 = 0\right\}~, \text{ and }~%\\
 (II): \left\{ \frac{\xi _5^2}{2}+\xi _5 \xi _9+\xi _7 \left(\xi _5+\xi _9\right) = 0 \right\} ~.
 %\end{split}
\end{align}
Along the locus $\{\xi_{3} = 0\} \subset (I)$, the W-boson $W_1$ in this chamber becomes massless. So this submanifold coincides with the hard wall which is the boundary of the Weyl chamber. The locus $\{\tfrac{1}{2}\xi_3 + \xi_9 = 0\}\subset (I)$ is not part of the K\"{a}hler chamber of resolution (1). The quadratic condition $(II)$ has two solutions: $\xi_5 =  -\xi_7-\xi_9 \pm \sqrt{\xi_7^2 + \xi_9^2}$. Both sign choices yield a negative value of $\xi_5$, which is inconsistent in this K\"{a}hler chamber. This implies that there are no magnetic walls in resolution (b) of the $E_{3}{}^{2,1}$ singularity. 

Away from any hard wall in this chamber, one of the three perturbative hypermultiplets $\cH_2$, $\cH_3$ or $\cH_6$ can become massless, respectively, at $\xi_5 = 0$, $\xi_7 = 0$ or $\xi_1 = 0$. These to flops of $\CC_5$, $\CC_7$ or $\CC_1$, which lead, respectively to gauge-theory resolutions (f), (g) or (a). Alternatively, away from any hard wall, the BPS instanton particle $\cI_2$ can become massless at $\xi_9 = 0$ (signaling a flop of $\CC_9$), indicating a traversible insantonic wall which leads to non-gauge-theoretic resolution (q$_3$).

\paragraph{RG Flow.} In this resolution, one can decouple the divisor $D_1$ by sending the volume of the curve $\CC_1$ to infinity, which is equivalent to sending the mass $m_2$ to $+\infty$. It is easy to see that this yields resolution (a) of the $E_{2}{}^{2,\fhalf}$ singularity shown in Figure \ref{fig:p1 res a final}. This is consistent with the fact that the effective Chern-Simons level changes from $k_{\text{eff}} = 1$ to $k_{\text{eff}} = 1-\frac{1}{2} = \frac{1}{2}$ (recall \eqref{eq:massive Dirac fermion collection U(1) half quantization}). An inspection of Figure \ref{fig:h1 crepant all} suggests that many such transitions are possible, including those involving non-gauge-theoretic phases.

\paragraph{The geometry $\leftrightarrow$ field theory map.} We can repeat the above analysis for  the remaining fourteen resolutions that admit vertical reductions, and match each of them to chambers of $SU(3)_{2}$ $N_{\text{f}}=2$ field theory (see Table \ref{tbl:chamber defs u3 nf2}). To do so, one may exploit the fact that map \eqref{eq:h1 res a mu nu} is constant across all resolutions. The result of this matching is outlined in Table \ref{tbl:map between H[1] geometry and field theory}.
\begin{table}[h]
%\hspace{-0.2in}
{\footnotesize
\begin{tabular}{|c|c|c|c|c|c|c|c|}\hline
 \text{Resolution} & \text{Chamber} & $\varphi _1{+}m_1$ & ${-}\varphi_1{+}\varphi _2{+}m_1$ & ${-}\varphi_2{+}m_1$ & $\varphi_1{+}m_2$ & ${-}\varphi_1{+}\varphi_2{+}m_2$ & ${-}\varphi_2{+}m_2$ \\ \hline
 (a)  & \text{ 11 } & \text{$\geq$ 0} & \text{$\geq$ 0} & \text{$<$ 0} & \text{$\geq$ 0} & \text{$\geq$ 0} & \text{$<$ 0} \\ \hline % res 1 = a
 (b) & \text{ 12 } & \text{$\geq$ 0} & \text{$\geq$ 0} & \text{$<$ 0} & \text{$\geq$ 0} & \text{$\geq$ 0} & \text{$\geq$ 0} \\ \hline % res 2 = b
 (c) & \text{ 10 } & \text{$\geq$ 0} & \text{$\geq$ 0} & \text{$<$ 0} & \text{$\geq$ 0} & \text{$<$ 0} & \text{$<$ 0} \\ \hline % res 3 = c
 (d) & \text{ 7 } & \text{$\geq$ 0} & \text{$<$ 0} & \text{$<$ 0} & \text{$\geq$ 0} & \text{$\geq$ 0} & \text{$<$ 0} \\ \hline % res 4 = d
 (e) & \text{ 15 } & \text{$\geq$ 0} & \text{$\geq$ 0} & \text{$\geq$ 0} & \text{$\geq$ 0} & \text{$\geq$ 0} & \text{$<$ 0} \\ \hline % res 5 = e
 (f) & \text{ 8} & \text{$\geq$ 0} & \text{$<$ 0} & \text{$<$ 0} & \text{$\geq$ 0} & \text{$\geq$ 0} & \text{$\geq$ 0} \\ \hline % res 8 = f
 (g) & \text{ 16 } & \text{$\geq$ 0} & \text{$\geq$ 0} & \text{$\geq$ 0} & \text{$\geq$ 0} & \text{$\geq$ 0} & \text{$\geq$ 0} \\ \hline % res 9 = g
 (h) & \text{ 9 } & \text{$\geq$ 0} & \text{$\geq$ 0} & \text{$<$ 0} & \text{$<$ 0} & \text{$<$ 0} & \text{$<$ 0} \\ \hline % res 11 = h
 (i) & \text{ 6 } & \text{$\geq$ 0} & \text{$<$ 0} & \text{$<$ 0} & \text{$\geq$ 0} & \text{$<$ 0} & \text{$<$ 0} \\ \hline % res 12 = i
 (j) & \text{ 14 } & \text{$\geq$ 0} & \text{$\geq$ 0} & \text{$\geq$ 0} & \text{$\geq$ 0} & \text{$<$ 0} & \text{$<$ 0} \\ \hline % res 13 = j
 (k) & \text{ 3} & \text{$<$ 0} & \text{$<$ 0} & \text{$<$ 0} & \text{$\geq$ 0} & \text{$\geq$ 0} & \text{$<$ 0} \\ \hline % res 15 = k
 (l) & \text{ 4} & \text{$<$ 0} & \text{$<$ 0} & \text{$<$ 0} & \text{$\geq$ 0} & \text{$\geq$ 0} & \text{$\geq$ 0} \\ \hline % res 19 = l
 (m) & \text{ 5 } & \text{$\geq$ 0} & \text{$<$ 0} & \text{$<$ 0} & \text{$<$ 0} & \text{$<$ 0} & \text{$<$ 0} \\ \hline % res 21 = m
 (n) & \text{ 13 } & \text{$\geq$ 0} & \text{$\geq$ 0} & \text{$\geq$ 0} & \text{$<$ 0} & \text{$<$ 0} & \text{$<$ 0} \\ \hline % res 22 = n
 (o) & \text{ 2 } & \text{$<$ 0} & \text{$<$ 0} & \text{$<$ 0} & \text{$\geq$ 0} & \text{$<$ 0} & \text{$<$ 0} \\ \hline % res 24 = o
 (p) & \text{ 1 } & \text{$<$ 0} & \text{$<$ 0} & \text{$<$ 0} & \text{$<$ 0} & \text{$<$ 0} & \text{$<$ 0} \\ \hline % res 27 = p
\end{tabular}
\caption{The map between the $16$ crepant resolutions of the $E_{3}{}^{2,1}$ singularity that admit a vertical reduction, and the $16$ chambers of the $SU(3)_{2}$ $N_{\text{f}}=2$ field theory, with field theory chamber definitions expressed as inequalities.\label{tbl:map between H[1] geometry and field theory}}
}
\end{table}
%%%%%%%%%

The triple-intersection numbers of all $30$ crepant resolutions (including the $14$ which are non-gauge-theoretic phases) are listed in Appendix \ref{sec:intnos H[1]}, and the expressions for the corresponding M-theory prepotentials are listed in Appendix \ref{sec:geo prepot H[1]}.

\subsection{The $E_{3}{}^{2,0}$ singularity and $SU(3)_{0}$ $N_{\text{f}} =2$ gauge theory}
The $E_{3}{}^{2,0}$ singularity of Figure \ref{fig:h4-sing} admits $24$ crepant resolutions. However, an $SL(2,\Z)$ transformation -- for example, by $ST^{-1}S^2$ -- transforms this singularity to the ``beetle singularity'' that was extensively analyzed in \cite{Closset:2018bjz}. An interesting feature of the beetle singularity is that there are both horizontal as well as vertical reductions, and these describe either the $SU(2)\times SU(2)$ theory, or the $SU(3)_{0}$ $N_{\text{f}}=2$ theory. As we have remarked before, this geometry can lead to a number of smaller geometries (see, for instance, Figure \ref{fig:RG flow chain}), including, in particular the $E_{2}{}^{\pm \frac{1}{2}}$ geometries, and the $E_{1}{}^{2,\ell}$ geometries for $\ell = -1,0,1$, but also geometries corresponding to non-gauge-theoretic phases. We refer the reader to \cite{Closset:2018bjz} for details.

\section*{Acknowledgements} 
\normalsize I would like to especially thank Cyril Closset and Michele Del Zotto for ongoing collaboration on related topics, and for their careful reading of an earlier draft. I thank Andr\'{e}s Collinucci, Nathan Haouzi, Radu Iona\c{s}, Daniel Lichtblau, Nicol\`{o} Piazzalunga, Martin Ro\v{c}ek, and Alessandro Tomasiello for interesting discussions and correspondence. I thank the referee for helpful suggestions. This work is supported in part by NSF grant PHY-1620628.

\appendix
%
%\section{The 5d $\mathcal{N}=1$ supersymmetry algebra and central charges}
%The most general (i.e. centrally extended) $\mathcal{N}$-extended Poincar\'{e} superalgebra in $d = 1+4$ dimensions has an $Sp(\mathcal{N}) \cong USp(2\mathcal{N})$ R-symmetry, and has the form \cite{Cremmer:1980gs,Ferrara:1999si}:
%\begin{align}
%\{Q_{\alpha}^A, Q_{\beta}^B\} &= (\gamma^{\mu}C)_{\alpha\beta} P_{\mu}\Omega^{AB} + (\gamma^{\mu}C)_{\alpha\beta} Z_{\mu}^{\circ[AB]} + C_{\alpha\beta} Z^{[AB]} + (\gamma^{\mu\nu}C)_{\alpha\beta} Z_{\mu\nu}^{(AB)}  ~,	\label{eq:5d superpoincare}
%\end{align}
%where $\alpha, \beta$ here are spinor indices (only for the purposes of this equation, and not to be confused with their use elsewhere in this paper), $\mu, \nu$ are five-dimensional spacetime indices, $C$ is the charge conjugation matrix, $\Omega$ denotes the symplectic form, and (crucially) $Z$'s denote \textit{real} central charges. The indices $A, B$ range over $A, B = 1, \ldots, 2\mathcal{N}$. Here $Z_{\mu}^{\circ[AB]}$ and $Z^{[AB]}$ are antisymmetric, and $Z_{\mu}^{\circ[AB]}$ is symplectic traceless: $\Omega_{AB}Z_{\mu}^{\circ[AB]} = 0$. The central charges $Z_{\mu}^{\circ[AB]}$ and $Z_{\mu\nu}^{(AB)}$ contribute to strings and membranes, respectively, whereas $Z^{[AB]}$ contributes to particle states that enter \eqref{eq:mass BPS particles} (see below). For the purposes of this paper, since we focus on 5d SCFTs, we restrict to $\mathcal{N}=1$ in \eqref{eq:5d superpoincare}.

\section{Field theory prepotentials and instanton masses\label{sec:field theory prepotentials}}
\normalsize In this section, we list the expressions for the field theory prepotential \eqref{eq:our prepot} evaluated for the models that we have analyzed in this paper, and also the instanton masses in each chamber. We begin by describing the procedure used to compute the instanton masses in field theory.

For the special case of $\GG = \prod_{i}SU(n_i)$, we follow \cite{Assel:2018rcw,Closset:2018bjz} and consider an auxiliary gauge group $\GG' = \prod_{i}U(n_i)$ obtained by replacing each $SU(n_i)$ gauge group factor with $U(n_i)$. Let us denote the $U(n_i)$ Coulomb branch vevs by $\phi_{i,(1)}, \phi_{i,(2)}, \ldots, \phi_{i,(n_i)}$. Then for each such gauge group $U(n_i)$ there are $n_i$ ``instanton states'' with masses given by the second derivatives of the prepotential:
      \begin{align}
        M(\mathcal{I}_{U(n_i)}^{(k)}) &= \left.\frac{\partial^2\FF}{\partial \phi_{i,(k)}^2}\right|_{U(n_i) \rightarrow SU(n_i)}	~, \quad \text{ for } k = 1, \ldots, n_i ~. \label{eq:BPS instanton mass}
      \end{align}
      Here the notation $U(n_i) \rightarrow SU(n_i)$ refers to the operation of imposing the ``traceless condition,'' to transform to $SU(n_i)$ variables, \textit{after} computing the second derivative. This entails the following substitution for every $U(n_i)$ factor in $\GG'$:
      \begin{align}
       	   \phi_{i,a} &= \varphi_{i,a} - \varphi_{i,a-1}, \quad a = 1, \ldots, n_i, \quad \text{ with } \varphi_{a,0} = \varphi_{a,n_i} = 0 ~. \label{eq:un to sun}
      \end{align}
      Here we have denoted the Coulomb vevs of $SU(n_i)$ by $\varphi_{i,1}, \ldots, \varphi_{i,n_i-1}$. The leading term is $M(\mathcal{I}) = h_{0} + \cdots$ corresponding to the familiar fact that instanton masses scale as $h_{0} = \frac{8\pi^2}{g^2}$. %The instanton particle with the lowest mass can be treated as the elementary instanton, with other instanton particles being interpreted as marginal bound states of the elementary instanton with other BPS particles (W-bosons or hypermultiplets) \cite{Aharony:1997bh}.

\paragraph{$U(3)_{k}$ field theory.} The fundamental Weyl chamber is defined by $\phi_1 \geq \phi_2 \geq \phi_3$.  The gauge theory prepotential \eqref{eq:our prepot} for pure 5d $U(3)_{k}$ gauge theory is: 
\begin{align}\label{eq:u3 k prepot}
\cF_{U(3)_{k}} &= \frac{1}{2}h_0(\phi_1^2 + \phi_2^2+ \phi_3^2) + \frac{k}{6}(\phi_1^3 + \phi_2^3 + \phi_3^2) + \frac{1}{6}\left[(\phi_1-\phi_2)^3 + (\phi_2-\phi_3)^3 + (\phi_1-\phi_3)^3\right] ~.
\end{align}

\paragraph{$SU(3)_{k_{\text{eff}}}$ + $N_{\text{f}} = 1$ field theory.} \normalsize The $U(3)_{k_{\text{eff}}}$ $N_{\text{f}}=1$ field theory is specified by 3 Coulomb vevs $(\phi_1, \phi_2, \phi_3)$, one inverse gauge coupling $h_{0}$, and one real mass $m \in \R$. There are 4 chambers inside the fundamental Weyl chamber given by $\phi_1 \geq \phi_2 \geq \phi_3$. These are defined by the allowed ranges of the arguments of the $\Theta$ functions in \eqref{eq:our prepot}. The instanton masses in various chambers, computed using \eqref{eq:BPS instanton mass}, are listed in Table \ref{tbl:u3 nf1 instantons} to facilitate calculations in the main text. Here $k$ denotes the bare CS level, given in terms of the effective CS level $k_{\text{eff}}$ for $N_{\text{f}}=1$ by
\begin{align}
k \equiv k_{\text{bare}} = k_{\text{eff}} + \frac{1}{2} ~,	
\end{align}
in the $U(1)_{-\frac{1}{2}}$ quantization scheme (cf. \eqref{eq:U(1) half quantization}).
\begin{table}[ht!]	
\centering
{\footnotesize
\begin{tabular}{|c|c|c|c|c|c|c|c|}\hline
  $\#$ & \begin{tabular}{c} $\phi_1 + m$ \\ \text{ or } \\ $\varphi_1{+}m$ \end{tabular} & \begin{tabular}{c} $\phi_2 + m$ \\ \text{ or } \\ $-\varphi_1{+}\varphi_2{+}m$ \end{tabular} & \begin{tabular}{c} $\phi_3 + m$ \\ \text{ or } \\ $-\varphi_2{+}m$ \end{tabular} & $M(\cI_1)$ & $M(\cI_2)$ &  $M(\cI_3)$ \\ \hline
 \text{1} & \text{$<$ 0} & \text{$<$ 0} & \text{$<$ 0} & $h_{0}{+}(k+3)\varphi_1$ &  \begin{tabular}{c}$h_{0}{+}(1-k)\varphi_1$\\${+}(1+k)\varphi_2$\end{tabular} & $h_{0}{+}(3-k)\varphi_2$ \\ \hline
  \text{2} & \text{$\geq$ 0} & \text{$<$ 0} & \text{$<$ 0} & \begin{tabular}{c}$h_0{-}m$\\${+}(k{+}2)\varphi_1$\end{tabular} &  \begin{tabular}{c}$h_0 + (1-k)\varphi_1$\\${+}(1+k)\varphi_2$\end{tabular} & $h_{0}{+}(3-k)\varphi_2$ \\ \hline
    \text{3} & \text{$\geq$ 0} & \text{$\geq$ 0} & \text{$<$ 0} & \begin{tabular}{c}$h_0{-}m$\\${+}(k{+}2)\varphi_1$\end{tabular} &  \begin{tabular}{c}$h_0{-}m$ \\ $+(2-k)\varphi_1$\\${+}k\varphi_2$\end{tabular} & $h_{0}{+}(3-k)\varphi_2$ \\ \hline
    \text{4} & \text{$\geq$ 0} & \text{$\geq$ 0} & \text{$\geq$ 0} & \begin{tabular}{c}$h_0{-}m$\\${+}(k{+}2)\varphi_1$\end{tabular} &  \begin{tabular}{c}$h_0{-}m$ \\ $+(2-k)\varphi_1$\\${+}k\varphi_2$\end{tabular} & \begin{tabular}{c}$h_0{-}m$\\${+}(4-k)\varphi_2$\end{tabular} \\ \hline
\end{tabular}
\caption{Chamber definitions of the $U(3)_{k_{\text{eff}}}$ or $SU(3)_{k_{\text{eff}}}$ gauge theory with $N_{\text{f}}=1$  and the corresponding instanton masses. The variables $(\phi_1, \phi_2, \phi_3)$ denote $U(3)$ vevs, while the variables $(\varphi_1, \varphi_2)$ denote $SU(3)$ vevs.\label{tbl:u3 nf1 instantons}}
}
\end{table}
%\\\\
%\noindent 
The $SU(3)_{k_{\text{eff}}}$ $N_{\text{f}}=1$ theory is specified by $2$ Coulomb vevs $\varphi_1, \varphi_2$, one inverse coupling $h_0$ and one real mass $m \in \R$. The fundamental Weyl chamber is given by $\{2\varphi_1 - \varphi_2 \geq 0\} \cap \{-\varphi_2 + 2\varphi_2 \geq 0\}$. 
%\\\\
%\noindent 
Below we list the prepotentials in the four field theory chambers.
\begin{align}
	   \hspace{-0.1in}\mathcal{F}_{SU(3)_{k,N_{\text{f}}=1}}^{\text{chamber 1}} &= \frac{4}{3}\varphi_1^3 + \frac{4}{3}\varphi_2^3 + \frac{(k-1)}{2}\varphi_1^2\varphi_2 - \frac{(k+1)}{2}\varphi_1\varphi_2^2 + h_{0}(\varphi_1^2 + \varphi_2^2 - \varphi_1\varphi_2)   \label{eq:ft prepot SU(3) k Nf1 chamber 1} ~.\\
	   \mathcal{F}_{SU(3)_{k,N_{\text{f}}=1}}^{\text{chamber 2}} &= \frac{7}{6}\varphi_1^3 + \frac{4}{3}\varphi_2^3 + \frac{(k-1)}{2}\varphi_1^2\varphi_2 - \frac{(k+1)}{2}\varphi_1\varphi_2^2 + \left(h_{0}-\frac{m}{2}\right)\varphi_1^2 + h_{0}\varphi_2^2\nonumber\\
	   & - h_{0}\varphi_{1}\varphi_{2} - \frac{m^2}{2}\varphi_1 - \frac{m^3}{6} \label{eq:ft prepot SU(3) k Nf1 chamber 2} ~.\\
	   \mathcal{F}_{SU(3)_{k,N_{\text{f}}=1}}^{\text{chamber 3}} &= \frac{4}{3}\varphi_1^3 + \frac{7}{6}\varphi_2^3 + \frac{(k-2)}{2}\varphi_1^2 \varphi_2 - \frac{k}{2}\varphi_1\varphi_2^2 + (h_{0}-m)\varphi_1^2 + \left(h_{0}-\frac{m}{2}\right)\varphi_2^2\nonumber\\
	   & + (m-h_{0})\varphi_1\varphi_2 - \frac{m^2}{2}\varphi_2 - \frac{m^3}{3}
	   \label{eq:ft prepot SU(3) k Nf1 chamber 3} ~.\\
	   \mathcal{F}_{SU(3)_{k,N_{\text{f}}=1}}^{\text{chamber 4}} &= \frac{4}{3}\varphi_1^3 + \frac{4}{3}\varphi_2^3 + \frac{(k-2)}{2}\varphi_1^2\varphi_2 -\frac{k}{2}\varphi_1\varphi_2^2 + (h_{0}-m)\varphi_1^2 + (h_{0}-m)\varphi_{2}^2 \nonumber\\
	   &+ (m-h_{0})\varphi_1\varphi_2 - \frac{m^3}{2} \label{eq:ft prepot SU(3) k Nf1 chamber 4} ~.
	\end{align}

\paragraph{$SU(3)_{k_{\text{eff}}}$ + $N_{\text{f}} = 2$ field theory.} \normalsize The $SU(3)_{k_{\text{eff}}}$ $N_{\text{f}}=2$ theory is likewise specified by 2 Coulomb vevs $(\varphi_1, \varphi_2)$, one gauge coupling $h_{0}$ and two real masses $m_{1}, m_{2} \in \R$. Table \ref{tbl:chamber defs u3 nf2} lists the $16$ chambers of the theory, while Table \ref{tbl:u3 nf2 instantons} lists the elementary instanton masses in the $16$ chambers.
\begin{table}[ht]	
\centering
{\footnotesize
\begin{tabular}{|c|c|c|c|c|c|c|}\hline
  \text{Chamber} & \begin{tabular}{c} $\phi_1 + m_1$ \\ \text{ or } \\ $\varphi_1{+}m_1$ \end{tabular} & \begin{tabular}{c} $\phi_2 + m_1$ \\ \text{ or } \\ $-\varphi_1{+}\varphi_2{+}m_1$ \end{tabular} & \begin{tabular}{c} $\phi_3 + m_1$ \\ \text{ or } \\ $-\varphi_2{+}m_1$ \end{tabular} & \begin{tabular}{c} $\phi_1 + m_2$ \\ \text{ or } \\ $\varphi_1{+}m_2$ \end{tabular} & \begin{tabular}{c} $\phi_2 + m_2$ \\ \text{ or } \\ $-\varphi_1{+}\varphi_2{+}m_2$ \end{tabular} &  \begin{tabular}{c} $\phi_3 + m_2$ \\ \text{ or } \\ $-\varphi_2{+}m_2$ \end{tabular} \\ \hline
 \text{1} & \text{$<$ 0} & \text{$<$ 0} & \text{$<$ 0} & \text{$<$ 0} & \text{$<$ 0} & \text{$<$ 0} \\ \hline
 \text{2} & \text{$<$ 0} & \text{$<$ 0} & \text{$<$ 0} & \text{$\geq$ 0} & \text{$<$ 0} & \text{$<$ 0} \\ \hline
 \text{3} & \text{$<$ 0} & \text{$<$ 0} & \text{$<$ 0} & \text{$\geq$ 0} & \text{$\geq$ 0} & \text{$<$ 0} \\ \hline
 \text{4} & \text{$<$ 0} & \text{$<$ 0} & \text{$<$ 0} & \text{$\geq$ 0} & \text{$\geq$ 0} & \text{$\geq$ 0} \\ \hline
 \text{5} & \text{$\geq$ 0} & \text{$<$ 0} & \text{$<$ 0} & \text{$<$ 0} & \text{$<$ 0} & \text{$<$ 0} \\ \hline
 \text{6} & \text{$\geq$ 0} & \text{$<$ 0} & \text{$<$ 0} & \text{$\geq$ 0} & \text{$<$ 0} & \text{$<$ 0} \\ \hline
 \text{7} & \text{$\geq$ 0} & \text{$<$ 0} & \text{$<$ 0} & \text{$\geq$ 0} & \text{$\geq$ 0} & \text{$<$ 0} \\ \hline
 \text{8} & \text{$\geq$ 0} & \text{$<$ 0} & \text{$<$ 0} & \text{$\geq$ 0} & \text{$\geq$ 0} & \text{$\geq$ 0} \\ \hline
 \text{9} & \text{$\geq$ 0} & \text{$\geq$ 0} & \text{$<$ 0} & \text{$<$ 0} & \text{$<$ 0} & \text{$<$ 0} \\ \hline
 \text{10} & \text{$\geq$ 0} & \text{$\geq$ 0} & \text{$<$ 0} & \text{$\geq$ 0} & \text{$<$ 0} & \text{$<$ 0} \\ \hline
 \text{11} & \text{$\geq$ 0} & \text{$\geq$ 0} & \text{$<$ 0} & \text{$\geq$ 0} & \text{$\geq$ 0} & \text{$<$ 0} \\ \hline
 \text{12} & \text{$\geq$ 0} & \text{$\geq$ 0} & \text{$<$ 0} & \text{$\geq$ 0} & \text{$\geq$ 0} & \text{$\geq$ 0} \\ \hline
 \text{13} & \text{$\geq$ 0} & \text{$\geq$ 0} & \text{$\geq$ 0} & \text{$<$ 0} & \text{$<$ 0} & \text{$<$ 0} \\ \hline
 \text{14} & \text{$\geq$ 0} & \text{$\geq$ 0} & \text{$\geq$ 0} & \text{$\geq$ 0} & \text{$<$ 0} & \text{$<$ 0} \\ \hline
 \text{15} & \text{$\geq$ 0} & \text{$\geq$ 0} & \text{$\geq$ 0} & \text{$\geq$ 0} & \text{$\geq$ 0} & \text{$<$ 0} \\ \hline
 \text{16} & \text{$\geq$ 0} & \text{$\geq$ 0} & \text{$\geq$ 0} & \text{$\geq$ 0} & \text{$\geq$ 0} & \text{$\geq$ 0} \\ \hline
\end{tabular}
\caption{Chamber definitions of the $U(3)_{k_{\text{eff}}}$ or $SU(3)_{k_{\text{eff}}}$ gauge theory with $N_{\text{f}}=2$ flavors. The variables $(\phi_1, \phi_2, \phi_3)$ denote $U(3)$ vevs, while the variables $(\varphi_1, \varphi_2)$ denote $SU(3)$ vevs.\label{tbl:chamber defs u3 nf2}}
}
\end{table}

\begin{table}[ht]	
\centering
{\footnotesize
\begin{tabular}{|c|c|c|c|}\hline
 \# & $M(\cI_1)$ & $M(\cI_2)$ & $M(\cI_3)$ \\ \hline
 1 & $h_0+(k+3) \varphi _1$ & $h_0+(1-k)\varphi _1+(k+1) \varphi _2$ & $h_0+(3-k) \varphi _2$ \\ \hline
 2 & $h_0+(k+2) \varphi _1-m_2$ & $h_0+(1-k) \varphi _1+(k+1) \varphi _2$ & $h_0+(3-k) \varphi _2$ \\ \hline
 3 & $h_0+(k+2) \varphi _1-m_2$ & $h_0+(2-k) \varphi _1+k \varphi _2-m_2$ & $h_0+(3-k) \varphi _2$ \\\hline
 4 & $h_0+(k+2) \varphi _1-m_2$ & $h_0+(2-k) \varphi _1+k \varphi _2-m_2$ & $h_0+(4-k) \varphi _2-m_2$ \\\hline
 5 & $h_0+(k+2) \varphi _1-m_1$ & $h_0+(1-k) \varphi _1+(k+1) \varphi _2$ & $h_0+(3-k) \varphi _2$ \\ \hline
 6 & $h_0+(k+1) \varphi _1-m_1-m_2$ & $h_0+(1-k) \varphi _1+(k+1) \varphi _2$ & $h_0+(3-k) \varphi _2$ \\\hline
 7 & $h_0+(k+1) \varphi _1-m_1-m_2$ & $h_0+(2-k) \varphi _1+k \varphi _2-m_2$ & $h_0+(3-k) \varphi _2$ \\\hline
 8 & $h_0+(k+1) \varphi _1-m_1-m_2$ & $h_0+(2-k) \varphi _1+k \varphi _2-m_2$ & $h_0+(4-k) \varphi _2-m_2$ \\\hline
 9 & $h_0+(k+2) \varphi _1-m_1$ & $h_0+(2-k) \varphi _1+k \varphi _2-m_1$ & $h_0+(3-k) \varphi _2$ \\\hline
 10 & $h_0+(k+1) \varphi _1-m_1-m_2$ & $h_0+(2-k) \varphi _1+k \varphi _2-m_1$ & $h_0+(3-k) \varphi _2$ \\\hline
 11 & $h_0+(k+1) \varphi _1-m_1-m_2$ & $h_0+(3-k) \varphi _1+(k-1) \varphi _2-m_1-m_2$ & $h_0+(3-k) \varphi _2$ \\\hline
 12 & $h_0+(k+1) \varphi _1-m_1-m_2$ & $h_0+(3-k) \varphi _1+(k-1) \varphi _2-m_1-m_2$ & $h_0+(4-k) \varphi _2-m_2$ \\\hline
 13 & $h_0+(k+2) \varphi _1-m_1$ & $h_0+(2-k) \varphi _1+k \varphi _2-m_1$ & $h_0+(4-k) \varphi _2-m_1$ \\\hline
 14 & $h_0+(k+1) \varphi _1-m_1-m_2$ & $h_0+(2-k) \varphi _1+k \varphi _2-m_1$ & $h_0+(4-k) \varphi _2-m_1$ \\\hline
 15 & $h_0+(k+1) \varphi _1-m_1-m_2$ & $h_0+(3-k) \varphi _1+(k-1) \varphi _2-m_1-m_2$ & $h_0+(4-k) \varphi _2-m_1$ \\\hline
 16 & $h_0+(k+1) \varphi _1-m_1-m_2$ & $h_0+(3-k) \varphi _1+(k-1) \varphi _2-m_1-m_2$ & $h_0+(5-k) \varphi _2-m_1-m_2$ \\\hline
\end{tabular}
\caption{Instanton masses in the $16$ chambers of the $SU(3)_{k_{\text{eff}}}$ $N_{\text{f}}=2$ field theory.}\label{tbl:u3 nf2 instantons}
}
\end{table}

Note that in Table \ref{tbl:u3 nf2 instantons}, the symbol $k$ denotes the bare CS level, which in this case is related to the effective CS level $k_{\text{eff}}$ by,
\begin{align}
	k \equiv k_{\text{bare}} = k_{\text{eff}} + 1 ~,
\end{align}
in the $U(1)_{-\frac{1}{2}}$ quantization scheme (cf. \eqref{eq:U(1) half quantization}). 

The prepotential \eqref{eq:our prepot} of the $SU(3)_{k_{\text{eff}}}$ $N_{\text{f}}=2$ theory is given by:
\begin{align}
   	\cF_{SU(3)_{k},N_{\text{f}}=2} &= h_{0}(\varphi_1^2 - \varphi_2\varphi_1 + \varphi_2^2) + \frac{4}{3}(\varphi_1^3 + \varphi_2^3) + \frac{1}{2}((k-1)\varphi_1^2\varphi_2 - (k+1)\varphi_1\varphi_2^2)\nonumber\\
   	&+ \frac{1}{6}\sum_{i=1}^{2}\big[\Theta\left(\varphi_1 + m_i\right)\left(\varphi_1 + m_i\right)^3 + \Theta\left(-\varphi_1 + \varphi_2 + m_i\right)\left(-\varphi_1 + \varphi_2 + m_i\right)^3\nonumber\\
   	&\qquad \qquad + \Theta\left(\varphi_2 + m_i\right)\left(\varphi_2 + m_i\right)^3\big] ~.\label{eq:ft prepot SU(3) k Nf2 chamber all}
\end{align}
The arguments of the $\Theta$ functions define various chambers, and are listed in Table \ref{tbl:chamber defs u3 nf2}. 

\section{Triple-intersection numbers of the $E_{3}{}^{2,1}$ geometry\label{sec:intnos H[1]}}
Using the GLSM approach, it is straightforward to compute triple-intersection numbers involving at least one compact divisor, as discussed in the main text. The results are listed below.
\paragraph{Resolutions that have a vertical reduction.} Resolutions (a)-(p) (see Figure \ref{fig:h1 crepant all}) have an allowed vertical reduction. The relevant triple-intersection numbers are:
{\footnotesize
\begin{align}% Resolution (1)
  {\rm \bf (a)} : \left\{\begin{array}{c@{~,\quad}c@{~,\quad}c@{~,\quad}c@{~,\quad}c@{~,}}
 \bE_1^3=6 & \bE_2^3=8 & D_1 \bE_1 \bE_2=0 & D_2 \bE_1 \bE_2=1 & D_3 \bE_1 \bE_2=0  \\
 \bE_1^2 \bE_2=-1 & \bE_1 \bE_2^2=-1 & D_1^2 \bE_1=-1 & D_1^2 \bE_2=0 & D_1 D_2 \bE_1=1 \\
 D_1 D_2 \bE_2=0 & D_1 D_3 \bE_1=0 & D_1 D_3 \bE_2=0 & D_2^2 \bE_1=-1 & D_2^2 \bE_2=0 \\
 D_2 D_3 \bE_1=0 & D_2 D_3 \bE_2=1 & D_3^2 \bE_1=0 & D_3^2 \bE_2=1 & D_1 \bE_1^2=-1 \\
 D_1 \bE_2^2=0 & D_2 \bE_1^2=-1 & D_2 \bE_2^2=-2 & D_3 \bE_1^2=0 & D_3 \bE_2^2=-3
\end{array}\right.
\end{align}}
\vspace{-0.1cm}
{\footnotesize
\begin{align}% Resolution (2)
  {\rm \bf (b)} : \left\{\begin{array}{c@{~,\quad}c@{~,\quad}c@{~,\quad}c@{~,\quad}c@{~,}}
  \bE_1^3=7 & \bE_2^3=8 & D_1 \bE_1 \bE_2=0 & D_2 \bE_1 \bE_2=1 & D_3 \bE_1 \bE_2=0 \\
 \bE_1^2 \bE_2=-1 & \bE_1 \bE_2^2=-1 & D_1^2 \bE_1=0 & D_1^2 \bE_2=0 & D_1 D_2 \bE_1=0 \\
 D_1 D_2 \bE_2=0 & D_1 D_3 \bE_1=0 & D_1 D_3 \bE_2=0 & D_2^2 \bE_1=0 & D_2^2 \bE_2=0 \\
 D_2 D_3 \bE_1=0 & D_2 D_3 \bE_2=1 & D_3^2 \bE_1=0 & D_3^2 \bE_2=1 & D_1 \bE_1^2=0 \\
 D_1 \bE_2^2=0 & D_2 \bE_1^2=-2 & D_2 \bE_2^2=-2 & D_3 \bE_1^2=0 & D_3 \bE_2^2=-3 
\end{array}\right.
\end{align}}
\vspace{-0.1cm}
{\footnotesize
\begin{align}% Resolution (3)
  {\rm \bf (c)} : \left\{\begin{array}{c@{~,\quad}c@{~,\quad}c@{~,\quad}c@{~,\quad}c@{~,}}
 \bE_1^3=7 & \bE_2^3=7 & D_1 \bE_1 \bE_2=1 & D_2 \bE_1 \bE_2=0 & D_3 \bE_1 \bE_2=0 \\
 \bE_1^2 \bE_2=-2 & \bE_1 \bE_2^2=0 & D_1^2 \bE_1=0 & D_1^2 \bE_2=-1 & D_1 D_2 \bE_1=0 \\
 D_1 D_2 \bE_2=1 & D_1 D_3 \bE_1=0 & D_1 D_3 \bE_2=0 & D_2^2 \bE_1=0 & D_2^2 \bE_2=-1 \\
 D_2 D_3 \bE_1=0 & D_2 D_3 \bE_2=1 & D_3^2 \bE_1=0 & D_3^2 \bE_2=1 & D_1 \bE_1^2=-2 \\
 D_1 \bE_2^2=-1 & D_2 \bE_1^2=0 & D_2 \bE_2^2=-1 & D_3 \bE_1^2=0 & D_3 \bE_2^2=-3 
\end{array}\right.
\end{align}}
\vspace{-0.1cm}
{\footnotesize
\begin{align}% Resolution (4)
  {\rm \bf (d)} : \left\{\begin{array}{c@{~,\quad}c@{~,\quad}c@{~,\quad}c@{~,\quad}c@{~,}}
 \bE_1^3=7 & \bE_2^3=7 & D_1 \bE_1 \bE_2=0 & D_2 \bE_1 \bE_2=1 & D_3 \bE_1 \bE_2=0 \\
 \bE_1^2 \bE_2=-2 & \bE_1 \bE_2^2=0 & D_1^2 \bE_1=-1 & D_1^2 \bE_2=0 & D_1 D_2 \bE_1=1 \\
 D_1 D_2 \bE_2=0 & D_1 D_3 \bE_1=0 & D_1 D_3 \bE_2=0 & D_2^2 \bE_1=-1 & D_2^2 \bE_2=0 \\
 D_2 D_3 \bE_1=0 & D_2 D_3 \bE_2=1 & D_3^2 \bE_1=0 & D_3^2 \bE_2=1 & D_1 \bE_1^2=-1 \\
 D_1 \bE_2^2=0 & D_2 \bE_1^2=-1 & D_2 \bE_2^2=-2 & D_3 \bE_1^2=0 & D_3 \bE_2^2=-3
\end{array}\right.
\end{align}}
\vspace{-0.1cm}
{\footnotesize
\begin{align}% Resolution (5)
  {\rm \bf (e)} : \left\{\begin{array}{c@{~,\quad}c@{~,\quad}c@{~,\quad}c@{~,\quad}c@{~,}}
 \bE_1^3=7 & \bE_2^3=8 & D_1 \bE_1 \bE_2=0 & D_2 \bE_1 \bE_2=1 & D_3 \bE_1 \bE_2=0 \\
 \bE_1^2 \bE_2=-1 & \bE_1 \bE_2^2=-1 & D_1^2 \bE_1=-1 & D_1^2 \bE_2=0 & D_1 D_2 \bE_1=1 \\
 D_1 D_2 \bE_2=0 & D_1 D_3 \bE_1=0 & D_1 D_3 \bE_2=0 & D_2^2 \bE_1=-1 & D_2^2 \bE_2=0 \\
 D_2 D_3 \bE_1=0 & D_2 D_3 \bE_2=1 & D_3^2 \bE_1=0 & D_3^2 \bE_2=1 & D_1 \bE_1^2=-1 \\
 D_1 \bE_2^2=0 & D_2 \bE_1^2=-1 & D_2 \bE_2^2=-2 & D_3 \bE_1^2=0 & D_3 \bE_2^2=-3
\end{array}\right.
\end{align}}
\vspace{-0.1cm}
{\footnotesize
\begin{align}% Resolution (8)
  {\rm \bf (f)} : \left\{\begin{array}{c@{~,\quad}c@{~,\quad}c@{~,\quad}c@{~,\quad}c@{~,}}
 \bE_1^3=8 & \bE_2^3=7 & D_1 \bE_1 \bE_2=0 & D_2 \bE_1 \bE_2=1 & D_3 \bE_1 \bE_2=0 \\
 \bE_1^2 \bE_2=-2 & \bE_1 \bE_2^2=0 & D_1^2 \bE_1=0 & D_1^2 \bE_2=0 & D_1 D_2 \bE_1=0 \\
 D_1 D_2 \bE_2=0 & D_1 D_3 \bE_1=0 & D_1 D_3 \bE_2=0 & D_2^2 \bE_1=0 & D_2^2 \bE_2=0 \\
 D_2 D_3 \bE_1=0 & D_2 D_3 \bE_2=1 & D_3^2 \bE_1=0 & D_3^2 \bE_2=1 & D_1 \bE_1^2=0 \\
 D_1 \bE_2^2=0 & D_2 \bE_1^2=-2 & D_2 \bE_2^2=-2 & D_3 \bE_1^2=0 & D_3 \bE_2^2=-3
\end{array}\right.
\end{align}}
\vspace{-0.1cm}
{\footnotesize
\begin{align}% Resolution (9)
  {\rm \bf (g)} : \left\{\begin{array}{c@{~,\quad}c@{~,\quad}c@{~,\quad}c@{~,\quad}c@{~,}}
 \bE_1^3=8 & \bE_2^3=8 & D_1 \bE_1 \bE_2=0 & D_2 \bE_1 \bE_2=1 & D_3 \bE_1 \bE_2=0 \\
 \bE_1^2 \bE_2=-1 & \bE_1 \bE_2^2=-1 & D_1^2 \bE_1=0 & D_1^2 \bE_2=0 & D_1 D_2 \bE_1=0 \\
 D_1 D_2 \bE_2=0 & D_1 D_3 \bE_1=0 & D_1 D_3 \bE_2=0 & D_2^2 \bE_1=0 & D_2^2 \bE_2=0 \\
 D_2 D_3 \bE_1=0 & D_2 D_3 \bE_2=1 & D_3^2 \bE_1=0 & D_3^2 \bE_2=1 & D_1 \bE_1^2=0 \\
 D_1 \bE_2^2=0 & D_2 \bE_1^2=-2 & D_2 \bE_2^2=-2 & D_3 \bE_1^2=0 & D_3 \bE_2^2=-3
\end{array}\right.
\end{align}}
\vspace{-0.1cm}
{\footnotesize
\begin{align}% Resolution (11)
  {\rm \bf (h)} : \left\{\begin{array}{c@{~,\quad}c@{~,\quad}c@{~,\quad}c@{~,\quad}c@{~,}}
 \bE_1^3=7 & \bE_2^3=8 & D_1 \bE_1 \bE_2=1 & D_2 \bE_1 \bE_2=0 & D_3 \bE_1 \bE_2=0 \\
 \bE_1^2 \bE_2=-2 & \bE_1 \bE_2^2=0 & D_1^2 \bE_1=0 & D_1^2 \bE_2=0 & D_1 D_2 \bE_1=0 \\
 D_1 D_2 \bE_2=0 & D_1 D_3 \bE_1=0 & D_1 D_3 \bE_2=1 & D_2^2 \bE_1=0 & D_2^2 \bE_2=0 \\
 D_2 D_3 \bE_1=0 & D_2 D_3 \bE_2=0 & D_3^2 \bE_1=0 & D_3^2 \bE_2=2 & D_1 \bE_1^2=-2 \\
 D_1 \bE_2^2=-2 & D_2 \bE_1^2=0 & D_2 \bE_2^2=0 & D_3 \bE_1^2=0 & D_3 \bE_2^2=-4
\end{array}\right.
\end{align}}
\vspace{-0.1cm}
{\footnotesize
\begin{align}% Resolution (12)
  {\rm \bf (i)} : \left\{\begin{array}{c@{~,\quad}c@{~,\quad}c@{~,\quad}c@{~,\quad}c@{~,}}
 \bE_1^3=8 & \bE_2^3=6 & D_1 \bE_1 \bE_2=1 & D_2 \bE_1 \bE_2=0 & D_3 \bE_1 \bE_2=0 \\
 \bE_1^2 \bE_2=-3 & \bE_1 \bE_2^2=1 & D_1^2 \bE_1=0 & D_1^2 \bE_2=-1 & D_1 D_2 \bE_1=0 \\
 D_1 D_2 \bE_2=1 & D_1 D_3 \bE_1=0 & D_1 D_3 \bE_2=0 & D_2^2 \bE_1=0 & D_2^2 \bE_2=-1 \\
 D_2 D_3 \bE_1=0 & D_2 D_3 \bE_2=1 & D_3^2 \bE_1=0 & D_3^2 \bE_2=1 & D_1 \bE_1^2=-2 \\
 D_1 \bE_2^2=-1 & D_2 \bE_1^2=0 & D_2 \bE_2^2=-1 & D_3 \bE_1^2=0 & D_3 \bE_2^2=-3
\end{array}\right.
\end{align}}
\vspace{-0.1cm}
{\footnotesize
\begin{align}% Resolution (13)
  {\rm \bf (j)} : \left\{\begin{array}{c@{~,\quad}c@{~,\quad}c@{~,\quad}c@{~,\quad}c@{~,}}
 \bE_1^3=8 & \bE_2^3=7 & D_1 \bE_1 \bE_2=1 & D_2 \bE_1 \bE_2=0 & D_3 \bE_1 \bE_2=0 \\
 \bE_1^2 \bE_2=-2 & \bE_1 \bE_2^2=0 & D_1^2 \bE_1=0 & D_1^2 \bE_2=-1 & D_1 D_2 \bE_1=0 \\
 D_1 D_2 \bE_2=1 & D_1 D_3 \bE_1=0 & D_1 D_3 \bE_2=0 & D_2^2 \bE_1=0 & D_2^2 \bE_2=-1 \\
 D_2 D_3 \bE_1=0 & D_2 D_3 \bE_2=1 & D_3^2 \bE_1=0 & D_3^2 \bE_2=1 & D_1 \bE_1^2=-2 \\
 D_1 \bE_2^2=-1 & D_2 \bE_1^2=0 & D_2 \bE_2^2=-1 & D_3 \bE_1^2=0 & D_3 \bE_2^2=-3 
\end{array}\right.
\end{align}}
\vspace{-0.1cm}
{\footnotesize
\begin{align}% Resolution (15)
  {\rm \bf (k)} : \left\{\begin{array}{c@{~,\quad}c@{~,\quad}c@{~,\quad}c@{~,\quad}c@{~,}}
 \bE_1^3=7 & \bE_2^3=8 & D_1 \bE_1 \bE_2=0 & D_2 \bE_1 \bE_2=1 & D_3 \bE_1 \bE_2=0 \\
 \bE_1^2 \bE_2=-2 & \bE_1 \bE_2^2=0 & D_1^2 \bE_1=-1 & D_1^2 \bE_2=0 & D_1 D_2 \bE_1=1 \\
 D_1 D_2 \bE_2=0 & D_1 D_3 \bE_1=0 & D_1 D_3 \bE_2=0 & D_2^2 \bE_1=-1 & D_2^2 \bE_2=0 \\
 D_2 D_3 \bE_1=0 & D_2 D_3 \bE_2=1 & D_3^2 \bE_1=0 & D_3^2 \bE_2=2 & D_1 \bE_1^2=-1 \\
 D_1 \bE_2^2=0 & D_2 \bE_1^2=-1 & D_2 \bE_2^2=-2 & D_3 \bE_1^2=0 & D_3 \bE_2^2=-4
\end{array}\right.
\end{align}}
\vspace{-0.1cm}
{\footnotesize
\begin{align}% Resolution (19)
  {\rm \bf (l)} : \left\{\begin{array}{c@{~,\quad}c@{~,\quad}c@{~,\quad}c@{~,\quad}c@{~,}}
 \bE_1^3=8 & \bE_2^3=8 & D_1 \bE_1 \bE_2=0 & D_2 \bE_1 \bE_2=1 & D_3 \bE_1 \bE_2=0 \\
 \bE_1^2 \bE_2=-2 & \bE_1 \bE_2^2=0 & D_1^2 \bE_1=0 & D_1^2 \bE_2=0 & D_1 D_2 \bE_1=0 \\
 D_1 D_2 \bE_2=0 & D_1 D_3 \bE_1=0 & D_1 D_3 \bE_2=0 & D_2^2 \bE_1=0 & D_2^2 \bE_2=0 \\
 D_2 D_3 \bE_1=0 & D_2 D_3 \bE_2=1 & D_3^2 \bE_1=0 & D_3^2 \bE_2=2 & D_1 \bE_1^2=0 \\
 D_1 \bE_2^2=0 & D_2 \bE_1^2=-2 & D_2 \bE_2^2=-2 & D_3 \bE_1^2=0 & D_3 \bE_2^2=-4
\end{array}\right.
\end{align}}
\vspace{-0.1cm}
{\footnotesize
\begin{align}% Resolution (21)
  {\rm \bf (m)} : \left\{\begin{array}{c@{~,\quad}c@{~,\quad}c@{~,\quad}c@{~,\quad}c@{~,}}
 \bE_1^3=8 & \bE_2^3=7 & D_1 \bE_1 \bE_2=1 & D_2 \bE_1 \bE_2=0 & D_3 \bE_1 \bE_2=0 \\
 \bE_1^2 \bE_2=-3 & \bE_1 \bE_2^2=1 & D_1^2 \bE_1=0 & D_1^2 \bE_2=0 & D_1 D_2 \bE_1=0 \\
 D_1 D_2 \bE_2=0 & D_1 D_3 \bE_1=0 & D_1 D_3 \bE_2=1 & D_2^2 \bE_1=0 & D_2^2 \bE_2=0 \\
 D_2 D_3 \bE_1=0 & D_2 D_3 \bE_2=0 & D_3^2 \bE_1=0 & D_3^2 \bE_2=2 & D_1 \bE_1^2=-2 \\
 D_1 \bE_2^2=-2 & D_2 \bE_1^2=0 & D_2 \bE_2^2=0 & D_3 \bE_1^2=0 & D_3 \bE_2^2=-4
\end{array}\right.
\end{align}}
\vspace{-0.1cm}
{\footnotesize
\begin{align}% Resolution (22)
  {\rm \bf (n)} : \left\{\begin{array}{c@{~,\quad}c@{~,\quad}c@{~,\quad}c@{~,\quad}c@{~,}}
 \bE_1^3=8 & \bE_2^3=8 & D_1 \bE_1 \bE_2=1 & D_2 \bE_1 \bE_2=0 & D_3 \bE_1 \bE_2=0 \\
 \bE_1^2 \bE_2=-2 & \bE_1 \bE_2^2=0 & D_1^2 \bE_1=0 & D_1^2 \bE_2=0 & D_1 D_2 \bE_1=0 \\
 D_1 D_2 \bE_2=0 & D_1 D_3 \bE_1=0 & D_1 D_3 \bE_2=1 & D_2^2 \bE_1=0 & D_2^2 \bE_2=0 \\
 D_2 D_3 \bE_1=0 & D_2 D_3 \bE_2=0 & D_3^2 \bE_1=0 & D_3^2 \bE_2=2 & D_1 \bE_1^2=-2 \\
 D_1 \bE_2^2=-2 & D_2 \bE_1^2=0 & D_2 \bE_2^2=0 & D_3 \bE_1^2=0 & D_3 \bE_2^2=-4
\end{array}\right.
\end{align}}
\vspace{-0.1cm}
{\footnotesize
\begin{align}% Resolution (24)
  {\rm \bf (o)} : \left\{\begin{array}{c@{~,\quad}c@{~,\quad}c@{~,\quad}c@{~,\quad}c@{~,}}
 \bE_1^3=8 & \bE_2^3=7 & D_1 \bE_1 \bE_2=1 & D_2 \bE_1 \bE_2=0 & D_3 \bE_1 \bE_2=0 \\
 \bE_1^2 \bE_2=-3 & \bE_1 \bE_2^2=1 & D_1^2 \bE_1=0 & D_1^2 \bE_2=-1 & D_1 D_2 \bE_1=0 \\
 D_1 D_2 \bE_2=1 & D_1 D_3 \bE_1=0 & D_1 D_3 \bE_2=0 & D_2^2 \bE_1=0 & D_2^2 \bE_2=-1 \\
 D_2 D_3 \bE_1=0 & D_2 D_3 \bE_2=1 & D_3^2 \bE_1=0 & D_3^2 \bE_2=2 & D_1 \bE_1^2=-2 \\
 D_1 \bE_2^2=-1 & D_2 \bE_1^2=0 & D_2 \bE_2^2=-1 & D_3 \bE_1^2=0 & D_3 \bE_2^2=-4
\end{array}\right.
\end{align}}
\vspace{-0.1cm}
{\footnotesize
\begin{align}% Resolution (27)
  {\rm \bf (p)} : \left\{\begin{array}{c@{~,\quad}c@{~,\quad}c@{~,\quad}c@{~,\quad}c}
 \bE_1^3=8 & \bE_2^3=8 & D_1 \bE_1 \bE_2=1 & D_2 \bE_1 \bE_2=0 & D_3 \bE_1 \bE_2=0~, \\
 \bE_1^2 \bE_2=-3 & \bE_1 \bE_2^2=1 & D_1^2 \bE_1=0 & D_1^2 \bE_2=0 & D_1 D_2 \bE_1=0~, \\
 D_1 D_2 \bE_2=0 & D_1 D_3 \bE_1=0 & D_1 D_3 \bE_2=1 & D_2^2 \bE_1=0 & D_2^2 \bE_2=0~, \\
 D_2 D_3 \bE_1=0 & D_2 D_3 \bE_2=0 & D_3^2 \bE_1=0 & D_3^2 \bE_2=3 & D_1 \bE_1^2=-2~, \\
 D_1 \bE_2^2=-2 & D_2 \bE_1^2=0 & D_2 \bE_2^2=0 & D_3 \bE_1^2=0 & D_3 \bE_2^2=-5~.
\end{array}\right.
\end{align}}
\paragraph{Resolutions without a gauge theory phase.} Resolutions (q$_{1}$)-(q$_{14}$) (see Figure \ref{fig:h1 crepant all}) do not admit vertical reductions, and thus have no gauge theory phases. For the sake of completeness, their triple-intersection numbers are listed below.
{\footnotesize
\begin{align}% Resolution (6)
  {\rm \bf (q_{1})} : \left\{\begin{array}{c@{~,\quad}c@{~,\quad}c@{~,\quad}c@{~,\quad}c@{~,}}
 \bE_1^3=7 & \bE_2^3=8 & D_1 \bE_1 \bE_2=0 & D_2 \bE_1 \bE_2=1 & D_3 \bE_1 \bE_2=0 \\
 \bE_1^2 \bE_2=-1 & \bE_1 \bE_2^2=-1 & D_1^2 \bE_1=0 & D_1^2 \bE_2=0 & D_1 D_2 \bE_1=1 \\
 D_1 D_2 \bE_2=0 & D_1 D_3 \bE_1=0 & D_1 D_3 \bE_2=0 & D_2^2 \bE_1=-1 & D_2^2 \bE_2=0 \\
 D_2 D_3 \bE_1=0 & D_2 D_3 \bE_2=1 & D_3^2 \bE_1=0 & D_3^2 \bE_2=1 & D_1 \bE_1^2=-2 \\
 D_1 \bE_2^2=0 & D_2 \bE_1^2=-1 & D_2 \bE_2^2=-2 & D_3 \bE_1^2=0 & D_3 \bE_2^2=-3
\end{array}\right.
\end{align}}
\vspace{-0.1cm}
{\footnotesize
\begin{align}% Resolution (7)
  {\rm \bf (q_{2})} : \left\{\begin{array}{c@{~,\quad}c@{~,\quad}c@{~,\quad}c@{~,\quad}c@{~,}}
 \bE_1^3=7 & \bE_2^3=9 & D_1 \bE_1 \bE_2=0 & D_2 \bE_1 \bE_2=0 & D_3 \bE_1 \bE_2=0 \\
 \bE_1^2 \bE_2=1 & \bE_1 \bE_2^2=1 & D_1^2 \bE_1=-1 & D_1^2 \bE_2=0 & D_1 D_2 \bE_1=1 \\
 D_1 D_2 \bE_2=0 & D_1 D_3 \bE_1=0 & D_1 D_3 \bE_2=0 & D_2^2 \bE_1=0 & D_2^2 \bE_2=1 \\
 D_2 D_3 \bE_1=0 & D_2 D_3 \bE_2=1 & D_3^2 \bE_1=0 & D_3^2 \bE_2=1 & D_1 \bE_1^2=-1 \\
 D_1 \bE_2^2=0 & D_2 \bE_1^2=-2 & D_2 \bE_2^2=-3 & D_3 \bE_1^2=0 & D_3 \bE_2^2=-3
\end{array}\right.
\end{align}}
\vspace{-0.1cm}
{\footnotesize
\begin{align}% Resolution (10)
  {\rm \bf (q_{3})} : \left\{\begin{array}{c@{~,\quad}c@{~,\quad}c@{~,\quad}c@{~,\quad}c@{~,}}
 \bE_1^3=8 & \bE_2^3=9 & D_1 \bE_1 \bE_2=0 & D_2 \bE_1 \bE_2=0 & D_3 \bE_1 \bE_2=0 \\
 \bE_1^2 \bE_2=1 & \bE_1 \bE_2^2=1 & D_1^2 \bE_1=0 & D_1^2 \bE_2=0 & D_1 D_2 \bE_1=0 \\
 D_1 D_2 \bE_2=0 & D_1 D_3 \bE_1=0 & D_1 D_3 \bE_2=0 & D_2^2 \bE_1=1 & D_2^2 \bE_2=1 \\
 D_2 D_3 \bE_1=0 & D_2 D_3 \bE_2=1 & D_3^2 \bE_1=0 & D_3^2 \bE_2=1 & D_1 \bE_1^2=0 \\
 D_1 \bE_2^2=0 & D_2 \bE_1^2=-3 & D_2 \bE_2^2=-3 & D_3 \bE_1^2=0 & D_3 \bE_2^2=-3
\end{array}\right.
\end{align}}
\vspace{-0.1cm}
{\footnotesize
\begin{align}% Resolution (14)
  {\rm \bf (q_{4})} : \left\{\begin{array}{c@{~,\quad}c@{~,\quad}c@{~,\quad}c@{~,\quad}c@{~,}}
 \bE_1^3=8 & \bE_2^3=7 & D_1 \bE_1 \bE_2=1 & D_2 \bE_1 \bE_2=0 & D_3 \bE_1 \bE_2=0 \\
 \bE_1^2 \bE_2=-2 & \bE_1 \bE_2^2=0 & D_1^2 \bE_1=1 & D_1^2 \bE_2=-1 & D_1 D_2 \bE_1=0 \\
 D_1 D_2 \bE_2=1 & D_1 D_3 \bE_1=0 & D_1 D_3 \bE_2=0 & D_2^2 \bE_1=0 & D_2^2 \bE_2=-1 \\
 D_2 D_3 \bE_1=0 & D_2 D_3 \bE_2=1 & D_3^2 \bE_1=0 & D_3^2 \bE_2=1 & D_1 \bE_1^2=-3 \\
 D_1 \bE_2^2=-1 & D_2 \bE_1^2=0 & D_2 \bE_2^2=-1 & D_3 \bE_1^2=0 & D_3 \bE_2^2=-3
\end{array}\right.
\end{align}}
\vspace{-0.1cm}
{\footnotesize
\begin{align}% Resolution (16)
  {\rm \bf (q_{5})} : \left\{\begin{array}{c@{~,\quad}c@{~,\quad}c@{~,\quad}c@{~,\quad}c@{~,}}
 \bE_1^3=8 & \bE_2^3=7 & D_1 \bE_1 \bE_2=0 & D_2 \bE_1 \bE_2=1 & D_3 \bE_1 \bE_2=0 \\
 \bE_1^2 \bE_2=-2 & \bE_1 \bE_2^2=0 & D_1^2 \bE_1=0 & D_1^2 \bE_2=0 & D_1 D_2 \bE_1=1 \\
 D_1 D_2 \bE_2=0 & D_1 D_3 \bE_1=0 & D_1 D_3 \bE_2=0 & D_2^2 \bE_1=-1 & D_2^2 \bE_2=0 \\
 D_2 D_3 \bE_1=0 & D_2 D_3 \bE_2=1 & D_3^2 \bE_1=0 & D_3^2 \bE_2=1 & D_1 \bE_1^2=-2 \\
 D_1 \bE_2^2=0 & D_2 \bE_1^2=-1 & D_2 \bE_2^2=-2 & D_3 \bE_1^2=0 & D_3 \bE_2^2=-3
\end{array}\right.
\end{align}}
\vspace{-0.1cm}
{\footnotesize
\begin{align}% Resolution (17)
  {\rm \bf (q_{6})} : \left\{\begin{array}{c@{~,\quad}c@{~,\quad}c@{~,\quad}c@{~,\quad}c@{~,}}
 \bE_1^3=8 & \bE_2^3=9 & D_1 \bE_1 \bE_2=0 & D_2 \bE_1 \bE_2=0 & D_3 \bE_1 \bE_2=0 \\
 \bE_1^2 \bE_2=1 & \bE_1 \bE_2^2=1 & D_1^2 \bE_1=-1 & D_1^2 \bE_2=0 & D_1 D_2 \bE_1=1 \\
 D_1 D_2 \bE_2=0 & D_1 D_3 \bE_1=0 & D_1 D_3 \bE_2=0 & D_2^2 \bE_1=0 & D_2^2 \bE_2=1 \\
 D_2 D_3 \bE_1=0 & D_2 D_3 \bE_2=1 & D_3^2 \bE_1=0 & D_3^2 \bE_2=1 & D_1 \bE_1^2=-1 \\
 D_1 \bE_2^2=0 & D_2 \bE_1^2=-2 & D_2 \bE_2^2=-3 & D_3 \bE_1^2=0 & D_3 \bE_2^2=-3
\end{array}\right.
\end{align}}
\vspace{-0.1cm}
{\footnotesize
\begin{align}% Resolution (18)
  {\rm \bf (q_{7})} : \left\{\begin{array}{c@{~,\quad}c@{~,\quad}c@{~,\quad}c@{~,\quad}c@{~,}}
 \bE_1^3=8 & \bE_2^3=9 & D_1 \bE_1 \bE_2=0 & D_2 \bE_1 \bE_2=0 & D_3 \bE_1 \bE_2=0 \\
 \bE_1^2 \bE_2=1 & \bE_1 \bE_2^2=1 & D_1^2 \bE_1=0 & D_1^2 \bE_2=0 & D_1 D_2 \bE_1=1 \\
 D_1 D_2 \bE_2=0 & D_1 D_3 \bE_1=0 & D_1 D_3 \bE_2=0 & D_2^2 \bE_1=0 & D_2^2 \bE_2=1 \\
 D_2 D_3 \bE_1=0 & D_2 D_3 \bE_2=1 & D_3^2 \bE_1=0 & D_3^2 \bE_2=1 & D_1 \bE_1^2=-2 \\
 D_1 \bE_2^2=0 & D_2 \bE_1^2=-2 & D_2 \bE_2^2=-3 & D_3 \bE_1^2=0 & D_3 \bE_2^2=-3
\end{array}\right.
\end{align}}
\vspace{-0.1cm}
{\footnotesize
\begin{align}% Resolution (20)
  {\rm \bf (q_{8})} : \left\{\begin{array}{c@{~,\quad}c@{~,\quad}c@{~,\quad}c@{~,\quad}c@{~,}}
 \bE_1^3=9 & \bE_2^3=9 & D_1 \bE_1 \bE_2=0 & D_2 \bE_1 \bE_2=0 & D_3 \bE_1 \bE_2=0 \\
 \bE_1^2 \bE_2=1 & \bE_1 \bE_2^2=1 & D_1^2 \bE_1=0 & D_1^2 \bE_2=0 & D_1 D_2 \bE_1=0 \\
 D_1 D_2 \bE_2=0 & D_1 D_3 \bE_1=0 & D_1 D_3 \bE_2=0 & D_2^2 \bE_1=1 & D_2^2 \bE_2=1 \\
 D_2 D_3 \bE_1=0 & D_2 D_3 \bE_2=1 & D_3^2 \bE_1=0 & D_3^2 \bE_2=1 & D_1 \bE_1^2=0 \\
 D_1 \bE_2^2=0 & D_2 \bE_1^2=-3 & D_2 \bE_2^2=-3 & D_3 \bE_1^2=0 & D_3 \bE_2^2=-3
\end{array}\right.
\end{align}}
\vspace{-0.1cm}
{\footnotesize
\begin{align}% Resolution (23)
  {\rm \bf (q_{9})} : \left\{\begin{array}{c@{~,\quad}c@{~,\quad}c@{~,\quad}c@{~,\quad}c@{~,}}
 \bE_1^3=8 & \bE_2^3=8 & D_1 \bE_1 \bE_2=1 & D_2 \bE_1 \bE_2=0 & D_3 \bE_1 \bE_2=0 \\
 \bE_1^2 \bE_2=-2 & \bE_1 \bE_2^2=0 & D_1^2 \bE_1=1 & D_1^2 \bE_2=0 & D_1 D_2 \bE_1=0 \\
 D_1 D_2 \bE_2=0 & D_1 D_3 \bE_1=0 & D_1 D_3 \bE_2=1 & D_2^2 \bE_1=0 & D_2^2 \bE_2=0 \\
 D_2 D_3 \bE_1=0 & D_2 D_3 \bE_2=0 & D_3^2 \bE_1=0 & D_3^2 \bE_2=2 & D_1 \bE_1^2=-3 \\
 D_1 \bE_2^2=-2 & D_2 \bE_1^2=0 & D_2 \bE_2^2=0 & D_3 \bE_1^2=0 & D_3 \bE_2^2=-4
\end{array}\right.
\end{align}}
\vspace{-0.1cm}
{\footnotesize
\begin{align}% Resolution (25)
  {\rm \bf (q_{10})} : \left\{\begin{array}{c@{~,\quad}c@{~,\quad}c@{~,\quad}c@{~,\quad}c@{~,}}
 \bE_1^3=9 & \bE_2^3=6 & D_1 \bE_1 \bE_2=1 & D_2 \bE_1 \bE_2=0 & D_3 \bE_1 \bE_2=0 \\
 \bE_1^2 \bE_2=-3 & \bE_1 \bE_2^2=1 & D_1^2 \bE_1=1 & D_1^2 \bE_2=-1 & D_1 D_2 \bE_1=0 \\
 D_1 D_2 \bE_2=1 & D_1 D_3 \bE_1=0 & D_1 D_3 \bE_2=0 & D_2^2 \bE_1=0 & D_2^2 \bE_2=-1 \\
 D_2 D_3 \bE_1=0 & D_2 D_3 \bE_2=1 & D_3^2 \bE_1=0 & D_3^2 \bE_2=1 & D_1 \bE_1^2=-3 \\
 D_1 \bE_2^2=-1 & D_2 \bE_1^2=0 & D_2 \bE_2^2=-1 & D_3 \bE_1^2=0 & D_3 \bE_2^2=-3
\end{array}\right.
\end{align}}
\vspace{-0.1cm}
{\footnotesize
\begin{align}% Resolution (26)
  {\rm \bf (q_{11})} : \left\{\begin{array}{c@{~,\quad}c@{~,\quad}c@{~,\quad}c@{~,\quad}c@{~,}}
 \bE_1^3=8 & \bE_2^3=8 & D_1 \bE_1 \bE_2=0 & D_2 \bE_1 \bE_2=1 & D_3 \bE_1 \bE_2=0 \\
 \bE_1^2 \bE_2=-2 & \bE_1 \bE_2^2=0 & D_1^2 \bE_1=0 & D_1^2 \bE_2=0 & D_1 D_2 \bE_1=1 \\
 D_1 D_2 \bE_2=0 & D_1 D_3 \bE_1=0 & D_1 D_3 \bE_2=0 & D_2^2 \bE_1=-1 & D_2^2 \bE_2=0 \\
 D_2 D_3 \bE_1=0 & D_2 D_3 \bE_2=1 & D_3^2 \bE_1=0 & D_3^2 \bE_2=2 & D_1 \bE_1^2=-2 \\
 D_1 \bE_2^2=0 & D_2 \bE_1^2=-1 & D_2 \bE_2^2=-2 & D_3 \bE_1^2=0 & D_3 \bE_2^2=-4
\end{array}\right.
\end{align}}
\vspace{-0.1cm}
{\footnotesize
\begin{align}% Resolution (28)
  {\rm \bf (q_{12})} : \left\{\begin{array}{c@{~,\quad}c@{~,\quad}c@{~,\quad}c@{~,\quad}c@{~,}}
 \bE_1^3=9 & \bE_2^3=7 & D_1 \bE_1 \bE_2=1 & D_2 \bE_1 \bE_2=0 & D_3 \bE_1 \bE_2=0 \\
 \bE_1^2 \bE_2=-3 & \bE_1 \bE_2^2=1 & D_1^2 \bE_1=1 & D_1^2 \bE_2=0 & D_1 D_2 \bE_1=0 \\
 D_1 D_2 \bE_2=0 & D_1 D_3 \bE_1=0 & D_1 D_3 \bE_2=1 & D_2^2 \bE_1=0 & D_2^2 \bE_2=0 \\
 D_2 D_3 \bE_1=0 & D_2 D_3 \bE_2=0 & D_3^2 \bE_1=0 & D_3^2 \bE_2=2 & D_1 \bE_1^2=-3 \\
 D_1 \bE_2^2=-2 & D_2 \bE_1^2=0 & D_2 \bE_2^2=0 & D_3 \bE_1^2=0 & D_3 \bE_2^2=-4
\end{array}\right.
\end{align}}
\vspace{-0.1cm}
{\footnotesize
\begin{align}% Resolution (29)
  {\rm \bf (q_{13})} : \left\{\begin{array}{c@{~,\quad}c@{~,\quad}c@{~,\quad}c@{~,\quad}c@{~,}}
 \bE_1^3=9 & \bE_2^3=7 & D_1 \bE_1 \bE_2=1 & D_2 \bE_1 \bE_2=0 & D_3 \bE_1 \bE_2=0 \\
 \bE_1^2 \bE_2=-3 & \bE_1 \bE_2^2=1 & D_1^2 \bE_1=1 & D_1^2 \bE_2=-1 & D_1 D_2 \bE_1=0 \\
 D_1 D_2 \bE_2=1 & D_1 D_3 \bE_1=0 & D_1 D_3 \bE_2=0 & D_2^2 \bE_1=0 & D_2^2 \bE_2=-1 \\
 D_2 D_3 \bE_1=0 & D_2 D_3 \bE_2=1 & D_3^2 \bE_1=0 & D_3^2 \bE_2=2 & D_1 \bE_1^2=-3 \\
 D_1 \bE_2^2=-1 & D_2 \bE_1^2=0 & D_2 \bE_2^2=-1 & D_3 \bE_1^2=0 & D_3 \bE_2^2=-4
\end{array}\right.
\end{align}}
\vspace{-0.1cm}
{\footnotesize
\begin{align}% Resolution (30)
  {\rm \bf (q_{14})} : \left\{\begin{array}{c@{~,\quad}c@{~,\quad}c@{~,\quad}c@{~,\quad}c}
 \bE_1^3=9 & \bE_2^3=8 & D_1 \bE_1 \bE_2=1 & D_2 \bE_1 \bE_2=0 & D_3 \bE_1 \bE_2=0~,\\
 \bE_1^2 \bE_2=-3 & \bE_1 \bE_2^2=1 & D_1^2 \bE_1=1 & D_1^2 \bE_2=0 & D_1 D_2 \bE_1=0~, \\
 D_1 D_2 \bE_2=0 & D_1 D_3 \bE_1=0 & D_1 D_3 \bE_2=1 & D_2^2 \bE_1=0 & D_2^2 \bE_2=0~, \\
 D_2 D_3 \bE_1=0 & D_2 D_3 \bE_2=0 & D_3^2 \bE_1=0 & D_3^2 \bE_2=3 & D_1 \bE_1^2=-3~, \\
 D_1 \bE_2^2=-2 & D_2 \bE_1^2=0 & D_2 \bE_2^2=0 & D_3 \bE_1^2=0 & D_3 \bE_2^2=-5~.
\end{array}\right.
\end{align}}

%%%%%%%%%%%

\section{M-theory prepotentials of the $E_{3}{}^{2,1}$ geometry\label{sec:geo prepot H[1]}}
In this appendix, we list the geometric prepotentials for all the crepant resolutions of the $E_{3}{}^{2,1}$ singularity (see Figure \ref{fig:h1 crepant all}), using  \eqref{eq:ess for H[1]}, \eqref{eq:geometric prepot}, and the triple-intersection numbers from Appendix \ref{sec:intnos H[1]}. %We list the results below.
\paragraph{Resolutions that have a vertical reduction.} (Resolutions (a)-(p) in Figure \ref{fig:h1 crepant all}.)
{{\footnotesize
\bea\nn
 	\cF_{\text{a}} &= \left(\frac{\mu _1^2}{2}-\mu _1 \mu _2+\frac{\mu _2^2}{2}\right) \nu
   _1+\left(\frac{\mu _1}{2}+\frac{\mu _2}{2}\right) \nu _1^2-\nu
   _1^3+\left(-\mu _2 \mu _3-\frac{\mu _3^2}{2}\right) \nu _2-\mu _2 \nu _1
   \nu _2+\frac{1}{2} \nu _1^2 \nu _2\\&\qquad+\left(\mu _2+\frac{3 \mu _3}{2}\right)
   \nu _2^2+\frac{1}{2} \nu _1 \nu _2^2-\frac{4 \nu _2^3}{3}~,\\
   \cF_{\text{b}} &= \mu _2 \nu _1^2-\frac{7 \nu _1^3}{6}+\left(-\mu _2 \mu _3-\frac{\mu
   _3^2}{2}\right) \nu _2-\mu _2 \nu _1 \nu _2+\frac{1}{2} \nu _1^2 \nu
   _2+\left(\mu _2+\frac{3 \mu _3}{2}\right) \nu _2^2+\frac{1}{2} \nu _1 \nu
   _2^2-\frac{4 \nu _2^3}{3}~,\\
   \cF_{\text{c}} &= \mu _1 \nu _1^2-\frac{7 \nu _1^3}{6}+\left(\frac{\mu _1^2}{2}-\mu _1 \mu
   _2+\frac{\mu _2^2}{2}-\mu _2 \mu _3-\frac{\mu _3^2}{2}\right) \nu _2-\mu
   _1 \nu _1 \nu _2+\nu _1^2 \nu _2+\left(\frac{\mu _1}{2}+\frac{\mu
   _2}{2}+\frac{3 \mu _3}{2}\right) \nu _2^2-\frac{7 \nu _2^3}{6}~,\\
   \cF_{\text{d}} &= \left(\frac{\mu _1^2}{2}-\mu _1 \mu _2+\frac{\mu _2^2}{2}\right) \nu
   _1+\left(\frac{\mu _1}{2}+\frac{\mu _2}{2}\right) \nu _1^2-\frac{7 \nu
   _1^3}{6}+\left(-\mu _2 \mu _3-\frac{\mu _3^2}{2}\right) \nu _2-\mu _2 \nu
   _1 \nu _2+\nu _1^2 \nu _2+\left(\mu _2+\frac{3 \mu _3}{2}\right) \nu
   _2^2-\frac{7 \nu _2^3}{6}~,\\
   \cF_{\text{e}} &= \left(\frac{\mu _1^2}{2}-\mu _1 \mu _2+\frac{\mu _2^2}{2}\right) \nu
   _1+\left(\frac{\mu _1}{2}+\frac{\mu _2}{2}\right) \nu _1^2-\frac{7 \nu
   _1^3}{6}+\left(-\mu _2 \mu _3-\frac{\mu _3^2}{2}\right) \nu _2-\mu _2 \nu
   _1 \nu _2+\frac{1}{2} \nu _1^2 \nu _2\nonumber\\&\qquad +\left(\mu _2+\frac{3 \mu
   _3}{2}\right) \nu _2^2+\frac{1}{2} \nu _1 \nu _2^2-\frac{4 \nu _2^3}{3}~,\\
   \cF_{\text{f}} &= \mu _2 \nu _1^2-\frac{4 \nu _1^3}{3}+\left(-\mu _2 \mu _3-\frac{\mu
   _3^2}{2}\right) \nu _2-\mu _2 \nu _1 \nu _2+\nu _1^2 \nu _2+\left(\mu
   _2+\frac{3 \mu _3}{2}\right) \nu _2^2-\frac{7 \nu _2^3}{6}~,\\
   \cF_{\text{g}} &= \mu _2 \nu _1^2-\frac{4 \nu _1^3}{3}+\left(-\mu _2 \mu _3-\frac{\mu
   _3^2}{2}\right) \nu _2-\mu _2 \nu _1 \nu _2+\frac{1}{2} \nu _1^2 \nu
   _2+\left(\mu _2+\frac{3 \mu _3}{2}\right) \nu _2^2+\frac{1}{2} \nu _1 \nu
   _2^2-\frac{4 \nu _2^3}{3}~,\\
   \cF_{\text{h}} &= \mu _1 \nu _1^2-\frac{7 \nu _1^3}{6}+\left(-\mu _1 \mu _3-\mu _3^2\right) \nu
   _2-\mu _1 \nu _1 \nu _2+\nu _1^2 \nu _2+\left(\mu _1+2 \mu _3\right) \nu
   _2^2-\frac{4 \nu _2^3}{3}~,\\
   \cF_{\text{i}} &= \mu _1 \nu _1^2-\frac{4 \nu _1^3}{3}+\left(\frac{\mu _1^2}{2}-\mu _1 \mu
   _2+\frac{\mu _2^2}{2}-\mu _2 \mu _3-\frac{\mu _3^2}{2}\right) \nu _2-\mu
   _1 \nu _1 \nu _2+\frac{3}{2} \nu _1^2 \nu _2+\left(\frac{\mu
   _1}{2}+\frac{\mu _2}{2}+\frac{3 \mu _3}{2}\right) \nu _2^2-\frac{1}{2} \nu
   _1 \nu _2^2-\nu _2^3~,\\
   \cF_{\text{j}} &= \mu _1 \nu _1^2-\frac{4 \nu _1^3}{3}+\left(\frac{\mu _1^2}{2}-\mu _1 \mu
   _2+\frac{\mu _2^2}{2}-\mu _2 \mu _3-\frac{\mu _3^2}{2}\right) \nu _2-\mu
   _1 \nu _1 \nu _2+\nu _1^2 \nu _2+\left(\frac{\mu _1}{2}+\frac{\mu
   _2}{2}+\frac{3 \mu _3}{2}\right) \nu _2^2-\frac{7 \nu _2^3}{6}~,\\
   \cF_{\text{k}} &= \left(\frac{\mu _1^2}{2}-\mu _1 \mu _2+\frac{\mu _2^2}{2}\right) \nu
   _1+\left(\frac{\mu _1}{2}+\frac{\mu _2}{2}\right) \nu _1^2-\frac{7 \nu
   _1^3}{6}+\left(-\mu _2 \mu _3-\mu _3^2\right) \nu _2-\mu _2 \nu _1 \nu
   _2+\nu _1^2 \nu _2+\left(\mu _2+2 \mu _3\right) \nu _2^2-\frac{4 \nu
   _2^3}{3} ~,\\
   \cF_{\text{l}} &= \mu _2 \nu _1^2-\frac{4 \nu _1^3}{3}+\left(-\mu _2 \mu _3-\mu _3^2\right) \nu
   _2-\mu _2 \nu _1 \nu _2+\nu _1^2 \nu _2+\left(\mu _2+2 \mu _3\right) \nu
   _2^2-\frac{4 \nu _2^3}{3} ~,\\
   \cF_{\text{m}} &= \mu _1 \nu _1^2-\frac{4 \nu _1^3}{3}+\left(-\mu _1 \mu _3-\mu _3^2\right) \nu
   _2-\mu _1 \nu _1 \nu _2+\frac{3}{2} \nu _1^2 \nu _2+\left(\mu _1+2 \mu
   _3\right) \nu _2^2-\frac{1}{2} \nu _1 \nu _2^2-\frac{7 \nu _2^3}{6}~,\\
   \cF_{\text{n}} &= \mu _1 \nu _1^2-\frac{4 \nu _1^3}{3}+\left(-\mu _1 \mu _3-\mu _3^2\right) \nu
   _2-\mu _1 \nu _1 \nu _2+\nu _1^2 \nu _2+\left(\mu _1+2 \mu _3\right) \nu
   _2^2-\frac{4 \nu _2^3}{3} ~,\\
   \cF_{\text{o}} &= \mu _1 \nu _1^2-\frac{4 \nu _1^3}{3}+\left(\frac{\mu _1^2}{2}-\mu _1 \mu
   _2+\frac{\mu _2^2}{2}-\mu _2 \mu _3-\mu _3^2\right) \nu _2-\mu _1 \nu _1
   \nu _2+\frac{3}{2} \nu _1^2 \nu _2+\left(\frac{\mu _1}{2}+\frac{\mu
   _2}{2}+2 \mu _3\right) \nu _2^2-\frac{1}{2} \nu _1 \nu _2^2-\frac{7 \nu
   _2^3}{6} ~,\\
   \cF_{\text{p}} &= \mu _1 \nu _1^2-\frac{4 \nu _1^3}{3}+\left(-\mu _1 \mu _3-\frac{3 \mu
   _3^2}{2}\right) \nu _2-\mu _1 \nu _1 \nu _2+\frac{3}{2} \nu _1^2 \nu
   _2+\left(\mu _1+\frac{5 \mu _3}{2}\right) \nu _2^2-\frac{1}{2} \nu _1 \nu
   _2^2-\frac{4 \nu _2^3}{3} ~.
\eea
}}%
\normalsize \paragraph{Resolutions without a gauge theory phase.} \hspace{-0.1in}(Resolutions (q$_{1}$)-(q$_{14}$) in Figure \ref{fig:h1 crepant all}.)
{\footnotesize{
 \bea\nn
    \cF_{q_{1}} &= \left(-\mu _1 \mu _2+\frac{\mu _2^2}{2}\right) \nu _1+\left(\mu _1+\frac{\mu
   _2}{2}\right) \nu _1^2-\frac{7 \nu _1^3}{6}+\left(-\mu _2 \mu _3-\frac{\mu
   _3^2}{2}\right) \nu _2-\mu _2 \nu _1 \nu _2+\frac{1}{2} \nu _1^2 \nu
   _2+\left(\mu _2+\frac{3 \mu _3}{2}\right) \nu _2^2\\&\qquad +\frac{1}{2} \nu _1 \nu
   _2^2-\frac{4 \nu _2^3}{3} ~,\\
   \cF_{q_{2}} &= \left(\frac{\mu _1^2}{2}-\mu _1 \mu _2\right) \nu _1+\left(\frac{\mu
   _1}{2}+\mu _2\right) \nu _1^2-\frac{7 \nu _1^3}{6}+\left(-\frac{\mu
   _2^2}{2}-\mu _2 \mu _3-\frac{\mu _3^2}{2}\right) \nu _2-\frac{1}{2} \nu
   _1^2 \nu _2+\left(\frac{3 \mu _2}{2}+\frac{3 \mu _3}{2}\right) \nu
   _2^2-\frac{1}{2} \nu _1 \nu _2^2-\frac{3 \nu _2^3}{2} ~,\\
   \cF_{q_{3}} &= -\frac{1}{2} \mu _2^2 \nu _1+\frac{3}{2} \mu _2 \nu _1^2-\frac{4 \nu
   _1^3}{3}+\left(-\frac{\mu _2^2}{2}-\mu _2 \mu _3-\frac{\mu _3^2}{2}\right)
   \nu _2-\frac{1}{2} \nu _1^2 \nu _2+\left(\frac{3 \mu _2}{2}+\frac{3 \mu
   _3}{2}\right) \nu _2^2-\frac{1}{2} \nu _1 \nu _2^2-\frac{3 \nu _2^3}{2} ~,\\
   \cF_{q_{4}} &= -\frac{1}{2} \mu _1^2 \nu _1+\frac{3}{2} \mu _1 \nu _1^2-\frac{4 \nu
   _1^3}{3}+\left(\frac{\mu _1^2}{2}-\mu _1 \mu _2+\frac{\mu _2^2}{2}-\mu _2
   \mu _3-\frac{\mu _3^2}{2}\right) \nu _2-\mu _1 \nu _1 \nu _2+\nu _1^2 \nu
   _2+\left(\frac{\mu _1}{2}+\frac{\mu _2}{2}+\frac{3 \mu _3}{2}\right) \nu
   _2^2-\frac{7 \nu _2^3}{6} ~,
\eea
\bea\nn
   \cF_{q_{5}} &= \left(-\mu _1 \mu _2+\frac{\mu _2^2}{2}\right) \nu _1+\left(\mu _1+\frac{\mu
   _2}{2}\right) \nu _1^2-\frac{4 \nu _1^3}{3}+\left(-\mu _2 \mu _3-\frac{\mu
   _3^2}{2}\right) \nu _2-\mu _2 \nu _1 \nu _2+\nu _1^2 \nu _2+\left(\mu
   _2+\frac{3 \mu _3}{2}\right) \nu _2^2-\frac{7 \nu _2^3}{6} ~,\\
   \cF_{q_{6}} &= \left(\frac{\mu _1^2}{2}-\mu _1 \mu _2\right) \nu _1+\left(\frac{\mu
   _1}{2}+\mu _2\right) \nu _1^2-\frac{4 \nu _1^3}{3}+\left(-\frac{\mu
   _2^2}{2}-\mu _2 \mu _3-\frac{\mu _3^2}{2}\right) \nu _2-\frac{1}{2} \nu
   _1^2 \nu _2+\left(\frac{3 \mu _2}{2}+\frac{3 \mu _3}{2}\right) \nu
   _2^2-\frac{1}{2} \nu _1 \nu _2^2-\frac{3 \nu _2^3}{2} ~,\\
   \cF_{q_{7}} &= -\mu _1 \mu _2 \nu _1+\left(\mu _1+\mu _2\right) \nu _1^2-\frac{4 \nu
   _1^3}{3}+\left(-\frac{\mu _2^2}{2}-\mu _2 \mu _3-\frac{\mu _3^2}{2}\right)
   \nu _2-\frac{1}{2} \nu _1^2 \nu _2+\left(\frac{3 \mu _2}{2}+\frac{3 \mu
   _3}{2}\right) \nu _2^2-\frac{1}{2} \nu _1 \nu _2^2-\frac{3 \nu _2^3}{2} ~,\\
   \cF_{q_{8}} &= -\frac{1}{2} \mu _2^2 \nu _1+\frac{3}{2} \mu _2 \nu _1^2-\frac{3 \nu
   _1^3}{2}+\left(-\frac{\mu _2^2}{2}-\mu _2 \mu _3-\frac{\mu _3^2}{2}\right)
   \nu _2-\frac{1}{2} \nu _1^2 \nu _2+\left(\frac{3 \mu _2}{2}+\frac{3 \mu
   _3}{2}\right) \nu _2^2-\frac{1}{2} \nu _1 \nu _2^2-\frac{3 \nu _2^3}{2} ~,
\eea
\bea\nn
   \cF_{q_{9}} &= -\frac{1}{2} \mu _1^2 \nu _1+\frac{3}{2} \mu _1 \nu _1^2-\frac{4 \nu
   _1^3}{3}+\left(-\mu _1 \mu _3-\mu _3^2\right) \nu _2-\mu _1 \nu _1 \nu
   _2+\nu _1^2 \nu _2+\left(\mu _1+2 \mu _3\right) \nu _2^2-\frac{4 \nu
   _2^3}{3} ~,\\
   \cF_{q_{10}} &= -\frac{1}{2} \mu _1^2 \nu _1+\frac{3}{2} \mu _1 \nu _1^2-\frac{3 \nu
   _1^3}{2}+\left(\frac{\mu _1^2}{2}-\mu _1 \mu _2+\frac{\mu _2^2}{2}-\mu _2
   \mu _3-\frac{\mu _3^2}{2}\right) \nu _2-\mu _1 \nu _1 \nu _2+\frac{3}{2}
   \nu _1^2 \nu _2+\left(\frac{\mu _1}{2}+\frac{\mu _2}{2}+\frac{3 \mu
   _3}{2}\right) \nu _2^2\\&\qquad -\frac{1}{2} \nu _1 \nu _2^2-\nu _2^3 ~,\\
   \cF_{q_{11}} &= \left(-\mu _1 \mu _2+\frac{\mu _2^2}{2}\right) \nu _1+\left(\mu _1+\frac{\mu
   _2}{2}\right) \nu _1^2-\frac{4 \nu _1^3}{3}+\left(-\mu _2 \mu _3-\mu
   _3^2\right) \nu _2-\mu _2 \nu _1 \nu _2+\nu _1^2 \nu _2+\left(\mu _2+2 \mu
   _3\right) \nu _2^2-\frac{4 \nu _2^3}{3} ~,\\
   \cF_{q_{12}} &= -\frac{1}{2} \mu _1^2 \nu _1+\frac{3}{2} \mu _1 \nu _1^2-\frac{3 \nu
   _1^3}{2}+\left(-\mu _1 \mu _3-\mu _3^2\right) \nu _2-\mu _1 \nu _1 \nu
   _2+\frac{3}{2} \nu _1^2 \nu _2+\left(\mu _1+2 \mu _3\right) \nu
   _2^2-\frac{1}{2} \nu _1 \nu _2^2-\frac{7 \nu _2^3}{6} ~,
\eea
\bea\nn
   \cF_{q_{13}} &= -\frac{1}{2} \mu _1^2 \nu _1+\frac{3}{2} \mu _1 \nu _1^2-\frac{3 \nu
   _1^3}{2}+\left(\frac{\mu _1^2}{2}-\mu _1 \mu _2+\frac{\mu _2^2}{2}-\mu _2
   \mu _3-\mu _3^2\right) \nu _2-\mu _1 \nu _1 \nu _2+\frac{3}{2} \nu _1^2
   \nu _2+\left(\frac{\mu _1}{2}+\frac{\mu _2}{2}+2 \mu _3\right) \nu
   _2^2\\&\qquad -\frac{1}{2} \nu _1 \nu _2^2-\frac{7 \nu _2^3}{6} ~,\\
   \cF_{q_{14}} &= -\frac{1}{2} \mu _1^2 \nu _1+\frac{3}{2} \mu _1 \nu _1^2-\frac{3 \nu
   _1^3}{2}+\left(-\mu _1 \mu _3-\frac{3 \mu _3^2}{2}\right) \nu _2-\mu _1
   \nu _1 \nu _2+\frac{3}{2} \nu _1^2 \nu _2+\left(\mu _1+\frac{5 \mu
   _3}{2}\right) \nu _2^2-\frac{1}{2} \nu _1 \nu _2^2-\frac{4 \nu _2^3}{3} ~.
 \eea
}}

%%%%%%%%%%%%%%%%%%%%%%%%%%%%%%%%%%%%%%
%%%%%%%%%%%%%%%%%%%%%%%%%%%%%%%%%%%%%%
%%%%%%%%%%%%%%%%%%%%%%%%%%%%%%%%%%%%%%
%%%%%%%%%%%%%%%%%%%%%%%%%%%%%%%%%%%%%%

%%%%%%%%%%%%%%%%%%%%%%%%%%%%%%%%%

\bibliographystyle{utphys}
\bibliography{paper4}{}

\end{document}